\documentclass[10pt,english,a4paper]{article}

\usepackage[a4paper,left=2cm,right=2cm,top=2.5cm,bottom=2.5cm]{geometry}
\usepackage{dcolumn}
\usepackage{bm}
\usepackage{graphicx}
\usepackage{amsmath}
\usepackage{amsfonts}
\usepackage{amssymb}
\usepackage{amsthm}
\usepackage{amstext}
\usepackage{amsbsy}
\usepackage{amsopn}
\usepackage{amscd}
\usepackage{amsxtra}

\usepackage{color}

\begin{document}
\title{Quantum control of molecular rotation}
\author{Christiane P. Koch\footnote{Theoretische Physik, Universit\"{a}t Kassel,
Heinrich-Plett-Stra{\ss}e 40, 34132 Kassel, Germany, christiane.koch@uni-kassel.de}, Mikhail Lemeshko\footnote{IST Austria (Institute of Science and Technology Austria), Am Campus 1, 3400 Klosterneuburg, Austria, mikhail.lemeshko@ist.ac.at}, Dominique Sugny\footnote{Laboratoire Interdisciplinaire Carnot de
Bourgogne (ICB), UMR 6303 CNRS-Universit\'e Bourgogne-Franche Comt\'e, 9 Av. A.
Savary, BP 47 870, F-21078 Dijon Cedex, France and Institute for Advanced Study, Technische Universit\"at M\"unchen, Lichtenbergstrasse 2 a, D-85748 Garching, Germany, dominique.sugny@u-bourgogne.fr}}

\maketitle

\begin{abstract}
  The angular momentum of molecules, or, equivalently, their rotation in three-dimensional space, is ideally suited for  quantum control. Molecular angular momentum is naturally quantized, time evolution is governed by a well-known Hamiltonian with only a few accurately known parameters, and transitions between rotational levels can be driven by external fields from various parts of the electromagnetic spectrum. Control over the rotational motion can be exerted in one-, two- and many-body scenarios, thereby  allowing to probe Anderson localization, target stereoselectivity of bimolecular reactions, or encode quantum information, to name just a few examples. The corresponding approaches to quantum control are pursued within separate, and typically disjoint, subfields of physics, including ultrafast science, cold collisions, ultracold gases, quantum information science, and condensed matter physics. It is the purpose of this review to present the various control phenomena, which all rely on the same underlying physics, within a unified framework. To this end, we recall the Hamiltonian for free rotations, assuming the rigid rotor approximation to be valid, and summarize the different ways for a rotor to interact with external electromagnetic fields. These interactions can be exploited for control --- from achieving alignment, orientation, or laser cooling in a one-body framework, steering bimolecular collisions, or realizing a quantum computer or quantum simulator in the many-body setting.
\end{abstract}

\maketitle

\section{Introduction}
Molecules, unlike atoms, are extended objects that possess a number of
different types of motion. In particular, the geometric arrangement of their
constituent atoms endows molecules with the basic capability to rotate
in three-dimensional space. Rotations can couple to vibrations of the
atomic nuclei as well as to the orbital and spin angular momentum of
the electrons. The resulting complexity of the energy level structure~\cite{HerzbergBook,BunkerBook,KreStwFrieColdMol, LemKreDoyKais13,
  LevebvreBrionField, KremsBook18} may be daunting. It offers, on the other hand,
a variety of knobs for control and thus is at the core of numerous
applications,
from the classic example of the
ammonia maser~\cite{GordonPR55} all the way to recent measurements
of the electron's electric dipole moment in a cryogenic molecular beam
of thorium monoxide~\cite{ACMEedm}.

A key advantage of rotational degrees of freedom is that they
occupy the low-energy part of the energy spectrum. Quantization of the rotational motion has been an early hallmark of quantum mechanics due to its connection to selection rules that govern all light-matter interactions~\cite{Zare:88}.
Nowadays, rotational states and rotational molecular dynamics feature
prominently in all active areas of Atomic, Molecular and Optical (AMO) physics research as well as in
neighbouring fields such as physical chemistry and quantum
information science. Control over the rotational motion is crucial
in one-body, two-body and many-body scenarios. For example, rotational
state-selective excitation could pave the way towards separating left- and right handed enantiomers of chiral molecules~\cite{EibenbergerPRL17,PerezAngewandte17}. Still within the one-body scenario, molecular rotation can serve as a testbed for a manifold of quantum phenomena including Bloch oscillations~\cite{floss2014,floss:2015}, Anderson localization~\cite{bitter:2016}, or quantum chaos~\cite{bitter:2017}.
Alignment and orientation of molecules in space is another
long-standing goal in the quantum control of molecular rotation. Control over alignment is by
now well understood~\cite{CaiFriCCCC01,Stapelfeldt:03, Seideman:05}, and the choice of specific
polarizations has allowed to extend it
to two and three spatial dimensions
\cite{larsen:2000,korobenko:2014,korech:2013,karras:2015}. Molecular orientation
is not yet  at the same stage of development, although
use of terahertz radiation~\cite{Nelson:11,Babilotte:16} and two-color
laser fields~\cite{de:2009} have recently boosted experimental progress.

While the dynamics leading to alignment and orientation can be
understood within the one-body scenario~\cite{Stapelfeldt:03,
  Seideman:05, CaiFriCCCC01}, the main motivation for these efforts was derived from the goal of stereoselectivity of chemical reactions~\cite{larsen99} which involve two-body interactions.
Another way to control the stereodynamics of a chemical reaction has been made possible by the impressive progress over the last decade in preparing molecules that are both internally and translationally cold~\cite{KreStwFrieColdMol, JinYeCRev12, LemKreDoyKais13}.
Controlling quantum stereodynamics of bimolecular reactions has thus become possible~\cite{MirandaNatPhys11}. Remarkably, quantization of molecular rotation governs reaction dynamics not only at low temperatures, but determines the rate even for reactions that can otherwise be described by a  classical theory~\cite{ShagamNatChem15}.

When molecules interact with their environment, a one-body or two-body
picture of the rotational dynamics becomes insufficient. The resulting
phenomena can broadly be classified into decoherence of the
rotational motion, for example via  intermolecular
collisions~\cite{RamakrishnaSeideman2005,Viellard:13}, and into
genuine many-body dynamics~\cite{Lemeshko_2016_book}. A recent
highlight of the latter is the understanding of the laser-induced
alignment of molecules inside helium
nanodroplets~\cite{Shepperson16}. The underlying picture of rotational
quasiparticles -- the `angulons'~\cite{SchmidtLem15, SchmidtLem16,
  LemeshkoDroplets16} represents a prototype for the role of
rotational states in many-body dynamics.
Ultracold molecules in optical lattices amount to one of the most promising platforms for studying many-particle physics in a fully controlled environment~\cite{MosesNatPhys17}. This includes both quantum simulation of condensed-matter models, see e.g.~\cite{gorshkovPRL11, YanNature13}, as well as the study of novel, previously unexplored phases of matter, e.g.~\cite{CooperPRL09, SyzranovNatComm14}.

These examples provide a first glimpse onto the prominent role
of molecular rotation in the various strands of current AMO and
quantum optics research.
Molecular rotation is a mature subject covered by several earlier
reviews~\cite{Seideman:05,Stapelfeldt:03,KreStwFrieColdMol,
  LemKreDoyKais13,ohshima:2010,fleischerrev:2012,pabst:2013}. Recently, a modern and very didactic introduction into molecular physics -- including molecular rotation -- became available~\cite{KremsBook18}. Rotational
degrees of freedom, however, occur in various contexts, and often different
languages are used to describe them. A unified treatment presenting the
tools and concepts used by the different communities to study the role
of molecular rotations in one-, two- and many-body phenomena,
and bridging, moreover, the gap to the quantum control framework~\cite{Brif:10,Glaser:15} is currently missing. The
present review seeks to fill this void.

The review is organized as follows. We start by summarizing the one-body
rotational structure for the three different classes of molecular rotors in
Sec.~\ref{sec:states}. Rotational dynamics and control over rotational
motion that can be understood within a one-body picture is
reviewed in Section~\ref{sec:align}, focussing on molecular alignment
and orientation, and Section~\ref{sec:onebodydyn}, dedicated to
dynamical phenomena in molecular rotation.
The role of rotations in two-body interactions is the
subject of Section~\ref{sec:collisions} whereas many-body scenarios
for control of molecular rotation are covered in
Sections~\ref{sec:environ},~\ref{sec:manybody}, and \ref{sec:QuantSim}. In particular, the
influence of the environment on molecular rotation is discussed
in Section~\ref{sec:environ}, and Sections~\ref{sec:manybody} and \ref{sec:QuantSim} showcase the use of molecular rotations in quantum information and quantum
simulation. The conclusions of this review are drawn in Section~\ref{sec:concl}.

\section{Molecular rotational structure and interaction with an electromagnetic field}
\label{sec:states}

The rotational structure of a molecule and the form of its interaction with an external electromagnetic field provide the basis for controlling its rotational dynamics. In this section, we summarize the basic concepts for each case of molecular rotor~\cite{Zare:88} and discuss the prospects to control its rotational motion.
A vast literature on quantum rotation of molecules exists from the point of view of molecular spectroscopy~\cite{LevebvreBrionField,TownesSchawlow,BernathBook}. The main goal of this review, on the other hand, is to describe rotations from the point of view of \textit{quantum dynamics and control}.

\subsection{Free rotation of a molecule}
\label{subsec:freerot}

First, let us consider the dynamics of an isolated molecule in its center-of-mass frame, neglecting its translational motion. Such a space-fixed reference frame is called the space-fixed frame. Internal molecular degrees of freedom can be classified as electronic, vibrational and rotational~\cite{KremsBook18}. The molecule is assumed to be in its vibronic, i.e., electronic and vibrational, ground state,
which is a good approximation for most small molecules at room temperature and below. Within this approximation, we can omit the discussion of the vibrational and electronic degrees of freedom and entirely focus on the rotational motion. In particular, the electromagnetic fields we consider are far-off-resonant from any electronic or vibrational transition.

\begin{figure}[tp]
\centering
      \includegraphics[width=0.5\linewidth]{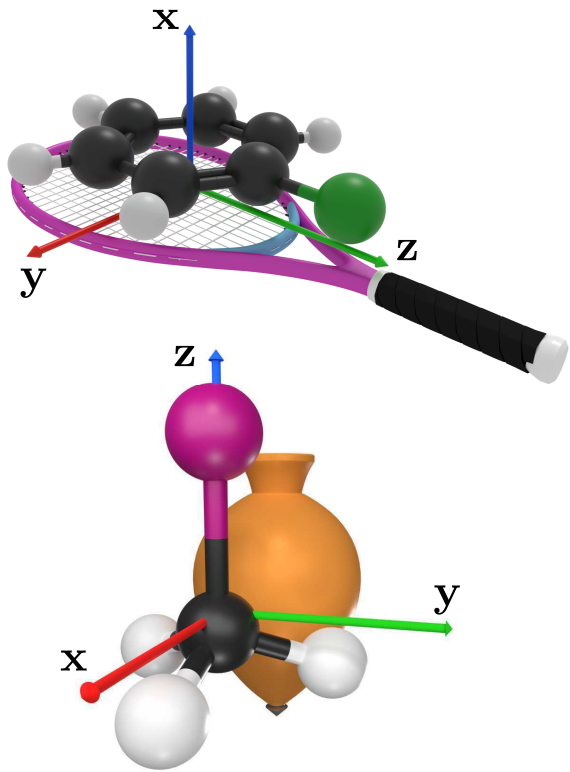}
\caption{(Color online) Schematic picture of the correspondence between molecules and classical objects. Two examples of rigid asymmetric tops, the chlorobenzene molecule, C$_6$H$_5$Cl, and a tennis racket, are represented in panel (a). Panel (b) displays a prolate symmetric top, the iodomethane molecule, CH$_3$I, and its classical analog. The body-fixed frame, $(\mathbf{x},\mathbf{y},\mathbf{z})$, is defined for each case.}
\label{fig0}
\end{figure}
Before entering into a description of the quantum rotational dynamics, we start the discussion by describing free rotation of a rigid body in classical mechanics~\cite{arnold,goldstein,landau}. Free rotation of a rigid body is defined by the position of the body-fixed frame, $(x,y,z)$, with respect to the space-fixed frame,  $(X,Y,Z)$. In a body-fixed frame, a linear transformation connects the angular momentum, $\mathbf{J}$, to the angular velocity, $\boldsymbol{\omega}$:
$$
\mathbf{J}=I\boldsymbol{\omega},
$$
where $I$ is a $3\times 3$ inertia matrix. The constant elements of $I$, known as the inertia elements, describe the mass distribution of the body with respect to the three axes of the frame. The elements of $I$, whose dimension is mass times length squared, can be expressed for a continuous body as:
$$
I_{jk}=\int_V\rho(\mathbf{r})(r^2\delta_{jk}-x_jx_k)d^3\mathbf{r},
$$
where $\rho$ is the mass density, $V$ the volume of the body and the coordinates of the position vector, $\mathbf{r}$, are denoted by $x_j$, $(x_1,x_2,x_3)\equiv (x,y,z)$. The matrix $I$ is a symmetric and real matrix such that a specific body-fixed frame, $(x,y,z)$, can be defined in which the inertia matrix is diagonal. The corresponding eigenvalues,  $(I_x,I_y,I_z)$, are positive and are called the moments of inertia. The moments of inertia can be physically interpreted as a measure of resistance of the body to rotational motion. The axes of the reference body-fixed frame, i.e., the eigenvectors of the inertia matrix, are the principal inertia directions of the body.

Rigid bodies can be classified according to the relative values of their moments of inertia. We adopt the convention $I_x\geq I_y\geq I_z$. We distinguish the following types:
\begin{equation*}
\begin{cases}
I_z< I_y< I_x:~\textrm{asymmetric top} \\
I_z< I_y=I_x:~\textrm{prolate symmetric top}\\
I_z=I_y< I_x:~\textrm{oblate symmetric top}\\
I_x=I_y=I_z:~\textrm{spherical top}\\
I_z=0,~I_x=I_y:~\textrm{linear top}.
\end{cases}
\end{equation*}
A standard example of a classical asymmetric top is the tennis racket as shown in Fig.~\ref{fig0}. For a tennis racket, the inertia axes are defined as follows. The axis $\mathbf{z}$ is along the handle of the racket, $\mathbf{y}$ lies in
the plane of the head of the racket and is orthogonal to $\mathbf{z}$, while $\mathbf{x}$ is orthogonal to the head of the racket. For a plane object, note that the moment of inertia about the axis normal to the plane is equal to the sum of the two other moments of inertia, i.e., here $I_x=I_y+I_z$. Most molecules are asymmetric tops, with the example of the Chlorobenzene molecule represented schematically in Fig.~\ref{fig0} (a). Figure~\ref{fig0}(b) shows the example of a body with a symmetry axis, i.e., a symmetric top. In this case, the two moments of inertia associated with directions orthogonal to the symmetry axis are equal. We distinguish the prolate and the oblate cases in which the moment of inertia of the symmetry axis is, respectively, the lowest and the largest moments. A prolate body has the shape of an American football, while an oblate body has a frisbee-like shape.  CH$_3$I and CHI$_3$ are two examples of prolate and oblate top molecules. A molecule with an additional symmetry axis such as CCl$_4$ or CH$_4$ is a spherical top for which all the moments of inertia are equal. Finally, linear molecules can be viewed as a limiting case of prolate symmetric tops where the lowest moment of inertia is assumed to be zero. Standard examples are any diatomic molecule or the CO$_2$ molecule.

\begin{figure}[tb]
\centering
     \includegraphics[width=0.5\linewidth]{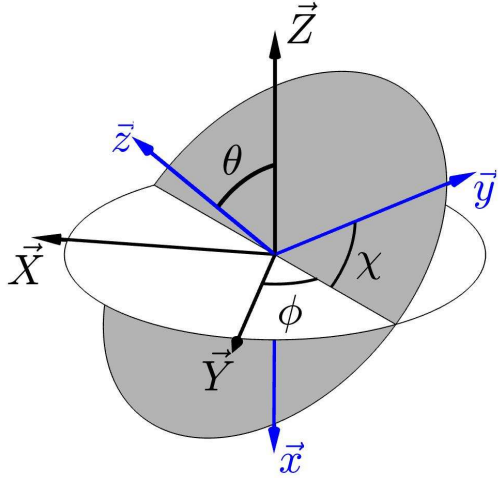}
\caption{(Color online) Definition of the Euler angles, $(\theta,\phi,\chi)$, which are used to describe the position of the body-fixed frame, $(\mathbf{x},\mathbf{y},\mathbf{z})$, in the space-fixed frame, $(\mathbf{X},\mathbf{Y},\mathbf{Z})$.}
\label{fig1}
\end{figure}
The relative motion of the space-fixed and body-fixed frames is characterized by the three Euler angles, $(\theta,\phi,\chi)$. A particular set of Euler angles (which is not unique) is defined in Fig.~\ref{fig1}. The angle $\theta$ is the angle between the axis $z$ of the body-fixed frame and the space-fixed axis $Z$. The angles $\phi$ and $\chi$ describe respectively the rotation of the body about the axes $Z$ and $z$.
The base-change matrix, $R(\theta,\phi,\chi)$, defined by the relation $(x,y,z)^\intercal=R~{(X,Y,Z)^\intercal}$, where $^\intercal$ means the transpose of a vector, can be expressed as:
\begin{equation}\label{mateuler}
R=\begin{pmatrix} \cos\phi\cos\chi\cos\theta-\sin\phi\sin\chi & \sin\phi\cos\chi\cos\theta+\cos\phi\sin\chi & -\sin\theta\cos\chi \\
-\cos\phi\sin\chi\cos\theta-\sin\phi\cos\chi & -\sin\phi\sin\chi\cos\theta-\cos\phi\cos\chi & \sin\theta\sin\chi \\
\sin\theta\cos\phi & \sin\theta\sin\phi & \cos\theta
\end{pmatrix},
\end{equation}
where the matrix elements of $R$ are the direction cosines, denoted $\cos\theta_{\gamma\Gamma}$ between the pairs of axes of the two frames, with $\gamma=x,y,z$ and $\Gamma=X,Y,Z$. They satisfy the relations:
\begin{equation}
\begin{cases}
\sum_\gamma\cos\theta_{\gamma\Gamma}\cos\theta_{\gamma\Gamma '}=\delta_{\Gamma\Gamma '}\,,\\
\sum_{\Gamma}\cos\theta_{\gamma\Gamma}\cos\theta_{\gamma '\Gamma}=\delta_{\gamma \gamma '}\,,
\end{cases}
\end{equation}
which can be deduced from the property that $R$ is a rotation matrix, $R^\intercal=R^{-1}$.

During a free rotation, the angular momentum $\mathbf{J}$ is constant in the space-fixed frame and usually chosen along the $Z$- axis of this frame. The
components of $\mathbf{J}$ can be expressed in the body-fixed frame as $J_x= -J\sin\theta\cos\chi$, $J_y = J \sin\theta\sin\chi$ and $J_z=J\cos\theta$ where $J = |\mathbf{J}|$. Note that the components of $\mathbf{J}$ correspond to the third column of the rotation matrix $R$ defined in Eq.~\eqref{mateuler}. The Euler equations $\dot{\mathbf{J}}=\mathbf{J}\times \boldsymbol{\Omega}$, with $\Omega_{k}=J_k/I_k$, $k=x,y,z$,  govern the dynamics of the angular momentum in the body-fixed frame. Explicit solutions of the Euler equations can be written in terms of Jacobi elliptic functions. Using the definition of the coordinates of $\mathbf{J}$ and of the angular velocities, it can be shown that the Euler angles satisfy the differential system~\cite{landau}:
\begin{equation*}
\begin{cases}
\dot{\theta}=J(\frac{1}{I_y}-\frac{1}{I_x})\sin\theta\sin\chi\cos\chi\,, \\
\dot{\phi}=J(\frac{1}{I_y}\sin^2\chi+\frac{1}{I_x}\cos^2\chi) \,,\\
\dot{\chi}=J(\frac{1}{I_z}-\frac{1}{I_y}\sin^2\chi-\frac{1}{I_x}\cos^2\chi)\cos\theta,
\end{cases}
\end{equation*}
whose solutions can be expressed as the sum of elliptic integrals of the first and third kinds~\cite{goldstein}.

\begin{figure}[tb]
\centering
     \includegraphics[width=0.5\linewidth]{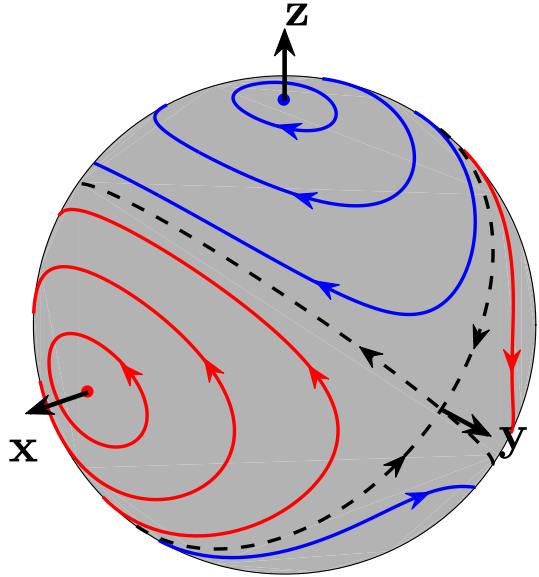}
\caption{(Color online) Trajectories of the angular momentum
of a rigid body in the body-fixed frame $(\mathbf{x},\mathbf{y},\mathbf{z})$. The blue (dark gray) and red (light gray)
lines represent respectively the rotating and oscillating trajectories of the angular momentum. The dashed black line is the separatrix.}
\label{fig2}
\end{figure}
From the general expression of the rotational kinetic energy, $T=I\boldsymbol{\omega}^2/2$, we deduce the field-free rotational Hamiltonian $H_0$,
\begin{equation}\label{eqhamfree}
H_0=\frac{J_x^2}{2I_x}+\frac{J_y^2}{2I_y}+\frac{J_z^2}{2I_z}.
\end{equation}
Equation~\eqref{eqhamfree} defines an ellipsoid in the $(J_x,J_y,J_z)$- space. Conservation of the angular momentum magnitude, $J^2=J_x^2+J_y^2+J_z^2$, implies that $\mathbf{J}$ follows trajectories lying on the intersections of the ellipsoid and the sphere. The two constants of motion, $H_0$ and $J$, define  Hamiltonian integrable dynamics. The corresponding classical phase space for asymmetric tops has a simple structure including a separatrix which is the
boundary between two families of trajectories, the rotating and
the oscillating ones. Each family of trajectories is distributed
around a stable fixed point, cf. Fig.~\ref{fig2}, and the separatrix connects the unstable fixed points.

\begin{figure}[tb]
\centering
      \includegraphics[width=0.5\linewidth]{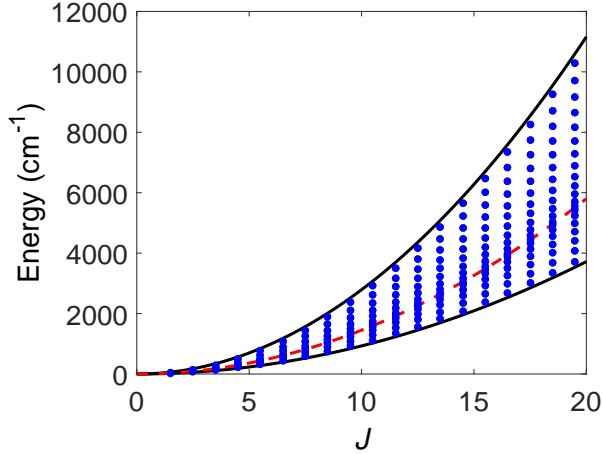}
\caption{(Color online) Energy momentum diagram of the water
molecule. The red dashed line depicts the position of the separatrix and
the black solid lines the boundary of the accessible EM. The blue dots indicate the position of the energy levels in the diagram. We use the convention $J=h(j+\frac{1}{2})$, with $h=1$ and $j$ is the quantum number defined in Eq.~\eqref{eqj}.}\label{fig3}
\end{figure}

As displayed in Fig.~\ref{fig3}, the energy-momentum diagram (EM) is
another useful way to visualize the global dynamics of an integrable
system. It corresponds to all the possible values of $H_0 = E$ as a function of $J$.
The position of the stable fixed points of the free rotation of
a rigid body, $E = J^2/(2I_x)$ and $E = J^2/(2I_z)$, delimits the boundary of
the EM and of the accessible phase space, while the separatrix is
defined by $E = J^2/(2I_y)$. The separatrix distinguishes the two
families of trajectories, namely the rotating and the oscillating
ones for $E > J^2/(2I_y)$ and $E < J^2/(2I_y)$, respectively.

After this brief review of the classical motion of a rigid body, we now consider a quantum description of the rotational dynamics. We introduce the rotational constants, $A$, $B$ and $C$, that are inversely proportional to the moments of inertia of the molecule, $A=\hbar/(4\pi I_x)$, $B=\hbar/(4\pi I_y)$ and $C=\hbar/(4\pi I_z)$. We adopt the standard convention where $A\leq B\leq C$. The quantum Hamiltonian is then given by $\hat{H}=A\hat{J}_x^2+B\hat{J}_y^2+C\hat{J}_z^2$, where $\hat{J}_{x,y,z}$ are operators acting on the infinite dimensional Hilbert space $\mathcal{H}$. Note that centrifugal terms due to the rotation-vibration interaction can also be added to $\hat{H}$~\cite{Zare:88}.

A connection between the classical and quantum dynamics of a rigid rotor can be established in the semi-classical limit. We refer the interested reader to~\cite{Child:91} for details. The Wigner representation can also be used to facilitate the analysis of molecular rotational dynamics~\cite{zhdanov:2015}.

A basis of the Hilbert space $\mathcal{H}$ is given by the wave functions $|j,k,m\rangle$~\cite{Zare:88} defined by
\begin{equation}\label{eqsymdj}
\langle \phi,\theta,\chi|j,k,m\rangle = \sqrt{\frac{2j+1}{4\pi}}D^{j*}_{m,k}(\phi,\theta,\chi)\,,
\end{equation}
with $j\geq 0$, $-j\leq k\leq j$ and $-j\leq m\leq j$. The coefficients $D^j_{mk}(\phi,\theta,\chi)$ can be derived from the rotation matrix $R$,
\begin{equation}
D^j_{mk}=\langle j,m|R(\phi,\theta,\chi)|j,k\rangle\,,
\end{equation}
where $|j, m\rangle $ are the eigenstates of the $\hat{\mathbf{J}}^2$ and $\hat{J}_Z$ operators, with the wavefunctions given by Spherical Harmonics, $\langle \theta, \phi | j, m \rangle = Y_{jm} (\theta, \phi)$. In the $|j,k,m\rangle$- basis, the components of the angular momentum satisfy:
\begin{equation}\label{eqj}
\begin{cases}
\hat{\mathbf{J}}^2|j,k,m\rangle =j(j+1)|j,k,m\rangle\,, \\
\hat{J}_z|j,k,m\rangle=k|j,k,m\rangle\,, \\
\hat{J}_Z|j,k,m\rangle = m|j,k,m\rangle\,.
\end{cases}
\end{equation}
Note that the matrix elements of $\hat{J}_X$ and $\hat{J}_Y$ (resp. $\hat{J}_x$ and $\hat{J}_y$) do not depend on $k$ (resp. $m$). For linear molecules, the Hamiltonian reads $\hat{H}=B\hat{\mathbf{J}}^2$ and the dynamics are restricted to the subspace with $k=0$, in which the angular momentum is orthogonal to the molecular axis. The energy levels of the form $Bj(j+1)$ are $(2j+1)$-degenerate and the eigenbasis is given by the kets $|j,m\rangle$. The second case concerns symmetric top molecules for which two of the rotational constants are equal. The energy levels, $E_{j,k}$, can be expressed respectively as $Aj(j+1)+(C-A)k^2$ and $Cj(j+1)+(A-C)k^2$ in the prolate and oblate cases with the eigenkets $|j,k,m\rangle$. In the general situation of an asymmetric top molecule, $k$ is not a good quantum number and the Hamiltonian matrix has a block-diagonal structure labelled by $j$ and $m$ in the $|j,k,m\rangle$- basis. The energy levels of the water molecule are represented in Fig.~\ref{fig3}. Since in the body-fixed frame, the Hamiltonian matrix does not depend on the quantum number $m$, the spectrum is composed of $2j+1$ levels which are $(2j+1)$-degenerate.

\subsection{Interaction between a molecule and an external electromagnetic field}
After this short description of the free rotational Hamiltonian, we introduce the terms describing the interaction with a electromagnetic field. We consider the semi-classical approximation in which the field is classical while the molecule is treated quantum-mechanically. Let $\hat{\boldsymbol\mu}$ be the electric dipole moment of the molecule. The dipole moment can be permanent for a polar molecule or induced by an intense electric field. The interaction term can be expressed as $H_I=-\hat{\boldsymbol \mu}\cdot\mathbf{E}(t)$ where $\mathbf{E}(t)$ is the electric field the molecule is subjected to at time $t$. In the dipole approximation, we assume that the electric field has no spatial variation across the extent of the molecule, i.e., $\mathbf{E}(\mathbf{r},t) \equiv \mathbf{E}(t)$. The dipole moment can be written as a power series expansion of $\mathbf{E}$:
\begin{equation}
\hat{\boldsymbol \mu}=\hat{\boldsymbol\mu}_0+\frac{1}{2}\hat{\boldsymbol{\alpha}}\mathbf{E}+\frac{1}{6}\hat{\boldsymbol\beta}\mathbf{E}^2+\cdots\,,
\end{equation}
where $\hat{\boldsymbol\mu}_0$ is the permanent dipole moment, and the polarizability, $\hat{\boldsymbol{\alpha}}$, and hyperpolarizabiliy, $\hat{\boldsymbol\beta}$, are given by tensors of rank 2 and 3, respectively.

The polarizability tensor is a $3\times 3$ diagonal matrix in the molecular frame for linear and symmetric molecules with the elements $(\alpha_\perp,\alpha_\perp,\alpha_\parallel)$ where $\alpha_\parallel$ and $\alpha_\perp$ are respectively the components parallel and perpendicular to the molecular or the symmetry axis. The difference between the two coordinates, i.e., the polarizability anisotropy, is denoted by $\Delta\alpha=\alpha_\parallel-\alpha_\perp$. $\Delta\alpha$ is positive for linear and prolate molecules and negative for oblate symmetric tops. The situation is more complex for asymmetric molecules. Only molecules with $D_2$, $C_{2v}$ or $D_{2h}$ symmetry have a diagonal polarizability tensor in the body-fixed frame with elements $\alpha_{xx}$, $\alpha_{yy}$ and $\alpha_{zz}$~\cite{cottonBook,yachmenev:2016,gershnabel:2018}. We will assume
the polarizability tensor to be diagonal in this section.

The electric field is defined in the space-fixed frame, and we denote its coordinates along the three directions of the frame by $(E_X,E_Y,E_Z)$. Its polarization is given by the time evolution of the three components, $E_X$, $E_Y$ and $E_Z$. Using the rotation matrix $R$ defined in Eq.~\eqref{mateuler}, the components in the molecular frame read
\begin{equation}
\begin{pmatrix} E_x \\ E_y \\ E_z \end{pmatrix}
= R \begin{pmatrix} E_X \\ E_Y \\ E_Z \end{pmatrix}\,.
\end{equation}
To second order, the interaction Hamiltonian can be written as
\begin{eqnarray*}
\hat{H}_I&=&-\sum_{K}\mu_0\cos\hat{\theta}_{zK}E_K 
 -\sum_{K}\frac{E_K^2}{2}(\Delta \alpha\cos^2\hat{\theta}_{zK}+\alpha_\perp) \\
& & -\sum_{K,K';K\neq K'}E_KE_{K'}\Delta\alpha\cos\hat{\theta}_{zK}\cos\hat{\theta}_{zK'}\,,
\end{eqnarray*}
with $K,K'=X,Y,Z$. We assume here that the permanent dipole moment points along the $z$- axis, which is usually the case for linear and symmetric top molecules. Several
different cases can then be distinguished. If the electric field is resonant with some of the rotational frequencies then $\hat{H}_I$ cannot be simplified. It is the situation encountered, e.g., with terahertz pulses. When the molecule is exposed to optical laser fields with a frequency much larger than the rotational ones, a high-frequency approximation can be used to separate different time scales and to simplify the expression of $\hat{H}_I$~\cite{keller:2000}. A heuristic derivation can be carried out as follows. We assume that the electric field can be expressed as
\begin{equation}
\mathbf{E}(t)=\sum_{K=X,Y,Z}\mathcal{E}_K(t)\cos(\omega t+\phi_K)\mathbf{u}_K,
\end{equation}
where $\omega$ is the carrier frequency, $\boldsymbol{\mathcal{E}}$ a slowly varying time-dependent vector of coordinates $(\mathcal{E}_X,\mathcal{E}_Y,\mathcal{E}_Z)$, $\phi_K$ the phases of the field which define its polarization state and $\vec{u}_K$ the unitary vector along the $K$-direction. For a non-resonant field, we introduce the time-averaged value $\langle\cdot \rangle $ over a time $\tau$ given for a function $a(t)$ by $\langle a\rangle=\frac{1}{\tau}\int_0^\tau a(t)dt$. The time $\tau$ satisfies $\frac{2\pi}{\omega}\ll \tau \ll T_{\textrm{rot}}$, where $T_{\textrm{rot}}$ is a typical time scale of the rotational dynamics. For linear and symmetric molecules, $T_{\textrm{rot}}$ is the rotational period. We then obtain:
\begin{equation}\label{eqtimeaverage}
\begin{cases}
\langle E_K\rangle =0\,, \\
\langle E_K^2\rangle = \frac{\mathcal{E}_K^2}{2}\,, \\
\langle E_KE_{K'}\rangle=\frac{\mathcal{E}_K\mathcal{E}_{K'}}{2}\cos(\phi_K-\phi_{K'}),~K\neq K'\,,
\end{cases}
\end{equation}
which leads to
\begin{eqnarray*}
& & \hat{H}_I=-\frac{1}{4}\sum_K\mathcal{E}_K^2(\Delta\alpha\cos^2\hat{\theta}_{zK}+\alpha_\perp)\\
& & -\frac{1}{2}\sum_{K,K';K\neq K'}\mathcal{E}_K\mathcal{E}_{K'}\cos(\phi_K-\phi_{K'})\Delta\alpha\cos\hat{\theta}_{zK}\cos\hat{\theta}_{zK'}.
\end{eqnarray*}
We observe that, for circular polarization in the $(X,Y)$- plane, the cross term with $\mathcal{E}_X\mathcal{E}_Y$ vanishes since the relative phase $\phi_X-\phi_Y=\frac{\pi}{2}$, while this term is maximum for linear polarization when the phase difference is equal to 0 or $\pi$. The same analysis can be conducted for the hyperpolarizability term or for more complex control fields. As an example, we consider a linearly polarized two-color laser field, interacting up to third order with a linear molecule. The non-zero components of the hyperpolarizability tensor are $\beta_\parallel=\beta_{zzz}$ and $\beta_\perp=\beta_{zxx}=\beta_{zyy}$. We obtain that
\begin{equation}
\hat{\boldsymbol \beta}\mathbf{E}^2=\begin{pmatrix}
2\beta_\perp E_xE_z \\
2\beta_\perp E_yE_z \\
\beta_\parallel E_z^2+\beta_\perp (E_x^2+E_y^2)
\end{pmatrix}
\end{equation}
and
\begin{equation}
\frac{\hat{\boldsymbol \beta}}{6}\mathbf{E}^3=\frac{1}{6}E(t)^3[(\beta_\parallel-3\beta_\perp)\cos^3\hat{\theta}+3\beta_\perp\cos\hat{\theta}]\,.
\end{equation}
The electric field can be expressed as
\begin{equation}
\mathbf{E}(t)=[\mathcal{E}_1\cos(\omega t)+\mathcal{E}_2\cos(2\omega t+\phi)]\mathbf{u}_Z\,,
\end{equation}
where $\phi$ is the relative phase between the two components and $\mathbf{u}_Z$ the unitary vector along the $Z$- direction. After averaging over the rapid oscillations of the field, the interaction Hamiltonian reads
\begin{eqnarray*}
\hat{H}_I&=&-\frac{1}{4}\left(
\Delta\alpha\cos^2\hat{\theta}+\alpha_\perp\right)
\left(\mathcal{E}_1^2+\mathcal{E}_2^2\right)\\
& & -\frac{\cos\phi}{8}\left[
\left(\beta_\parallel-3\beta_\perp\right)\cos^3\hat{\theta}+
3\beta_\perp\cos\hat{\theta}\right]\mathcal{E}_1^2\mathcal{E}_2\,.
\end{eqnarray*}

The last technical point concerns the construction of the matrix associated with the operator $\hat{H}_I$ in the $|j,k,m\rangle$- basis. An efficient approach consists in deriving the direction cosines in terms of the coefficients $D^j_{m,k}$. For instance, we have $\cos\hat{\theta}_{z,Z}=\cos\hat{\theta}= D^1_{0,0}(0,\hat{\theta},0)$. The matrix elements can then be computed by using Eq.~\eqref{eqsymdj}
and the sum rule,
\begin{eqnarray*}
\int D_{m_1k_1}^{j_1}D_{m_2k_2}^{j_2}D_{m_3k_3}^{j_3}d\Omega= 
8\pi^2
\begin{pmatrix} j_1 & j_2 & j_3 \\ m_1 & m_2 & m_3\end{pmatrix}
\begin{pmatrix} j_1 & j_2 & j_3 \\ k_1 & k_2 & k_3\end{pmatrix}\,,
\end{eqnarray*}
where $\Omega(\theta,\phi,\chi)$ is the integration volume and the 3-$j$ symbols have been introduced. The 3-$j$ symbols are coefficients allowing to add angular momenta in quantum mechanics~\cite{Zare:88}.

\subsection{Molecular rotational dynamics}
The time evolution of molecular rotational dynamics is governed by different differential equations according to the experimental conditions. At zero temperature, the rotational state of an isolated molecule in the gas phase is described by a state vector, $|\psi(t)\rangle$, which satisfies the Schr\"odinger equation,
\begin{equation}
i\hbar \frac{d}{dt}|\psi(t)\rangle = [\hat{H}_0+\hat{H}_I]|\psi(t)\rangle\,.
\end{equation}
The initial state is usually the level of lowest energy, i.e. $|0,0,0\rangle$. At non-zero temperature, the Liouville-von Neumann equation,
\begin{equation}\label{eq:LvN}
i\hbar\frac{\partial \hat{\rho}}{\partial t}=[\hat{H}_0+\hat{H}_I,\hat{\rho}]\,,
\end{equation}
governs the dynamics of a gas at low pressure,
where $\hat{\rho}$ is the density matrix of the rotational system. For a gas at temperature $T$, the initial state $\hat{\rho}_0$ is given by the canonical density operator. For a linear molecule, this state can be expressed as
\begin{equation}
\hat{\rho}(0)=\frac{1}{Z}\sum_{j,m}g_Je^{-Bj(j+1)/(k_BT)}|j,m\rangle\langle j,m|\,,
\end{equation}
where $Z=\sum_{j,m}g_Je^{-Bj(j+1)/(k_BT)}$ is the partition function and $k_B$ the Boltzmann constant. The term $g_J$ is the nuclear spin degeneracy factor. For homonuclear molecules, this factor depends on the parity of $J$. For instance, $g_J$ is equal to 6 and 3, respectively, for even and odd $J$-states of $N_2$. Equation~\eqref{eq:LvN}
is typically valid for times shorter than the relaxation time due to molecular collisions.  A more complete description is given when including decoherence and dissipation, for example in terms of a Lindblad-type equation,
\begin{equation}\label{eq:Lindblad}
\frac{\partial \hat{\rho}}{\partial t}=\frac{1}{i\hbar}[\hat{H}_0+\hat{H}_I,\hat{\rho}]+\mathcal{L}(\hat{\rho})\,,
\end{equation}
where $\mathcal{L}$ is a linear operator modelling relaxation processes such as collisions~\cite{RamakrishnaSeideman2005}. Equation~\eqref{eq:Lindblad} assumes the Markov approximation, i.e., where the quantum system evolves without memory and does not influence the state of its environment.

When analyzing quantum rotational dynamics, typical observables are expressed in terms of rotation angles. For example, a simple and useful description of a molecule's orientation
in space is given, to first order, by the expectation values of the direction cosines: $\langle \cos\theta_{\gamma\Gamma}\rangle =\textrm{Tr}[\rho\cos\theta_{\gamma\Gamma}]$. Note that only a coarse-grained information of the rotational state is provided by these quantities~\cite{ramakrishna:2013}. By definition, the different expectation values belong to the interval $[-1,1]$. A molecule is said to be oriented along a given axis of the space-fixed frame if one of the expectation values satisfies $|\langle\cos\theta_{\gamma\Gamma}\rangle|\simeq 1$. This concept of orientation can be generalized to three dimensions with the corresponding three direction cosines. Further information about the rotational probability density can be gained from the second-order moments,  $\langle\cos^2\theta_{\gamma\Gamma}\rangle$. These quantities are essential for symmetric molecules with no permanent dipole moment. An example is given by homonuclear diatomic molecules such as N$_2$ or O$_2$. In this case, the notion of orientation is replaced by that of alignment in which the axes direction is not considered. For linear molecules, the degree of alignment is quantified through the quantum averages, $\langle \cos^2\theta_{z\Gamma}\rangle$, where $\theta_{z\Gamma}$ is the angle between the molecular axis and the axes of the space-fixed frame. They satisfy the relation $\sum_\Gamma \langle \cos^2\theta_{z\Gamma}\rangle =1$. At ambient temperature, the three expectation values are for symmetry reasons equal to 1/3. The molecule is said to be aligned along the $X$- direction if $\langle \cos^2\theta_{zX}\rangle\simeq 1$. When $\langle \cos^2\theta_{zX}\rangle\simeq 0$, the molecular axis is delocalized in the $(Y,Z)$- plane, leading to planar alignment~\cite{lapert:2009,Hoque:11}.

\subsection{Geometric description of rotational dynamics}

\begin{figure}[tb]
\centering
      \includegraphics[width=0.5\linewidth]{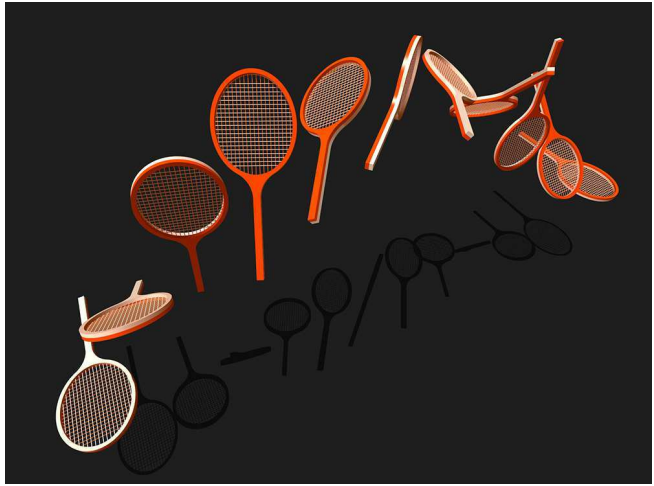}
\caption{(Color online) Schematic description of the tennis racket effect. Note the $\pi$- flip of the head of the racket when the handle makes a
rotation.}\label{figracket}
\end{figure}
The geometric study of the rotation of a rigid body is a fundamental subject both in classical and quantum mechanics~\cite{landau,landauQM}. The starting point is the integrable case~\cite{arnold}, that is, the dynamics of the free asymmetric top and symmetric top in a constant external field. The two systems are known as the Euler top and the Lagrange top in mathematics and  have been studied for many centuries~\cite{cushman}. Non-trivial effects are still investigated by mathematicians and theoretical physicists. An example is given for asymmetric tops by the tennis racket effect~\cite{cushman,vandamme:2017,ashbaugh:1991}. This classical geometric phenomenon occurs in the free rotation of any asymmetric rigid body. In particular, it describes what happens when a tennis racket is
tossed into the air while imparting a rotation about the unstable axis. In addition to the $2\pi$-rotation about this axis, the racket performs an unexpected $\pi$-flip about its handle. A schematic description of this effect is given in Fig.~\ref{figracket}.

In some specific cases, the rotational energies and wave functions of a molecule in an electric field can be derived analytically.
For a linear molecule subject to combined orienting and aligning interactions,
this is made possibly by use of supersymmetric quantum mechanics~\cite{schmidt:2015,schmidt:2014,lemeshko:2011}. For symmetric top molecules, quasi-solvability has been investigated by~\cite{schatz:2018} using the quantum Hamiltonian-Jacobi theory. The properties of polar paramagnetic molecules subject to congruent electric, magnetic and optical fields were derived by~\cite{sharma:2015}. A systematic analysis of the non-adiabatic and adiabatic regimes is performed in~\cite{mirahmadi:2018} with both analytical and numerical computations.

A numerical approach allows to investigate the role of vibrational motion for rotational dynamics, for example in strong  electric fields inducing adiabatic dynamics~\cite{gonzales:2004}.
An adiabatic rotor approach was proposed by~\cite{gonzales:2005} to account for the different dynamical effects of the rovibrational coupling. For the example of the LiCs molecule, a complete study of the coupled ro-vibrational motion has been performed by~\cite{gonzales:2006}.

A topic of renewed interest today is to find signatures of classical dynamics in the quantum regime. A large number of studies have shown the advantage of classical analysis for revealing the properties of quantum molecular spectra. This aspect has been investigated for the free molecular rotation~\cite{Child:91,cuisset:2012,harter:1984,sadovskii:1990,hamraoui:2018,child:2007,vandamme:SR}, but also for molecules subject to a constant electric field~\cite{kozin:2003,arango:2004,arango:2005}. Take Hamiltonian monodromy~\cite{cushman} as an example, which is the simplest topological obstruction to the existence of global action angle variables in classical integrable systems. Monodromy has a quantum counterpart which prevents the existence of global good quantum numbers in the rotational spectrum~\cite{EfstathiouBook}. Monodromy in the spectrum of a quantum symmetric top molecule in an electric field was described by~\cite{kozin:2003}. The dynamics of linear molecules in tilted electric fields were investigated by~\cite{arango:2004,arango:2005}. Note that a geometric framework can also be derived for non-integrable case in terms of location and  bifurcation of relative equilibria in the energy-momentum diagram of the molecule~\cite{arango:2008}.

\subsection{Controllability of rotational dynamics}\label{sec:contra}
Controllability of quantum dynamics is of fundamental as well of practical importance since it determines the extent to which
a quantum system can be manipulated and brought to a desired target state~\cite{alessandro_book}. The mathematical theory of controllability is by now well established for finite-dimensional closed quantum systems~\cite{albertini:2003,altafini:2002,dirr:2012}. In the case of an infinite-dimensional Hilbert space, the problem is much more intricate from a mathematical point of view, even for a quantum system with a discrete spectrum~\cite{beauchard:2005,chambrion:2009,boscain:2012,boscain:2014,boscain:2015}.

For infinite-dimensional systems, two concepts of controllability are introduced, namely exact and approximate controllability. In the approximate controllability version, the system is not steered exactly to the target state, but to an arbitrary small neighborhood of the final state. Note that in a finite-dimensional setting, the notions of exact and approximate controllability coincide. Exact controllability seems rather exceptional in quantum control and very difficult to prove rigorously~\cite{beauchard:2005}. Approximate controllability is much more common and many different results have been obtained using, e.g., Galerkin techniques, i.e., finite-dimensional approximations of the Hilbert space~\cite{chambrion:2009,boscain:2012,boscain:2014,boscain:2015}. A crucial aspect of the proof is based on the structure of the energy levels, which has to be as different as possible from the spectrum of a quantum harmonic oscillator. It can be shown mathematically that the latter is not controllable~\cite{rouchon:2004}. This is also easily understood intuitively since an electromagnetic field driving a harmonic oscillator cannot distinguish different transitions, such as the transition between $|0\rangle$ and  $|1\rangle$ and that between $|1\rangle$ and $|2\rangle$.
Rotational dynamics with its infinite-dimensional Hilbert space and the anharmonic discrete spectrum of the associated Hamiltonian is an ideal physical example for testing controllability techniques. Approximate controllability in the infinite-dimensional setting has been shown for linear molecules~\cite{boscain:2012,boscain:2014,boscain:2015}. However, controllability of the dynamics of symmetric and asymmetric top molecules remains an open mathematical question.

Another approach to establishing controllability consists in selecting a finite-dimensional subspace of the physical Hilbert space~\cite{RanganPRL04}. For  control of molecular rotations, this subspace is spanned by the kets $\{|j,k,m\rangle\}$ with $j\leq j_{\textrm{max}}$. Such a dimensionality reduction can be physically justified by the finite temperature of the system and the fact that  electromagnetic fields only transfer a finite amount of energy to the molecule,  confining the system to a finite-dimensional subspace~\cite{turinici:2004,turinici:2007,sugny:2004,sugny:2005,sugny:2005b}. The parameter $j_{\textrm{max}}$ depends on the temperature, on the frequency bandwidth and the  energy of the pulse that the molecule is subjected to. With such a physically motivated assumption, controllability can be shown, using the tools of the finite-dimensional setting.

If the molecule is controlled by two independent electric fields along two orthogonal directions, then the system is usually completely controllable, that is any target state can be reached from any initial state of the Hilbert space~\cite{alessandro_book}. There are some exceptions to this general rule due to  symmetries of the molecular system. A simple example of a non-completely controllable system is a decoupled one. In this case, the Hamiltonian matrix is block diagonal and the Hilbert
space can be decomposed into a direct sum of at least two orthogonal subspaces. In this situation, the relevant concept is simultaneous controllability, that is the possibility to control uncoupled quantum systems by the same external field. A standard example is encountered in a linear molecule driven by a linearly polarized laser pulse at non-zero temperature. This system is not completely controllable since $m$ is a good quantum number~\cite{sugny:2005b}. The Hilbert space can be written as the sum of subspaces $\mathcal{H}_m$. By symmetry, the dynamics in $\mathcal{H}_m$ and $\mathcal{H}_{-m}$ are the same so that only the positive $m$ values have to be taken into account in a controllability analysis~\cite{sugny:2005b}. Simultaneous controllability can be proven for this class of examples in the finite and infinite dimensional setting~\cite{boscain:2014}.

\section{Alignment and orientation}
\label{sec:align}

Laser-induced alignment and orientation are important
examples for the control of molecular dynamics. A one-body picture is sufficient to analyze and understand it.
A noticeable degree of alignment or orientation is crucial in a variety of applications ranging from chemical reaction dynamics~\cite{Stapelfeldt:03,LevineBook} to nanoscale design~\cite{seideman:1997c} and attosecond electron dynamics~\cite{Haessler:2010}. Molecular alignment and orientation can also be viewed as crucial prerequisites in laser control of molecular dynamics before exploring more complex scenarios. For instance, the enhancement of the yield of a chemical reaction often requires the control of the spatial orientation of the molecules. In the case of a molecule driven by a linearly polarized laser field, alignment (corresponding to a double-headed arrow, $\updownarrow$) is defined by an increased probability distribution along the polarization axis compared to the direction perpendicular to it. Orientation, on the other hand, corresponds to a selected direction in space, i.e., a single-headed arrow, $\uparrow$. In such a way, orientation implies alignment, however, the reverse is not the case~\cite{Stapelfeldt:03,Seideman:05}.

Molecular alignment and orientation can be produced under two different regimes, namely the adiabatic and the sudden ones. We introduce
the characteristic time-scale of the free rotation of the molecule, $T_{\textrm{rot}}$, which is
typically of the order of 1 to 100$\,$ps for small molecules. The sudden mechanism is based on short and intense laser pulses of duration $T_s$
with  $T_s\ll T_{\textrm{rot}}$. The time $T_s$ is of the order of tens or hundreds of femtoseconds. A specific rotational wave packet is created and field-free transient alignment and orientation are observed as long as the coherence of the process is preserved. In the adiabatic limit, on the other hand, pulses with a duration $T_a\gg T_{\textrm{rot}}$ are used. In practice, this has been obtained with pulse durations of a few nanoseconds or few hundred picoseconds. The adiabatic transfer is induced by slowly turning on the laser field with respect to the rotational period. The molecule is then adiabatically transferred to rotationally excited states
during the pulse and goes back to its initial state when the field is turned off. In this case, alignment and orientation are not generated under field-free conditions, except for the case where the field is abruptly switched off~\cite{seideman:2001,underwood:2003}.

In this section, we review  recent theoretical and experimental progress in the generation of alignment and orientation, from the standard one-dimensional alignment to recent extensions to two and three spatial dimensions. The experimental detection of alignment and orientation will be also discussed, as well as recent applications of molecular alignment. For a
more detailed description of the concepts and applications, we refer the interested reader to
~\cite{Stapelfeldt:03,Seideman:05,LemKreDoyKais13,CaiFriCCCC01,ohshima:2010,fleischerrev:2012,pabst:2013} and references therein.

\subsection{Generation of 1D alignment and orientation}

\subsubsection*{Alignment by static fields and linearly polarized nanosecond and femtosecond laser pulses}

Confining molecular rotation to a specific direction is a long-standing goal in physics. Pioneering works date back to the early 1960's. Two techniques based on static electric fields were used to orient molecules~\cite{Stapelfeldt:03,LemKreDoyKais13}. The oldest method selects molecules in a single rotational state by using an electrical
hexapole~\cite{brooks:1979,baugh:1994,stolte:1988,cho:1991}.
The second one consists in applying a strong, static and homogeneous electric field to orient the molecules~\cite{loesch:1990,wu:1994,friedrich:1991}. This approach requires that the interaction energy, i.e. the dipole moment times the field strength, is larger than the rotational energy. These strategies have an intrinsic limitation due to the maximum strength of electrostatic fields that can be applied. Intense laser fields
came into play only about 30 years later with the advance of optical technology.
It does thus not come as a surprise that the first results on the spatial manipulation of molecules in the adiabatic regime with intense nanosecond laser fields were obtained theoretically~\cite{zon:1975,friedrich:1995,seideman:1995,seideman:1997a,seideman:1997b,seideman:1997c,seideman:1999b,friedrich:1995b,friedrich:1999}. This theoretical activity encouraged experimental developments on molecular alignment for which the first direct evidences were reported in the late 1990s~\cite{sakai:1999,larsen99,larsen:1999,sugita:2000,normand:1992,baumfalk:2001}.

At the same time, new alignment techniques under field-free conditions based on the application of femtosecond laser pulses were proposed in different theoretical studies~\cite{averbukh:2001,ortigoso:1999,seideman:1999,machholm:1999,machholm:2001b}. Molecular alignment and orientation in combined electrostatic and pulsed laser fields were described by~\cite{cai:2001b}. Field-free molecular alignment was then observed by using femtosecond laser pulses for linear molecules~\cite{rosca:2001,rosca:2002a,rosca:2002b} and for asymmetric top molecules~\cite{peronne:2003}.
Several theoretical
studies, focussing on  alignment dynamics, explored the best way to align linear molecules in the sudden excitation regime with one or a series of femtosecond laser pulses. The revival structure of alignment dynamics was described by~\cite{seideman:1999,ortigoso:1999}. The roles of permanent dipole moment and polarizability terms in the alignment process were discussed by~\cite{dion:1999}. Enhancement of molecular alignment by a train of short laser pulses was shown by~\cite{leibscher:2003,leibscher:2004}. The possibility to maintain molecular alignment for arbitrarily long time by using an appropriate periodic sequence of laser pulses was proposed in~\cite{ortigoso:2004,ortigoso:2010}.
The alignment mechanism was described in terms of the coherent control of rotational wave packets~\cite{spanner:2004}. Phase-shaped femtosecond laser pulses can be used to control the degree of alignment~\cite{renard:2005,renard:2004,hertz:2007}. The alignment was optimized experimentally by a feedback closed loop procedure~\cite{suzuki:2008}. The corresponding rotational wave packet can be reconstructed through the interference with a replica of the laser pulse~\cite{hasegawa:2008}. Enhancement of alignment by a combination of a short and a long laser pulses polarized in the same direction was shown by~\cite{poulsen:2006,guerin:2008}.

The intermediate alignment regime in which the duration of the laser pulse is of the same order as the rotational period, i.e.,  inbetween the adiabatic and sudden limits, is explored theoretically by~\cite{ortigoso:1999,seideman:1999,torres:2005,mirahmadi:2018}. Experimentally, this regime has been explored with the OCS molecule by~\cite{trippel:2014}. A wave packet of field-dressed states is generated and its time evolution observed during and after the laser pulse. Contrary to the adiabatic regime, the degree of alignment presents oscillations due to the propagation of the quantum system in the field-dressed potential of the molecule. Recently, it was pointed out how the coupling between the rotational angular momenta and the nuclear spins through the electric quadrupole interaction can decrease the observed degree of alignment in the sudden regime~\cite{thomas:2018}. Finally, note that field-free molecular alignment can be strongly enhanced by a cavity as shown by~\cite{benko:2015}.

\subsubsection*{Adiabatic orientation by two-color and static fields}

In the non-resonant case, the carrier frequency of the laser field is very large compared to rotational frequencies. As shown in Eq.~\eqref{eqtimeaverage}, the interaction between the molecule and the external field through the permanent dipole moment averages to zero. Therefore, only the polarizability term plays a role in the rotational dynamics which leads to molecular alignment. Molecular orientation of polar molecules can only be generated in the adiabatic regime if the inversion symmetry of the alignment process is broken.  Different control strategies have been studied. One option is to use a two-color laser field and the asymmetry caused by the hyperpolarizability interaction~\cite{kanai:2001}. Parity breaking can also be achieved from a resonance between the frequency of the laser field and a vibrational transition~\cite{guerin:2002}. Hyperpolarizability-induced orientation was demonstrated experimentally by~\cite{Oda:10}. A recent theoretical
study has shown how to improve molecular orientation by optimizing the relative delay and intensities of the two-color excitation~\cite{mun:2018}. Note that field-free molecular orientation can also be achieved by slowly ramping up the laser field and then abruptly switching it off~\cite{muramatsu}.

Another control protocol consists in combining an electrostatic field with an intense non resonant laser field~\cite{friedrich:1999b}. This procedure was first applied experimentally by~\cite{sakai:2003,bukh:2006}. The same control strategy was used in several experiments~\cite{Goban:08,nielsen:2012,trippel:2015,hansen:2013,kienitz:2016} and studied theoretically by~\cite{hartelt} and~\cite{Omiste:16,thesing:2017} for symmetric and asymmetric top molecules, respectively. Non-adiabatic effects  were revealed to play a role in molecular orientation by~\cite{omiste:2012}. In particular, this work provides the field parameters under which the dynamics are adiabatic. Control of molecular orientation of the LiCs molecule in static electric fields through radiative rotational transitions was studied by~\cite{gonzales:2007}.

Orientation and alignment drastically decrease with temperature. As temperature increases, the initial state of the dynamics becomes more and more a statistical mixture of rotational states which tends to misalign or misorient the molecule. This difficulty can be overcome by considering state-selected molecules. This can be achieved in  a seeded supersonic expansion. Due to collisions with the buffer gas, molecular rotation is cooled down to typically a few kelvin so that only the lowest rotational states are populated. A strong inhomogeneous static electric field is applied in a second step to select polar molecules in a single or a few individual rotational quantum states. The produced state-selected molecules are then used to generate very high laser induced alignment and orientation. This idea was realized  experimentally in a series of studies~\cite{holmegaard:2009,filsinger:2009,Ghafur:09,mun:2014,luo:2015} with state-selected molecules. The control fields were theoretically optimized by~\cite{rouzee:2009} to improve the efficiency of the alignment or orientation process.
\subsubsection*{Orientation by two-color femtosecond lasers and THz field}
As in the adiabatic regime, the
generation of molecular orientation by two-color femtosecond laser pulses is
based on the light-matter interaction involving the molecular polarizability and hyperpolarizability terms.
This method was proposed and studied theoretically by~\cite{Tehini:08,vrakking:1997} and
demonstrated experimentally by~\cite{de:2009,kraus:14}. A comparison between this mechanism and a two-color laser induced orientation based on ionization depletion is made by~\cite{spanner:2012}. In the second process, the pulse ionizes selectively molecules according to their orientation angles. An experimental observation of the transition between these two mechanisms was done in~\cite{znakovskaya:2014}. Other theoretical studies have investigated the conditions required to maximize the achieved degree of orientation. \cite{zeng:2010} showed that field-free orientation can be strongly enhanced if a second in-phase or antiphase dual-color laser pulse is applied at the revival or half-revival time. \cite{yun:2011} observed that the parity of excited rotational states is a crucial parameter in the enhancement of molecular orientation. Phase-dependence of molecular orientation in a two-color excitation process was investigated by~\cite{qin:2014}. Alignment enhanced orientation obtained from a delay between single and two-color laser pulses was also demonstrated by~\cite{zhang:2011,Tehini:2012} and observed experimentally by~\cite{ren:2014b}. Field-free orientation by other control protocols with short pulses of different colors was studied~\cite{zhang:2011b,chen:2010,zhdanov:2008}. The efficiency of a multicolor laser field with a superposition of the fundamental wave and its harmonics was analyzed by~\cite{zhang:2011b}. Two circularly polarized pulses of different wavelengths can also be used to orient molecules~\cite{chen:2010}. \cite{zhdanov:2008} showed that molecular orientation can be generated by a three-color laser field even at room temperature.

Orientation by THz fields has been the subject of an intense theoretical activity to determine the feasibility as well as the efficiency of the control process. The typical time scale of a THz pulse is of the same order as the rotational period, leading to direct one photon transitions between rotational states. A THz field with a highly asymmetrical temporal shape imparts a sudden momentum kick through the permanent dipole moment and thereby orients the molecule. However, due to limitations to current pulse shaping techniques, only half-cycle and few-cycle pulses can be generated experimentally. The efficiency of the kick mechanism and its robustness against temperature effects were studied theoretically~\cite{dion:1999,Lapert:12,machholm:1999,machholm:2001,chuan:2013,ortigoso:2012,matos:2003,sugny:2004b}. The combination of THz excitation with a far-off-resonant laser field was investigated by~\cite{daems:2005b,gershnabel:2006,gershnabel:2006b,Kitano:11,shu:2008,hu:2009}. A high degree of orientation can also be generated by a train of half-cycle pulses~\cite{sugny:2004,dion:2005}. Experimental demonstrations have been
achieved more recently both for linear~\cite{Nelson:11,jones:2014}, symmetric~\cite{Babilotte:16} and asymmetric top molecules~\cite{damari:2016}. More complex control scenarios have been explored theoretically using, to mention a few, a two-color laser field and a THz pulse~\cite{li:2013} or a THz few-cycle field with a specific phase~\cite{shu:2009,qin:2012}. \cite{fleischer:2012} have shown theoretically and experimentally that quantum coherences in a rotational system can be manipulated by two properly delayed THz pulses.

\subsubsection*{Optimal control for alignment and orientation}

Optimization techniques have been
widely used to theoretically design control fields maximizing orientation and alignment. The controllability aspects have been described in Sec.~\ref{sec:contra}. Manipulating molecular rotation by laser fields represents an ideal control problem because the Hamiltonian is known with very high precision---only the rotational constants and the permanent and induced dipole moments are needed for computing the dynamics. A very good agreement between theory and experiment is usually observed. Therefore, open-loop optimization process, which is only based on the knowledge of the quantum system~\cite{Glaser:15}, is expected to result in improved performance. A pioneering study on the subject was carried out by~\cite{rabitz:1990}, who explored the question of controllability as well as the numerical design of optimal pulses. Also, note that a parallel can be made with spin dynamics in which optimal control fields are known to be experimentally very efficient~\cite{Glaser:15,kobzar:2005,skinner:2012,lapert:2010}.

To simplify the description of the control process, we consider the optimization of the degree of orientation of a linear molecule by a linearly polarized electric field. The goal is to maximize at a given time $t$ the expectation value $\langle \cos\hat{\theta}\rangle$, where $\theta$ is the angle between the molecular axis and the polarization direction. These arguments can be generalized to molecular alignment and to the control of symmetric and asymmetric top molecules. From a theoretical point of view, the main difficulty for the optimization procedure comes from the infinite dimension of the Hilbert space of the quantum system. The operator $\cos\hat{\theta}$ has a continuous spectrum and one cannot define
a target state that saturates
the maximum orientation. This target state, i.e., the eigenstate of $\cos\hat{\theta}$ with the largest (or smallest) eigenvalue, does not belong to the Hilbert space and is physically not relevant
since it can only be expressed as an infinite linear combination of spherical harmonics. An alternative strategy to define a target state consists in selecting a finite-dimensional subspace of the physical Hilbert space. A standard choice is to fix a maximum value $j_{\textrm{opt}}$ of the angular momentum and to consider the space $\mathcal{H}_{\textrm{opt}}$ spanned by the states $|j,m\rangle$ with $j\leq j_{\textrm{opt}}$. The projection of $\cos\hat{\theta}$ onto $\mathcal{H}_{\textrm{opt}}$ has a discrete spectrum and a target state maximizing the orientation can thus be defined~\cite{sugny:2004}. Note that, in order to avoid numerical artefacts, the numerical computations should be performed in a larger Hilbert space with $j>j_{\textrm{opt}}$. Another option is to optimize $\langle \cos\hat{\theta}\rangle $ directly but with a penalization on the pulse energy to avoid populating high $j$- levels, which would require a very intense laser field~\cite{abe:2011,salomon:2005}.

Different optimization methods have been used in the literature for maximizing molecular orientation and alignment. Pulses with a fixed duration were derived with numerical optimal control algorithms either from a target state or the maximization of the expectation value of an observable~\cite{salomon:2005,nakajima:2012,nakagami:2008b,hoki:2001}. The control time was optimized in~\cite{NdongJMO14}. Different extensions of the standard algorithms were proposed to account for some specific experimental constraints
such as the requirement of a zero-area field~\cite{Sugny:14,shu:2016}, spectral constraints~\cite{lapert:2009b}, fields with specific shape constraints~\cite{abe:2011,Yoshida:14,liao:2013,yoshida:2015}, and different polarizations~\cite{lapert:2009,abe:2012}. Polarizability and hyperpolarizability terms lead to a non-linear interaction between the molecular system and the control field. Different optimization procedures were proposed for this kind of dynamics~\cite{nakagami:2008,lapert:2008,Ndong:2013}. Other situations with isotope selectivity~\cite{nakajima:2016} and control by microwave pulses~\cite{chi:2008} have also been explored. Due to the computational complexity of the quantum dynamics, optimal control of asymmetric top molecules is not at the same stage of development, even if some recent studies have started investigating this aspect~\cite{artamonov:2010,coudert:2017,coudert:2018}. Molecular rotation can also be studied from the perspective of quantum tracking control in which the goal is to steer the expectation values of specific observables along desired paths in time. Analytical and singularity-free expressions
for fields capable of controlling the orientation of a planar molecule were derived by~\cite{magann:2018}.

In all these examples, the derived optimal pulses are generally very efficient but cannot be directly implemented experimentally with the current state-of-the-art pulse shaping techniques. In view of experimental applications, another approach, based on a closed-form expression for the control field which then depends on a finite number of parameters, was developed~\cite{skinner2010}.
The parametrization of the field is chosen to satisfy specific experimental constraints or limitations, corresponding to those of the pulse shapers. At the second step of the optimization procedure, the optimal values of the parameters can be determined by using global optimization procedures such as genetic algorithms in the time or frequency domain~\cite{benhaj:2002,rouzee:2011,hertz:2007,leibscher:2003,leibscher:2004,dion:2005,dion:2002}. The diversity of the potential optimal control solutions is explored by~\cite{shir:2008}. This approach leads to efficient and experimentally relevant control pulses.

Experimental pulse shaping devices, for laser pulses,
operate in the frequency domain by tailoring the spectral phase and amplitude of the field over a given bandwidth~\cite{weiner:2000}. More precisely, the spectral phase and amplitude are two piecewise constant functions with a number of pixels typically equal to 64, 128 or 256. These pixels are equally spaced in the frequency interval defined by the bandwidth. The shaping of THz pulses is not yet at the same stage of development, although recent years have seen impressive experimental development, see e.g.~\cite{ahn:2003,amico:2009,vidal:2014,gingras:2018} to mention but a few. These studies show that the shape of the generated THz waveform can be optimized to some extent. Moreover, the central frequency can be tuned and the width of the spectrum can be modified.

\subsection{Generation of unidirectional rotation and 3D alignment via polarization control}

A much more challenging goal than 1D alignment is to generate 3D alignment or more complicated dynamics through the control of laser field polarization either in the sudden or in the adiabatic regime. In 3D alignment, the three molecular axes are forced to be aligned along given fixed axes in space. The idea behind this research program is to make progress towards full control of the rotational motion.

\begin{figure}[tb]
\centering
      \includegraphics[width=0.5\linewidth]{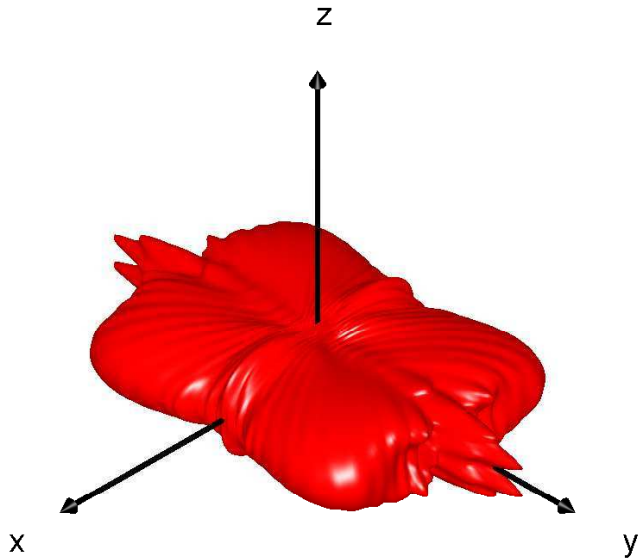}
\caption{(Color online) Angular distribution of the CO$_2$ molecule after interaction with two short pulses polarized in orthogonal directions. Note the planar alignment of the molecule in the $(x,y)$- plane. Reprinted with permission from~\cite{Hoque:11}.}\label{figplanar}
\end{figure}

We start by discussing  sudden control of molecular rotation. Since the duration of field-free alignment of a linear molecule along a given fixed direction is intrinsically limited by the rotational dynamics, a less demanding objective is to permanently confine the molecular axis in a plane. Such a planar alignment has no intrinsic limit as shown by~\cite{lapert:2009}. Since, for linear molecules, the angular momentum is orthogonal to the molecular axis, this control is equivalent to the alignment of the angular momentum. It was successfully achieved in the experiment by~\cite{Hoque:11}, applying two short laser pulses properly delayed and polarized in orthogonal directions. The simulated angular distribution of the CO$_2$ molecule after interaction with the laser pulses is displayed in Fig.~\ref{figplanar}.
As proposed theoretically by~\cite{feischer:2009} and realized experimentally by~\cite{kitano:2009}, orientation of the angular momentum can be obtained with the same control sequence, but using two pulses with linear polarization tilted by $45^\circ$ with respect to each other. Such an arrangement leads to an asymmetric distribution of the rotational population in states with positive and negative values of the quantum number $m$. This excitation induces the rotation of the molecules in a preferential direction, clockwise or counter-clockwise. The same control strategy can even induce molecular orientation in specific asymmetric chiral molecules whose polarizability tensor is non-diagonal in the molecular frame~\cite{YachmenevPRL16,TutunnikovJPCL18,gershnabel:2018}. Other control procedures can be used to generate unidirectional rotation of the molecular axis such as
a spinning linear polarization~\cite{karras:2015} and a chiral train of laser pulses~\cite{zhdanovich:2011,floss:2012b}. The rotational Doppler effect can be used to measure the rotation of spinning molecules as proposed and shown by~\cite{steinitz:2014,korech:2013,karras:2015}. A field-free two-direction alignment alternation can be induced by short elliptic laser pulses~\cite{daems:2005,maan:2016}.

Unidirectional rotation can be produced in the adiabatic regime with a much larger efficiency~\cite{karczmarek:1999}. This was first experimentally observed by~\cite{villeneuve:2000}, introducing the concept of the optical centrifuge. Spinning of molecules to very high angular momentum states is induced by a $\sim$ 100 picosecond long laser pulse with a rotating polarization axis. Such an excitation can be sufficient to break molecular bonds. Dissociation and multiple collisions processes were used for the identification of the formation of superrotors~\cite{villeneuve:2000,yuan:2011}. These new molecular objects were further studied, measured and controlled
by~\cite{korobenko:2014} from a coherent detection technique using a probe beam. The dynamics and the physical properties of these structures were then studied in a series of papers. The magneto-optical properties of paramagnetic superrotors were investigated by~\cite{milner:2015}. Ultrafast magnetization of a dense molecular gas was achieved by~\cite{MilnerPRL17}. \cite{steinitz:2012} showed numerically that the collisions in a dense gas of unidirectionally rotating molecules can induce macroscopic vortex gas flows. A two-dimensional centrifuge was proposed as a tool to produce long-lived alignment~\cite{milner:2016}. A complete theoretical description of the generation of asymmetric superrotors is given by~\cite{omiste:2018}.

Linearly polarized pulses can be used to align the most polarizable axis of asymmetric-top molecules along the polarization direction~\cite{holmegaard:2007,rouzee:2006,peronne:2004}. However, the ultimate goal in the control of the rotation of such molecules is 3D alignment or orientation where all three molecular axes are forced to be aligned or oriented along the three directions of the space-fixed frame. Such a degree of control requires to go beyond linearly polarized laser fields. Three-dimensional alignment was achieved for the first time experimentally with an elliptically polarized nanosecond pulse by~\cite{larsen:2000} and by~\cite{viftrup:2007,viftrup:2009} combining nanosecond and femtosecond laser pulses polarized in orthogonal directions. Three-dimensional orientation was observed by~\cite{tanji:2005}, using a combination of an electrostatic and an elliptically polarized laser field. Field-free 3D molecular alignment was first demonstrated by~\cite{lee:2006} with two time-delayed, orthogonally polarized femtosecond laser pulses. The same observation was made by~\cite{rouzee:2008} with a short elliptically polarized field and by~\cite{ren:2014} with a sequence of impulsive laser pulses with different ellipticities. Optimal control techniques have been used to theoretically study the extent to which asymmetric-top molecules can be aligned along the three directions at the same time~\cite{artamonov:2010}. Laser field-free 3D orientation of asymmetric-top molecules was achieved by~\cite{takei:2016} with combined weak electrostatic and elliptically polarized laser fields. The first all-optical field-free orientation was demonstrated with two-color fields~\cite{kang:2018}.

In addition to rotational degrees of freedom, it is also interesting to control the molecular torsion, that is the angle between the different groups of atoms around a molecular bond. The control of torsional angles is a crucial step for the development of emerging technologies such as molecular electronics. The first theoretical proposal was brought forward by~\cite{ramakrishna:2007}. Adiabatic torsional quantum control was studied by~\cite{coudert:2011,ortigoso:2013}. Modeling of torsional control was explored by~\cite{coudert:2015,grohmann:2017} to analyze the number of internal molecular coordinates to account for in this dynamical process as well as the validity of reduced dimensionality models. \cite{flosstorsion:2012} showed that laser-induced torsional alignment depends on the nuclear spin of the molecule, which leads to selective control of nuclear spin isomers. The interplay between the torsional and the rotational motion of aligned molecules was studied theoretically by~\cite{omiste:2017,grohmann:2018}. Torsional motion was found to have a strong impact on the rotational dynamics. Control over the torsional motion has been demonstrated experimentally and theoretically by~\cite{madsen:2009,hansen:2012} where a nanosecond elliptically polarized laser pulse is used to produce 3D alignment of the molecules while a linearly polarized short pulse initiates torsion. A pair of delayed linearly polarized laser pulses was used by~\cite{christensen:2014} to generate torsional motion in substituted benzene rings.

\subsection{Detection of alignment and orientation}

The experimental measurement and quantification of molecular alignment and orientation is important to estimate the efficiency of the control process and to probe the angular distribution of the generated rotational wavepacket. Both destructive and non-destructive measurement techniques
have been developed.

The first strategy to estimate the molecular alignment is based on the
substantial modification of rotational eigenstates when the molecule is subject to a strong laser pulse. Rotational spectroscopy can therefore be used to observe and measure alignment~\cite{kim:1996,kim:1997,kim:1998}. Another way to quantify alignment is to use a destructive technique
based on an intense probe, which breaks the aligned molecule, and
the measurement of the angular distribution of the corresponding fragments~\cite{larsen:1999,sakai:1999,larsen:1998}. More precisely, a pulse ionizes the molecule, which leads to a Coulomb explosion~\cite{AminiJCP17}. The produced ions are then detected by measuring the three-dimensional velocity of each fragment, either the full 3D vector or its 2D projection on a detector screen where the third projection of the ion's velocity can be extracted from the time of flight. For a linear molecule and a sufficiently fast dissociation process, the velocities of the fragments are parallel to the molecular axis and the measured distribution reflects therefore the laboratory-frame molecular orientation. This measurement can be performed in one or several directions. Direct imaging of rotational wave-packet dynamics of linear molecules was achieved by~\cite{dooley:2003}. The ion produced by Coulomb exploding the molecule is detected in a series of measurements carried out at different times. This information allows for the reconstruction of the molecular angular distribution function. Note that recent progress has been made in the Coulomb explosion imaging techniques~\cite{UnderwoodRSI15,ChristensenPRA16,AminiJCP17}. Rotating molecular wave packets have been visualized recently by~\cite{mizuse:2015,lin:2015}.

A non-intrusive observation of molecular alignment can be performed through a weak probe which is aimed at measuring the optical properties of the gas of aligned molecules. It can be shown that the macroscopic measure of birefringence is related to the degree of alignment~\cite{renard:2003,renard:2004b}. This technique can be generalized to measure the alignment along two orthogonal directions~\cite{hertz:2007b}. Molecular dichroism can be used in a comparable way as birefringence to estimate the alignment of the molecules~\cite{lavorel:2016}. Furthermore, degenerate four-wave mixing has been employed for measuring the degree of alignment~\cite{ren:2012}.

Non-intrusive methods have  also been developed to measure molecular orientation. When a molecular sample is oriented by laser excitation, a free-induction decay (FID) is emitted by the gas~\cite{Babilotte:16,Nelson:11}. The FID can be connected to the degree of orientation of the molecule. In particular, it can be shown that the emitted THz field is proportional to the time derivative of $\langle \cos\theta\rangle$~\cite{Babilotte:16,Nelson:11}. The observed experimental signal can be reproduced with good accuracy by numerical simulations~\cite{Babilotte:16}. It was recently shown that FID leads to a decay of molecular orientation which is not observed for non-polar molecules~\cite{damari:2017}.

High-order harmonic generation (HHG) from aligned molecules was first observed experimentally by~\cite{velotta:2001}. Theoretically, the H$_2^+$ model system was studied in this respect by~\cite{hay:2002}. The role of orbital symmetry in HHG was discussed by~\cite{nalda:2004}. It is shown that the minimum high-harmonic yield is obtained when the molecular axis is parallel to the polarization direction.
HHG was then studied theoretically by~\cite{Ramakrishna:07}, showing in particular that information about the molecule and its rotational dynamics can be extracted from the HHG spectra~\cite{ramakrishna:2008,Ramakrishna:10}. HHG was also experimentally analyzed in different studies~\cite{levesque:2007,yoshii:2008,rupenyan:2012}. The characteristics of oriented wave packets were studied via HHG by~\cite{frumker:2012b,frumker:2012a,Kraus:12}. Complex revival dynamics of rotational wave packets are measured via Coulomb imaging and HHG by~\cite{weber:2013}. HHG from spinning molecules was also observed~\cite{faucher:2016,He:2018,prost:2017}. The rotational Doppler shift in the harmonics frequency was observed by~\cite{faucher:2016,prost:2017}. HHG has been recently used to probe the spinning dynamics in real time~\cite{He:2018}.

Imaging polyatomic molecular structure can also be achieved with laser-induced electron diffraction technique as proposed by~\cite{peters:2011}. The approach is based on measuring the photo-electron diffraction pattern which encodes information about the geometry of the molecule's nuclei.

\subsection{Applications of molecular alignment}

Molecular alignment results in a highly peaked angular
distribution of the rotational wavepacket. It is generally used as a preparation step for further interactions or chemical processes. The decisive role of molecular alignment has been shown in a growing number of applications such as molecular imaging and selectivity, the control of molecular scattering and ionization, to name a few. One can also make use of molecular alignment for probing collisional relaxation and enhancing the interaction of a molecule with a surface. Since a large number of applications of laser aligned molecules have emerged during the past few years, only the recent developments will be reviewed in this work and we refer the reader to earlier reviews discussing these aspects~\cite{Stapelfeldt:03,Seideman:05,LemKreDoyKais13}. Note that the rotational dynamics in an environment is discussed in Section~\ref{sec:environ}.

During the past two decades, alignment has been shown to be crucial
in chemical reactions and stereochemistry,
for studying molecular structure~\cite{torres:2007,itatani:2004}, HHG~\cite{kanai:2005,kanai:2007,hay:2002}, as well as in nanoscale design~\cite{seideman:1997c} and quantum computing~\cite{shapiro:2003}.
In a molecular gas, it was shown experimentally that field-free molecular alignment can be used to track collisional relaxation~\cite{Viellard:13,karras:2014}. Strong alignment is also a way to tailor the dipole force of a molecule by tuning the effective molecular polarization~\cite{purcell:2009}.

Molecular images can be obtained from high-order harmonics generated by aligned molecules. This technique was used to obtain photoelectron angular distributions in ionization~\cite{le:2009}. Imaging isolated molecules is made possible through diffraction with external sources of femtosecond x-ray pulses~\cite{ho:2008,kupper:2014} or femtosecond electron pulses~\cite{blaga:2012,hensley:2012,boll:2013,reckenthaeler:2009}. Aligned molecules were recently imaged by diffraction with internal electron pulses, a technique called laser-induced electron diffraction (LIED)~\cite{Wolter:2016}.

A molecular movie of bond breaking in a small linear molecule was realized by measuring molecular frame photoelectron angular distributions~\cite{bisgaard:2009}. Nodal planes of orbitals in strong-field ionization were imaged by~\cite{holmegaard:2010}. In addition, charge migration in molecular cations was measured on an attosecond time scale by using high-order harmonic spectroscopy of oriented molecules~\cite{kraus:2015}.

The crucial role of molecular alignment on molecular scattering was shown in several studies~\cite{gershnabel:2010,gershnabel:2010b,gershnabel:2011,kim:2016}. The degree of alignment modifies the dipole force felt by the molecules, and therefore the scattering of molecules in external fields can be controlled.
Molecular alignment can also help to understand the complex structures of molecular attosecond transient-absorption spectra~\cite{baekhoj:2016}. Rotationally aligned wave packets can be used to probe the structure and dynamics of molecular clusters~\cite{galinis:2014}. The principle of a nanoscale molecular switch driven by molecular alignment was proposed by~\cite{reuter:2008}. The conductance of the system switches according to the position of the molecule with respect to the field polarization axis. Alignment-dependent ionization of linear molecules was studied by~\cite{petretti:2010} and measured by~\cite{pavicic:2007,xie:2014}.

Another
much discussed application of alignment is molecular selectivity. Laser control of molecular alignment was first applied to selective manipulation of multicomponent isotopic mixtures~\cite{fleischer:2006}. Spin-selective alignment of ortho- and para molecular spin isomers at room temperature was experimentally demonstrated by~\cite{fleisher:2007}. Selective rotational excitation of molecules using a sequence of ultrashort laser pulses has experimentally been explored by~\cite{zhdanovich:2012}, leading to the discrimination of  isotopologues and of para- and ortho-isomers. Nuclear spin-selective alignment for molecules with four different nuclear spin isomers has been investigated by~\cite{grohmann:2011}. Isotope-selective ionization with a specific train of femtosecond laser pulses  was proposed by~\cite{akagi:2015}.

\section{Dynamical phenomena in the rotation of molecules}
\label{sec:onebodydyn}

For understanding and controlling the dynamics of molecules, it is useful to prepare the molecules in a well-defined initial state. Two basic strategies exist---cooling a trapped sample of molecules to such low temperatures that eventually all excitations are frozen out, or selecting a specific internal state with an external field. The rotational structure of the molecules is important for both strategies. We first briefly review progress in trapping, cooling and state preparation, differentiating between neutral molecules and molecular ions in
Sections~\ref{subsec:coldneutrals} and~\ref{subsec:coldions},
respectively, and then spotlight possible control scenarios starting
from molecules in a single quantum state.

\subsection{Rotationally cold neutral molecules}
\label{subsec:coldneutrals}

Laser cooling is a key tool for preparing samples of ultracold neutral  atoms~\cite{MetcalfBook}. One route to creating samples of trapped, state-selected ultracold molecules has thus been to first laser cool atoms, then associate atoms to molecules using magnetic field control of Feshbach resonances~\cite{ChinRMP10} and transfer these molecules to a desired final state using stimulated Raman adiabatic passage~\cite{OspelkausNatPhys08,LangPRL08,DanzlSci08,NiSci08,OspelkausPRL10}.
In addition to making  molecules using a magnetic field,  photoassociation, i.e.,  using a laser field to create the bond, has also been shown to allow for the prepration of ultracold molecules in their absolute ground state~\cite{KermanPRL04b,DeiglmayrPRL08,ViteauSci08,ManaiPRL12}. State selectivity is either afforded by the specific level structure of the molecule, as in the case of RbCs~\cite{KermanPRL04b} and LiCs~\cite{DeiglmayrPRL08}, or can be enforced using rovibrational laser cooling based on broadband optical pumping~\cite{ViteauSci08,ManaiPRL12}.
While this progress is impressive and has allowed, for example, to observe quantum state controlled chemical reactions~\cite{OspelkausSci10}, the strategy to associate molecules from ultracold atoms is limited to molecules consisting of atomic species that are straightforward to cool, i.e., alkali and alkaline earth atoms.

In contrast, direct laser cooling of molecules should be applicable to other molecular species but has long been thought impossible. Since laser cooling relies on scattering many photons, it requires a closed cooling cycle between two (or a few) energy levels. The complex internal level structure of molecules seems to impede isolating a closed cooling cycle. Certain molecules, however, in particular alkaline-earth monohydrides and diatomic molecules with a similar electronic structure, possess a level structure that is favorable to laser cooling \cite{DiRosaEPJD04}. Strontium monofluoride~\cite{ShumanPRL09, ShumanNature10} and yttrium (II) oxide~\cite{HummonPRL13} have thus been laser cooled to temperatures of a few millikelvin.

Rotational state control, for example using microwave fields~\cite{OspelkausPRL10}, is an important tool for preparing cold molecules. Mixing rotational states with microwave fields provides a way to close loss channels and thereby enhance cooling efficiency~\cite{YeoPRL15}. While the complex internal level structure of molecules implies a potential hurdle not only for laser cooling, but also for magneto-optical trapping, \cite{BarryNature14} were able to demonstrate three dimensional trapping of strontium monofluoride. In this study, it was key to properly account for the rotational level structure in the design of the trap. Laser cooling has recently been extended to calcium and ytterbium monofluoride~\cite{TruppeNatPhys17,AndereggPRL17,AndereggNatPhys18,LimPRL18}, species which are candidates for quantum simulation and for measuring the electron's dipole moment. Chemical substitution rules allow to further extend the range of molecules that can directly be laser cooled to polyatomics~\cite{IsaevPRL16}. Thus, for triatomic SrOH, the reduction of temperature in one dimension to the submillikelvin range  has recently been achieved by~\cite{KozyryevPRL17}.

One alternative to laser cooling is evaporative cooling which often also makes use of the rotational structure. Microwave-forced evaporative cooling has been demonstrated for neutral hydroxyl  molecules~\cite{StuhlNature12}.
A direct observation of collisions between trapped, naturally occurring molecules -- O$_2$ molecules -- has recently been reported by \cite{Segev2019}.
Another alternative is provided by sisyphus cooling where the reduction of energy and the export of entropy are separated~\cite{ZeppenfeldPRA09}. The latter comes with the advantage of being directly applicable to polyatomic molecules~\cite{ZeppenfeldNature12,GloecknerPRL15}.
This technique has so far produced the largest number of state-selectively prepared neutral molecules at submillikelvin temperatures~\cite{PrehnPRL16}.

\subsection{Rotationally cold molecular ions}
\label{subsec:coldions}

The key tool to preparing molecular cations in a well-defined initial state is
sympathetic cooling with laser-cooled atomic cations forming a Coulomb crystal~\cite{WillitschIRPC12}. For molecular anions, much less progress has been made so far, although direct laser cooling is predicted to be feasible at least for some species~\cite{YzombardPRL15}. Sympathetic cooling of cations relies on elastic collisions of a molecular ion with an atomic ion which are governed by the long-range Coulomb interaction. This transfers the molecule's kinetic energy to the atom from where it is removed by laser cooling~\cite{MolhavePRA00}. However, due to the frequency mismatch between the normal modes of the Coulomb crystal and the rovibrational transitions of the molecule, vibrations and, most notably, rotations are not cooled~\cite{WillitschIRPC12}. One remedy consists in rotational laser cooling where either a small number of carefully chosen rovibrational transitions~\cite{VogeliusPRL02,StaanumNatPhys10,SchneiderNatPhys10} or broadband optical pumping~\cite{LienNatCommun14} allow for the accumulation of population in the rotational ground state. Due to the finite number of states to be addressed, rotational laser cooling may even be used to realize  controlled population transfer into a selected hyperfine state~\cite{BresselPRL12}. Cooling rotations may also be brought about by a buffer gas~\cite{HansenNature14}. Alternatively, molecular ions can be created state-selectively in the first place, by photoionization into a specific rovibronic state~\cite{TongPRL10}. Care then needs to be taken to avoid rotational excitation during the subsequent sympathetic cooling~\cite{Berglund2019}.
Rotational state selectivity can also be achieved by projective measurements~\cite{VogeliusJPB06,ChouNat2017}. While most experiments have been carried out for diatomic molecular ions, sympathetic cooling can also be employed for larger molecular ions~\cite{OstendorfPRL06,HojbjerrePRA08}. The state-selective preparation of polyatomic molecular ions in a trap remains, however, an open goal to date~\cite{PattersonPRA18}.

A key application of trapped and translationally as well as rotationally cold molecular ions is chemical reaction dynamics~\cite{WillitschIRPC12}. In particular, Coulomb-crystallized ions have allowed to observe single-ion reactions~\cite{DrewsenPRL04,StaanumPRL08} and ion-neutral reactions at ultralow energy as recently reviewed by~\cite{ZhangChapter18}. Precision spectroscopy represents another important area of application for state-selectively prepared molecular ions~\cite{CalvinJPCL18}. In particular, cold molecular ions could be used for mass spectroscopy~\cite{SchillerPRA05,AlighanbariNatPhys18},  in optical clocks~\cite{SchillerPRL14}, and to experimentally measure molecular parameters~\cite{BerglundNJP15}. The latter proposal  specifically exploits the rotational dynamics of a state-selectively prepared, trapped molecular ion in a Ramsey-type interferometer to determine the polarizability anisotropy.

\subsection{Rotating molecules as a testbed for exploring quantum phenomena}

Experiments on the rotational dynamics of laser-kicked molecules represent a new testing ground where fundamental quantum phenomena can be studied. These observations have also implications for the understanding of laser controlled molecular processes.
A first example is given by rotational echoes. Since its discovery in magnetic resonance in the 1950s, echo has become a well-known phenomenon which has been observed in many different situations, both in classical and quantum physics~\cite{lin:2016}. Recently, a new type of echo based on rotational dynamics has been proposed and demonstrated experimentally: Echoes occur in an ensemble of molecules due to an excitation by a pair of time-delayed laser kicks. The second pulse is able to reverse the flow of time, thereby recreating the initial event. Orientation and alignment echoes were observed by~\cite{karras_echo}. Fractional echoes were measured by~\cite{karras:20016,lin:2017}. The phenomenon of rotational echoes has been thoroughly described by~\cite{lin:2016}. Rephasing of centrifugal distortion by rotational echoes was shown by~\cite{rosenberg:2017}. The dynamics of rotational echoes and their dependence on the delay and intensity of the excitation pulses were studied experimentally and theoretically by~\cite{rosenberg:2018}. Orientational echoes induced by THz fields were observed by~\cite{lu:2016}.

The periodically kicked quantum rotor is another common example in physics where quantum localization phenomena such as Anderson localization or Bloch oscillations can be observed. Rotational dynamics excited by a periodic train of short laser pulses are a perfect testing ground in which such properties can be exhibited both theoretically~\cite{floss:2013,floss2014} and experimentally~\cite{floss:2015,bitter:2016}. \cite{floss:2012} showed that a periodic train of laser pulses can also be used for selective rotational excitation in a molecular mixture. The effect of random deviations of the train period was studied by~\cite{kamalov:2015}.

Finally, chaotic dynamics of molecular rotors subject to a periodic sequence of ultrashort laser pulses have been  investigated. By controlling the initial wave packet, the rotational distribution and the energy of the final state can be modified as shown by~\cite{bitter:2017}. Laser-induced molecular alignment in presence of chaotic rotational dynamics is analyzed by~\cite{floss:2017} for asymmetric top molecules subject to a static electric field. Numerical computations suggest that molecular alignment is robust against rotational chaos. Such studies allow to reveal the profound connection between the classical and the quantum regime.

\subsection{Chiral-sensitive rotational dynamics}

\begin{figure*}[tb]
\centering
  \includegraphics[width=0.85\linewidth]{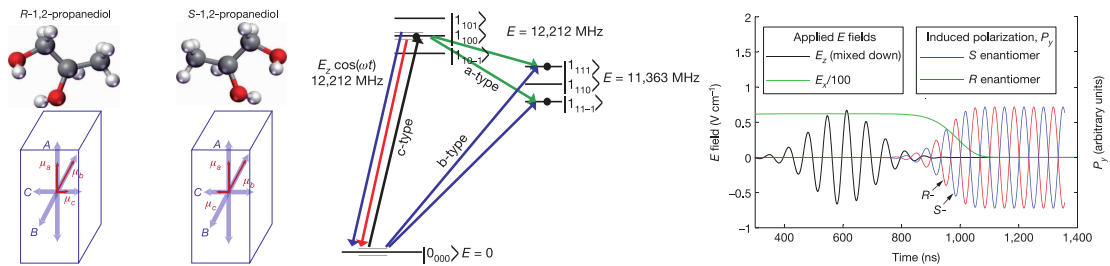}%
  \caption{Chiral-sensitive three-wave mixing microwave spectroscopy: Right-handed and left-handed enantiomers of chiral molecules with $C_1$ symmetry possess three orthogonal electric dipole moment components, identical in magnitude but differing in sign for one of the components. When rotational transitions are driven using two of the components and fluorescence is collected along the third direction, the different sign accumulated in the three-wave mixing results in a relative phase of $\pi$ between the two enantiomers. Reprinted with permission from~\cite{PattersonNat13}.}
  \label{fig:3WM}
\end{figure*}
Molecules without any improper axis of rotation, more specifically chiral molecules belonging to the $C_1$
symmetry point group, offer a particularly interesting avenue for
quantum control. These molecules possess a permanent dipole moment
with non-zero components along all three principal axes of inertia. They are
necessarily chiral such that the two stereoisomers, also termed
enantiomers, are non-superposable mirror images of each other. While
enantiomers have the same rotational constants and the same magnitude
of dipole moment components, the product of the three dipole moment
components differs in sign~\cite{PattersonNat13,HirotaPJA12}. This
difference in sign can be probed by microwave three-wave mixing
spectroscopy~\cite{PattersonNat13,PattersonPRL13}, cf. Fig.~\ref{fig:3WM},
independent of the orientation of the molecules.
To this end, three rotational levels have to be chosen that are connected by electric dipole transitions corresponding to the three orthogonal dipole moments~\cite{PattersonNat13,HirotaPJA12}. Two of the transitions are driven by microwave fields, possibly chirped~\cite{ShubertACIE2014,ShubertJPCL16}, whereas the free-induction decay on the third transition is collected, showing a phase shift of $\pi$ between left-handed and right-handed molecules~\cite{PattersonNat13}. The free induction decay can be optimized by accounting for the spatial degeneracy when choosing the pulse flip angles~\cite{LehmannJCP18}. Using group theory, one can show that indeed three orthogonal polarization directions are required to  pick up the desired phase shift in three-wave mixing~\cite{LeibscherJCP19}.

The first application of the microwave three-wave mixing technique is to detect
enantiomeric excess in a mixture containing molecules of both handednesses~\cite{PattersonPRL13}.
Second, the phase shift picked up in the  three-wave mixing provides a handle for coherent control with the goal of realizing enantiomer-selective excitation and
separation~\cite{KralPRL01,KralPRL03,LiPRL07,JacobJCP12}. The idea is to exploit the enantio-sensitive phase for constructive, respectively destructive, interference when driving an additional transition. First experiments have confirmed the basic feasibility of such enantiomer-specific rotational state transfer~\cite{EibenbergerPRL17,PerezAngewandte17}. Theoretical proposals assuming zero temperature predict a contrast of 100\% between enantiomers for a variety of three-wave mixing schemes, including ones relying on adiabatic passage~\cite{KralPRL01} or shortcuts to adiabaticity~\cite{VitanovPRL19}. However, in the experiments,
enantiomer-selective enrichment of population in a specific rotational state is limited by thermal population. An alternative approach, more suitable to high temperatures, has been brought forward by \cite{YachmenevPRL16}.  Their proposal relies on the enantiomers having a different sign in the off-diagonal elements of the polarizability tensor. This property can be exploited to excite a unidirectional rotation of two enantiomers with a $\pi$ phaseshift,  very similar to that in microwave three-wave mixing.
Furthermore, it is possible to extend chiral-sensitive three-wave mixing spectroscopy from purely rotational to rovibrational transitions~\cite{LeibscherJCP19}.

The close connection between the handedness of chiral molecules and their rotational structure has further applications. It allows for enantioselective optical orientation by using twisted polarization~\cite{TutunnikovJPCL18} and can be exploited to determine  individual components of the optical activity pseudo-tensor of chiral molecules in rotational spectroscopy~\cite{CameronPRA2016}.
Finally, electric-dipole based, enantiomer-selective excitation is not limited to rotational transitions. \cite{Ordonez18} provide a unified view of chiral-sensitive electric dipole based excitation schemes, nicely connecting rotational three-wave mixing with other chiral signatures such as photoelectron circular dichroism~\cite{LuxAngewandte12}.

\section{Rotational states in two-body collisions}
\label{sec:collisions}

The rotational structure of a molecule is crucial for understanding two-body
interactions that are probed when the molecule is made to collide with
an atom or another molecule. In addition to revealing details of the interaction potential, collisions are at the core of many cooling protocols as mediators of thermal equilibration. A key question is whether elastic or inelastic collisions dominate. This determines the feasibility of e.g. sympathetic cooling.
One important source of inelasticity are transitions between the rotational states during the collision. First, we briefly describe collision
physics and the role of rotational states and then focus on orbiting resonances
as a novel phenomenon that is particularly promising for quantum
control.

\subsection{Role of rotational states in molecular collisions}

Collision physics is usually studied by solving the two-body Schr\"odinger equation~\cite{LevineBook}. Cold collision studies have by and large used the time-independent framework of scattering theory~\cite{KremsIRPC05, KremsPCCP08, KreStwFrieColdMol, LemKreDoyKais13,KlosChapter18,QuemenerChapter18}. It starts from the differential and total cross sections which are the observables measured in the experiment. When calculated theoretically, they reflect
the underlying theoretical model and thus connect experiment and theory~\cite{KlosChapter18}. The theoretical model includes, in particular, the interaction potential and couplings to external fields.
The cross sections can be expressed in terms of the $S$-matrix or the scattering amplitude for each partial wave which, in turn, are obtained by calculating the scattering wavefunction or its logarithmic derivative~\cite{KlosChapter18}. For scatterers with internal degrees of freedom such as rotation, the scattering wavefunction is expanded into a complete orthonormal set of product wavefunctions for the relative motion and internal degrees of freedom. While the radial wavefunctions are determined by integration,\footnote{Note that this integration over the radial coordinate is typically referred to as propagation in the scattering literature which is not to be confused with propagation as in time evolution.}
eigenbases for the other degrees of freedom, i.e., the end-over-end rotation and internal motion, are employed. This results in so-called coupled channels equations where each channel is characterized by a set of quantum numbers for the end-over-end rotation and internal motion. A very didactic introduction into how to choose the basis functions, derive and solve the resulting coupled channels equations has recently been provided by~\cite{KremsBook18}.

Coupling between the channels may be due to the interparticle interaction or external fields. For example, channels corresponding to different rotational states may be coupled due to an anisotropy of the interaction potential or due to an electric field~\cite{KremsPCCP08,LemKreDoyKais13,KremsBook18}. Collisions changing only the internal rotational state may be observed when the colliding molecule can be treated as a rigid rotor. This is the case, e.g., for hydroxyl anions
for which absolute scattering rate coefficients for rotational state changing collisions with helium have been measured~\cite{HauserNatPhys15}. For open-shell molecules or molecules with hyperfine structure, rotations are coupled to electronic and nuclear spins such that state-changing collisions become even more likely~\cite{KremsPCCP08}. External field control may provide a means to protect the molecule from state-changing collisions and suppress inelasticities~\cite{KremsPCCP08,LemKreDoyKais13,KremsBook18}.
For example, microwave coupling of rotational states leads to long-range repulsive interactions between ultracold polar molecules~\cite{GorshkovPRL08,karman:2018,MicheliPRA07} and allow to control the molecular scattering length~\cite{Lassabliere:2018}.

\subsection{Orbiting resonances in cold collisions}

The kinetic energy of the end-over-end rotation
involved in a collision of two particles turns into a centrifugal barrier upon a partial wave expansion. The potential barrier gives rise to quasi-bound states called orbiting or shape resonances, depending on the diffuseness of the resonance wavefunction.\footnote{Whereas the two terms are often used interchangeably in the physics literature, the physical chemistry community distinguishes orbiting and shape resonances according to their energy being above or below the height of the rotational barrier, respectively. According to this definition, orbiting resonances are much more diffuse than shape resonances.} They correspond to a temporary trapping of probability amplitude of the particles' scattering wavefunction at short interparticle separations. Orbiting and shape resonances can thus be viewed as quantization of the scattering motion. If the collision is reactive, this leads to an enhanced reaction rate, an effect that becomes particularly pronounced at low collision energy. Cold collisions thus provide an ideal testbed for observing this quantum phenomenon.

The position of the rotational barrier coincides with the long-range part of the interaction potential, at least for partial waves with low angular momentum which are most relevant in cold collisions. Orbiting and shape resonances can therefore very accurately be described  just in terms of the van der Waals coefficient of the long-range potential in addition to  the reduced mass of the scattering complex \cite{GaoPRA09,LondonoPRA10}.

\begin{figure}[tb]
  \centering
  \includegraphics[width=0.85\linewidth]{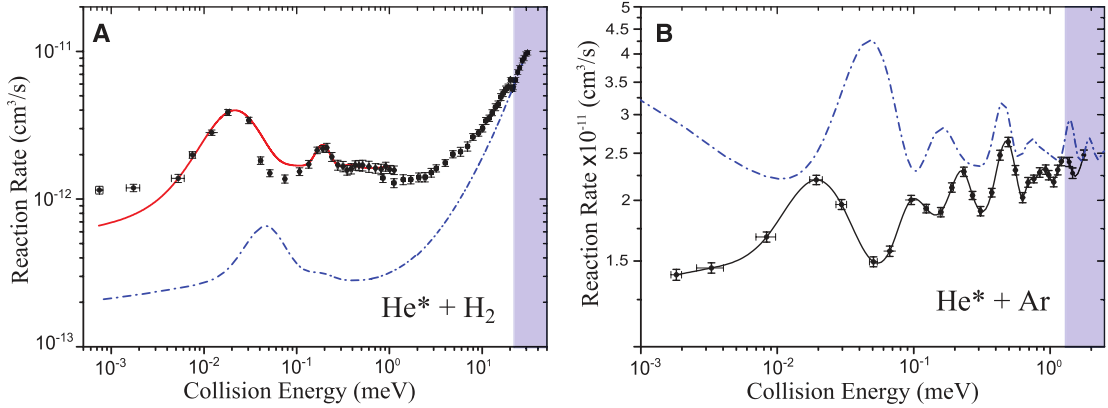}
  \caption{Experimental observation of shape resonances in Penning ionization reactions. Reprinted with permission from \cite{HensonSci12}.}
  \label{fig:shaperes}
\end{figure}
Experimental observation of shape resonances has been made possible by combining Penning ionization reactions with the merged beam technique~\cite{HensonSci12,Lavert-OfirNatChem2014,JankunasJCP15}. The latter provides the capability to tune the collision energy over several orders of magnitude, down to $k_B E$ in the milli-kelvin regime~\cite{ShagamJPCC13,OsterwalderEPJTI2015}.
Penning ionization occurs when an atom in a  metastable state has enough internal energy to ionize its collision partner~\cite{SiskaRMP93}. Orbiting resonances have been observed in collisions with both atoms and molecules~\cite{HensonSci12}, see Fig.~\ref{fig:shaperes}. They are sensitive to tiny changes in the effective interaction from one molecular  isotopologue to another~\cite{Lavert-OfirNatChem2014} and to the quantum state of the internal rotation in the case of a molecular collision partner~\cite{KleinNatPhys17,ShagamNatChem15}. The anisotropy of the interaction that occurs in the Penning ionization of molecules, even when it is rather small, plays a crucial role for shape resonances at low energy~\cite{KleinNatPhys17}. This can be rationalized in terms of adiabatic potential energy curves obtained when separating the relative motion's rotation and vibration~\cite{PawlakJCP15,PawlakJPCA17}. The character of the shape resonances, i.e., the corresponding angular momentum quantum number, is determined by comparison with scattering calculations. This is somewhat unsatisfactory since even for diatomics, the best, currently available potential energy curves are not sufficiently accurate to correctly predict the scattering length and thus the exact position and partial wave character of orbiting and shape
resonances~\cite{LondonoPRA10}. An alternative would be provided by combining Penning ionization in merged beams with photoassociation spectroscopy~\cite{SkomorowskiJPCA16} where the rotational progression of the photoassociation spectrum would allow for identifying the resonance character.

Recently, low-lying shape resonances have also been observed using pulsed-field-ionization zero-kinetic-energy photoelectron spectroscopy for the simplest examples of collisions involving identical fermions and bosons, namely H$^+$ and H and D$^+$ and D, respectively~\cite{BeyerPRX18}.

\subsection{Orbiting resonances in photoassociation}

The scattering wavefunction's enhanced probability amplitude at short interparticle separation can also be exploited in light-induced reactions. One such example is photoassociation in an ultracold  gas where two colliding particles are excited by laser light into a bound level of a higher lying electronic state~\cite{JonesRMP06,KochChemRev12}. The initial state for the photoassociation reaction is the thermal ensemble of the trapped particles. An orbiting resonance becomes visible if the resonance position matches the trap temperature.  The corresponding signature is an enhancement of the photoassociation rate, first observed for rubidium by~\cite{BoestenPRL96} which has a low-lying orbiting resonance at about $300\,\mu$K$\cdot k_B$.

For most colliding particles, the positions for the lowest orbiting resonances lie at higher energies, corresponding to a few or a few tens of milli-kelvin. However, manipulation of the thermal cloud with an external field, for example a static electric field~\cite{ChakrabortyJPB11} or a strong off-resonant laser field~\cite{GonzalezPRA12}, allows for shifting the position of a resonance to match the trap temperature. When driving the photoassociation reaction in the presence of these additional fields, rate enhancements of up to several orders of magnitude are predicted~\cite{ChakrabortyJPB11,GonzalezPRA12}. In both theoretical proposals, very high field strengths are required. The generation of the required intensities in case of non-resonant light is challenging but within currently existing experimental capability~\cite{GonzalezPRA12}. Note that the interaction Hamiltonian describing the coupling to the non-resonant light is identical to the one used for laser-induced alignment~\cite{Stapelfeldt:03}. However, since the field is applied to a scattering complex instead of a tightly bound molecule, the dependence of the polarizability anisotropy on interparticle distance becomes important for the dynamics, resulting in strong hybridization of the rovibrational motion~\cite{GonzalezPRA12}.

\subsection{Control of collisions}

The strong hybridization of the rovibrational motion that non-resonant light exerts upon a collision complex discussed above is not only useful to enhance reaction rates but can also be used to control the collision itself. On the one hand, the non-resonant light modifies existing resonances in their position and width~\cite{GonzalezPRA12,TomzaPRL14} and also creates new resonances~\cite{TomzaPRL14}. These resonances, if present, dominate cold collisions. On the other hand, non-resonant light is predicted to tune the effective interaction strength in a cold collision, such as the scattering length for $s$-wave collisions~\cite{CrubellierPRA17} and the $p$-wave scattering volume~\cite{Crubellier18b}. Thanks to the long-range nature of the collision dynamics, this control is well described by a universal asymptotic model, extending the approach of \cite{LondonoPRA10} to the presence of non-resonant light~\cite{CrubellierNJP15}. For field-dressed orbiting resonances, the resonance properties were found to scale approximately linearly in the field intensity up to fairly large intensities, allowing for  a perturbative single-channel approach~\cite{CrubellierNJP15b}.

In the examples discussed above, the angular momentum associated with the rotation of a collision complex provides a handle to control the collision. Direct control over this rotation in case of reactive collisions or, in other words, control over the stereodynamics of a bimolecular reaction, has been a long-standing goal in chemical physics. One way to realize such control at extremely low temperature is provided by tight confinement. It allows polar molecules to approach each other only in 'side-by-side' collisions and lets Fermi statistics play out in full,  suppressing the atom-exchange reaction that would otherwise occur in a trapped ultracold gas of KRb molecules~\cite{MirandaNatPhys11}.  In this experiment, the dipole moments of the polar molecules were aligned using an external electric field. Magnetic fields provide another means for controlling chemical reaction stereodynamics~\cite{AoizPCCP15}.
Analytical models capture the essential physics of rotationally inelastic collisions in electric~\cite{LemeshkoJCP08,LemeshkoJPCA09} as well as  magnetic fields~\cite{LemeshkoPRA09}, allowing to rationalize the dynamics in terms of Fraunhofer scattering.
In a recent experiment, fixing the magnetic dipole moment of one of the collision partners with a tunable magnetic field which in turn determines the projection of the total angular momentum onto the interparticle axis has allowed to vary the branching ratio between Penning and associative ionization~\cite{GordonPRL17}, in line with an earlier theoretical prediction~\cite{ArangoPRL06}.

While full control of collisions remains a wide open goal, significant progress has been achieved in the control of so-called half-collisions.  A prime example is photodissociation where a laser pulse induces the breaking of a chemical bond and thereby initiates the collision~\cite{SchinkeBook}. This problem is much more amenable to quantum control  since the starting point is a single, coherent quantum state rather than the scattering continuum~\cite{KochChemRev12,KochChapter18}. Photodissociation using shaped  femtosecond laser pulses  has thus been one of the early success stories of quantum control~\cite{ShapiroBook2,BrixnerCPC03,WollenhauptAnnuRevPhysChem05}. It results, however, in energetically broad and quasi-classical continuum wavepackets. Progress in the control of half-collisions in the distinctly quantum regime has been more recent.
For example, theoretical proposals suggested to dissociate weakly bound molecules into low energy scattering states using a magnetic field~\cite{GneitingPRA10} or a non-resonant laser pulse~\cite{LemeshkoPRL09}.
The latter introduces a centrifugal term which expels the highest vibrational level from the potential that binds it~\cite{LemeshkoPRL09}. Experimentally,
manipulation of a half-collision process at the level of rotational states, producing well-defined quantum continuum states at low energies where the  traditional quasiclassical model of photodissociation fails, has been demonstrated by~\cite{McDonaldNat16}.
Precise quantum-state control of the molecules has also allowed for controlling photodissociation into low energy continuum states using a magnetic field~\cite{McDonaldPRL18} and for observing the cross-over from the distinctly quantum mechanical to the quasiclassical regime~\cite{KondovPRL18}.

\section{Rotational dynamics in an environment}
\label{sec:environ}

In `realistic' experiments, molecular rotations naturally take place in the presence of some kind of external bath, be it electromagnetic field noise, a solvent, or a crystal surface. Since rotations
occupy the low-energy part of the energy spectrum (frequencies of $1-100$~GHz or wavelengths of $3-300$~mm), they can be easily altered by an interaction with the surrounding medium. Controlling the effects of the external environment on molecular rotational structure is crucial for using molecules in quantum simulation and computation, as described in Sec.~\ref{sec:manybody}. Furthermore, molecular reactivity strongly depends on the relative orientation of molecules with respect to one another. Therefore, understanding how an external environment (e.g.\ solvent, surface, or solid-state matrix) affects molecular rotations is crucial to control chemical reactions at the quantum level under realistic experimental conditions. This, in turn, opens a new route for controlled chemistry~\cite{KremsBook18, KremsPCCP08}.

\subsection{Blackbody radiation}
An external environment that is ubiquitous and almost unavoidable in   experiment is  electromagnetic field noise. For example, blackbody radiation is present in essentially any laboratory is and, for polar molecules,
its effect on molecular states becomes crucial at milli-kelvin temperatures. \cite{VanhaeckeMolPhys07} studied the effect of blackbody radiation on cold molecules and showed theoretically that blackbody radiation can induce transitions between molecular rovibrational states, thereby causing limitations to precision measurements. \cite{HoekstraPRL07} experimentally studied rotational pumping of cold OH and OD radical that were Stark-decelerated and electrostatically trapped. They detected the transfer of  the molecular population from the ground rotational state, $J=3/2$, to the first excited rotational state, $J=5/2$, due to room-temperature blackbody radiation. \cite{WolfNature16} observed quantum jumps in a trapped  molecular ion, MgH$^+$, induced by thermal blackbody radiation. \cite{VogeliusPRL02} showed that blackbody radiation can be exploited as a resource to  laser-cool  translationally cold but internally hot molecules, as also discussed in Sec.~\ref{subsec:coldions}.
The scheme they introduced relies on optical pumping of the rovibrational population into the dark state, where blackbody radiation enhances the population decay. It was experimentally demonstrated for a trapped MgH$^+$ ion~\cite{VogeliusPRL02}. \cite{NabanitaPCCP13} theoretically studied the possibility to achieve blackbody-radiation-assisted rotational cooling for several diatomic molecules.

\subsection{Rotational relaxation due to collisions}
\label{sec:relcoll}

Molecular alignment experiments described in Sec.~\ref{sec:align} are usually performed in low-pressure gas ensembles, where the effects of interparticle collisions are negligible. Most chemical processes, on the other hand, take place in some kind of a dissipative environment, such as a dense gas or liquid. Studying how a dissipative environment affects molecular rotation through collisions is crucial to understand and control chemical reactivity.

The theory of alignment of single molecules by laser pulses was described in Sec.~\ref{sec:align}. \cite{RamakrishnaSeideman2005,  ramakrishna:2006jcp} extended this theory to the case where a dissipative environment is present. They considered the case of a Markovian environment, that is an environment without memory, or, in other words, a bath so large that its state cannot be altered by the presence of a molecule. Based on  the density matrix formalism, they predicted an exponential decay of molecular alignment due to the bath. \cite{pelzer:2007} theoretically studied the possibility to apply optimal control protocols to molecular alignment in dissipative media. \cite{hartmann:2012} developed quantum and classical approaches to study laser-induced molecular alignment under dissipative conditions. Compared to earlier studies, their quantum model explicitly accounts for dephasing and reorienting elastic collisions with the medium. \cite{zhdanov:2011} have theoretically shown the possibility to achieve transient molecular alignment even in dense dissipative media by making use of aligned dark states. \cite{viellard:2008,Viellard:13} experimentally studied the decay of field-free alignment of CO$_2$ molecules due to collisions with noble gases. \cite{OwschimikowJCP10} measured cross-sections for rotational decoherence in N$_2$ through the decay of laser-induced alignment. \cite{ZhangJCP18} demonstrated the possibility to achieve field-free molecular alignment in the presence of collisional relaxation for symmetric-top molecules such as ethane and C$_2$H$_6$. \cite{tenney:2016} measured the impulsive alignment in  a dense thermal sample of asymmetric-top molecules (SO$_2$) and demonstrated the possibility to observe revivals. \cite{karras:2014}  probed  molecular alignment to track collisional relaxation of CO$_2$ molecules. \cite{milner:2014} used an optical centrifuge to bring the molecules to high rotational angular momenta in order to study collisional decoherence as a function of molecular rotational state. They have observed that in the range of molecular rotational states between $J=8$ and $J=66$ the collisional relaxation rate changes by over one order of magnitude. Such molecules in extremely large angular momentum states $J$ -- so-called  `superrotors' --  were found to be much more resilient to collisional relaxation. In a later work, \cite{MilnerPRX15} studied the dynamics of the collisional relaxation of the rotation of superrotors in a molecular ensemble. The possibility to control molecular gas hydrodynamics through laser-induced rotational excitation and subsequent collisional relaxation was examined theoretically by~\cite{ZahedpourPRL14}. \cite{KhodorkovskyNatCom15} theoretically studied collisional dynamics and equilibration in a gas of such molecular superrotors. Rotational relaxation can be investigated using rotational echoes as recently shown by~\cite{zhang2019}.

\subsection{Rotation of molecules in helium droplets}
The settings described in Sec.~\ref{sec:relcoll} correspond to collisions with  a dilute bath under thermal conditions. In such a case, most of the undergoing  processes can be understood as a combination of two-body collisions averaged over the Boltzmann distribution. However, what happens if the environment is quantum and dense and collective (many-particle) effects start playing a role? A good example of such an environment is a quantum liquid such as superfluid helium.

Trapping molecules inside nanosized droplets of superfluid helium$-4$ has been used as a technique for molecular spectroscopy for over two decades~\cite{ToenniesARPC98, ToenniesAngChem04, StienkemeierJPB06, SzalewiczIRPC08, ToenniesMolPhys13, MudrichIRPC14}. The main motivation driving the field was to isolate molecular species in a cold ($\sim 0.4$~K) environment and record spectra free of collisional and Doppler broadening. Furthermore, trapping single molecules inside droplets of superfluid helium allows to study reactive species that are unstable in the gas phase.

It has been shown that, in general, superfluid helium does not strongly broaden molecular spectral lines, although there are a few exceptions, see e.g.~\cite{Slipchenko2005, MorrisonJPCA13, Cherepanov17}. However, molecule--helium interactions can alter molecular moments of inertia such that molecules rotate slower inside a superfluid. The helium-induced change in moments of inertia ranges from a few percent for light molecules, such as H$_2$O or HF, to a factor of $3-5$ for heavier species, such as CS$_2$ or N$_2$O~\cite{ToenniesAngChem04, LemeshkoDroplets16}. Semiclassically, this effect can be explained as follows. Through the molecule-helium interactions, the molecule distorts helium around it, thereby forming a non-superfluid shell~\cite{GrebenevOCS, CallegariPRL99}. If the molecule is rotating slow enough, that is, if its rotational kinetic energy is much smaller compared to molecule-helium and helium-helium interactions (or, equivalently, if the response time of the bath is much shorter compared to the rotational timescale), the non-superfluid shell co-rotates with it. Therefore this effect is most pronounced for heavy, slowly rotating molecules.

Renormalization of  molecular moments of inertia in superfluids has been studied by numerical techniques based on  variational, path-integral, reptation, and  diffusion quantum Monte Carlo algorithms~\cite{SzalewiczIRPC08, RodriguesIRPC16} and density functional theory~\cite{AncilottoIRPC17}. Considering a finite-size system of a molecule and $\sim 10^2-10^3$ He atoms allowed to reproduce the numerical values of the molecular moments of inertia for several species, in good agreement with experiment.

An alternative approach to understand molecular rotation in superfluids is based on the recently introduced angulon quasiparticles~\cite{SchmidtLem15, LemeshkoDroplets16}. In the quasiparticle language, the renormalization of molecular moments of inertia is a phenomenon similar to renormalization of the effective mass for electrons moving in solids~\cite{Devreese15}. The angulon theory allows to describe strong renormalization for heavy molecules by constructing a quantum many-body wavefunction similar to the co-rotating non-superfluid shell described above. On the other hand, weak renormalization observed for light molecules has been described  in terms of  the `rotational Lamb shift' -- differential renormalization of molecular states due to virtual phonon excitations carrying angular momentum.

Since the main motivation behind the field of molecules in helium nanodroplets was to obtain more insight into molecular structure, most of the studies were performed in equilibrium conditions.  That is, after having been trapped in a droplet, the molecules had enough time to thermalize with the surrounding superfluid and then were probed using  a weak spectroscopic field. This implies that the magnitude of the external field changes very slowly, on timescales much longer than both the molecular rotational period and the timescale of the molecule--helium interactions. Therefore, the field cannot bring the system out of equilibrium.

\begin{figure}[tb]
\centering
    \includegraphics[width=0.5\linewidth]{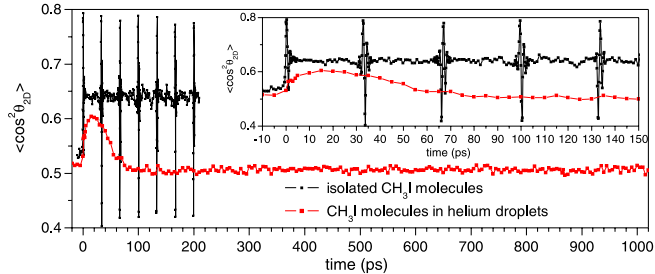}
\caption{(Color online) Time dependence of the alignment cosine of CH$_3$I molecules after a 450 fs laser pulse, in helium droplets (red) compared to the gas phase (black). Reprinted with permission from \cite{pentlehner:2013}.}
\label{StapelfeldtExp}
\end{figure}
A few years ago, \cite{pentlehner:2013} pioneered experiments on far-from-equilibrium rotational dynamics of molecules in a superfluid. In their experiments, molecules were aligned using a short laser pulse, non-adiabatic both with respect to the molecular rotational timescale (as described in Sec.~\ref{sec:align}), and to the timescale of molecule--helium interactions. In these first experiment an unexpected behavior was observed: Not only did the revivals not form, but also a new slower timescale emerged, see Fig.~\ref{StapelfeldtExp}. For example, one can see that the first maximum of alignment is reached at two orders of magnitude larger times compared to the gas-phase data.

The behavior presented in Fig.~\ref{StapelfeldtExp} cannot be explained by the effect of a Markovian environment as described in Sec.~\ref{sec:relcoll}~\cite{RamakrishnaSeideman2005}. Moreover, such slow dynamics  was found to be universal, i.e.\ featured by different molecular species (linear rotor, symmetric top, asymmetric top) and for various intensities of the aligning  laser. Interestingly, a completely `normal' behavior of molecules was observed for an excitation by long laser pulses, adiabatic both with respect to molecular rotation and to molecule-helium interactions~\cite{PentlehnerPRA13}. In the follow-up experiments,
it was shown that the degree of field-free molecular alignment inside helium droplets can be enhanced using a combination of laser pulses~\cite{ChristiansenPRA15} and that alignment can be achieved using near-adiabatic laser pulses~\cite{Shepperson17, SheppersonPRA18}. \cite{Shepperson16} have shown that strong short laser pulses can induce detachment of molecules from the surrounding superfluid shell. At lower laser intensities, however, it is still possible to observe revivals of the rotational wavepacket even inside a superfluid and explain the observations within the angulon theory.

Large molecules have also been coherently manipulated inside superfluid helium nanodroplets. For example, a comparison between alignment of 1,4-diiodobenzene molecules in free space and superfluid helium has been carried out by~\cite{ChristiansenPRA16}. \cite{ChatterleyPRL17} demonstrated complete three-dimensional alignment of 3,5-dichloroiodobenzene molecules using elliptically polarized laser pulses. \cite{PickeringPRL18} have shown the possibility to align van der Waals molecular complexes -- in this case (CS$_2$)$_2$ -- and thereby determine the dimer structure. It was found that the superfluid helium environment can stabilize complexes unstable in gas phase. As of 2018, however, a fully satisfactory theory describing quantum dynamics of molecules in quantum solvents -- even as ``simple'' as helium -- is still to be developed.

\subsection{Rotation of molecules in liquid phase}
Superfluid helium represents a nice model system to study molecule-solvent interactions at the quantum level. However, most real chemistry occurs in thermal solutions whose properties are often substantially more complex compared to superfluid helium. An important question is whether one can exploit quantum coherences of rotational motion in order to manipulate chemical reactions in such systems. \cite{moskun_rotational_2006} studied ICN molecules which were photodissociated with a short laser pulse forming highly rotationally excited CN rotors inside a thermal solution such as water or alcohols. They found that in such CN fragments the rotational coherence persists for several rotational periods before it decays due to strong interactions with a room-temperature solvent. Furthermore, they observed effects going beyond linear response. That is, in these experiments the solvent is not merely acting as a `sink' for energy, but is strongly coupled to the molecules such that molecules can actually alter the solvent state. As a result, the molecular coherence decays much slower compared to the exponential decay expected from the linear response approach.  \cite{tao_molecular_2006} performed a theoretical study of the linear response breakdown and have shown that the mismatch of timescales between molecular rotational dynamics and  that of the solvent evolution is a key factor leading to non-linear response. \cite{ohkubo:2004} studied the dynamics of molecular alignment in a liquid phase using molecular dynamics simulations.

\subsection{Relaxation due to surfaces}
In several applications, such as heterogeneous catalysis, molecules interact with a solid-state  surface. It is therefore of crucial importance to understand how molecular rotations are affected by the presence of a surface. \cite{BuhmannPRA08} theoretically studied rovibrational heating of cold molecules placed in the vicinity of various surfaces at finite temperature, from the perspective of quantum electrodynamics. Exchange of virtual photons between a molecule and a surface, accompanied by rotational and vibrational excitations, has a dissipative component (or ``heating''), whose magnitude depends on the molecule-surface distance and temperature. The theory allowed to determine optimal distances for trapping of molecules in a vicinity of different surfaces which would maximize their lifetime.

Recently it became possible to use scanning tunnelling microscopy (STM) to probe rotational motion of molecules adsorbed on surfaces such as  graphene~\cite{NattererPRL13} and metal surfaces~\cite{LiPRL13}. As opposed to conventional microwave spectroscopy,  rotational resonances in STM are observed as peaks in the electron current tunneling through the molecules. This paves the way to understanding the microscopic mechanisms of molecule-surface interactions and, ultimately, to controlling molecular rotational dynamics on surfaces, which would have applications in catalysis.

\section{Rotations in quantum information processing and  many-body physics}
\label{sec:manybody}

Experimental progress in cooling and trapping molecules enabled their applications in quantum information processing (QIP). During the recent years there has been a lot of theoretical and experimental progress in engineering quantum information registers -- qubits -- using various atomic, optical, and solid-state systems. An advantage of using molecules is that their rotational states are long-lived, as opposed to, e.g., Rydberg states of atoms~\cite{SaffmanMolmerRMP10}. Furthermore,  cold controlled molecules can be isolated extremely well from the external environment, which is challenging to achieve e.g.\ with solid-state qubits~\cite{LaddNature10}.

\subsection{Molecules in optical lattices}

\begin{figure}[b]
\centering
     \includegraphics[width=0.5\linewidth]{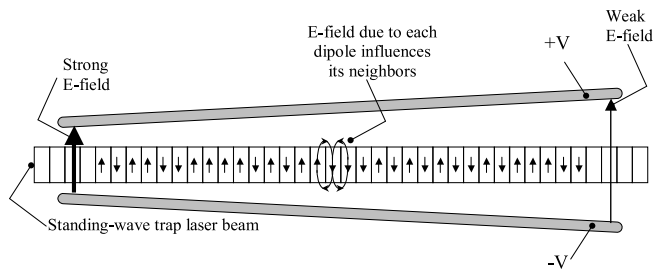}
\caption{First proposal of a quantum computer based on ultracold molecules in an optical lattice. Reprinted with permission from \cite{DeMillePRL02}.}
\label{DeMilleQIP}
\end{figure}
The first proposal to use cold molecules in optical lattices as qubits was put forward by~\cite{DeMillePRL02}. In this proposal, schematically illustrated in Fig.~\ref{DeMilleQIP}, the $\vert 0 \rangle$ and $\vert 1 \rangle$ qubit states are mapped onto the orientational states of a cold polar molecule, pointing along or against an external electric field, respectively. Transitions between the qubit states are driven by a microwave or an optical Raman field. At the same time, an electric field gradient makes it possible to individually address single molecules since their transition frequencies are off-resonant with respect to one another. The qubits are coupled one to another through electric dipole interactions. Microwave pulses that execute the Hadamard and CNOT gates for such a setup were derived by~\cite{arai:2015}. The proposal of DeMille triggered the development of a new research direction in the field of quantum information processing with a focus on ultracold polar molecules, see also Sec.~\ref{subsec:coldneutrals}.

The experiments on ultracold gases allow for a lot of controllability which gave the researchers hope that a molecule-based quantum computer can indeed be realized. However, a careful analysis of the system is crucial in order to evaluate possible sources of decoherence. One of the obvious decoherence sources is the trapping field of an optical lattice. \cite{KotochigovaPRA06} performed an exhaustive numerical analysis and have shown that it is indeed possible to control dipole-dipole interactions between molecules even in the presence of an optical lattice. Their main finding is that laser trapping frequencies can be chosen such that the trapping potential is nearly independent of the molecular rotational state and decoherence due to light scattering is reasonably low.

Later, \cite{BomblePRA10} performed numerical simulations of various quantum algorithms for an ultracold NaCs molecule in an optical lattice, in the presence of a static electric field. \cite{YelinPRA06} suggested a scheme to achieve robust control over molecule--molecule interactions, which is necessary for QIP with polar molecules in optical lattices. The main idea is to use pulsed fields to switch from the molecular states with a small dipole moment to the states with a large dipole moment. The scheme is thereby based on an effect analogous to the Rydberg blockade in atomic architectures~\cite{SaffmanMolmerRMP10}. In the follow-up work, \cite{KuznetsovaPRA08} performed a detailed analysis of phase gate architectures with polar molecules. \cite{charron:2007} proposed a scheme to generate entanglement between polar molecules using a sequence of laser pulses acting between different vibrational states. \cite{YuPCCP18} devised a technique to perform optimal control of orientation and entanglement of two planar molecular rotors coupled via dipole-dipole interactions. \cite{MilmanPRL07} derived Bell-type inequalities for non-locality and entanglement between two polar molecules in an optical lattice. The idea is based on measuring the correlations between the spacial orientation of the two molecules. \cite{PalaoPRL02,TeschPRL02} proposed QIP schemes where qubit states are encoded in molecular vibronic states. \cite{ShioyaMolPhys07} studied the possibility to realize quantum gates by mapping two qubits on the rotational and vibrational states of a single molecule. \cite{MishimaChemPhys09} numerically studied the possibility to implement Deutsch-Jozsa algorithm on rotational states of two polar molecules. \cite{KuznetsovaPRA16} proposed a technique for non-destructive readout of the rotational states in a one-dimensional cold-molecule array using a single Rydberg atom. \cite{WeiJCP11, ZhuJCP13, QiCPC16} developed various techniques to generate entanglement and to implement quantum logic gates using molecules in pendular states, created using a far-off-resonant laser beam (see Sec.~\ref{sec:align}). In the proposal, the qubit states are mapped onto the two lowest pendular states, $\vert \tilde{0}, 0 \rangle$ and $\vert \tilde{1}, 0 \rangle$. The coupling between the qubits is achieved through dipole--dipole interactions. \cite{HerreraNJP14} studied the ways to generate entanglement between open-shell $^2\Sigma$ molecules dressed by a far-off-resonant infrared field. \cite{KarraJCP16} analyzed the prospects of QIP with polar paramagnetic $^2\Sigma$ molecules in congruent electric and magnetic fields.

While most QIP proposals were based on  molecules with the simplest possible structure (rigid linear rotors), additional degrees of freedom featured by more complex species allow for more versatility and controllability. For example, \cite{WeiJCP11b} have shown that it is possible to generate entanglement with symmetric-top molecules, which exhibit a first-order Stark effect.
The qubits can be encoded in different $J$-, $M_J$- \cite{WeiJCP11b} or $K$-states \cite{Yu_2019}.
\cite{ZhangSciRep17}, in turn, studied bipartite quantum correlations of polar symmetric-top molecules in pendular states.

\subsection{Alternative schemes}
\cite{LeePRA05} proposed an alternative scheme to perform QIP using molecular levels. There, the $\vert 0 \rangle$ and $\vert 1 \rangle$ qubit states are mapped onto the scattering state of two ultracold atoms (per site in an optical lattice) and on the bound molecular pair, respectively. Switching between the two states is achieved using a Raman association/dissociation pulse. The dipole-dipole interaction arises when the neighbouring qubits are both brought to a bound molecular state. \cite{KuznetsovaPRA10} suggested to use dipole-dipole interacting molecular states to construct phase gates, while the atomic hyperfine states are used to initialize and store quantum information. \cite{Ortner2011} proposed another QIP scheme, based on self-assembled crystals of polar molecules (see also Sec.~\ref{sec:QuantSim}). The repulsive dipole-dipole interaction between the molecules is counteracted by a harmonic trapping potential which results in formation of stable Wigner-crystal-like structures, similar to those mentioned in Sec.~\ref{subsec:coldions}. The qubit states are, in turn, encoded into the stable spin states of the ground molecular state. The intermolecular interactions are mediated by phonons, in a similar way as it was previously realized with trapped atomic ions~\cite{HaffnerPhysRep08}. \cite{LeePRL04} studied (experimentally and theoretically) QIP with rotational wavepackets in a thermal ensemble of molecules, whose translational motion is not confined by any kind of trap or optical lattice. They have shown that  manipulating the revivals of the molecular wavepacket can be mapped onto qubit manipulation. \cite{HalversonJCP18} considered an even more different setup: instead of molecules in optical lattices, they proposed to create an array of endohedral fullerenes -- C$_{60}$ doped with freely rotating HF molecules. They developed a variational approach to reveal the entanglement entropy in an ensemble of such species.

\subsection{Trapped molecular ions}
As already discussed in Sec.~\ref{subsec:coldions}, it has recently become experimentally feasible to cool, trap, and control molecular ions~\cite{WillitschIRPC12, TongPRL10, StaanumNatPhys10, SchneiderNatPhys10, WolfNature16}. The advantages of this system for quantum information processing are  long lifetimes of cold-ion samples (hours, sometimes even days), the possibility to use phonons to introduce coupling between different ions, as well as  the fact that molecular ions can be coupled to atomic ions, which can be controlled in experiment extremely well~\cite{HaffnerPhysRep08}. In state-of-the-art experiments it is  possible to prepare and manipulate pure quantum states of a single molecular ion~\cite{ChouNat2017}. \cite{Mur-Petit2013} overviewed the possibilities to use trapped molecular ions for QIP, where the Zeeman  states of the molecular rotational levels act as qubit states. \cite{ShiNJP13} proposed to perform quantum logic spectroscopy, where an atomic ion is used to read out the state of a molecular ion qubit.
\cite{HudsonPRA18} have recently suggested a  chip-based molecular ion quantum processor with rotational states representing one option to encode qubits.

\subsection{Hybrid quantum systems}
Over the years, different platforms for quantum simulation and computation (based on atoms, ions, molecules, superconducting circuits, quantum defects in solids, etc.) were developing in parallel. A significant amount of attention has been paid to designing `hybrid' platforms that would combine the advantages of different systems~\cite{WallquistPS09}. \cite{RablPRL06, AndreNatPhys06} considered a hybrid system consisting of an ensemble of molecules coupled to a superconducting stripline cavity via microwave Raman processes. In this setup, long-living molecular rotational states act as a quantum memory, microwave photons in the stripline cavity transfer quantum information, while the computation itself is performed using charge qubits.  The coupling between the molecules and the stripline cavity is achieved through the large electric dipole moment of polar molecules, by making use of the fact that the rotational level splitting is in the same (GHz) frequency range as the resonance frequencies of stripline cavities. The microwave field of the cavity, in turn, couples to a cooper pair box (`a charge qubit'), which represents a superconducting island connected via Josephson junctions to a grounded reservoir. Cooper pairs can tunnel between the reservoir and the island, and the number of cooper pairs on the island determines the state of the qubit. \cite{SchusterPRA11} extended this idea to a hybrid system of trapped molecular ions coupled to a superconducting microwave cavity. The advantage of molecular ions is that their trapping is achieved through charge, and is therefore independent of their internal state. \cite{TordrupPRA08} proposed a hybrid scheme where quantum information is encoded in rotational excitations of a molecular ensemble, coupled to a Cooper-pair box. The scheme allows for a linear scaling of the number of qubits with the number of rotational molecular states involved.

\section{Quantum simulation with rotational states in dipolar gases}\label{sec:QuantSim}

In state-of-the-art experiments, cold molecules can be prepared in a single preordained quantum state. Moreover, their  orientation in space and mutual interactions can be fine-tuned using electromagnetic fields, which allows to control quantum dynamics. The final state of the many-body ensemble, on the other hand, can be detected at the single-particle level~\cite{OspelkausPRL10, MirandaNatPhys11, JinYeCRev12, LemKreDoyKais13}. In this section we survey recent progress in the application of controlled molecular rotations  to studying quantum many-particle physics.

\begin{figure}[b]
\centering
      \includegraphics[width=0.5\linewidth]{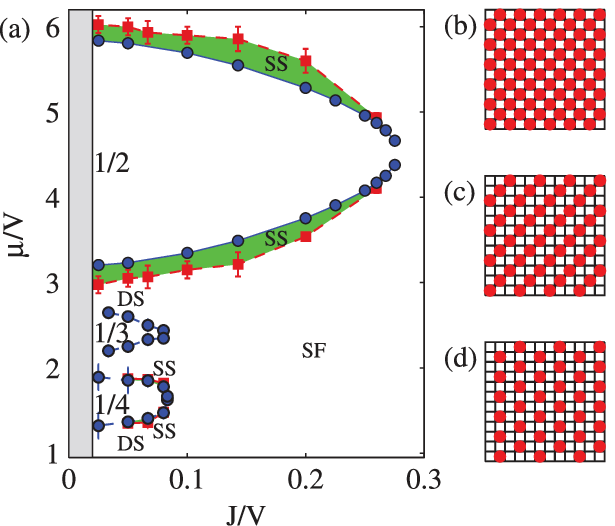}
\caption{(a) Phase diagram of dipolar bosons on a two-dimensional optical lattice. Supersolid (SS), superfluid (SF), as well as devil's staircase (DS) are indicated. (b)-(d) Examples of Mott solids with densities of 1/2, 1/3, and 1/4, respectively. Reprinted with permission from~\cite{CapogrossoSansonePRL10}.}
\label{DipolarPhase}
\end{figure}

As opposed to atoms, polar molecules feature long-range and anisotropic interactions, which paves the way to realizing exotic models of many-body physics, inaccessible in ``traditional'' condensed-matter systems. The many-body phenomena observed in molecular ensembles depend on the intermolecular interactions, which, in turn, depend on the relative orientation of molecules with respect to one another.  During the recent years, there has been a tremendous amount of work on many-body physics with ultracold molecules in optical lattices and traps~\cite{LemKreDoyKais13}. Most of these proposals make use of the long-range dipole-dipole interactions (in a general sense), which can in principle be realized with other dipolar systems, such as ultracold magnetic atoms  or magnetic defects in solids. In what follows we restrict ourselves to the schemes that make an explicit use of molecular rotational structure.

\subsection{Quantum phases of polar molecules in optical lattices}
\cite{GoralPRL02} studied many-body phases of dipolar bosons in an optical lattice. Using a variational approach based on Gutzwiller ansatz, they revealed the ground-state phase diagram, which was shown to include superfluid, supersolid, Mott insulator, checkerboard, and collapse phases. \cite{CapogrossoSansonePRL10} revealed the phase diagram of bosonic polar molecules on a 2D square lattice interacting via repulsive dipole-dipole interactions, using Monte Carlo simulations based on the worm  algorithm. The phase diagram they uncovered is shown in Fig.~\ref{DipolarPhase} and includes supersolid and superfluid phases, as well a devil's staircase of Mott solids,  where the density pattern is commensurate with the lattice at rational fillings (for example, with every second, third, or fourth lattice site occupied, Fig.~\ref{DipolarPhase}(b--d)). \cite{PolletPRL10} performed a similar calculation for a triangular lattice and focused on the formation of the supersolid phase. \cite{HePRA11} used dynamical mean-field theory to study the supersolid phase of cold fermionic polar molecules on a 2D optical lattice. \cite{SowinskiPRL12} studied an extended Bose-Hubbard model realized with polar molecules using  exact diagonalization and the multiscale entanglement renormalization ansatz. They have shown that taking into account the occupation-dependent tunneling and pair tunneling terms is important since they can destroy insulating phases and lead to novel quantum phases.

\cite{MicheliNatPhys06, BrennenNJP07}  showed that with open-shell $^2\Sigma$ molecules in an optical lattice one can engineer a variety of spin-model Hamiltonians, including those with topological properties.  \cite{BuchlerZollerPRL07} demonstrated that if molecules are confined in two dimensions, one can use electrostatic and microwave fields to engineer the strength and shape of the intermolecular potential. \cite{GorshkovPRL08} proposed a technique to enhance elastic collision rates and suppress inelastic collision rates between cold polar molecules. They effectively generate a repulsive van der Waals interaction using a combination of electrostatic and microwave fields. This, in turn, results in various self-assembled many-body phases. \cite{BuchlerNatPhys07} showed that using molecules in optical lattices dressed by microwave fields one can realize Hubbard models with strong nearest-neighbor three-body interactions, which paves the way to studying exotic quantum phases. \cite{SchmidtPRL08} studied the emergence of solid and supersolid phases in ultracold molecules using quantum Monte Carlo simulations, exact diagonalization, and a semiclassical approach. \cite{pupillo:2008,OrtnerNJP09} proposed to simulate extended Hubbard models using self-assembled crystals of polar molecules. \cite{ZhouPRA11} described a scheme to engineer effective long-range interactions in self-assembled crystals of polar molecules, mediated by lattice phonons.

\subsection{Quantum magnetism}
\cite{gorshkovPRL11, GorshkovPRA11}  showed that it is possible to engineer highly-tunable generalizations of the $t-J$ model, featuring spin-exchange, density-density, and density-spin interactions through long-range anisotropic dipole-dipole coupling. In such a setting, molecules are trapped in a two-dimensional optical lattice, where they hop between the sites and interact with one another via anisotropic dipole-dipole as well as hyperfine interactions. The electric dipole-dipole interaction term is unique for polar molecules and its strength can be controlled by applying an external electric field. In order to simulate magnetic systems, one chooses two rotational states dressed by microwave fields to represent `spin-up' and `spin-down'. Then, by tuning the strength of the electrostatic and microwave fields one is able to independently control the magnitude of the resulting dipole-dipole interactions. Such interactions can be not only of the density-density kind, i.e.\ taking place between molecules in  the same `spin' (rotational) states, but also of the `spin-flip' type, that is hopping of a spin (rotational) excitation between the lattice sites.  \cite{WeimerMolPhys13} has shown the possibility to engineer many-body spin interactions through digital (stroboscopic) quantum simulation with polar molecules. In particular, the realization of the Kitaev toric code was suggested. \cite{GorshkovMolPhys13} showed that exotic spin models, including the Kitaev honeycomb model, can be realized.  \cite{LinPRA10} predicted a ferroelectric phase, arising due to dipole-dipole interactions in an ensemble of ultracold polar molecules. \cite{KunsPRA11} have shown that molecules can realize $d-$wave superfluid phases. \cite{LemeshkoPRL12} proposed a technique to prepare spin crystals of ultracold molecules non-adiabatically, bypassing the usual Kibble-Zurek scaling. \cite{ManmanaPRA17} studied the phase diagram of the $t - J_\perp$ chain with long-range interactions realized with ultracold polar molecules. They found an enlarged superconducting phase.

\subsection{Topological phases}

The rich rotational structure of molecules paves the way to the realization of topological phases. For example, \cite{LevinsenPRA11} discussed the emergence of an exotic  topological $p_x + ip_y$ superfluid phase in a 2D gas of polar molecules dressed by a circularly-polarized microwave field. \cite{YaoPRL13} showed the possibility to realize fractional Chern insulator states in a two-dimensional array of molecules in an optical lattice. Realization of the Kitaev honeycomb  \cite{GorshkovMolPhys13} and toric code \cite{WeimerMolPhys13} models has been suggested. \cite{YaoNatPhys18} proposed to realize dipolar spin liquid and dipolar Heisenberg antiferromagnet using the two lowest molecular rotational states dressed with electric and microwave fields. \cite{ManmanaPRB13} demonstrated the possibility to engineer symmetry-protected topological phases in spin ladders using polar molecules in optical lattices. These phases were shown to survive even in the presence of long-range dipole-dipole interactions.

\subsection{Quantum transport and impurity physics}
One of the simplest many-body models consists of a single quantum particle (in solid-state physics it is usually an electron or a localized spin), coupled to a quantum bath, such as phonons, an electron liquid or a spin ensemble. These so-called `impurity problems' can be considered as an elementary building block of strongly correlated systems of condensed matter physics. The intricate internal structure featured by molecules paves the way to realize novel impurity models, inaccessible in conventional solid-state physics.

\cite{HerreraPRA11, XiangPRA12} studied realizations of Holstein polaron and exciton physics with molecules in optical lattices. In such a setting, molecules are considered to be trapped in an optical lattice, with a unit filling (one molecule per site). In the presence of a weak electric field, it is possible to isolate an effective two-level system, e.g.\ consisting of the ground, $J=0$, and the first excited, $J=1$, molecular rotational states with a given $M$. Molecules interact via dipole-dipole interaction, which enables hopping of a rotational excitation between the neighboring sites. Such an excitation can be effectively considered as a quasiparticle -- a polaron or an exciton, which can be localized or delocalized in space (the latter corresponds to collective rotational excitations in an ensemble of molecules). On the other hand,  molecules in an optical lattice are never completely pinned in space and can oscillate around their equilibrium positions. Such small displacements (that correspond to collective vibrational excitations or `phonons'  in the molecular lattice) alter the distances between the neighboring molecules and thereby the dipole-dipole interactions between them. Change in the dipole-dipole interaction, in turn, affects the hopping magnitude of the rotational excitation. This leads to an effective coupling between the `particle' (rotational excitation) and the phonon bath (vibrations in the molecular lattice), inherent to the polaron models. \cite{HerreraPRA10} considered a two-species mixture of ultracold molecules in an optical lattice, showing that it maps onto rotational excitons interacting with disordered impurities with tunable disorder. \cite{PeresRiosNJP10} studied collective spin excitations in an ensemble of $^2\Sigma$ molecules, which can be controlled with external electric and magnetic fields. \cite{HerreraPRL13} proposed to use ultracold molecules to realize a polaron model with mixed  breathing-mode and Su-Schrieffer-Heeger couplings. They have shown that the model exhibits two sharp transitions, in contrast to one featured by the standard  Su-Schrieffer-Heeger model. \cite{KwasigrochPRA14}  theoretically demonstrated many-body localization of rotational excitations of molecules in a lattice following a microwave pulse. Quantum walk and Anderson localization of rotational excitations in disordered ensembles of polar molecules was studied by~\cite{XuNJP15}. \cite{LemeshkoDroplets16, Lemeshko_2016_book, BikashPRA16} have shown that using cold molecules coupled to a many-body environment, one can realize a novel `angulon' model, which allows to study angular momentum dynamics in quantum many-body systems, with possible applications to solid-state magnetism.

\subsection{More exotic developments}
In addition to dipole-dipole coupling, ultracold molecules feature other types of interactions that are challenging to realize with cold atoms. For example, \cite{ByrdPRL12, BhongalePRL13} suggested to make use of quadrupole-quadrupole interactions between ultracold molecules to obtain novel many-body phases. On one hand, quadrupole-quadrupole interactions are substantially weaker compared to dipole-dipole forces between polar molecules. On the other hand, however, it is  experimentally easier to achieve dense ensembles of homonuclear molecules through Feshbach association, as opposed to preparing dense dipolar gases. The peculiar symmetry of quadrupole-quadrupole interactions leads to new phases of matter. While most of cold-molecule proposals are concerned with linear-rotor molecules, using more complex species  allows to realize more complex and exotic phases of matter. For example, \cite{WallAnnPhys13, WallNJP15} proposed to make use of the complex internal structure of symmetric top molecules to realize non-conventional magnetism models. In particular,  linear Stark effect inherent to symmetric-top molecules allows to engineer level crossings between rotational states with the same $J$ and different $M$ values using a combination of microwave and electrostatic fields~\cite{WallNJP15}. This results in new molecule-molecule interaction terms, which in the context of spin models can be seen as terms not conserving magnetization.

Novel multichannel Hubbard models with non-reactive molecules in optical lattices were also proposed~\cite{DocajPRL16, WallPRA17}. In these models, the on-site interaction parameter, $U$, of the Hubbard model is replaced by multichannel interaction, due to the rich internal structure of molecular species. This term arises from short-range physics and is therefore also present for homonuclear species in the absence of external fields. \cite{SundarSciRep18} proposed to realize synthetic dimensions using polar molecules. In addition to real spatial dimensions (1D, 2D, or 3D), additional synthetic dimensions can be mapped onto the internal molecular rotational states. Hopping in  synthetic dimensions is realized using microwave fields transferring the population between molecular rotational levels. The controllability of molecular dynamics in both real and synthetic dimensions allows to access rich physics, including synthetic gauge fields and topological phases to many-body localization.

\subsection{Experimental realizations}
For several years, the field of quantum simulation with ultracold molecules was largely driven by theory. Recent experimental developments have, however, already allowed to observe some of the predicted phenomena. \cite{YanNature13}  experimentally realized a lattice spin model using a many-body system of molecules on an optical lattice and demonstrated the presence of dipolar spin-exchange interactions. \cite{HazzardPRL14} used Ramsey spectroscopy to experimentally probe quantum dynamics of a disordered sample of polar molecules on an optical lattice. Theory based on the cluster expansion technique was found to be in good agreement with the measurements. This paves the way for exploring quantum many-body dynamics of molecules in the regimes inaccessible to theoretical models.

In current experiments, it is challenging to achieve a high density of polar molecules, close to unit filling in an optical lattice, and bring the molecules close to quantum degeneracy. \cite{HazzardPRL13} have theoretically shown that, even in non-degenerate samples well below unit filling, one can observe intriguing physics of far-from-equilibrium magnetism, already pushing the limits accessible to theory.

Although creating high-density samples of ultracold molecules with full quantum control represents a formidable challenge, there has been a lot of experimental progress  during the last years. For example, \cite{ReichsollnerPRL17} created a  low-entropy gas of heteronuclear bosonic  molecules (RbCs) in an optical lattice, with the lattice filling exceeding 30\%. Fermionic NaK molecules have been created by~\cite{ParkPRL15} and their rotational and hyperfine states have been controlled using microwave fields~\cite{WillPRL16}. NaRb molecules have been prepared as well~\cite{GuoPRL16}. \cite{MosesSci15} reported on the creation of a low-entropy gas of polar KRb molecules in a 3D optical lattice at a filling fraction as high as 25\%. \cite{CoveyNatComm16} have experimentally prepared an ensemble of polar molecules in an optical lattice, such that the lattice sites are either empty or occupied by a Bose-Fermi atomic pair, and have studied the production of ultracold molecules in the lattice. Finally, a Fermi-degenerate quantum gas of ultracold polar molecules (KRb) has been created recently~\cite{DeMarco18}. This gives us hope that the exciting quantum many-body models described in this section can be experimentally realized in the nearest future.

\section{Summary and outlook}
\label{sec:concl}

In this review, we presented the recent advances, both theoretical and experimental, in controlling molecular rotations, together with the various avenues for exploiting this control at the forefront of current research in quantum physics. Hallmarks include the control and the study of superrotors, the manipulation of molecular torsion angles,
the observation of orbiting resonances in cold collisions,
enantiomer-selectivity via quantum pathway interference in three-wave mixing rotational spectroscopy of chiral molecules, quantum simulation of long-range interacting many-body systems as well as ultracold molecules in optical lattices.

Quantum control of molecular rotation is challenging and promising at the same time. Quantization is a basic prerequisite for quantum control, which is naturally met for molecular rotation.  However, since the spectra of molecular rotations are unbounded, the question as to how much control is fundamentally achievable still remains open for polyatomic, notably asymmetric top molecules. In the same direction, it would be interesting to identify both the best control mechanism to orient or align molecules with respect to the experimental conditions, and the most sensitive way to measure the efficiency of the control process.

Another important control problem is the preparation of molecules in a single rotational state. While diatomic, i.e., linear top molecules have been successfully brought into a single rotational state by laser cooling, state-selective laser excitation, and projective measurements, no general accessible route seems available for polyatomic molecules. In favorable cases, small linear and symmetric top molecules can be selected in a single rotational quantum state by passage through inhomogeneous static fields in, e.g., an electrostatic deflector or an electrical hexapole. This severely hampers the efficiency of coherent control schemes, for example for the enantioselective excitation of chiral molecules. There, even at temperatures as low as a few kelvin, many rotational states are thermally populated.

Coherent control of bimolecular reactions has been a long-standing goal, with laser alignment of the collision partners envisioned as a possible route towards it. For diatomic molecules, preparing the molecules in very tight confinement, possible at the extremely low temperatures of the nano-kelvin range, has provided an alternative. Whether this approach can be extended to polyatomic molecules and more complex reactions is still an open question.

Finally, ultracold molecules in optical lattices potentially represent a versatile platform for quantum simulation of both conventional and exotic many-body models. We hope that future experimental advances will allow to achieve higher densities of ultracold molecules, as well as to cool and trap polyatomic molecules in optical lattices. This would bring us even closer to the realisation of molecular quantum simulators which, in turn, might alter the way we study condensed matter systems.

Quantum control of molecular rotation, as exemplified in this review,
is at the core of current research endeavors, aiming to, on the one hand,  shed light on fundamental questions of quantum control and, on the other hand, to realize  novel applications in AMO physics, physical chemistry and quantum information science.\\

\noindent \textbf{Acknowledgments}\\
We would like to thank (in alphabetical order) O.~Atabek, B.~Friedrich, R.~Gonzalez-Ferez, M.~Leibscher, A.~B.~Magann, Y.~Ohtsuki, G.~Pupillo, H.~A.~Rabitz, Y.~Shikano and T.~Zelevinsky for comments on the manuscript.
C.P.K. would like to thank the Weizmann Institute for Science for hospitality.
We are grateful for financial support by a Rosi and Max Varon Visiting Professorship (CPK), from the Austrian Science
Fund (FWF), under project No. P29902-N27 (ML), and from the PICS program (DS) as well as the ANR-DFG research program COQS (CPK and DS, grant numbers ANR-15-CE30-0023-01 and Ko 2301/11-1). The work of D. S. has been done with the support of the Technische Universit\"at M\"unchen – Institute for Advanced Study, funded by the German Excellence Initiative and the European Union Seventh Framework Programme under grant agreement 291763.


\begin{thebibliography}{607}
\expandafter\ifx\csname natexlab\endcsname\relax\def\natexlab#1{#1}\fi
\expandafter\ifx\csname bibnamefont\endcsname\relax
  \def\bibnamefont#1{#1}\fi
\expandafter\ifx\csname bibfnamefont\endcsname\relax
  \def\bibfnamefont#1{#1}\fi
\expandafter\ifx\csname citenamefont\endcsname\relax
  \def\citenamefont#1{#1}\fi
\expandafter\ifx\csname url\endcsname\relax
  \def\url#1{\texttt{#1}}\fi
\expandafter\ifx\csname urlprefix\endcsname\relax\def\urlprefix{URL }\fi
\providecommand{\bibinfo}[2]{#2}
\providecommand{\eprint}[2][]{\url{#2}}

\bibitem{abe:2011}
\bibinfo{author}{\bibnamefont{Abe}, \bibfnamefont{H.}}, and
  \bibinfo{author}{\bibfnamefont{Y.}~\bibnamefont{Ohtsuki}},
  \bibinfo{year}{2011}, \bibinfo{journal}{Phys. Rev. A}
  \textbf{\bibinfo{volume}{83}}, \bibinfo{pages}{053410}.

\bibitem{abe:2012}
\bibinfo{author}{\bibnamefont{Abe}, \bibfnamefont{H.}}, and
  \bibinfo{author}{\bibfnamefont{Y.}~\bibnamefont{Ohtsuki}},
  \bibinfo{year}{2012}, \bibinfo{journal}{Chem. Phys.}
  \textbf{\bibinfo{volume}{400}}, \bibinfo{pages}{13}.

\bibitem{ahn:2003}
\bibinfo{author}{\bibnamefont{Ahn}, \bibfnamefont{J.}},
  \bibinfo{author}{\bibfnamefont{A.~V.} \bibnamefont{Efimov}},
  \bibinfo{author}{\bibfnamefont{R.~D.} \bibnamefont{Averitt}}, and
  \bibinfo{author}{\bibfnamefont{A.~J.} \bibnamefont{Taylor}},
  \bibinfo{year}{2003}, \bibinfo{journal}{Opt. Express}
  \textbf{\bibinfo{volume}{11}}, \bibinfo{pages}{2486}.

\bibitem{akagi:2015}
\bibinfo{author}{\bibnamefont{Akagi}, \bibfnamefont{H.}},
  \bibinfo{author}{\bibfnamefont{T.}~\bibnamefont{Kasajima}},
  \bibinfo{author}{\bibfnamefont{T.}~\bibnamefont{Kumada}},
  \bibinfo{author}{\bibfnamefont{R.}~\bibnamefont{Itakura}},
  \bibinfo{author}{\bibfnamefont{A.}~\bibnamefont{Yokoyama}},
  \bibinfo{author}{\bibfnamefont{H.}~\bibnamefont{Hasegawa}}, and
  \bibinfo{author}{\bibfnamefont{Y.}~\bibnamefont{Ohshima}},
  \bibinfo{year}{2015}, \bibinfo{journal}{Phys. Rev. A}
  \textbf{\bibinfo{volume}{91}}, \bibinfo{pages}{063416}.

\bibitem{albertini:2003}
\bibinfo{author}{\bibnamefont{Albertini}, \bibfnamefont{F.}}, and
  \bibinfo{author}{\bibfnamefont{D.}~\bibnamefont{D'Alessandro}},
  \bibinfo{year}{2003}, \bibinfo{journal}{IEEE Trans. Autom. Control}
  \textbf{\bibinfo{volume}{48}}, \bibinfo{pages}{1399}.

\bibitem{AlighanbariNatPhys18}
\bibinfo{author}{\bibnamefont{Alighanbari}, \bibfnamefont{S.}},
  \bibinfo{author}{\bibfnamefont{M.~G.} \bibnamefont{Hansen}},
  \bibinfo{author}{\bibfnamefont{V.~I.} \bibnamefont{Korobov}}, and
  \bibinfo{author}{\bibfnamefont{S.}~\bibnamefont{Schiller}},
  \bibinfo{year}{2018}, \bibinfo{journal}{Nature Phys.}
  \textbf{\bibinfo{volume}{14}}, \bibinfo{pages}{555}.

\bibitem{altafini:2002}
\bibinfo{author}{\bibnamefont{Altafini}, \bibfnamefont{C.}},
  \bibinfo{year}{2002}, \bibinfo{journal}{J. Math. Phys.}
  \textbf{\bibinfo{volume}{43}}, \bibinfo{pages}{2051}.

\bibitem{AminiJCP17}
\bibinfo{author}{\bibnamefont{Amini}, \bibfnamefont{K.}},
  \bibinfo{author}{\bibfnamefont{R.}~\bibnamefont{Boll}},
  \bibinfo{author}{\bibfnamefont{A.}~\bibnamefont{Lauer}},
  \bibinfo{author}{\bibfnamefont{M.}~\bibnamefont{Burt}},
  \bibinfo{author}{\bibfnamefont{J.~W.~L.} \bibnamefont{Lee}},
  \bibinfo{author}{\bibfnamefont{L.}~\bibnamefont{Christensen}},
  \bibinfo{author}{\bibfnamefont{F.}~\bibnamefont{Brauβe}},
  \bibinfo{author}{\bibfnamefont{T.}~\bibnamefont{Mullins}},
  \bibinfo{author}{\bibfnamefont{E.}~\bibnamefont{Savelyev}},
  \bibinfo{author}{\bibfnamefont{U.}~\bibnamefont{Ablikim}},
  \bibinfo{author}{\bibfnamefont{N.}~\bibnamefont{Berrah}},
  \bibinfo{author}{\bibfnamefont{C.}~\bibnamefont{Bomme}}, \emph{et~al.},
  \bibinfo{year}{2017}, \bibinfo{journal}{J. Chem. Phys.}
  \textbf{\bibinfo{volume}{147}}(\bibinfo{number}{1}), \bibinfo{pages}{013933}.

\bibitem{AncilottoIRPC17}
\bibinfo{author}{\bibnamefont{Ancilotto}, \bibfnamefont{F.}},
  \bibinfo{author}{\bibfnamefont{M.}~\bibnamefont{Barranco}},
  \bibinfo{author}{\bibfnamefont{F.}~\bibnamefont{Coppens}},
  \bibinfo{author}{\bibfnamefont{J.}~\bibnamefont{Eloranta}},
  \bibinfo{author}{\bibfnamefont{N.}~\bibnamefont{Halberstadt}},
  \bibinfo{author}{\bibfnamefont{A.}~\bibnamefont{Hernando}},
  \bibinfo{author}{\bibfnamefont{D.}~\bibnamefont{Mateo}}, and
  \bibinfo{author}{\bibfnamefont{M.}~\bibnamefont{Pi}}, \bibinfo{year}{2017},
  \bibinfo{journal}{Int. Rev. in Phys. Chem.}
  \textbf{\bibinfo{volume}{36}}(\bibinfo{number}{4}), \bibinfo{pages}{621}.

\bibitem{AndereggNatPhys18}
\bibinfo{author}{\bibnamefont{Anderegg}, \bibfnamefont{L.}},
  \bibinfo{author}{\bibfnamefont{B.~L.} \bibnamefont{Augenbraun}},
  \bibinfo{author}{\bibfnamefont{Y.}~\bibnamefont{Bao}},
  \bibinfo{author}{\bibfnamefont{S.}~\bibnamefont{Burchesky}},
  \bibinfo{author}{\bibfnamefont{L.~W.} \bibnamefont{Cheuk}},
  \bibinfo{author}{\bibfnamefont{W.}~\bibnamefont{Ketterle}}, and
  \bibinfo{author}{\bibfnamefont{J.~M.} \bibnamefont{Doyle}},
  \bibinfo{year}{2018}, \bibinfo{journal}{Nature Phys.}
  \textbf{\bibinfo{volume}{14}}, \bibinfo{pages}{890}.

\bibitem{AndereggPRL17}
\bibinfo{author}{\bibnamefont{Anderegg}, \bibfnamefont{L.}},
  \bibinfo{author}{\bibfnamefont{B.~L.} \bibnamefont{Augenbraun}},
  \bibinfo{author}{\bibfnamefont{E.}~\bibnamefont{Chae}},
  \bibinfo{author}{\bibfnamefont{B.}~\bibnamefont{Hemmerling}},
  \bibinfo{author}{\bibfnamefont{N.~R.} \bibnamefont{Hutzler}},
  \bibinfo{author}{\bibfnamefont{A.}~\bibnamefont{Ravi}},
  \bibinfo{author}{\bibfnamefont{A.}~\bibnamefont{Collopy}},
  \bibinfo{author}{\bibfnamefont{J.}~\bibnamefont{Ye}},
  \bibinfo{author}{\bibfnamefont{W.}~\bibnamefont{Ketterle}}, and
  \bibinfo{author}{\bibfnamefont{J.~M.} \bibnamefont{Doyle}},
  \bibinfo{year}{2017}, \bibinfo{journal}{Phys. Rev. Lett.}
  \textbf{\bibinfo{volume}{119}}, \bibinfo{pages}{103201}.

\bibitem{AndreNatPhys06}
\bibinfo{author}{\bibnamefont{Andr{\'e}}, \bibfnamefont{A.}},
  \bibinfo{author}{\bibfnamefont{D.}~\bibnamefont{DeMille}},
  \bibinfo{author}{\bibfnamefont{J.~M.} \bibnamefont{Doyle}},
  \bibinfo{author}{\bibfnamefont{M.~D.} \bibnamefont{Lukin}},
  \bibinfo{author}{\bibfnamefont{S.~E.} \bibnamefont{Maxwell}},
  \bibinfo{author}{\bibfnamefont{P.}~\bibnamefont{Rabl}},
  \bibinfo{author}{\bibfnamefont{R.~J.} \bibnamefont{Schoelkopf}}, and
  \bibinfo{author}{\bibfnamefont{P.}~\bibnamefont{Zoller}},
  \bibinfo{year}{2006}, \bibinfo{journal}{Nature Physics}
  \textbf{\bibinfo{volume}{2}}, \bibinfo{pages}{636}.

\bibitem{AoizPCCP15}
\bibinfo{author}{\bibnamefont{Aoiz}, \bibfnamefont{F.~J.}},
  \bibinfo{author}{\bibfnamefont{M.}~\bibnamefont{Brouard}},
  \bibinfo{author}{\bibfnamefont{S.~D.~S.} \bibnamefont{Gordon}},
  \bibinfo{author}{\bibfnamefont{B.}~\bibnamefont{Nichols}},
  \bibinfo{author}{\bibfnamefont{S.}~\bibnamefont{Stolte}}, and
  \bibinfo{author}{\bibfnamefont{V.}~\bibnamefont{Walpole}},
  \bibinfo{year}{2015}, \bibinfo{journal}{Phys. Chem. Chem. Phys.}
  \textbf{\bibinfo{volume}{17}}, \bibinfo{pages}{30210}.

\bibitem{arai:2015}
\bibinfo{author}{\bibnamefont{Arai}, \bibfnamefont{K.}}, and
  \bibinfo{author}{\bibfnamefont{Y.}~\bibnamefont{Ohtsuki}},
  \bibinfo{year}{2015}, \bibinfo{journal}{Phys. Rev. A}
  \textbf{\bibinfo{volume}{92}}, \bibinfo{pages}{062302}.

\bibitem{arango:2008}
\bibinfo{author}{\bibnamefont{Arango}, \bibfnamefont{C.~A.}}, and
  \bibinfo{author}{\bibfnamefont{G.~S.} \bibnamefont{Ezra}},
  \bibinfo{year}{2008}, \bibinfo{journal}{Int. J. Bifurcat. Chaos}
  \textbf{\bibinfo{volume}{18}}, \bibinfo{pages}{1127}.

\bibitem{arango:2004}
\bibinfo{author}{\bibnamefont{Arango}, \bibfnamefont{C.~A.}},
  \bibinfo{author}{\bibfnamefont{W.~W.} \bibnamefont{Kennerly}}, and
  \bibinfo{author}{\bibfnamefont{G.~S.} \bibnamefont{Ezra}},
  \bibinfo{year}{2004}, \bibinfo{journal}{Chem. Phys. Lett.}
  \textbf{\bibinfo{volume}{392}}, \bibinfo{pages}{486}.

\bibitem{arango:2005}
\bibinfo{author}{\bibnamefont{Arango}, \bibfnamefont{C.~A.}},
  \bibinfo{author}{\bibfnamefont{W.~W.} \bibnamefont{Kennerly}}, and
  \bibinfo{author}{\bibfnamefont{G.~S.} \bibnamefont{Ezra}},
  \bibinfo{year}{2005}, \bibinfo{journal}{J. Chem. Phys.}
  \textbf{\bibinfo{volume}{122}}, \bibinfo{pages}{184303}.

\bibitem{ArangoPRL06}
\bibinfo{author}{\bibnamefont{Arango}, \bibfnamefont{C.~A.}},
  \bibinfo{author}{\bibfnamefont{M.}~\bibnamefont{Shapiro}}, and
  \bibinfo{author}{\bibfnamefont{P.}~\bibnamefont{Brumer}},
  \bibinfo{year}{2006}, \bibinfo{journal}{Phys. Rev. Lett.}
  \textbf{\bibinfo{volume}{97}}, \bibinfo{pages}{193202}.

\bibitem{arnold}
\bibinfo{author}{\bibnamefont{Arnol'd}, \bibfnamefont{V.~I.}},
  \bibinfo{year}{1989}, \emph{\bibinfo{title}{Mathematical Methods of Classical
  Mechanics}} (\bibinfo{publisher}{Springer-Verlag, New York}).

\bibitem{artamonov:2010}
\bibinfo{author}{\bibnamefont{Artamonov}, \bibfnamefont{M.}}, and
  \bibinfo{author}{\bibfnamefont{T.}~\bibnamefont{Seideman}},
  \bibinfo{year}{2010}, \bibinfo{journal}{Phys. Rev. A}
  \textbf{\bibinfo{volume}{82}}, \bibinfo{pages}{023413}.

\bibitem{ashbaugh:1991}
\bibinfo{author}{\bibnamefont{Ashbaugh}, \bibfnamefont{M.~S.}},
  \bibinfo{author}{\bibfnamefont{C.~C.} \bibnamefont{Chiconc}}, and
  \bibinfo{author}{\bibfnamefont{R.~H.} \bibnamefont{Cushman}},
  \bibinfo{year}{1991}, \bibinfo{journal}{J. Dyn. Diff. Eq.}
  \textbf{\bibinfo{volume}{3}}, \bibinfo{pages}{67}.

\bibitem{averbukh:2001}
\bibinfo{author}{\bibnamefont{Averbukh}, \bibfnamefont{I.~S.}}, and
  \bibinfo{author}{\bibfnamefont{R.}~\bibnamefont{Arvieu}},
  \bibinfo{year}{2001}, \bibinfo{journal}{Phys. Rev. Lett.}
  \textbf{\bibinfo{volume}{87}}, \bibinfo{pages}{163601}.

\bibitem{Babilotte:16}
\bibinfo{author}{\bibnamefont{Babilotte}, \bibfnamefont{P.}},
  \bibinfo{author}{\bibfnamefont{K.}~\bibnamefont{Hamraoui}},
  \bibinfo{author}{\bibfnamefont{F.}~\bibnamefont{Billard}},
  \bibinfo{author}{\bibfnamefont{E.}~\bibnamefont{Hertz}},
  \bibinfo{author}{\bibfnamefont{B.}~\bibnamefont{Lavorel}},
  \bibinfo{author}{\bibfnamefont{O.}~\bibnamefont{Faucher}}, and
  \bibinfo{author}{\bibfnamefont{D.}~\bibnamefont{Sugny}},
  \bibinfo{year}{2016}, \bibinfo{journal}{Phys. Rev. A}
  \textbf{\bibinfo{volume}{94}}, \bibinfo{pages}{043403}.

\bibitem{baekhoj:2016}
\bibinfo{author}{\bibnamefont{B\ae{}kh\o{}j}, \bibfnamefont{J.~E.}}, and
  \bibinfo{author}{\bibfnamefont{L.~B.} \bibnamefont{Madsen}},
  \bibinfo{year}{2016}, \bibinfo{journal}{Phys. Rev. A}
  \textbf{\bibinfo{volume}{94}}, \bibinfo{pages}{043414}.

\bibitem{BarryNature14}
\bibinfo{author}{\bibnamefont{Barry}, \bibfnamefont{J.~F.}},
  \bibinfo{author}{\bibfnamefont{D.~J.} \bibnamefont{McCarron}},
  \bibinfo{author}{\bibfnamefont{E.~B.} \bibnamefont{Norrgard}},
  \bibinfo{author}{\bibfnamefont{M.~H.} \bibnamefont{Steinecker}}, and
  \bibinfo{author}{\bibfnamefont{D.}~\bibnamefont{DeMille}},
  \bibinfo{year}{2014}, \bibinfo{journal}{Nature}
  \textbf{\bibinfo{volume}{512}}, \bibinfo{pages}{7514}.

\bibitem{baugh:1994}
\bibinfo{author}{\bibnamefont{Baugh}, \bibfnamefont{D.~A.}},
  \bibinfo{author}{\bibfnamefont{D.~Y.} \bibnamefont{Kim}},
  \bibinfo{author}{\bibfnamefont{V.~A.} \bibnamefont{Cho}},
  \bibinfo{author}{\bibfnamefont{L.~C.} \bibnamefont{Pipes}},
  \bibinfo{author}{\bibfnamefont{J.~C.} \bibnamefont{Petteway}}, and
  \bibinfo{author}{\bibfnamefont{C.~D.} \bibnamefont{Fuglesang}},
  \bibinfo{year}{1994}, \bibinfo{journal}{Chemical Physics Letters}
  \textbf{\bibinfo{volume}{219}}(\bibinfo{number}{3}), \bibinfo{pages}{207 }.

\bibitem{baumfalk:2001}
\bibinfo{author}{\bibnamefont{Baumfalk}, \bibfnamefont{R.}},
  \bibinfo{author}{\bibfnamefont{N.~H.} \bibnamefont{Nahler}}, and
  \bibinfo{author}{\bibfnamefont{U.}~\bibnamefont{Buck}}, \bibinfo{year}{2001},
  \bibinfo{journal}{J. Chem. Phys.}
  \textbf{\bibinfo{volume}{114}}(\bibinfo{number}{11}), \bibinfo{pages}{4755}.

\bibitem{beauchard:2005}
\bibinfo{author}{\bibnamefont{Beauchard}, \bibfnamefont{K.}},
  \bibinfo{year}{2005}, \bibinfo{journal}{J. Math. Pures et Appl.}
  \textbf{\bibinfo{volume}{84}}, \bibinfo{pages}{851}.

\bibitem{benhaj:2002}
\bibinfo{author}{\bibnamefont{Ben Haj-Yedder}, \bibfnamefont{A.}},
  \bibinfo{author}{\bibfnamefont{A.}~\bibnamefont{Auger}},
  \bibinfo{author}{\bibfnamefont{C.~M.} \bibnamefont{Dion}},
  \bibinfo{author}{\bibfnamefont{E.}~\bibnamefont{Canc\`es}},
  \bibinfo{author}{\bibfnamefont{A.}~\bibnamefont{Keller}},
  \bibinfo{author}{\bibfnamefont{C.}~\bibnamefont{Le~Bris}}, and
  \bibinfo{author}{\bibfnamefont{O.}~\bibnamefont{Atabek}},
  \bibinfo{year}{2002}, \bibinfo{journal}{Phys. Rev. A}
  \textbf{\bibinfo{volume}{66}}, \bibinfo{pages}{063401}.

\bibitem{benko:2015}
\bibinfo{author}{\bibnamefont{Benko}, \bibfnamefont{C.}},
  \bibinfo{author}{\bibfnamefont{L.}~\bibnamefont{Hua}},
  \bibinfo{author}{\bibfnamefont{T.~K.} \bibnamefont{Allison}},
  \bibinfo{author}{\bibfnamefont{F.~m.~c.} \bibnamefont{Labaye}}, and
  \bibinfo{author}{\bibfnamefont{J.}~\bibnamefont{Ye}}, \bibinfo{year}{2015},
  \bibinfo{journal}{Phys. Rev. Lett.} \textbf{\bibinfo{volume}{114}},
  \bibinfo{pages}{153001}.

\bibitem{BerglundNJP15}
\bibinfo{author}{\bibnamefont{Berglund}, \bibfnamefont{J.~M.}},
  \bibinfo{author}{\bibfnamefont{M.}~\bibnamefont{Drewsen}}, and
  \bibinfo{author}{\bibfnamefont{C.~P.} \bibnamefont{Koch}},
  \bibinfo{year}{2015}, \bibinfo{journal}{New J. Phys.}
  \textbf{\bibinfo{volume}{17}}, \bibinfo{pages}{025007}.

\bibitem{Berglund2019}
\bibinfo{author}{\bibnamefont{Berglund}, \bibfnamefont{J.~M.}},
  \bibinfo{author}{\bibfnamefont{M.}~\bibnamefont{Drewsen}}, and
  \bibinfo{author}{\bibfnamefont{C.~P.} \bibnamefont{Koch}},
  \bibinfo{year}{2019}, \bibinfo{journal}{arXiv:1905.02130} .

\bibitem{BernathBook}
\bibinfo{author}{\bibnamefont{Bernath}, \bibfnamefont{P.~F.}},
  \bibinfo{year}{2005}, \emph{\bibinfo{title}{Spectra of Atoms and Molecules}}
  (\bibinfo{publisher}{Oxford University Press, UK}), \bibinfo{edition}{2}
  edition.

\bibitem{BeyerPRX18}
\bibinfo{author}{\bibnamefont{Beyer}, \bibfnamefont{M.}}, and
  \bibinfo{author}{\bibfnamefont{F.}~\bibnamefont{Merkt}},
  \bibinfo{year}{2018}, \bibinfo{journal}{Phys. Rev. X}
  \textbf{\bibinfo{volume}{8}}, \bibinfo{pages}{031085}.

\bibitem{BhongalePRL13}
\bibinfo{author}{\bibnamefont{Bhongale}, \bibfnamefont{S.~G.}},
  \bibinfo{author}{\bibfnamefont{L.}~\bibnamefont{Mathey}},
  \bibinfo{author}{\bibfnamefont{E.}~\bibnamefont{Zhao}},
  \bibinfo{author}{\bibfnamefont{S.}~\bibnamefont{Yelin}}, and
  \bibinfo{author}{\bibfnamefont{M.}~\bibnamefont{Lemeshko}},
  \bibinfo{year}{2013}, \bibinfo{journal}{Phys. Rev. Lett.}
  \textbf{\bibinfo{volume}{110}}, \bibinfo{pages}{155301}.

\bibitem{bisgaard:2009}
\bibinfo{author}{\bibnamefont{Bisgaard}, \bibfnamefont{C.~Z.}},
  \bibinfo{author}{\bibfnamefont{O.~J.} \bibnamefont{Clarkin}},
  \bibinfo{author}{\bibfnamefont{G.}~\bibnamefont{Wu}},
  \bibinfo{author}{\bibfnamefont{A.~M.~D.} \bibnamefont{Lee}},
  \bibinfo{author}{\bibfnamefont{O.}~\bibnamefont{Ge{\ss}ner}},
  \bibinfo{author}{\bibfnamefont{C.~C.} \bibnamefont{Hayden}}, and
  \bibinfo{author}{\bibfnamefont{A.}~\bibnamefont{Stolow}},
  \bibinfo{year}{2009}, \bibinfo{journal}{Science}
  \textbf{\bibinfo{volume}{323}}(\bibinfo{number}{5920}),
  \bibinfo{pages}{1464}.

\bibitem{bitter:2016}
\bibinfo{author}{\bibnamefont{Bitter}, \bibfnamefont{M.}}, and
  \bibinfo{author}{\bibfnamefont{V.}~\bibnamefont{Milner}},
  \bibinfo{year}{2016}, \bibinfo{journal}{Phys. Rev. Lett.}
  \textbf{\bibinfo{volume}{117}}, \bibinfo{pages}{144104}.

\bibitem{bitter:2017}
\bibinfo{author}{\bibnamefont{Bitter}, \bibfnamefont{M.}}, and
  \bibinfo{author}{\bibfnamefont{V.}~\bibnamefont{Milner}},
  \bibinfo{year}{2017}, \bibinfo{journal}{Phys. Rev. Lett.}
  \textbf{\bibinfo{volume}{118}}, \bibinfo{pages}{034101}.

\bibitem{blaga:2012}
\bibinfo{author}{\bibnamefont{Blaga}, \bibfnamefont{C.~I.}},
  \bibinfo{author}{\bibfnamefont{J.}~\bibnamefont{Xu}},
  \bibinfo{author}{\bibfnamefont{A.~D.} \bibnamefont{DiChiara}},
  \bibinfo{author}{\bibfnamefont{E.}~\bibnamefont{Sistrunk}},
  \bibinfo{author}{\bibfnamefont{K.}~\bibnamefont{Zhang}},
  \bibinfo{author}{\bibfnamefont{P.}~\bibnamefont{Agostini}},
  \bibinfo{author}{\bibfnamefont{T.~A.} \bibnamefont{Miller}},
  \bibinfo{author}{\bibfnamefont{L.~F.} \bibnamefont{DiMauro}}, and
  \bibinfo{author}{\bibfnamefont{C.~D.} \bibnamefont{Lin}},
  \bibinfo{year}{2012}, \bibinfo{journal}{Nature}
  \textbf{\bibinfo{volume}{483}}, \bibinfo{pages}{194}.

\bibitem{BoestenPRL96}
\bibinfo{author}{\bibnamefont{Boesten}, \bibfnamefont{H.~M.~J.~M.}},
  \bibinfo{author}{\bibfnamefont{C.~C.} \bibnamefont{Tsai}},
  \bibinfo{author}{\bibfnamefont{B.~J.} \bibnamefont{Verhaar}}, and
  \bibinfo{author}{\bibfnamefont{D.~J.} \bibnamefont{Heinzen}},
  \bibinfo{year}{1996}, \bibinfo{journal}{Phys. Rev. Lett.}
  \textbf{\bibinfo{volume}{77}}, \bibinfo{pages}{5194}.

\bibitem{boll:2013}
\bibinfo{author}{\bibnamefont{Boll}, \bibfnamefont{R.}},
  \bibinfo{author}{\bibfnamefont{D.}~\bibnamefont{Anielski}},
  \bibinfo{author}{\bibfnamefont{C.}~\bibnamefont{Bostedt}},
  \bibinfo{author}{\bibfnamefont{J.~D.} \bibnamefont{Bozek}},
  \bibinfo{author}{\bibfnamefont{L.}~\bibnamefont{Christensen}},
  \bibinfo{author}{\bibfnamefont{R.}~\bibnamefont{Coffee}},
  \bibinfo{author}{\bibfnamefont{S.}~\bibnamefont{De}},
  \bibinfo{author}{\bibfnamefont{P.}~\bibnamefont{Decleva}},
  \bibinfo{author}{\bibfnamefont{S.~W.} \bibnamefont{Epp}},
  \bibinfo{author}{\bibfnamefont{B.}~\bibnamefont{Erk}},
  \bibinfo{author}{\bibfnamefont{L.}~\bibnamefont{Foucar}},
  \bibinfo{author}{\bibfnamefont{F.}~\bibnamefont{Krasniqi}}, \emph{et~al.},
  \bibinfo{year}{2013}, \bibinfo{journal}{Phys. Rev. A}
  \textbf{\bibinfo{volume}{88}}, \bibinfo{pages}{061402}.

\bibitem{BomblePRA10}
\bibinfo{author}{\bibnamefont{Bomble}, \bibfnamefont{L.}},
  \bibinfo{author}{\bibfnamefont{P.}~\bibnamefont{Pellegrini}},
  \bibinfo{author}{\bibfnamefont{P.}~\bibnamefont{Ghesqui\`ere}}, and
  \bibinfo{author}{\bibfnamefont{M.}~\bibnamefont{Desouter-Lecomte}},
  \bibinfo{year}{2010}, \bibinfo{journal}{Phys. Rev. A}
  \textbf{\bibinfo{volume}{82}}, \bibinfo{pages}{062323}.

\bibitem{boscain:2012}
\bibinfo{author}{\bibnamefont{Boscain}, \bibfnamefont{U.}},
  \bibinfo{author}{\bibfnamefont{M.}~\bibnamefont{Caponigro}},
  \bibinfo{author}{\bibfnamefont{T.}~\bibnamefont{Chambrion}}, and
  \bibinfo{author}{\bibfnamefont{M.}~\bibnamefont{Sigalotti}},
  \bibinfo{year}{2012}, \bibinfo{journal}{Comm. Math. Phys.}
  \textbf{\bibinfo{volume}{311}}, \bibinfo{pages}{423}.

\bibitem{boscain:2014}
\bibinfo{author}{\bibnamefont{Boscain}, \bibfnamefont{U.}},
  \bibinfo{author}{\bibfnamefont{M.}~\bibnamefont{Caponigro}}, and
  \bibinfo{author}{\bibfnamefont{M.}~\bibnamefont{Sigalotti}},
  \bibinfo{year}{2014}, \bibinfo{journal}{J. Diff. Eq.}
  \textbf{\bibinfo{volume}{256}}, \bibinfo{pages}{3524}.

\bibitem{boscain:2015}
\bibinfo{author}{\bibnamefont{Boscain}, \bibfnamefont{U.}},
  \bibinfo{author}{\bibfnamefont{J.-P.} \bibnamefont{Gauthier}},
  \bibinfo{author}{\bibfnamefont{F.}~\bibnamefont{Rossi}}, and
  \bibinfo{author}{\bibfnamefont{M.}~\bibnamefont{Sigalotti}},
  \bibinfo{year}{2015}, \bibinfo{journal}{Comm. Math. Phys.}
  \textbf{\bibinfo{volume}{333}}, \bibinfo{pages}{1225}.

\bibitem{BrennenNJP07}
\bibinfo{author}{\bibnamefont{Brennen}, \bibfnamefont{G.~K.}},
  \bibinfo{author}{\bibfnamefont{A.}~\bibnamefont{Micheli}}, and
  \bibinfo{author}{\bibfnamefont{P.}~\bibnamefont{Zoller}},
  \bibinfo{year}{2007}, \bibinfo{journal}{New J. Phys.}
  \textbf{\bibinfo{volume}{9}}(\bibinfo{number}{5}), \bibinfo{pages}{138}.

\bibitem{BresselPRL12}
\bibinfo{author}{\bibnamefont{Bressel}, \bibfnamefont{U.}},
  \bibinfo{author}{\bibfnamefont{A.}~\bibnamefont{Borodin}},
  \bibinfo{author}{\bibfnamefont{J.}~\bibnamefont{Shen}},
  \bibinfo{author}{\bibfnamefont{M.}~\bibnamefont{Hansen}},
  \bibinfo{author}{\bibfnamefont{I.}~\bibnamefont{Ernsting}}, and
  \bibinfo{author}{\bibfnamefont{S.}~\bibnamefont{Schiller}},
  \bibinfo{year}{2012}, \bibinfo{journal}{Phys. Rev. Lett.}
  \textbf{\bibinfo{volume}{108}}, \bibinfo{pages}{183003}.

\bibitem{Brif:10}
\bibinfo{author}{\bibnamefont{Brif}, \bibfnamefont{C.}},
  \bibinfo{author}{\bibfnamefont{R.}~\bibnamefont{Chakrabarti}}, and
  \bibinfo{author}{\bibfnamefont{H.}~\bibnamefont{Rabitz}},
  \bibinfo{year}{2010}, \bibinfo{journal}{New J. Phys.}
  \textbf{\bibinfo{volume}{12}}, \bibinfo{pages}{075008}.

\bibitem{BrixnerCPC03}
\bibinfo{author}{\bibnamefont{Brixner}, \bibfnamefont{T.}}, and
  \bibinfo{author}{\bibfnamefont{G.}~\bibnamefont{Gerber}},
  \bibinfo{year}{2003}, \bibinfo{journal}{ChemPhysChem}
  \textbf{\bibinfo{volume}{4}}(\bibinfo{number}{5}), \bibinfo{pages}{418}.

\bibitem{brooks:1979}
\bibinfo{author}{\bibnamefont{Brooks}, \bibfnamefont{P.~R.}},
  \bibinfo{author}{\bibfnamefont{J.~S.} \bibnamefont{McKillop}}, and
  \bibinfo{author}{\bibfnamefont{H.~G.} \bibnamefont{Pippin}},
  \bibinfo{year}{1979}, \bibinfo{journal}{Chem. Phys. Lett.}
  \textbf{\bibinfo{volume}{66}}, \bibinfo{pages}{144}.

\bibitem{BuchlerZollerPRL07}
\bibinfo{author}{\bibnamefont{B{\"u}chler}, \bibfnamefont{H.~P.}},
  \bibinfo{author}{\bibfnamefont{E.}~\bibnamefont{Demler}},
  \bibinfo{author}{\bibfnamefont{M.}~\bibnamefont{Lukin}},
  \bibinfo{author}{\bibfnamefont{A.}~\bibnamefont{Micheli}},
  \bibinfo{author}{\bibfnamefont{N.}~\bibnamefont{Prokof'ev}},
  \bibinfo{author}{\bibfnamefont{G.}~\bibnamefont{Pupillo}}, and
  \bibinfo{author}{\bibfnamefont{P.}~\bibnamefont{Zoller}},
  \bibinfo{year}{2007}, \bibinfo{journal}{Phys. Rev. Lett.}
  \textbf{\bibinfo{volume}{98}}, \bibinfo{pages}{060404}.

\bibitem{BuchlerNatPhys07}
\bibinfo{author}{\bibnamefont{{B{\"u}chler~\textit{et al.}}},
  \bibfnamefont{H.~P.}}, \bibinfo{year}{2007}, \bibinfo{journal}{Nature
  Physics} \textbf{\bibinfo{volume}{3}}, \bibinfo{pages}{726}.

\bibitem{bukh:2006}
\bibinfo{author}{\bibnamefont{Buck}, \bibfnamefont{U.}}, and
  \bibinfo{author}{\bibfnamefont{M.}~\bibnamefont{Farnik}},
  \bibinfo{year}{2006}, \bibinfo{journal}{Int. Rev. Phys. Chem.}
  \textbf{\bibinfo{volume}{25}}, \bibinfo{pages}{583}.

\bibitem{BuhmannPRA08}
\bibinfo{author}{\bibnamefont{Buhmann}, \bibfnamefont{S.~Y.}},
  \bibinfo{author}{\bibfnamefont{M.~R.} \bibnamefont{Tarbutt}},
  \bibinfo{author}{\bibfnamefont{S.}~\bibnamefont{Scheel}}, and
  \bibinfo{author}{\bibfnamefont{E.~A.} \bibnamefont{Hinds}},
  \bibinfo{year}{2008}, \bibinfo{journal}{Phys. Rev. A}
  \textbf{\bibinfo{volume}{78}}, \bibinfo{pages}{052901}.

\bibitem{BunkerBook}
\bibinfo{author}{\bibnamefont{Bunker}, \bibfnamefont{P.~R.}}, and
  \bibinfo{author}{\bibfnamefont{P.}~\bibnamefont{Jensen}},
  \bibinfo{year}{2012}, \bibinfo{journal}{NRC Research Press Canada}
  \bibinfo{note}{2nd ed.}

\bibitem{ByrdPRL12}
\bibinfo{author}{\bibnamefont{Byrd}, \bibfnamefont{J.~N.}},
  \bibinfo{author}{\bibfnamefont{J.~A.} \bibnamefont{Montgomery}}, and
  \bibinfo{author}{\bibfnamefont{R.}~\bibnamefont{C\^ot\'e}},
  \bibinfo{year}{2012}, \bibinfo{journal}{Phys. Rev. Lett.}
  \textbf{\bibinfo{volume}{109}}, \bibinfo{pages}{083003}.

\bibitem{CaiFriCCCC01}
\bibinfo{author}{\bibnamefont{Cai}, \bibfnamefont{L.}}, and
  \bibinfo{author}{\bibfnamefont{B.}~\bibnamefont{Friedrich}},
  \bibinfo{year}{2001}, \bibinfo{journal}{Collect. Czech. Chem. Commun.}
  \textbf{\bibinfo{volume}{66}}, \bibinfo{pages}{991}.

\bibitem{cai:2001b}
\bibinfo{author}{\bibnamefont{Cai}, \bibfnamefont{L.}},
  \bibinfo{author}{\bibfnamefont{J.}~\bibnamefont{Marango}}, and
  \bibinfo{author}{\bibfnamefont{B.}~\bibnamefont{Friedrich}},
  \bibinfo{year}{2001}, \bibinfo{journal}{Phys. Rev. Lett.}
  \textbf{\bibinfo{volume}{86}}, \bibinfo{pages}{775}.

\bibitem{CallegariPRL99}
\bibinfo{author}{\bibnamefont{Callegari}, \bibfnamefont{C.}},
  \bibinfo{author}{\bibfnamefont{A.}~\bibnamefont{Conjusteau}},
  \bibinfo{author}{\bibfnamefont{I.}~\bibnamefont{Reinhard}},
  \bibinfo{author}{\bibfnamefont{K.~K.} \bibnamefont{Lehmann}},
  \bibinfo{author}{\bibfnamefont{G.}~\bibnamefont{Scoles}}, and
  \bibinfo{author}{\bibfnamefont{F.}~\bibnamefont{Dalfovo}},
  \bibinfo{year}{1999}, \bibinfo{journal}{Phys. Rev. Lett.}
  \textbf{\bibinfo{volume}{83}}, \bibinfo{pages}{5058}.

\bibitem{CalvinJPCL18}
\bibinfo{author}{\bibnamefont{Calvin}, \bibfnamefont{A.~T.}}, and
  \bibinfo{author}{\bibfnamefont{K.~R.} \bibnamefont{Brown}},
  \bibinfo{year}{2018}, \bibinfo{journal}{J. Phys. Chem. Lett.}
  \textbf{\bibinfo{volume}{9}}(\bibinfo{number}{19}), \bibinfo{pages}{5797}.

\bibitem{CameronPRA2016}
\bibinfo{author}{\bibnamefont{Cameron}, \bibfnamefont{R.~P.}},
  \bibinfo{author}{\bibfnamefont{J.~B.} \bibnamefont{G\"otte}}, and
  \bibinfo{author}{\bibfnamefont{S.~M.} \bibnamefont{Barnett}},
  \bibinfo{year}{2016}, \bibinfo{journal}{Phys. Rev. A}
  \textbf{\bibinfo{volume}{94}}, \bibinfo{pages}{032505}.

\bibitem{CapogrossoSansonePRL10}
\bibinfo{author}{\bibnamefont{Capogrosso-Sansone}, \bibfnamefont{B.}},
  \bibinfo{author}{\bibfnamefont{C.}~\bibnamefont{Trefzger}},
  \bibinfo{author}{\bibfnamefont{M.}~\bibnamefont{Lewenstein}},
  \bibinfo{author}{\bibfnamefont{P.}~\bibnamefont{Zoller}}, and
  \bibinfo{author}{\bibfnamefont{G.}~\bibnamefont{Pupillo}},
  \bibinfo{year}{2010}, \bibinfo{journal}{Phys. Rev. Lett.}
  \textbf{\bibinfo{volume}{104}}, \bibinfo{pages}{125301}.

\bibitem{ChakrabortyJPB11}
\bibinfo{author}{\bibnamefont{Chakraborty}, \bibfnamefont{D.}},
  \bibinfo{author}{\bibfnamefont{J.}~\bibnamefont{Hazra}}, and
  \bibinfo{author}{\bibfnamefont{B.}~\bibnamefont{Deb}}, \bibinfo{year}{2011},
  \bibinfo{journal}{J. Phys. B} \textbf{\bibinfo{volume}{44}},
  \bibinfo{pages}{095201}.

\bibitem{chambrion:2009}
\bibinfo{author}{\bibnamefont{Chambrion}, \bibfnamefont{T.}},
  \bibinfo{author}{\bibfnamefont{P.}~\bibnamefont{Mason}},
  \bibinfo{author}{\bibfnamefont{M.}~\bibnamefont{Sigalotti}}, and
  \bibinfo{author}{\bibfnamefont{U.}~\bibnamefont{Boscain}},
  \bibinfo{year}{2009}, \bibinfo{journal}{Ann. Inst. H. Poincar\'e}
  \textbf{\bibinfo{volume}{26}}, \bibinfo{pages}{329}.

\bibitem{charron:2007}
\bibinfo{author}{\bibnamefont{Charron}, \bibfnamefont{E.}},
  \bibinfo{author}{\bibfnamefont{P.}~\bibnamefont{Milman}},
  \bibinfo{author}{\bibfnamefont{A.}~\bibnamefont{Keller}}, and
  \bibinfo{author}{\bibfnamefont{O.}~\bibnamefont{Atabek}},
  \bibinfo{year}{2007}, \bibinfo{journal}{Phys. Rev. A}
  \textbf{\bibinfo{volume}{75}}, \bibinfo{pages}{033414}.

\bibitem{ChatterleyPRL17}
\bibinfo{author}{\bibnamefont{Chatterley}, \bibfnamefont{A.~S.}},
  \bibinfo{author}{\bibfnamefont{B.}~\bibnamefont{Shepperson}}, and
  \bibinfo{author}{\bibfnamefont{H.}~\bibnamefont{Stapelfeldt}},
  \bibinfo{year}{2017}, \bibinfo{journal}{Phys. Rev. Lett.}
  \textbf{\bibinfo{volume}{119}}, \bibinfo{pages}{073202}.

\bibitem{chen:2010}
\bibinfo{author}{\bibnamefont{Chen}, \bibfnamefont{C.}},
  \bibinfo{author}{\bibfnamefont{J.}~\bibnamefont{Wu}}, and
  \bibinfo{author}{\bibfnamefont{H.}~\bibnamefont{Zeng}}, \bibinfo{year}{2010},
  \bibinfo{journal}{Phys. Rev. A} \textbf{\bibinfo{volume}{82}},
  \bibinfo{pages}{033409}.

\bibitem{Cherepanov17}
\bibinfo{author}{\bibnamefont{Cherepanov}, \bibfnamefont{I.}}, and
  \bibinfo{author}{\bibfnamefont{M.}~\bibnamefont{Lemeshko}},
  \bibinfo{year}{2017}, \bibinfo{journal}{Phys. Rev. Materials}
  \textbf{\bibinfo{volume}{1}}, \bibinfo{pages}{035602}.

\bibitem{chi:2008}
\bibinfo{author}{\bibnamefont{Chi}, \bibfnamefont{F.}}, and
  \bibinfo{author}{\bibfnamefont{J.}~\bibnamefont{Chen}}, \bibinfo{year}{2008},
  \bibinfo{journal}{Chemical Physics Letters}
  \textbf{\bibinfo{volume}{465}}(\bibinfo{number}{4}), \bibinfo{pages}{299 }.

\bibitem{Child:91}
\bibinfo{author}{\bibnamefont{Child}, \bibfnamefont{M.~S.}},
  \bibinfo{year}{1991}, \emph{\bibinfo{title}{Semiclassical mechanics with
  molecular applications}} (\bibinfo{publisher}{Clarendon Press, Oxford}).

\bibitem{child:2007}
\bibinfo{author}{\bibnamefont{Child}, \bibfnamefont{M.~S.}},
  \bibinfo{year}{2007}, \bibinfo{journal}{Adv. Chem. Phys.}
  \textbf{\bibinfo{volume}{136}}, \bibinfo{pages}{39}.

\bibitem{ChinRMP10}
\bibinfo{author}{\bibnamefont{Chin}, \bibfnamefont{C.}},
  \bibinfo{author}{\bibfnamefont{R.}~\bibnamefont{Grimm}},
  \bibinfo{author}{\bibfnamefont{P.}~\bibnamefont{Julienne}}, and
  \bibinfo{author}{\bibfnamefont{E.}~\bibnamefont{Tiesinga}},
  \bibinfo{year}{2010}, \bibinfo{journal}{Rev. Mod. Phys.}
  \textbf{\bibinfo{volume}{82}}(\bibinfo{number}{2}), \bibinfo{pages}{1225}.

\bibitem{cho:1991}
\bibinfo{author}{\bibnamefont{Cho}, \bibfnamefont{V.~A.}}, and
  \bibinfo{author}{\bibfnamefont{R.~B.} \bibnamefont{Bernstein}},
  \bibinfo{year}{1991}, \bibinfo{journal}{J. Phys. Chem.}
  \textbf{\bibinfo{volume}{95}}, \bibinfo{pages}{8129}.

\bibitem{ChouNat2017}
\bibinfo{author}{\bibnamefont{Chou}, \bibfnamefont{C.}},
  \bibinfo{author}{\bibfnamefont{C.}~\bibnamefont{Kurz}},
  \bibinfo{author}{\bibfnamefont{D.~B.} \bibnamefont{Hume}},
  \bibinfo{author}{\bibfnamefont{P.~N.} \bibnamefont{Plessow}},
  \bibinfo{author}{\bibfnamefont{D.~R.} \bibnamefont{Leibrandt}}, and
  \bibinfo{author}{\bibfnamefont{D.}~\bibnamefont{Leibfried}},
  \bibinfo{year}{2017}, \bibinfo{journal}{Nature}
  \textbf{\bibinfo{volume}{545}}, \bibinfo{pages}{203}.

\bibitem{ChristensenPRA16}
\bibinfo{author}{\bibnamefont{Christensen}, \bibfnamefont{L.}},
  \bibinfo{author}{\bibfnamefont{L.}~\bibnamefont{Christiansen}},
  \bibinfo{author}{\bibfnamefont{B.}~\bibnamefont{Shepperson}}, and
  \bibinfo{author}{\bibfnamefont{H.}~\bibnamefont{Stapelfeldt}},
  \bibinfo{year}{2016}, \bibinfo{journal}{Phys. Rev. A}
  \textbf{\bibinfo{volume}{94}}, \bibinfo{pages}{023410}.

\bibitem{christensen:2014}
\bibinfo{author}{\bibnamefont{Christensen}, \bibfnamefont{L.}},
  \bibinfo{author}{\bibfnamefont{J.~H.} \bibnamefont{Nielsen}},
  \bibinfo{author}{\bibfnamefont{C.~B.} \bibnamefont{Brandt}},
  \bibinfo{author}{\bibfnamefont{C.~B.} \bibnamefont{Madsen}},
  \bibinfo{author}{\bibfnamefont{L.~B.} \bibnamefont{Madsen}},
  \bibinfo{author}{\bibfnamefont{C.~S.} \bibnamefont{Slater}},
  \bibinfo{author}{\bibfnamefont{A.}~\bibnamefont{Lauer}},
  \bibinfo{author}{\bibfnamefont{M.}~\bibnamefont{Brouard}},
  \bibinfo{author}{\bibfnamefont{M.~P.} \bibnamefont{Johansson}},
  \bibinfo{author}{\bibfnamefont{B.}~\bibnamefont{Shepperson}}, and
  \bibinfo{author}{\bibfnamefont{H.}~\bibnamefont{Stapelfeldt}},
  \bibinfo{year}{2014}, \bibinfo{journal}{Phys. Rev. Lett.}
  \textbf{\bibinfo{volume}{113}}, \bibinfo{pages}{073005}.

\bibitem{ChristiansenPRA16}
\bibinfo{author}{\bibnamefont{Christiansen}, \bibfnamefont{L.}},
  \bibinfo{author}{\bibfnamefont{J.~H.} \bibnamefont{Nielsen}},
  \bibinfo{author}{\bibfnamefont{L.}~\bibnamefont{Christensen}},
  \bibinfo{author}{\bibfnamefont{B.}~\bibnamefont{Shepperson}},
  \bibinfo{author}{\bibfnamefont{D.}~\bibnamefont{Pentlehner}}, and
  \bibinfo{author}{\bibfnamefont{H.}~\bibnamefont{Stapelfeldt}},
  \bibinfo{year}{2016}, \bibinfo{journal}{Phys. Rev. A}
  \textbf{\bibinfo{volume}{93}}, \bibinfo{pages}{023411}.

\bibitem{ChristiansenPRA15}
\bibinfo{author}{\bibnamefont{Christiansen}, \bibfnamefont{L.}},
  \bibinfo{author}{\bibfnamefont{J.~H.} \bibnamefont{Nielsen}},
  \bibinfo{author}{\bibfnamefont{D.}~\bibnamefont{Tobias}},
  \bibinfo{author}{\bibfnamefont{V.}~\bibnamefont{Pentlehner}},
  \bibinfo{author}{\bibfnamefont{J.~G.} \bibnamefont{Underwood}}, and
  \bibinfo{author}{\bibfnamefont{H.}~\bibnamefont{Stapelfeldt}},
  \bibinfo{year}{2015}, \bibinfo{journal}{Phys. Rev. A}
  \textbf{\bibinfo{volume}{92}}, \bibinfo{pages}{1050}.

\bibitem{CooperPRL09}
\bibinfo{author}{\bibnamefont{Cooper}, \bibfnamefont{N.~R.}}, and
  \bibinfo{author}{\bibfnamefont{G.~V.} \bibnamefont{Shlyapnikov}},
  \bibinfo{year}{2009}, \bibinfo{journal}{Phys. Rev. Lett.}
  \textbf{\bibinfo{volume}{103}}, \bibinfo{pages}{155302}.

\bibitem{cottonBook}
\bibinfo{author}{\bibnamefont{Cotton}, \bibfnamefont{F.~A.}},
  \bibinfo{year}{1990}, \emph{\bibinfo{title}{Chemical Applications of Group
  Theory}} (\bibinfo{publisher}{John Wiley and Sons, New York}).

\bibitem{coudert:2015}
\bibinfo{author}{\bibnamefont{Coudert}, \bibfnamefont{L.~H.}},
  \bibinfo{year}{2015}, \bibinfo{journal}{Phys. Rev. A}
  \textbf{\bibinfo{volume}{91}}, \bibinfo{pages}{013402}.

\bibitem{coudert:2017}
\bibinfo{author}{\bibnamefont{Coudert}, \bibfnamefont{L.~H.}},
  \bibinfo{year}{2017}, \bibinfo{journal}{J. Chem. Phys.}
  \textbf{\bibinfo{volume}{146}}(\bibinfo{number}{2}), \bibinfo{pages}{024303}.

\bibitem{coudert:2018}
\bibinfo{author}{\bibnamefont{Coudert}, \bibfnamefont{L.~H.}},
  \bibinfo{year}{2018}, \bibinfo{journal}{J. Chem. Phys.}
  \textbf{\bibinfo{volume}{148}}(\bibinfo{number}{9}), \bibinfo{pages}{094306}.

\bibitem{coudert:2011}
\bibinfo{author}{\bibnamefont{Coudert}, \bibfnamefont{L.~H.}},
  \bibinfo{author}{\bibfnamefont{L.~F.} \bibnamefont{Pacios}}, and
  \bibinfo{author}{\bibfnamefont{J.}~\bibnamefont{Ortigoso}},
  \bibinfo{year}{2011}, \bibinfo{journal}{Phys. Rev. Lett.}
  \textbf{\bibinfo{volume}{107}}, \bibinfo{pages}{113004}.

\bibitem{CoveyNatComm16}
\bibinfo{author}{\bibnamefont{Covey}, \bibfnamefont{J.~P.}},
  \bibinfo{author}{\bibfnamefont{S.~A.} \bibnamefont{Moses}},
  \bibinfo{author}{\bibfnamefont{M.}~\bibnamefont{G{\"a}rttner}},
  \bibinfo{author}{\bibfnamefont{A.}~\bibnamefont{Safavi-Naini}},
  \bibinfo{author}{\bibfnamefont{M.~T.} \bibnamefont{Miecnikowski}},
  \bibinfo{author}{\bibfnamefont{Z.}~\bibnamefont{Fu}},
  \bibinfo{author}{\bibfnamefont{J.}~\bibnamefont{Schachenmayer}},
  \bibinfo{author}{\bibfnamefont{P.~S.} \bibnamefont{Julienne}},
  \bibinfo{author}{\bibfnamefont{A.~M.} \bibnamefont{Rey}},
  \bibinfo{author}{\bibfnamefont{D.~S.} \bibnamefont{Jin}}, and
  \bibinfo{author}{\bibfnamefont{J.}~\bibnamefont{Ye}}, \bibinfo{year}{2016},
  \bibinfo{journal}{Nature Communications} \textbf{\bibinfo{volume}{7}},
  \bibinfo{pages}{11279 EP }.

\bibitem{CrubellierNJP15}
\bibinfo{author}{\bibnamefont{Crubellier}, \bibfnamefont{A.}},
  \bibinfo{author}{\bibfnamefont{R.}~\bibnamefont{Gonz\'alez-F\'erez}},
  \bibinfo{author}{\bibfnamefont{C.~P.} \bibnamefont{Koch}}, and
  \bibinfo{author}{\bibfnamefont{E.}~\bibnamefont{Luc-Koenig}},
  \bibinfo{year}{2015}{\natexlab{a}}, \bibinfo{journal}{New J. Phys.}
  \textbf{\bibinfo{volume}{17}}, \bibinfo{pages}{045020}.

\bibitem{CrubellierNJP15b}
\bibinfo{author}{\bibnamefont{Crubellier}, \bibfnamefont{A.}},
  \bibinfo{author}{\bibfnamefont{R.}~\bibnamefont{Gonz\'alez-F\'erez}},
  \bibinfo{author}{\bibfnamefont{C.~P.} \bibnamefont{Koch}}, and
  \bibinfo{author}{\bibfnamefont{E.}~\bibnamefont{Luc-Koenig}},
  \bibinfo{year}{2015}{\natexlab{b}}, \bibinfo{journal}{New J. Phys.}
  \textbf{\bibinfo{volume}{17}}, \bibinfo{pages}{045022}.

\bibitem{CrubellierPRA17}
\bibinfo{author}{\bibnamefont{Crubellier}, \bibfnamefont{A.}},
  \bibinfo{author}{\bibfnamefont{R.}~\bibnamefont{Gonz\'alez-F\'erez}},
  \bibinfo{author}{\bibfnamefont{C.~P.} \bibnamefont{Koch}}, and
  \bibinfo{author}{\bibfnamefont{E.}~\bibnamefont{Luc-Koenig}},
  \bibinfo{year}{2017}, \bibinfo{journal}{Phys. Rev. A}
  \textbf{\bibinfo{volume}{95}}, \bibinfo{pages}{023405}.

\bibitem{Crubellier18b}
\bibinfo{author}{\bibnamefont{Crubellier}, \bibfnamefont{A.}},
  \bibinfo{author}{\bibfnamefont{R.}~\bibnamefont{Gonz\'alez-F\'erez}},
  \bibinfo{author}{\bibfnamefont{C.~P.} \bibnamefont{Koch}}, and
  \bibinfo{author}{\bibfnamefont{E.}~\bibnamefont{Luc-Koenig}},
  \bibinfo{year}{2019}, \bibinfo{journal}{Phys. Rev. A}
  \textbf{\bibinfo{volume}{99}}, \bibinfo{pages}{032710},
  \bibinfo{note}{arXiv:1807.05787}.

\bibitem{cuisset:2012}
\bibinfo{author}{\bibnamefont{Cuisset}, \bibfnamefont{A.}},
  \bibinfo{author}{\bibfnamefont{O.}~\bibnamefont{Pirali}}, and
  \bibinfo{author}{\bibfnamefont{D.~A.} \bibnamefont{Sadovski\'{\i}}},
  \bibinfo{year}{2012}, \bibinfo{journal}{Phys. Rev. Lett.}
  \textbf{\bibinfo{volume}{109}}, \bibinfo{pages}{094101}.

\bibitem{cushman}
\bibinfo{author}{\bibnamefont{Cushman}, \bibfnamefont{R.~H.}}, and
  \bibinfo{author}{\bibfnamefont{L.}~\bibnamefont{Bates}},
  \bibinfo{year}{1997}, \emph{\bibinfo{title}{Global Aspects of Classical
  Integrable Systems}} (\bibinfo{publisher}{Birkhauser, Basel}).

\bibitem{daems:2005}
\bibinfo{author}{\bibnamefont{Daems}, \bibfnamefont{D.}},
  \bibinfo{author}{\bibfnamefont{S.}~\bibnamefont{Gu\'erin}},
  \bibinfo{author}{\bibfnamefont{E.}~\bibnamefont{Hertz}},
  \bibinfo{author}{\bibfnamefont{H.~R.} \bibnamefont{Jauslin}},
  \bibinfo{author}{\bibfnamefont{B.}~\bibnamefont{Lavorel}}, and
  \bibinfo{author}{\bibfnamefont{O.}~\bibnamefont{Faucher}},
  \bibinfo{year}{2005}{\natexlab{a}}, \bibinfo{journal}{Phys. Rev. Lett.}
  \textbf{\bibinfo{volume}{95}}, \bibinfo{pages}{063005}.

\bibitem{daems:2005b}
\bibinfo{author}{\bibnamefont{Daems}, \bibfnamefont{D.}},
  \bibinfo{author}{\bibfnamefont{S.}~\bibnamefont{Gu\'erin}},
  \bibinfo{author}{\bibfnamefont{D.}~\bibnamefont{Sugny}}, and
  \bibinfo{author}{\bibfnamefont{H.~R.} \bibnamefont{Jauslin}},
  \bibinfo{year}{2005}{\natexlab{b}}, \bibinfo{journal}{Phys. Rev. Lett.}
  \textbf{\bibinfo{volume}{94}}, \bibinfo{pages}{153003}.

\bibitem{alessandro_book}
\bibinfo{author}{\bibnamefont{D'Alessandro}, \bibfnamefont{D.}},
  \bibinfo{year}{2008}, \emph{\bibinfo{title}{Introduction to Quantum Control
  and Dynamics}} (\bibinfo{publisher}{Chapman and Hall, Boca Raton}).

\bibitem{damari:2016}
\bibinfo{author}{\bibnamefont{Damari}, \bibfnamefont{R.}},
  \bibinfo{author}{\bibfnamefont{S.}~\bibnamefont{Kallush}}, and
  \bibinfo{author}{\bibfnamefont{S.}~\bibnamefont{Fleischer}},
  \bibinfo{year}{2016}, \bibinfo{journal}{Phys. Rev. Lett.}
  \textbf{\bibinfo{volume}{117}}, \bibinfo{pages}{103001}.

\bibitem{damari:2017}
\bibinfo{author}{\bibnamefont{Damari}, \bibfnamefont{R.}},
  \bibinfo{author}{\bibfnamefont{D.}~\bibnamefont{Rosenberg}}, and
  \bibinfo{author}{\bibfnamefont{S.}~\bibnamefont{Fleischer}},
  \bibinfo{year}{2017}, \bibinfo{journal}{Phys. Rev. Lett.}
  \textbf{\bibinfo{volume}{119}}, \bibinfo{pages}{033002}.

\bibitem{amico:2009}
\bibinfo{author}{\bibnamefont{D'Amico}, \bibfnamefont{C.}},
  \bibinfo{author}{\bibfnamefont{M.}~\bibnamefont{Tondusson}},
  \bibinfo{author}{\bibfnamefont{J.}~\bibnamefont{Degert}}, and
  \bibinfo{author}{\bibfnamefont{E.}~\bibnamefont{Freysz}},
  \bibinfo{year}{2009}, \bibinfo{journal}{Opt. Express}
  \textbf{\bibinfo{volume}{17}}, \bibinfo{pages}{592}.

\bibitem{DanzlSci08}
\bibinfo{author}{\bibnamefont{Danzl}, \bibfnamefont{J.~G.}},
  \bibinfo{author}{\bibfnamefont{E.}~\bibnamefont{Haller}},
  \bibinfo{author}{\bibfnamefont{M.}~\bibnamefont{Gustavsson}},
  \bibinfo{author}{\bibfnamefont{M.~J.} \bibnamefont{Mark}},
  \bibinfo{author}{\bibfnamefont{R.}~\bibnamefont{Hart}},
  \bibinfo{author}{\bibfnamefont{N.}~\bibnamefont{Bouloufa}},
  \bibinfo{author}{\bibfnamefont{O.}~\bibnamefont{Dulieu}},
  \bibinfo{author}{\bibfnamefont{H.}~\bibnamefont{Ritsch}}, and
  \bibinfo{author}{\bibfnamefont{H.-C.} \bibnamefont{N\"agerl}},
  \bibinfo{year}{2008}, \bibinfo{journal}{Science}
  \textbf{\bibinfo{volume}{321}}(\bibinfo{number}{5892}),
  \bibinfo{pages}{1062}.

\bibitem{de:2009}
\bibinfo{author}{\bibnamefont{De}, \bibfnamefont{S.}},
  \bibinfo{author}{\bibfnamefont{I.}~\bibnamefont{Znakovskaya}},
  \bibinfo{author}{\bibfnamefont{D.}~\bibnamefont{Ray}},
  \bibinfo{author}{\bibfnamefont{F.}~\bibnamefont{Anis}},
  \bibinfo{author}{\bibfnamefont{N.~G.} \bibnamefont{Johnson}},
  \bibinfo{author}{\bibfnamefont{I.~A.} \bibnamefont{Bocharova}},
  \bibinfo{author}{\bibfnamefont{M.}~\bibnamefont{Magrakvelidze}},
  \bibinfo{author}{\bibfnamefont{B.~D.} \bibnamefont{Esry}},
  \bibinfo{author}{\bibfnamefont{C.~L.} \bibnamefont{Cocke}},
  \bibinfo{author}{\bibfnamefont{I.~V.} \bibnamefont{Litvinyuk}}, and
  \bibinfo{author}{\bibfnamefont{M.~F.} \bibnamefont{Kling}},
  \bibinfo{year}{2009}, \bibinfo{journal}{Phys. Rev. Lett.}
  \textbf{\bibinfo{volume}{103}}, \bibinfo{pages}{153002}.

\bibitem{NabanitaPCCP13}
\bibinfo{author}{\bibnamefont{Deb}, \bibfnamefont{N.}},
  \bibinfo{author}{\bibfnamefont{B.~R.} \bibnamefont{Heazlewood}},
  \bibinfo{author}{\bibfnamefont{M.~T.} \bibnamefont{Bell}}, and
  \bibinfo{author}{\bibfnamefont{T.~P.} \bibnamefont{Softley}},
  \bibinfo{year}{2013}, \bibinfo{journal}{Phys. Chem. Chem. Phys.}
  \textbf{\bibinfo{volume}{15}}, \bibinfo{pages}{14270}.

\bibitem{DeiglmayrPRL08}
\bibinfo{author}{\bibnamefont{Deiglmayr}, \bibfnamefont{J.}},
  \bibinfo{author}{\bibfnamefont{A.}~\bibnamefont{Grochola}},
  \bibinfo{author}{\bibfnamefont{M.}~\bibnamefont{Repp}},
  \bibinfo{author}{\bibfnamefont{K.}~\bibnamefont{M\"{o}rtlbauer}},
  \bibinfo{author}{\bibfnamefont{C.}~\bibnamefont{Gl\"{u}ck}},
  \bibinfo{author}{\bibfnamefont{J.}~\bibnamefont{Lange}},
  \bibinfo{author}{\bibfnamefont{O.}~\bibnamefont{Dulieu}},
  \bibinfo{author}{\bibfnamefont{R.}~\bibnamefont{Wester}}, and
  \bibinfo{author}{\bibfnamefont{M.}~\bibnamefont{Weidem\"{u}ller}},
  \bibinfo{year}{2008}, \bibinfo{journal}{Phys. Rev. Lett.}
  \textbf{\bibinfo{volume}{101}}, \bibinfo{pages}{133004}.

\bibitem{DeMillePRL02}
\bibinfo{author}{\bibnamefont{DeMille}, \bibfnamefont{D.}},
  \bibinfo{year}{2002}, \bibinfo{journal}{Phys. Rev. Lett.}
  \textbf{\bibinfo{volume}{88}}, \bibinfo{pages}{067901}.

\bibitem{Devreese15}
\bibinfo{author}{\bibnamefont{Devreese}, \bibfnamefont{J.~T.}},
  \bibinfo{year}{2015}, \bibinfo{journal}{arXiv:1012.4576v6} .

\bibitem{DiRosaEPJD04}
\bibinfo{author}{\bibnamefont{Di~Rosa}, \bibfnamefont{M.~D.}},
  \bibinfo{year}{2004}, \bibinfo{journal}{The European Physical Journal D -
  Atomic, Molecular, Optical and Plasma Physics}
  \textbf{\bibinfo{volume}{31}}(\bibinfo{number}{2}), \bibinfo{pages}{395}.

\bibitem{dion:2002}
\bibinfo{author}{\bibnamefont{Dion}, \bibfnamefont{C.~M.}},
  \bibinfo{author}{\bibfnamefont{A.}~\bibnamefont{Ben Haj-Yedder}},
  \bibinfo{author}{\bibfnamefont{E.}~\bibnamefont{Canc\`es}},
  \bibinfo{author}{\bibfnamefont{C.}~\bibnamefont{Le~Bris}},
  \bibinfo{author}{\bibfnamefont{A.}~\bibnamefont{Keller}}, and
  \bibinfo{author}{\bibfnamefont{O.}~\bibnamefont{Atabek}},
  \bibinfo{year}{2002}, \bibinfo{journal}{Phys. Rev. A}
  \textbf{\bibinfo{volume}{65}}, \bibinfo{pages}{063408}.

\bibitem{dion:2005}
\bibinfo{author}{\bibnamefont{Dion}, \bibfnamefont{C.~M.}},
  \bibinfo{author}{\bibfnamefont{A.}~\bibnamefont{Keller}}, and
  \bibinfo{author}{\bibfnamefont{O.}~\bibnamefont{Atabek}},
  \bibinfo{year}{2005}, \bibinfo{journal}{Phys. Rev. A}
  \textbf{\bibinfo{volume}{72}}, \bibinfo{pages}{023402}.

\bibitem{dion:1999}
\bibinfo{author}{\bibnamefont{Dion}, \bibfnamefont{C.~M.}},
  \bibinfo{author}{\bibfnamefont{A.}~\bibnamefont{Keller}},
  \bibinfo{author}{\bibfnamefont{O.}~\bibnamefont{Atabek}}, and
  \bibinfo{author}{\bibfnamefont{A.~D.} \bibnamefont{Bandrauk}},
  \bibinfo{year}{1999}, \bibinfo{journal}{Phys. Rev. A}
  \textbf{\bibinfo{volume}{59}}, \bibinfo{pages}{1382}.

\bibitem{DocajPRL16}
\bibinfo{author}{\bibnamefont{Do{\c{c}}aj}, \bibfnamefont{A.}},
  \bibinfo{author}{\bibfnamefont{M.~L.} \bibnamefont{Wall}},
  \bibinfo{author}{\bibfnamefont{R.}~\bibnamefont{Mukherjee}}, and
  \bibinfo{author}{\bibfnamefont{K.~R.~A.} \bibnamefont{Hazzard}},
  \bibinfo{year}{2016}, \bibinfo{journal}{Phys. Rev. Lett.}
  \textbf{\bibinfo{volume}{116}}, \bibinfo{pages}{135301}.

\bibitem{dooley:2003}
\bibinfo{author}{\bibnamefont{Dooley}, \bibfnamefont{P.~W.}},
  \bibinfo{author}{\bibfnamefont{I.~V.} \bibnamefont{Litvinyuk}},
  \bibinfo{author}{\bibfnamefont{K.~F.} \bibnamefont{Lee}},
  \bibinfo{author}{\bibfnamefont{D.~M.} \bibnamefont{Rayner}},
  \bibinfo{author}{\bibfnamefont{M.}~\bibnamefont{Spanner}},
  \bibinfo{author}{\bibfnamefont{D.~M.} \bibnamefont{Villeneuve}}, and
  \bibinfo{author}{\bibfnamefont{P.~B.} \bibnamefont{Corkum}},
  \bibinfo{year}{2003}, \bibinfo{journal}{Phys. Rev. A}
  \textbf{\bibinfo{volume}{68}}, \bibinfo{pages}{023406}.

\bibitem{DrewsenPRL04}
\bibinfo{author}{\bibnamefont{Drewsen}, \bibfnamefont{M.}},
  \bibinfo{author}{\bibfnamefont{A.}~\bibnamefont{Mortensen}},
  \bibinfo{author}{\bibfnamefont{R.}~\bibnamefont{Martinussen}},
  \bibinfo{author}{\bibfnamefont{P.}~\bibnamefont{Staanum}}, and
  \bibinfo{author}{\bibfnamefont{J.~L.} \bibnamefont{S\o{}rensen}},
  \bibinfo{year}{2004}, \bibinfo{journal}{Phys. Rev. Lett.}
  \textbf{\bibinfo{volume}{93}}, \bibinfo{pages}{243201}.

\bibitem{EfstathiouBook}
\bibinfo{author}{\bibnamefont{Efstathiou}, \bibfnamefont{K.}},
  \bibinfo{year}{2004}, \emph{\bibinfo{title}{Metamorphoses of Hamiltonian
  Systems with Symmetry}} (\bibinfo{publisher}{Lecture Notes in Mathematics
  Series-LNM 1864, Springer-Verlag, Heidelberg}).

\bibitem{jones:2014}
\bibinfo{author}{\bibnamefont{Egodapitiya}, \bibfnamefont{K.~N.}},
  \bibinfo{author}{\bibfnamefont{S.}~\bibnamefont{Li}}, and
  \bibinfo{author}{\bibfnamefont{R.~R.} \bibnamefont{Jones}},
  \bibinfo{year}{2014}, \bibinfo{journal}{Phys. Rev. Lett.}
  \textbf{\bibinfo{volume}{112}}, \bibinfo{pages}{103002}.

\bibitem{EibenbergerPRL17}
\bibinfo{author}{\bibnamefont{Eibenberger}, \bibfnamefont{S.}},
  \bibinfo{author}{\bibfnamefont{J.}~\bibnamefont{Doyle}}, and
  \bibinfo{author}{\bibfnamefont{D.}~\bibnamefont{Patterson}},
  \bibinfo{year}{2017}, \bibinfo{journal}{Phys. Rev. Lett.}
  \textbf{\bibinfo{volume}{118}}, \bibinfo{pages}{123002}.

\bibitem{faucher:2016}
\bibinfo{author}{\bibnamefont{Faucher}, \bibfnamefont{O.}},
  \bibinfo{author}{\bibfnamefont{E.}~\bibnamefont{Prost}},
  \bibinfo{author}{\bibfnamefont{E.}~\bibnamefont{Hertz}},
  \bibinfo{author}{\bibfnamefont{F.}~\bibnamefont{Billard}},
  \bibinfo{author}{\bibfnamefont{B.}~\bibnamefont{Lavorel}},
  \bibinfo{author}{\bibfnamefont{A.~A.} \bibnamefont{Milner}},
  \bibinfo{author}{\bibfnamefont{V.~A.} \bibnamefont{Milner}},
  \bibinfo{author}{\bibfnamefont{J.}~\bibnamefont{Zyss}}, and
  \bibinfo{author}{\bibfnamefont{I.~S.} \bibnamefont{Averbukh}},
  \bibinfo{year}{2016}, \bibinfo{journal}{Phys. Rev. A}
  \textbf{\bibinfo{volume}{94}}, \bibinfo{pages}{051402}.

\bibitem{filsinger:2009}
\bibinfo{author}{\bibnamefont{Filsinger}, \bibfnamefont{F.}},
  \bibinfo{author}{\bibfnamefont{J.}~\bibnamefont{K{\"u}pper}},
  \bibinfo{author}{\bibfnamefont{G.}~\bibnamefont{Meijer}},
  \bibinfo{author}{\bibfnamefont{L.}~\bibnamefont{Holmegaard}},
  \bibinfo{author}{\bibfnamefont{J.~H.} \bibnamefont{Nielsen}},
  \bibinfo{author}{\bibfnamefont{I.}~\bibnamefont{Nevo}},
  \bibinfo{author}{\bibfnamefont{J.~L.} \bibnamefont{Hansen}}, and
  \bibinfo{author}{\bibfnamefont{H.}~\bibnamefont{Stapelfeldt}},
  \bibinfo{year}{2009}, \bibinfo{journal}{J. Chem. Phys.}
  \textbf{\bibinfo{volume}{131}}(\bibinfo{number}{6}), \bibinfo{pages}{064309}.

\bibitem{fleischer:2006}
\bibinfo{author}{\bibnamefont{Fleischer}, \bibfnamefont{S.}},
  \bibinfo{author}{\bibfnamefont{I.~S.} \bibnamefont{Averbukh}}, and
  \bibinfo{author}{\bibfnamefont{Y.}~\bibnamefont{Prior}},
  \bibinfo{year}{2006}, \bibinfo{journal}{Phys. Rev. A}
  \textbf{\bibinfo{volume}{74}}, \bibinfo{pages}{041403}.

\bibitem{fleisher:2007}
\bibinfo{author}{\bibnamefont{Fleischer}, \bibfnamefont{S.}},
  \bibinfo{author}{\bibfnamefont{I.~S.} \bibnamefont{Averbukh}}, and
  \bibinfo{author}{\bibfnamefont{Y.}~\bibnamefont{Prior}},
  \bibinfo{year}{2007}, \bibinfo{journal}{Phys. Rev. Lett.}
  \textbf{\bibinfo{volume}{99}}, \bibinfo{pages}{093002}.

\bibitem{fleischer:2012}
\bibinfo{author}{\bibnamefont{Fleischer}, \bibfnamefont{S.}},
  \bibinfo{author}{\bibfnamefont{R.~W.} \bibnamefont{Field}}, and
  \bibinfo{author}{\bibfnamefont{K.~A.} \bibnamefont{Nelson}},
  \bibinfo{year}{2012}{\natexlab{a}}, \bibinfo{journal}{Phys. Rev. Lett.}
  \textbf{\bibinfo{volume}{109}}, \bibinfo{pages}{123603}.

\bibitem{fleischerrev:2012}
\bibinfo{author}{\bibnamefont{Fleischer}, \bibfnamefont{S.}},
  \bibinfo{author}{\bibfnamefont{Y.}~\bibnamefont{Khodorkovsky}},
  \bibinfo{author}{\bibfnamefont{E.}~\bibnamefont{Gershnabel}},
  \bibinfo{author}{\bibfnamefont{Y.}~\bibnamefont{Prior}}, and
  \bibinfo{author}{\bibfnamefont{I.~S.} \bibnamefont{Averbukh}},
  \bibinfo{year}{2012}{\natexlab{b}}, \bibinfo{journal}{Isr. J. Chem.}
  \textbf{\bibinfo{volume}{52}}, \bibinfo{pages}{414}.

\bibitem{feischer:2009}
\bibinfo{author}{\bibnamefont{Fleischer}, \bibfnamefont{S.}},
  \bibinfo{author}{\bibfnamefont{Y.}~\bibnamefont{Khodorkovsky}},
  \bibinfo{author}{\bibfnamefont{Y.}~\bibnamefont{Prior}}, and
  \bibinfo{author}{\bibfnamefont{I.~S.} \bibnamefont{Averbukh}},
  \bibinfo{year}{2009}, \bibinfo{journal}{New J. of Phys.}
  \textbf{\bibinfo{volume}{11}}, \bibinfo{pages}{105039}.

\bibitem{Nelson:11}
\bibinfo{author}{\bibnamefont{Fleischer}, \bibfnamefont{S.}},
  \bibinfo{author}{\bibfnamefont{Y.}~\bibnamefont{Zhou}},
  \bibinfo{author}{\bibfnamefont{R.~W.} \bibnamefont{Field}}, and
  \bibinfo{author}{\bibfnamefont{K.~A.} \bibnamefont{Nelson}},
  \bibinfo{year}{2011}, \bibinfo{journal}{Phys. Rev. Lett.}
  \textbf{\bibinfo{volume}{107}}, \bibinfo{pages}{163603}.

\bibitem{floss:2012b}
\bibinfo{author}{\bibnamefont{Flo\ss{}}, \bibfnamefont{J.}}, and
  \bibinfo{author}{\bibfnamefont{I.~S.} \bibnamefont{Averbukh}},
  \bibinfo{year}{2012}{\natexlab{a}}, \bibinfo{journal}{Phys. Rev. A}
  \textbf{\bibinfo{volume}{86}}, \bibinfo{pages}{063414}.

\bibitem{floss:2012}
\bibinfo{author}{\bibnamefont{Flo\ss{}}, \bibfnamefont{J.}}, and
  \bibinfo{author}{\bibfnamefont{I.~S.} \bibnamefont{Averbukh}},
  \bibinfo{year}{2012}{\natexlab{b}}, \bibinfo{journal}{Phys. Rev. A}
  \textbf{\bibinfo{volume}{86}}, \bibinfo{pages}{021401}.

\bibitem{floss2014}
\bibinfo{author}{\bibnamefont{Flo\ss{}}, \bibfnamefont{J.}}, and
  \bibinfo{author}{\bibfnamefont{I.~S.} \bibnamefont{Averbukh}},
  \bibinfo{year}{2014}, \bibinfo{journal}{Phys. Rev. Lett.}
  \textbf{\bibinfo{volume}{113}}, \bibinfo{pages}{043002}.

\bibitem{floss:2017}
\bibinfo{author}{\bibnamefont{Flo\ss{}}, \bibfnamefont{J.}}, and
  \bibinfo{author}{\bibfnamefont{P.}~\bibnamefont{Brumer}},
  \bibinfo{year}{2017}, \bibinfo{journal}{J. Chem. Phys.}
  \textbf{\bibinfo{volume}{146}}(\bibinfo{number}{12}),
  \bibinfo{pages}{124313}.

\bibitem{floss:2013}
\bibinfo{author}{\bibnamefont{Flo\ss{}}, \bibfnamefont{J.}},
  \bibinfo{author}{\bibfnamefont{S.}~\bibnamefont{Fishman}}, and
  \bibinfo{author}{\bibfnamefont{I.~S.} \bibnamefont{Averbukh}},
  \bibinfo{year}{2013}, \bibinfo{journal}{Phys. Rev. A}
  \textbf{\bibinfo{volume}{88}}, \bibinfo{pages}{023426}.

\bibitem{flosstorsion:2012}
\bibinfo{author}{\bibnamefont{Floss}, \bibfnamefont{J.}},
  \bibinfo{author}{\bibfnamefont{T.}~\bibnamefont{Grohmann}},
  \bibinfo{author}{\bibfnamefont{M.}~\bibnamefont{Leibscher}}, and
  \bibinfo{author}{\bibfnamefont{T.}~\bibnamefont{Seideman}},
  \bibinfo{year}{2012}, \bibinfo{journal}{J. Chem. Phys.}
  \textbf{\bibinfo{volume}{136}}(\bibinfo{number}{8}), \bibinfo{pages}{084309}.

\bibitem{floss:2015}
\bibinfo{author}{\bibnamefont{Flo\ss{}}, \bibfnamefont{J.}},
  \bibinfo{author}{\bibfnamefont{A.}~\bibnamefont{Kamalov}},
  \bibinfo{author}{\bibfnamefont{I.~S.} \bibnamefont{Averbukh}}, and
  \bibinfo{author}{\bibfnamefont{P.~H.} \bibnamefont{Bucksbaum}},
  \bibinfo{year}{2015}, \bibinfo{journal}{Phys. Rev. Lett.}
  \textbf{\bibinfo{volume}{115}}, \bibinfo{pages}{203002}.

\bibitem{friedrich:1999}
\bibinfo{author}{\bibnamefont{Friedrich}, \bibfnamefont{B.}}, and
  \bibinfo{author}{\bibnamefont{Herschbach}},
  \bibinfo{year}{1999}{\natexlab{a}}, \bibinfo{journal}{J. Phys. Chem. A}
  \textbf{\bibinfo{volume}{103}}(\bibinfo{number}{49}), \bibinfo{pages}{10280}.

\bibitem{friedrich:1995}
\bibinfo{author}{\bibnamefont{Friedrich}, \bibfnamefont{B.}}, and
  \bibinfo{author}{\bibfnamefont{D.}~\bibnamefont{Herschbach}},
  \bibinfo{year}{1995}{\natexlab{a}}, \bibinfo{journal}{Phys. Rev. Lett.}
  \textbf{\bibinfo{volume}{74}}, \bibinfo{pages}{4623}.

\bibitem{friedrich:1995b}
\bibinfo{author}{\bibnamefont{Friedrich}, \bibfnamefont{B.}}, and
  \bibinfo{author}{\bibfnamefont{D.}~\bibnamefont{Herschbach}},
  \bibinfo{year}{1995}{\natexlab{b}}, \bibinfo{journal}{J. Phys. Chem.}
  \textbf{\bibinfo{volume}{99}}(\bibinfo{number}{42}), \bibinfo{pages}{15686}.

\bibitem{friedrich:1999b}
\bibinfo{author}{\bibnamefont{Friedrich}, \bibfnamefont{B.}}, and
  \bibinfo{author}{\bibfnamefont{D.}~\bibnamefont{Herschbach}},
  \bibinfo{year}{1999}{\natexlab{b}}, \bibinfo{journal}{J. Chem. Phys.}
  \textbf{\bibinfo{volume}{111}}(\bibinfo{number}{14}), \bibinfo{pages}{6157}.

\bibitem{friedrich:1991}
\bibinfo{author}{\bibnamefont{Friedrich}, \bibfnamefont{B.}}, and
  \bibinfo{author}{\bibfnamefont{D.~R.} \bibnamefont{Herschbach}},
  \bibinfo{year}{1991}, \bibinfo{journal}{Z. Phys. D: At. Mol. Clusters}
  \textbf{\bibinfo{volume}{18}}, \bibinfo{pages}{153}.

\bibitem{frumker:2012b}
\bibinfo{author}{\bibnamefont{Frumker}, \bibfnamefont{E.}},
  \bibinfo{author}{\bibfnamefont{C.~T.} \bibnamefont{Hebeisen}},
  \bibinfo{author}{\bibfnamefont{N.}~\bibnamefont{Kajumba}},
  \bibinfo{author}{\bibfnamefont{J.~B.} \bibnamefont{Bertrand}},
  \bibinfo{author}{\bibfnamefont{H.~J.} \bibnamefont{W\"orner}},
  \bibinfo{author}{\bibfnamefont{M.}~\bibnamefont{Spanner}},
  \bibinfo{author}{\bibfnamefont{D.~M.} \bibnamefont{Villeneuve}},
  \bibinfo{author}{\bibfnamefont{A.}~\bibnamefont{Naumov}}, and
  \bibinfo{author}{\bibfnamefont{P.~B.} \bibnamefont{Corkum}},
  \bibinfo{year}{2012}{\natexlab{a}}, \bibinfo{journal}{Phys. Rev. Lett.}
  \textbf{\bibinfo{volume}{109}}, \bibinfo{pages}{113901}.

\bibitem{frumker:2012a}
\bibinfo{author}{\bibnamefont{Frumker}, \bibfnamefont{E.}},
  \bibinfo{author}{\bibfnamefont{N.}~\bibnamefont{Kajumba}},
  \bibinfo{author}{\bibfnamefont{J.~B.} \bibnamefont{Bertrand}},
  \bibinfo{author}{\bibfnamefont{H.~J.} \bibnamefont{W\"orner}},
  \bibinfo{author}{\bibfnamefont{C.~T.} \bibnamefont{Hebeisen}},
  \bibinfo{author}{\bibfnamefont{P.}~\bibnamefont{Hockett}},
  \bibinfo{author}{\bibfnamefont{M.}~\bibnamefont{Spanner}},
  \bibinfo{author}{\bibfnamefont{S.}~\bibnamefont{Patchkovskii}},
  \bibinfo{author}{\bibfnamefont{G.~G.} \bibnamefont{Paulus}},
  \bibinfo{author}{\bibfnamefont{D.~M.} \bibnamefont{Villeneuve}},
  \bibinfo{author}{\bibfnamefont{A.}~\bibnamefont{Naumov}}, and
  \bibinfo{author}{\bibfnamefont{P.~B.} \bibnamefont{Corkum}},
  \bibinfo{year}{2012}{\natexlab{b}}, \bibinfo{journal}{Phys. Rev. Lett.}
  \textbf{\bibinfo{volume}{109}}, \bibinfo{pages}{233904}.

\bibitem{galinis:2014}
\bibinfo{author}{\bibnamefont{Galinis}, \bibfnamefont{G.}},
  \bibinfo{author}{\bibfnamefont{C.}~\bibnamefont{Cacho}},
  \bibinfo{author}{\bibfnamefont{R.~T.} \bibnamefont{Chapman}},
  \bibinfo{author}{\bibfnamefont{A.~M.} \bibnamefont{Ellis}},
  \bibinfo{author}{\bibfnamefont{M.}~\bibnamefont{Lewerenz}},
  \bibinfo{author}{\bibfnamefont{L.~G.} \bibnamefont{Mendoza~Luna}},
  \bibinfo{author}{\bibfnamefont{R.~S.} \bibnamefont{Minns}},
  \bibinfo{author}{\bibfnamefont{M.}~\bibnamefont{Mladenovi\ifmmode~\acute{c}\else
  \'{c}\fi{}}}, \bibinfo{author}{\bibfnamefont{A.}~\bibnamefont{Rouz\'ee}},
  \bibinfo{author}{\bibfnamefont{E.}~\bibnamefont{Springate}},
  \bibinfo{author}{\bibfnamefont{I.~C.~E.} \bibnamefont{Turcu}},
  \bibinfo{author}{\bibfnamefont{M.~J.} \bibnamefont{Watkins}}, \emph{et~al.},
  \bibinfo{year}{2014}, \bibinfo{journal}{Phys. Rev. Lett.}
  \textbf{\bibinfo{volume}{113}}, \bibinfo{pages}{043004}.

\bibitem{GaoPRA09}
\bibinfo{author}{\bibnamefont{Gao}, \bibfnamefont{B.}}, \bibinfo{year}{2009},
  \bibinfo{journal}{Phys. Rev. A} \textbf{\bibinfo{volume}{80}},
  \bibinfo{pages}{012702}.

\bibitem{gershnabel:2010}
\bibinfo{author}{\bibnamefont{Gershnabel}, \bibfnamefont{E.}}, and
  \bibinfo{author}{\bibfnamefont{I.~S.} \bibnamefont{Averbukh}},
  \bibinfo{year}{2010}{\natexlab{a}}, \bibinfo{journal}{Phys. Rev. A}
  \textbf{\bibinfo{volume}{82}}, \bibinfo{pages}{033401}.

\bibitem{gershnabel:2010b}
\bibinfo{author}{\bibnamefont{Gershnabel}, \bibfnamefont{E.}}, and
  \bibinfo{author}{\bibfnamefont{I.~S.} \bibnamefont{Averbukh}},
  \bibinfo{year}{2010}{\natexlab{b}}, \bibinfo{journal}{Phys. Rev. Lett.}
  \textbf{\bibinfo{volume}{104}}, \bibinfo{pages}{153001}.

\bibitem{gershnabel:2018}
\bibinfo{author}{\bibnamefont{Gershnabel}, \bibfnamefont{E.}}, and
  \bibinfo{author}{\bibfnamefont{I.~S.} \bibnamefont{Averbukh}},
  \bibinfo{year}{2018}, \bibinfo{journal}{Phys. Rev. Lett.}
  \textbf{\bibinfo{volume}{120}}, \bibinfo{pages}{083204}.

\bibitem{gershnabel:2006b}
\bibinfo{author}{\bibnamefont{Gershnabel}, \bibfnamefont{E.}},
  \bibinfo{author}{\bibfnamefont{I.~S.} \bibnamefont{Averbukh}}, and
  \bibinfo{author}{\bibfnamefont{R.~J.} \bibnamefont{Gordon}},
  \bibinfo{year}{2006}{\natexlab{a}}, \bibinfo{journal}{Phys. Rev. A}
  \textbf{\bibinfo{volume}{74}}, \bibinfo{pages}{053414}.

\bibitem{gershnabel:2006}
\bibinfo{author}{\bibnamefont{Gershnabel}, \bibfnamefont{E.}},
  \bibinfo{author}{\bibfnamefont{I.~S.} \bibnamefont{Averbukh}}, and
  \bibinfo{author}{\bibfnamefont{R.~J.} \bibnamefont{Gordon}},
  \bibinfo{year}{2006}{\natexlab{b}}, \bibinfo{journal}{Phys. Rev. A}
  \textbf{\bibinfo{volume}{73}}, \bibinfo{pages}{061401}.

\bibitem{gershnabel:2011}
\bibinfo{author}{\bibnamefont{Gershnabel}, \bibfnamefont{E.}},
  \bibinfo{author}{\bibfnamefont{M.}~\bibnamefont{Shapiro}}, and
  \bibinfo{author}{\bibfnamefont{I.~S.} \bibnamefont{Averbukh}},
  \bibinfo{year}{2011}, \bibinfo{journal}{J. Chem. Phys.}
  \textbf{\bibinfo{volume}{135}}(\bibinfo{number}{19}),
  \bibinfo{pages}{194310}.

\bibitem{Ghafur:09}
\bibinfo{author}{\bibnamefont{Ghafur}, \bibfnamefont{O.}},
  \bibinfo{author}{\bibfnamefont{A.}~\bibnamefont{Rouzee}},
  \bibinfo{author}{\bibfnamefont{A.}~\bibnamefont{Gijsbertsen}},
  \bibinfo{author}{\bibfnamefont{W.~K.} \bibnamefont{Siu}},
  \bibinfo{author}{\bibfnamefont{S.}~\bibnamefont{Stolte}}, and
  \bibinfo{author}{\bibfnamefont{M.~J.~J.} \bibnamefont{Vrakking}},
  \bibinfo{year}{2009}, \bibinfo{journal}{Nature Phys.}
  \textbf{\bibinfo{volume}{5}}(\bibinfo{number}{4}), \bibinfo{pages}{289}.

\bibitem{gingras:2018}
\bibinfo{author}{\bibnamefont{Gingras}, \bibfnamefont{L.}},
  \bibinfo{author}{\bibfnamefont{W.}~\bibnamefont{Cui}},
  \bibinfo{author}{\bibfnamefont{A.~W.} \bibnamefont{Schiff-Kearn}},
  \bibinfo{author}{\bibfnamefont{J.-M.} \bibnamefont{M\'enard}}, and
  \bibinfo{author}{\bibfnamefont{D.~G.} \bibnamefont{Cooke}},
  \bibinfo{year}{2018}, \bibinfo{journal}{Optics Express}
  \textbf{\bibinfo{volume}{26}}, \bibinfo{pages}{13876}.

\bibitem{Glaser:15}
\bibinfo{author}{\bibnamefont{Glaser}, \bibfnamefont{S.~J.}},
  \bibinfo{author}{\bibfnamefont{U.}~\bibnamefont{Boscain}},
  \bibinfo{author}{\bibfnamefont{T.}~\bibnamefont{Calarco}},
  \bibinfo{author}{\bibfnamefont{C.~P.} \bibnamefont{Koch}},
  \bibinfo{author}{\bibfnamefont{W.}~\bibnamefont{K{\"o}ckenberger}},
  \bibinfo{author}{\bibfnamefont{R.}~\bibnamefont{Kosloff}},
  \bibinfo{author}{\bibfnamefont{I.}~\bibnamefont{Kuprov}},
  \bibinfo{author}{\bibfnamefont{B.}~\bibnamefont{Luy}},
  \bibinfo{author}{\bibfnamefont{S.}~\bibnamefont{Schirmer}},
  \bibinfo{author}{\bibfnamefont{T.}~\bibnamefont{Schulte-Herbr{\"u}ggen}},
  \bibinfo{author}{\bibfnamefont{D.}~\bibnamefont{Sugny}}, and
  \bibinfo{author}{\bibfnamefont{F.~K.} \bibnamefont{Wilhelm}},
  \bibinfo{year}{2015}, \bibinfo{journal}{Eur. Phys. J. D}
  \textbf{\bibinfo{volume}{69}}(\bibinfo{number}{12}), \bibinfo{pages}{279}.

\bibitem{GloecknerPRL15}
\bibinfo{author}{\bibnamefont{Gl\"ockner}, \bibfnamefont{R.}},
  \bibinfo{author}{\bibfnamefont{A.}~\bibnamefont{Prehn}},
  \bibinfo{author}{\bibfnamefont{B.~G.~U.} \bibnamefont{Englert}},
  \bibinfo{author}{\bibfnamefont{G.}~\bibnamefont{Rempe}}, and
  \bibinfo{author}{\bibfnamefont{M.}~\bibnamefont{Zeppenfeld}},
  \bibinfo{year}{2015}, \bibinfo{journal}{Phys. Rev. Lett.}
  \textbf{\bibinfo{volume}{115}}, \bibinfo{pages}{233001}.

\bibitem{GneitingPRA10}
\bibinfo{author}{\bibnamefont{Gneiting}, \bibfnamefont{C.}}, and
  \bibinfo{author}{\bibfnamefont{K.}~\bibnamefont{Hornberger}},
  \bibinfo{year}{2010}, \bibinfo{journal}{Phys. Rev. A}
  \textbf{\bibinfo{volume}{81}}, \bibinfo{pages}{013423}.

\bibitem{Goban:08}
\bibinfo{author}{\bibnamefont{Goban}, \bibfnamefont{A.}},
  \bibinfo{author}{\bibfnamefont{S.}~\bibnamefont{Minemoto}}, and
  \bibinfo{author}{\bibfnamefont{H.}~\bibnamefont{Sakai}},
  \bibinfo{year}{2008}, \bibinfo{journal}{Phys. Rev. Lett.}
  \textbf{\bibinfo{volume}{101}}, \bibinfo{pages}{013001}.

\bibitem{goldstein}
\bibinfo{author}{\bibnamefont{Goldstein}, \bibfnamefont{H.}},
  \bibinfo{year}{1950}, \emph{\bibinfo{title}{Classical mechanics}}
  (\bibinfo{publisher}{Addison-Wesley, Reading, M.A.}).

\bibitem{gonzales:2006}
\bibinfo{author}{\bibnamefont{Gonz\'alez-F\'erez}, \bibfnamefont{M., Mayle}},
  and \bibinfo{author}{\bibfnamefont{P.}~\bibnamefont{Schmelcher}},
  \bibinfo{year}{2006}, \bibinfo{journal}{Chem. Phys.}
  \textbf{\bibinfo{volume}{329}}, \bibinfo{pages}{203}.

\bibitem{GonzalezPRA12}
\bibinfo{author}{\bibnamefont{Gonz\'alez-F\'erez}, \bibfnamefont{R.}}, and
  \bibinfo{author}{\bibfnamefont{C.~P.} \bibnamefont{Koch}},
  \bibinfo{year}{2012}, \bibinfo{journal}{Phys. Rev. A}
  \textbf{\bibinfo{volume}{86}}, \bibinfo{pages}{063420}.

\bibitem{gonzales:2004}
\bibinfo{author}{\bibnamefont{Gonz\'alez-F\'erez}, \bibfnamefont{R.}}, and
  \bibinfo{author}{\bibfnamefont{P.}~\bibnamefont{Schmelcher}},
  \bibinfo{year}{2004}, \bibinfo{journal}{Phys. Rev. A}
  \textbf{\bibinfo{volume}{69}}, \bibinfo{pages}{023402}.

\bibitem{gonzales:2005}
\bibinfo{author}{\bibnamefont{Gonz\'alez-F\'erez}, \bibfnamefont{R.}}, and
  \bibinfo{author}{\bibfnamefont{P.}~\bibnamefont{Schmelcher}},
  \bibinfo{year}{2005}, \bibinfo{journal}{Phys. Rev. A}
  \textbf{\bibinfo{volume}{71}}, \bibinfo{pages}{033416}.

\bibitem{GoralPRL02}
\bibinfo{author}{\bibnamefont{G\'oral}, \bibfnamefont{K.}},
  \bibinfo{author}{\bibfnamefont{L.}~\bibnamefont{Santos}}, and
  \bibinfo{author}{\bibfnamefont{M.}~\bibnamefont{Lewenstein}},
  \bibinfo{year}{2002}, \bibinfo{journal}{Phys. Rev. Lett.}
  \textbf{\bibinfo{volume}{88}}, \bibinfo{pages}{170406}.

\bibitem{GordonPR55}
\bibinfo{author}{\bibnamefont{Gordon}, \bibfnamefont{J.}},
  \bibinfo{author}{\bibfnamefont{H.}~\bibnamefont{Zeiger}}, and
  \bibinfo{author}{\bibfnamefont{C.}~\bibnamefont{Townes}},
  \bibinfo{year}{1955}, \bibinfo{journal}{Phys. Rev.}
  \textbf{\bibinfo{volume}{99}}(\bibinfo{number}{4}), \bibinfo{pages}{1264}.

\bibitem{GordonPRL17}
\bibinfo{author}{\bibnamefont{Gordon}, \bibfnamefont{S.~D.~S.}},
  \bibinfo{author}{\bibfnamefont{J.}~\bibnamefont{Zou}},
  \bibinfo{author}{\bibfnamefont{S.}~\bibnamefont{Tanteri}},
  \bibinfo{author}{\bibfnamefont{J.}~\bibnamefont{Jankunas}}, and
  \bibinfo{author}{\bibfnamefont{A.}~\bibnamefont{Osterwalder}},
  \bibinfo{year}{2017}, \bibinfo{journal}{Phys. Rev. Lett.}
  \textbf{\bibinfo{volume}{119}}, \bibinfo{pages}{053001}.

\bibitem{GorshkovPRA11}
\bibinfo{author}{\bibnamefont{Gorshkov}, \bibfnamefont{A.}},
  \bibinfo{author}{\bibfnamefont{S.}~\bibnamefont{Manmana}},
  \bibinfo{author}{\bibfnamefont{G.}~\bibnamefont{Chen}},
  \bibinfo{author}{\bibfnamefont{E.}~\bibnamefont{Demler}},
  \bibinfo{author}{\bibfnamefont{M.~D.} \bibnamefont{Lukin}}, and
  \bibinfo{author}{\bibfnamefont{A.~M.} \bibnamefont{Rey}},
  \bibinfo{year}{2011}{\natexlab{a}}, \bibinfo{journal}{Phys. Rev. A}
  \textbf{\bibinfo{volume}{84}}, \bibinfo{pages}{033619}.

\bibitem{GorshkovMolPhys13}
\bibinfo{author}{\bibnamefont{Gorshkov}, \bibfnamefont{A.~V.}},
  \bibinfo{author}{\bibfnamefont{K.~R.} \bibnamefont{Hazzard}}, and
  \bibinfo{author}{\bibfnamefont{A.~M.} \bibnamefont{Rey}},
  \bibinfo{year}{2013}, \bibinfo{journal}{Molecular Physics}
  \textbf{\bibinfo{volume}{111}}(\bibinfo{number}{12-13}),
  \bibinfo{pages}{1908}.

\bibitem{gorshkovPRL11}
\bibinfo{author}{\bibnamefont{Gorshkov}, \bibfnamefont{A.~V.}},
  \bibinfo{author}{\bibfnamefont{S.~R.} \bibnamefont{Manmana}},
  \bibinfo{author}{\bibfnamefont{G.}~\bibnamefont{Chen}},
  \bibinfo{author}{\bibfnamefont{J.}~\bibnamefont{Ye}},
  \bibinfo{author}{\bibfnamefont{E.}~\bibnamefont{Demler}},
  \bibinfo{author}{\bibfnamefont{M.~D.} \bibnamefont{Lukin}}, and
  \bibinfo{author}{\bibfnamefont{A.~M.} \bibnamefont{Rey}},
  \bibinfo{year}{2011}{\natexlab{b}}, \bibinfo{journal}{Phys. Rev. Lett.}
  \textbf{\bibinfo{volume}{107}}, \bibinfo{pages}{115301}.

\bibitem{GorshkovPRL08}
\bibinfo{author}{\bibnamefont{Gorshkov}, \bibfnamefont{A.~V.}},
  \bibinfo{author}{\bibfnamefont{P.}~\bibnamefont{Rabl}},
  \bibinfo{author}{\bibfnamefont{G.}~\bibnamefont{Pupillo}},
  \bibinfo{author}{\bibfnamefont{A.}~\bibnamefont{Micheli}},
  \bibinfo{author}{\bibfnamefont{P.}~\bibnamefont{Zoller}},
  \bibinfo{author}{\bibfnamefont{M.~D.} \bibnamefont{Lukin}}, and
  \bibinfo{author}{\bibfnamefont{H.~P.} \bibnamefont{B\"uchler}},
  \bibinfo{year}{2008}, \bibinfo{journal}{Phys. Rev. Lett.}
  \textbf{\bibinfo{volume}{101}}, \bibinfo{pages}{073201}.

\bibitem{GrebenevOCS}
\bibinfo{author}{\bibnamefont{Grebenev}, \bibfnamefont{S.}},
  \bibinfo{author}{\bibfnamefont{M.}~\bibnamefont{Hartmann}},
  \bibinfo{author}{\bibfnamefont{M.}~\bibnamefont{Havenith}},
  \bibinfo{author}{\bibfnamefont{B.}~\bibnamefont{Sartakov}},
  \bibinfo{author}{\bibfnamefont{J.~P.} \bibnamefont{Toennies}}, and
  \bibinfo{author}{\bibfnamefont{A.~F.} \bibnamefont{Vilesov}},
  \bibinfo{year}{2000}, \bibinfo{journal}{J. Chem. Phys.}
  \textbf{\bibinfo{volume}{112}}, \bibinfo{pages}{4485}.

\bibitem{grohmann:2011}
\bibinfo{author}{\bibnamefont{Grohmann}, \bibfnamefont{T.}}, and
  \bibinfo{author}{\bibfnamefont{M.}~\bibnamefont{Leibscher}},
  \bibinfo{year}{2011}, \bibinfo{journal}{J. Chem. Phys.}
  \textbf{\bibinfo{volume}{134}}(\bibinfo{number}{20}),
  \bibinfo{pages}{204316}.

\bibitem{grohmann:2017}
\bibinfo{author}{\bibnamefont{Grohmann}, \bibfnamefont{T.}},
  \bibinfo{author}{\bibfnamefont{M.}~\bibnamefont{Leibscher}}, and
  \bibinfo{author}{\bibfnamefont{T.}~\bibnamefont{Seideman}},
  \bibinfo{year}{2017}, \bibinfo{journal}{Phys. Rev. Lett.}
  \textbf{\bibinfo{volume}{118}}, \bibinfo{pages}{203201}.

\bibitem{grohmann:2018}
\bibinfo{author}{\bibnamefont{Grohmann}, \bibfnamefont{T.}},
  \bibinfo{author}{\bibfnamefont{T.}~\bibnamefont{Seideman}}, and
  \bibinfo{author}{\bibfnamefont{M.}~\bibnamefont{Leibscher}},
  \bibinfo{year}{2018}, \bibinfo{journal}{J. Chem. Phys.}
  \textbf{\bibinfo{volume}{148}}(\bibinfo{number}{9}), \bibinfo{pages}{094304}.

\bibitem{guerin:2008}
\bibinfo{author}{\bibnamefont{Gu\'erin}, \bibfnamefont{S.}},
  \bibinfo{author}{\bibfnamefont{A.}~\bibnamefont{Rouz\'ee}}, and
  \bibinfo{author}{\bibfnamefont{E.}~\bibnamefont{Hertz}},
  \bibinfo{year}{2008}, \bibinfo{journal}{Phys. Rev. A}
  \textbf{\bibinfo{volume}{77}}, \bibinfo{pages}{041404}.

\bibitem{guerin:2002}
\bibinfo{author}{\bibnamefont{Gu\'erin}, \bibfnamefont{S.}},
  \bibinfo{author}{\bibfnamefont{L.~P.} \bibnamefont{Yatsenko}},
  \bibinfo{author}{\bibfnamefont{H.~R.} \bibnamefont{Jauslin}},
  \bibinfo{author}{\bibfnamefont{O.}~\bibnamefont{Faucher}}, and
  \bibinfo{author}{\bibfnamefont{B.}~\bibnamefont{Lavorel}},
  \bibinfo{year}{2002}, \bibinfo{journal}{Phys. Rev. Lett.}
  \textbf{\bibinfo{volume}{88}}, \bibinfo{pages}{233601}.

\bibitem{GuoPRL16}
\bibinfo{author}{\bibnamefont{Guo}, \bibfnamefont{M.}},
  \bibinfo{author}{\bibfnamefont{B.}~\bibnamefont{Zhu}},
  \bibinfo{author}{\bibfnamefont{B.}~\bibnamefont{Lu}},
  \bibinfo{author}{\bibfnamefont{X.}~\bibnamefont{Ye}},
  \bibinfo{author}{\bibfnamefont{F.}~\bibnamefont{Wang}},
  \bibinfo{author}{\bibfnamefont{R.}~\bibnamefont{Vexiau}},
  \bibinfo{author}{\bibfnamefont{N.}~\bibnamefont{Bouloufa-Maafa}},
  \bibinfo{author}{\bibfnamefont{G.}~\bibnamefont{Qu\'em\'ener}},
  \bibinfo{author}{\bibfnamefont{O.}~\bibnamefont{Dulieu}}, and
  \bibinfo{author}{\bibfnamefont{D.}~\bibnamefont{Wang}}, \bibinfo{year}{2016},
  \bibinfo{journal}{Phys. Rev. Lett.} \textbf{\bibinfo{volume}{116}},
  \bibinfo{pages}{205303}.

\bibitem{HaffnerPhysRep08}
\bibinfo{author}{\bibnamefont{H{\"a}ffner}, \bibfnamefont{H.}},
  \bibinfo{author}{\bibfnamefont{C.~F.} \bibnamefont{Roos}}, and
  \bibinfo{author}{\bibfnamefont{R.}~\bibnamefont{Blatt}},
  \bibinfo{year}{2008}, \bibinfo{journal}{Physics Reports}
  \textbf{\bibinfo{volume}{469}}, \bibinfo{pages}{155}.

\bibitem{HalversonJCP18}
\bibinfo{author}{\bibnamefont{Halverson}, \bibfnamefont{T.}},
  \bibinfo{author}{\bibfnamefont{D.}~\bibnamefont{Iouchtchenko}}, and
  \bibinfo{author}{\bibfnamefont{P.-N.} \bibnamefont{Roy}},
  \bibinfo{year}{2018}, \bibinfo{journal}{J. Chem. Phys.}
  \textbf{\bibinfo{volume}{148}}(\bibinfo{number}{7}), \bibinfo{pages}{074112}.

\bibitem{hamraoui:2018}
\bibinfo{author}{\bibnamefont{Hamraoui}, \bibfnamefont{K.}},
  \bibinfo{author}{\bibfnamefont{L.}~\bibnamefont{Van~Damme}},
  \bibinfo{author}{\bibfnamefont{P.}~\bibnamefont{Marde\ifmmode \check{s}\else
  \v{s}\fi{}i\ifmmode~\acute{c}\else \'{c}\fi{}}}, and
  \bibinfo{author}{\bibfnamefont{D.}~\bibnamefont{Sugny}},
  \bibinfo{year}{2018}, \bibinfo{journal}{Phys. Rev. A}
  \textbf{\bibinfo{volume}{97}}, \bibinfo{pages}{032118}.

\bibitem{HansenNature14}
\bibinfo{author}{\bibnamefont{Hansen}, \bibfnamefont{A.~K.}},
  \bibinfo{author}{\bibfnamefont{O.~O.} \bibnamefont{Versolato}},
  \bibinfo{author}{\bibfnamefont{{\L}.}~\bibnamefont{K{\l}osowski}},
  \bibinfo{author}{\bibfnamefont{S.~B.} \bibnamefont{Kristensen}},
  \bibinfo{author}{\bibfnamefont{A.}~\bibnamefont{Gingell}},
  \bibinfo{author}{\bibfnamefont{M.}~\bibnamefont{Schwarz}},
  \bibinfo{author}{\bibfnamefont{A.}~\bibnamefont{Windberger}},
  \bibinfo{author}{\bibfnamefont{J.}~\bibnamefont{Ullrich}},
  \bibinfo{author}{\bibfnamefont{J.~R.~C.} \bibnamefont{L\'opez-Urrutia}}, and
  \bibinfo{author}{\bibfnamefont{M.}~\bibnamefont{Drewsen}},
  \bibinfo{year}{2014}, \bibinfo{journal}{Nature}
  \textbf{\bibinfo{volume}{508}}, \bibinfo{pages}{76}.

\bibitem{hansen:2012}
\bibinfo{author}{\bibnamefont{Hansen}, \bibfnamefont{J.~L.}},
  \bibinfo{author}{\bibfnamefont{J.~H.} \bibnamefont{Nielsen}},
  \bibinfo{author}{\bibfnamefont{C.~B.} \bibnamefont{Madsen}},
  \bibinfo{author}{\bibfnamefont{A.~T.} \bibnamefont{Lindhardt}},
  \bibinfo{author}{\bibfnamefont{M.~P.} \bibnamefont{Johansson}},
  \bibinfo{author}{\bibfnamefont{T.}~\bibnamefont{Skrydstrup}},
  \bibinfo{author}{\bibfnamefont{L.~B.} \bibnamefont{Madsen}}, and
  \bibinfo{author}{\bibfnamefont{H.}~\bibnamefont{Stapelfeldt}},
  \bibinfo{year}{2012}, \bibinfo{journal}{J. Chem. Phys.}
  \textbf{\bibinfo{volume}{136}}(\bibinfo{number}{20}),
  \bibinfo{pages}{204310}.

\bibitem{hansen:2013}
\bibinfo{author}{\bibnamefont{Hansen}, \bibfnamefont{J.~L.}},
  \bibinfo{author}{\bibfnamefont{J.~J.} \bibnamefont{Omiste}},
  \bibinfo{author}{\bibfnamefont{J.~H.} \bibnamefont{Nielsen}},
  \bibinfo{author}{\bibfnamefont{D.}~\bibnamefont{Pentlehner}},
  \bibinfo{author}{\bibfnamefont{J.}~\bibnamefont{K{\"u}pper}},
  \bibinfo{author}{\bibfnamefont{R.}~\bibnamefont{Gonz{\'a}lez-F{\'e}rez}}, and
  \bibinfo{author}{\bibfnamefont{H.}~\bibnamefont{Stapelfeldt}},
  \bibinfo{year}{2013}, \bibinfo{journal}{J. Chem. Phys.}
  \textbf{\bibinfo{volume}{139}}(\bibinfo{number}{23}),
  \bibinfo{pages}{234313}.

\bibitem{hartelt}
\bibinfo{author}{\bibnamefont{H\"artelt}, \bibfnamefont{M.}}, and
  \bibinfo{author}{\bibfnamefont{B.}~\bibnamefont{Friedrich}},
  \bibinfo{year}{2008}, \bibinfo{journal}{J. Chem. Phys.}
  \textbf{\bibinfo{volume}{128}}(\bibinfo{number}{22}),
  \bibinfo{pages}{224313}.

\bibitem{harter:1984}
\bibinfo{author}{\bibnamefont{Harter}, \bibfnamefont{W.~G.}}, and
  \bibinfo{author}{\bibfnamefont{C.~W.} \bibnamefont{Patterson}},
  \bibinfo{year}{1984}, \bibinfo{journal}{J. Chem. Phys.}
  \textbf{\bibinfo{volume}{80}}, \bibinfo{pages}{4241}.

\bibitem{hartmann:2012}
\bibinfo{author}{\bibnamefont{Hartmann}, \bibfnamefont{J.-M.}}, and
  \bibinfo{author}{\bibfnamefont{C.}~\bibnamefont{Boulet}},
  \bibinfo{year}{2012}, \bibinfo{journal}{J. Chem. Phys.}
  \textbf{\bibinfo{volume}{136}}(\bibinfo{number}{18}),
  \bibinfo{pages}{184302}.

\bibitem{hasegawa:2008}
\bibinfo{author}{\bibnamefont{Hasegawa}, \bibfnamefont{H.}}, and
  \bibinfo{author}{\bibfnamefont{Y.}~\bibnamefont{Ohshima}},
  \bibinfo{year}{2008}, \bibinfo{journal}{Phys. Rev. Lett.}
  \textbf{\bibinfo{volume}{101}}, \bibinfo{pages}{053002}.

\bibitem{Haessler:2010}
\bibinfo{author}{\bibnamefont{Hassler}, \bibfnamefont{S.}},
  \bibinfo{author}{\bibfnamefont{J.}~\bibnamefont{Caillat}},
  \bibinfo{author}{\bibfnamefont{W.}~\bibnamefont{Boutu}},
  \bibinfo{author}{\bibfnamefont{C.}~\bibnamefont{Giovanetti-Teixeira}},
  \bibinfo{author}{\bibfnamefont{T.}~\bibnamefont{Ruchon}},
  \bibinfo{author}{\bibfnamefont{T.}~\bibnamefont{Auguste}},
  \bibinfo{author}{\bibfnamefont{Z.}~\bibnamefont{Diveki}},
  \bibinfo{author}{\bibfnamefont{P.}~\bibnamefont{Breger}},
  \bibinfo{author}{\bibfnamefont{A.}~\bibnamefont{Maquet}},
  \bibinfo{author}{\bibfnamefont{B.}~\bibnamefont{Carr\'e}},
  \bibinfo{author}{\bibfnamefont{R.}~\bibnamefont{Ta\"ieb}}, and
  \bibinfo{author}{\bibfnamefont{P.}~\bibnamefont{Sali\`eres}},
  \bibinfo{year}{2010}, \bibinfo{journal}{Nature Phys.}
  \textbf{\bibinfo{volume}{6}}, \bibinfo{pages}{200}.

\bibitem{HauserNatPhys15}
\bibinfo{author}{\bibnamefont{Hauser}, \bibfnamefont{D.}},
  \bibinfo{author}{\bibfnamefont{S.}~\bibnamefont{Lee}},
  \bibinfo{author}{\bibfnamefont{F.}~\bibnamefont{Carelli}},
  \bibinfo{author}{\bibfnamefont{S.}~\bibnamefont{Spieler}},
  \bibinfo{author}{\bibfnamefont{O.}~\bibnamefont{Lakhmanskaya}},
  \bibinfo{author}{\bibfnamefont{E.~S.} \bibnamefont{Endres}},
  \bibinfo{author}{\bibfnamefont{S.~S.} \bibnamefont{Kumar}},
  \bibinfo{author}{\bibfnamefont{F.}~\bibnamefont{Gianturco}}, and
  \bibinfo{author}{\bibfnamefont{R.}~\bibnamefont{Wester}},
  \bibinfo{year}{2015}, \bibinfo{journal}{Nat. Phys.}
  \textbf{\bibinfo{volume}{11}}, \bibinfo{pages}{467}.

\bibitem{hay:2002}
\bibinfo{author}{\bibnamefont{Hay}, \bibfnamefont{N.}},
  \bibinfo{author}{\bibfnamefont{R.}~\bibnamefont{Velotta}},
  \bibinfo{author}{\bibfnamefont{M.}~\bibnamefont{Lein}},
  \bibinfo{author}{\bibfnamefont{R.}~\bibnamefont{de~Nalda}},
  \bibinfo{author}{\bibfnamefont{E.}~\bibnamefont{Heesel}},
  \bibinfo{author}{\bibfnamefont{M.}~\bibnamefont{Castillejo}}, and
  \bibinfo{author}{\bibfnamefont{J.~P.} \bibnamefont{Marangos}},
  \bibinfo{year}{2002}, \bibinfo{journal}{Phys. Rev. A}
  \textbf{\bibinfo{volume}{65}}, \bibinfo{pages}{053805}.

\bibitem{HazzardPRL14}
\bibinfo{author}{\bibnamefont{Hazzard}, \bibfnamefont{K.~R.~A.}},
  \bibinfo{author}{\bibfnamefont{B.}~\bibnamefont{Gadway}},
  \bibinfo{author}{\bibfnamefont{M.}~\bibnamefont{Foss-Feig}},
  \bibinfo{author}{\bibfnamefont{B.}~\bibnamefont{Yan}},
  \bibinfo{author}{\bibfnamefont{S.~A.} \bibnamefont{Moses}},
  \bibinfo{author}{\bibfnamefont{J.~P.} \bibnamefont{Covey}},
  \bibinfo{author}{\bibfnamefont{N.~Y.} \bibnamefont{Yao}},
  \bibinfo{author}{\bibfnamefont{M.~D.} \bibnamefont{Lukin}},
  \bibinfo{author}{\bibfnamefont{J.}~\bibnamefont{Ye}},
  \bibinfo{author}{\bibfnamefont{D.~S.} \bibnamefont{Jin}}, and
  \bibinfo{author}{\bibfnamefont{A.~M.} \bibnamefont{Rey}},
  \bibinfo{year}{2014}, \bibinfo{journal}{Phys. Rev. Lett.}
  \textbf{\bibinfo{volume}{113}}, \bibinfo{pages}{195302}.

\bibitem{HazzardPRL13}
\bibinfo{author}{\bibnamefont{Hazzard}, \bibfnamefont{K.~R.~A.}},
  \bibinfo{author}{\bibfnamefont{S.~R.} \bibnamefont{Manmana}},
  \bibinfo{author}{\bibfnamefont{M.}~\bibnamefont{Foss-Feig}}, and
  \bibinfo{author}{\bibfnamefont{A.~M.} \bibnamefont{Rey}},
  \bibinfo{year}{2013}, \bibinfo{journal}{Phys. Rev. Lett.}
  \textbf{\bibinfo{volume}{110}}(\bibinfo{number}{7}), \bibinfo{pages}{075301}.

\bibitem{HePRA11}
\bibinfo{author}{\bibnamefont{He}, \bibfnamefont{L.}}, and
  \bibinfo{author}{\bibfnamefont{W.}~\bibnamefont{Hofstetter}},
  \bibinfo{year}{2011}, \bibinfo{journal}{Phys. Rev. A}
  \textbf{\bibinfo{volume}{83}}, \bibinfo{pages}{053629}.

\bibitem{He:2018}
\bibinfo{author}{\bibnamefont{He}, \bibfnamefont{L.}},
  \bibinfo{author}{\bibfnamefont{P.}~\bibnamefont{Lan}},
  \bibinfo{author}{\bibfnamefont{A.-T.} \bibnamefont{Le}},
  \bibinfo{author}{\bibfnamefont{B.}~\bibnamefont{Wang}},
  \bibinfo{author}{\bibfnamefont{B.}~\bibnamefont{Wang}},
  \bibinfo{author}{\bibfnamefont{X.}~\bibnamefont{Zhu}},
  \bibinfo{author}{\bibfnamefont{P.}~\bibnamefont{Lu}}, and
  \bibinfo{author}{\bibfnamefont{C.~D.} \bibnamefont{Lin}},
  \bibinfo{year}{2018}, \bibinfo{journal}{Phys. Rev. Lett.}
  \textbf{\bibinfo{volume}{121}}, \bibinfo{pages}{163201}.

\bibitem{hensley:2012}
\bibinfo{author}{\bibnamefont{Hensley}, \bibfnamefont{C.~J.}},
  \bibinfo{author}{\bibfnamefont{J.}~\bibnamefont{Yang}}, and
  \bibinfo{author}{\bibfnamefont{M.}~\bibnamefont{Centurion}},
  \bibinfo{year}{2012}, \bibinfo{journal}{Phys. Rev. Lett.}
  \textbf{\bibinfo{volume}{109}}, \bibinfo{pages}{133202}.

\bibitem{HensonSci12}
\bibinfo{author}{\bibnamefont{Henson}, \bibfnamefont{A.~B.}},
  \bibinfo{author}{\bibfnamefont{S.}~\bibnamefont{Gersten}},
  \bibinfo{author}{\bibfnamefont{Y.}~\bibnamefont{Shagam}},
  \bibinfo{author}{\bibfnamefont{J.}~\bibnamefont{Narevicius}}, and
  \bibinfo{author}{\bibfnamefont{E.}~\bibnamefont{Narevicius}},
  \bibinfo{year}{2012}, \bibinfo{journal}{Science}
  \textbf{\bibinfo{volume}{338}}(\bibinfo{number}{6104}), \bibinfo{pages}{234}.

\bibitem{HerreraNJP14}
\bibinfo{author}{\bibnamefont{Herrera}, \bibfnamefont{F.}},
  \bibinfo{author}{\bibfnamefont{Y.}~\bibnamefont{Cao}},
  \bibinfo{author}{\bibfnamefont{S.}~\bibnamefont{Kais}}, and
  \bibinfo{author}{\bibfnamefont{K.~B.} \bibnamefont{Whaley}},
  \bibinfo{year}{2014}, \bibinfo{journal}{New Journal of Physics}
  \textbf{\bibinfo{volume}{16}}(\bibinfo{number}{7}), \bibinfo{pages}{075001}.

\bibitem{HerreraPRA11}
\bibinfo{author}{\bibnamefont{Herrera}, \bibfnamefont{F.}}, and
  \bibinfo{author}{\bibfnamefont{R.~V.} \bibnamefont{Krems}},
  \bibinfo{year}{2011}, \bibinfo{journal}{Phys. Rev. A}
  \textbf{\bibinfo{volume}{84}}, \bibinfo{pages}{051401(R)}.

\bibitem{HerreraPRA10}
\bibinfo{author}{\bibnamefont{Herrera}, \bibfnamefont{F.}},
  \bibinfo{author}{\bibfnamefont{M.}~\bibnamefont{Litinskaya}}, and
  \bibinfo{author}{\bibfnamefont{R.~V.} \bibnamefont{Krems}},
  \bibinfo{year}{2010}, \bibinfo{journal}{Phys. Rev. A}
  \textbf{\bibinfo{volume}{82}}, \bibinfo{pages}{033428}.

\bibitem{HerreraPRL13}
\bibinfo{author}{\bibnamefont{Herrera}, \bibfnamefont{F.}},
  \bibinfo{author}{\bibfnamefont{K.~W.} \bibnamefont{Madison}},
  \bibinfo{author}{\bibfnamefont{R.~V.} \bibnamefont{Krems}}, and
  \bibinfo{author}{\bibfnamefont{M.}~\bibnamefont{Berciu}},
  \bibinfo{year}{2013}, \bibinfo{journal}{Phys. Rev. Lett.}
  \textbf{\bibinfo{volume}{110}}, \bibinfo{pages}{223002}.

\bibitem{hertz:2007b}
\bibinfo{author}{\bibnamefont{Hertz}, \bibfnamefont{E.}},
  \bibinfo{author}{\bibfnamefont{D.}~\bibnamefont{Daems}},
  \bibinfo{author}{\bibfnamefont{S.}~\bibnamefont{Gu\'erin}},
  \bibinfo{author}{\bibfnamefont{H.~R.} \bibnamefont{Jauslin}},
  \bibinfo{author}{\bibfnamefont{B.}~\bibnamefont{Lavorel}}, and
  \bibinfo{author}{\bibfnamefont{O.}~\bibnamefont{Faucher}},
  \bibinfo{year}{2007}{\natexlab{a}}, \bibinfo{journal}{Phys. Rev. A}
  \textbf{\bibinfo{volume}{76}}, \bibinfo{pages}{043423}.

\bibitem{hertz:2007}
\bibinfo{author}{\bibnamefont{Hertz}, \bibfnamefont{E.}},
  \bibinfo{author}{\bibfnamefont{A.}~\bibnamefont{Rouz\'ee}},
  \bibinfo{author}{\bibfnamefont{S.}~\bibnamefont{Gu\'erin}},
  \bibinfo{author}{\bibfnamefont{B.}~\bibnamefont{Lavorel}}, and
  \bibinfo{author}{\bibfnamefont{O.}~\bibnamefont{Faucher}},
  \bibinfo{year}{2007}{\natexlab{b}}, \bibinfo{journal}{Phys. Rev. A}
  \textbf{\bibinfo{volume}{75}}, \bibinfo{pages}{031403}.

\bibitem{HerzbergBook}
\bibinfo{author}{\bibnamefont{Herzberg}, \bibfnamefont{G.}},
  \bibinfo{year}{1989}, \emph{\bibinfo{title}{Molecular Spectra and Molecular
  Structure}} (\bibinfo{publisher}{Krieger}).

\bibitem{HirotaPJA12}
\bibinfo{author}{\bibnamefont{Hirota}, \bibfnamefont{E.}},
  \bibinfo{year}{2012}, \bibinfo{journal}{Proc. Jpn. Acad., Ser. B}
  \textbf{\bibinfo{volume}{88}}(\bibinfo{number}{3}), \bibinfo{pages}{120}.

\bibitem{ho:2008}
\bibinfo{author}{\bibnamefont{Ho}, \bibfnamefont{P.~J.}}, and
  \bibinfo{author}{\bibfnamefont{R.}~\bibnamefont{Santra}},
  \bibinfo{year}{2008}, \bibinfo{journal}{Phys. Rev. A}
  \textbf{\bibinfo{volume}{78}}, \bibinfo{pages}{053409}.

\bibitem{HoekstraPRL07}
\bibinfo{author}{\bibnamefont{Hoekstra}, \bibfnamefont{S.}},
  \bibinfo{author}{\bibfnamefont{J.~J.} \bibnamefont{Gilijamse}},
  \bibinfo{author}{\bibfnamefont{B.}~\bibnamefont{Sartakov}},
  \bibinfo{author}{\bibfnamefont{N.}~\bibnamefont{Vanhaecke}},
  \bibinfo{author}{\bibfnamefont{L.}~\bibnamefont{Scharfenberg}},
  \bibinfo{author}{\bibfnamefont{S.~Y.~T.} \bibnamefont{van~de Meerakker}}, and
  \bibinfo{author}{\bibfnamefont{G.}~\bibnamefont{Meijer}},
  \bibinfo{year}{2007}, \bibinfo{journal}{Phys. Rev. Lett.}
  \textbf{\bibinfo{volume}{98}}, \bibinfo{pages}{133001}.

\bibitem{HojbjerrePRA08}
\bibinfo{author}{\bibnamefont{H\o{}jbjerre}, \bibfnamefont{K.}},
  \bibinfo{author}{\bibfnamefont{D.}~\bibnamefont{Offenberg}},
  \bibinfo{author}{\bibfnamefont{C.~Z.} \bibnamefont{Bisgaard}},
  \bibinfo{author}{\bibfnamefont{H.}~\bibnamefont{Stapelfeldt}},
  \bibinfo{author}{\bibfnamefont{P.~F.} \bibnamefont{Staanum}},
  \bibinfo{author}{\bibfnamefont{A.}~\bibnamefont{Mortensen}}, and
  \bibinfo{author}{\bibfnamefont{M.}~\bibnamefont{Drewsen}},
  \bibinfo{year}{2008}, \bibinfo{journal}{Phys. Rev. A}
  \textbf{\bibinfo{volume}{77}}, \bibinfo{pages}{030702}.

\bibitem{hoki:2001}
\bibinfo{author}{\bibnamefont{Hoki}, \bibfnamefont{K.}}, and
  \bibinfo{author}{\bibfnamefont{Y.}~\bibnamefont{Fujimura}},
  \bibinfo{year}{2001}, \bibinfo{journal}{Chemical Physics}
  \textbf{\bibinfo{volume}{267}}(\bibinfo{number}{1}), \bibinfo{pages}{187 }.

\bibitem{holmegaard:2010}
\bibinfo{author}{\bibnamefont{Holmegaard}, \bibfnamefont{L.}},
  \bibinfo{author}{\bibfnamefont{J.~L.} \bibnamefont{Hansen}},
  \bibinfo{author}{\bibfnamefont{L.}~\bibnamefont{Kalhoj}},
  \bibinfo{author}{\bibfnamefont{S.~L.} \bibnamefont{Kragh}},
  \bibinfo{author}{\bibfnamefont{H.}~\bibnamefont{Stapelfeldt}},
  \bibinfo{author}{\bibfnamefont{F.}~\bibnamefont{Filsinger}},
  \bibinfo{author}{\bibfnamefont{J.}~\bibnamefont{Kupper}},
  \bibinfo{author}{\bibfnamefont{G.}~\bibnamefont{Meijer}},
  \bibinfo{author}{\bibfnamefont{D.}~\bibnamefont{Dimitrovski}},
  \bibinfo{author}{\bibfnamefont{M.}~\bibnamefont{Abu-samha}},
  \bibinfo{author}{\bibfnamefont{C.}~\bibnamefont{Martiny}}, and
  \bibinfo{author}{\bibfnamefont{L.~B.} \bibnamefont{Madsen}},
  \bibinfo{year}{2010}, \bibinfo{journal}{Nat. Phys.}
  \textbf{\bibinfo{volume}{6}}, \bibinfo{pages}{428}.

\bibitem{holmegaard:2009}
\bibinfo{author}{\bibnamefont{Holmegaard}, \bibfnamefont{L.}},
  \bibinfo{author}{\bibfnamefont{J.~H.} \bibnamefont{Nielsen}},
  \bibinfo{author}{\bibfnamefont{I.}~\bibnamefont{Nevo}},
  \bibinfo{author}{\bibfnamefont{H.}~\bibnamefont{Stapelfeldt}},
  \bibinfo{author}{\bibfnamefont{F.}~\bibnamefont{Filsinger}},
  \bibinfo{author}{\bibfnamefont{J.}~\bibnamefont{K\"upper}}, and
  \bibinfo{author}{\bibfnamefont{G.}~\bibnamefont{Meijer}},
  \bibinfo{year}{2009}, \bibinfo{journal}{Phys. Rev. Lett.}
  \textbf{\bibinfo{volume}{102}}, \bibinfo{pages}{023001}.

\bibitem{holmegaard:2007}
\bibinfo{author}{\bibnamefont{Holmegaard}, \bibfnamefont{L.}},
  \bibinfo{author}{\bibfnamefont{S.~S.} \bibnamefont{Viftrup}},
  \bibinfo{author}{\bibfnamefont{V.}~\bibnamefont{Kumarappan}},
  \bibinfo{author}{\bibfnamefont{C.~Z.} \bibnamefont{Bisgaard}},
  \bibinfo{author}{\bibfnamefont{H.}~\bibnamefont{Stapelfeldt}},
  \bibinfo{author}{\bibfnamefont{E.}~\bibnamefont{Hamilton}}, and
  \bibinfo{author}{\bibfnamefont{T.}~\bibnamefont{Seideman}},
  \bibinfo{year}{2007}, \bibinfo{journal}{Phys. Rev. A}
  \textbf{\bibinfo{volume}{75}}, \bibinfo{pages}{051403}.

\bibitem{Hoque:11}
\bibinfo{author}{\bibnamefont{Hoque}, \bibfnamefont{M.~Z.}},
  \bibinfo{author}{\bibfnamefont{M.}~\bibnamefont{Lapert}},
  \bibinfo{author}{\bibfnamefont{E.}~\bibnamefont{Hertz}},
  \bibinfo{author}{\bibfnamefont{F.}~\bibnamefont{Billard}},
  \bibinfo{author}{\bibfnamefont{D.}~\bibnamefont{Sugny}},
  \bibinfo{author}{\bibfnamefont{B.}~\bibnamefont{Lavorel}}, and
  \bibinfo{author}{\bibfnamefont{O.}~\bibnamefont{Faucher}},
  \bibinfo{year}{2011}, \bibinfo{journal}{Phys. Rev. A}
  \textbf{\bibinfo{volume}{84}}.

\bibitem{hu:2009}
\bibinfo{author}{\bibnamefont{Hu}, \bibfnamefont{W.-H.}},
  \bibinfo{author}{\bibfnamefont{C.-C.} \bibnamefont{Shu}},
  \bibinfo{author}{\bibfnamefont{Y.-C.} \bibnamefont{Han}},
  \bibinfo{author}{\bibfnamefont{K.-J.} \bibnamefont{Yuan}}, and
  \bibinfo{author}{\bibfnamefont{S.-L.} \bibnamefont{Cong}},
  \bibinfo{year}{2009}, \bibinfo{journal}{Chemical Physics Letters}
  \textbf{\bibinfo{volume}{474}}(\bibinfo{number}{1}), \bibinfo{pages}{222 }.

\bibitem{HudsonPRA18}
\bibinfo{author}{\bibnamefont{Hudson}, \bibfnamefont{E.~R.}}, and
  \bibinfo{author}{\bibfnamefont{W.~C.} \bibnamefont{Campbell}},
  \bibinfo{year}{2018}, \bibinfo{journal}{Phys. Rev. A(R)}
  \textbf{\bibinfo{volume}{98}}, \bibinfo{pages}{040302}.

\bibitem{HummonPRL13}
\bibinfo{author}{\bibnamefont{Hummon}, \bibfnamefont{M.~T.}},
  \bibinfo{author}{\bibfnamefont{M.}~\bibnamefont{Yeo}},
  \bibinfo{author}{\bibfnamefont{B.~K.} \bibnamefont{Stuhl}},
  \bibinfo{author}{\bibfnamefont{A.~L.} \bibnamefont{Collopy}},
  \bibinfo{author}{\bibfnamefont{Y.}~\bibnamefont{Xia}}, and
  \bibinfo{author}{\bibfnamefont{J.}~\bibnamefont{Ye}}, \bibinfo{year}{2013},
  \bibinfo{journal}{Phys. Rev. Lett.} \textbf{\bibinfo{volume}{110}},
  \bibinfo{pages}{143001}.

\bibitem{IsaevPRL16}
\bibinfo{author}{\bibnamefont{Isaev}, \bibfnamefont{T.~A.}}, and
  \bibinfo{author}{\bibfnamefont{R.}~\bibnamefont{Berger}},
  \bibinfo{year}{2016}, \bibinfo{journal}{Phys. Rev. Lett.}
  \textbf{\bibinfo{volume}{116}}, \bibinfo{pages}{063006}.

\bibitem{itatani:2004}
\bibinfo{author}{\bibnamefont{Itatani}, \bibfnamefont{J.}},
  \bibinfo{author}{\bibfnamefont{J.}~\bibnamefont{Levesque}},
  \bibinfo{author}{\bibfnamefont{D.}~\bibnamefont{Zeidler}},
  \bibinfo{author}{\bibfnamefont{H.}~\bibnamefont{Niikura}},
  \bibinfo{author}{\bibfnamefont{H.}~\bibnamefont{Pepin}},
  \bibinfo{author}{\bibfnamefont{J.~C.} \bibnamefont{Kieffer}},
  \bibinfo{author}{\bibfnamefont{P.~B.} \bibnamefont{Corkum}}, and
  \bibinfo{author}{\bibfnamefont{D.~M.} \bibnamefont{Villeneuve}},
  \bibinfo{year}{2004}, \bibinfo{journal}{Nature}
  \textbf{\bibinfo{volume}{432}}, \bibinfo{pages}{867}.

\bibitem{JacobJCP12}
\bibinfo{author}{\bibnamefont{Jacob}, \bibfnamefont{A.}}, and
  \bibinfo{author}{\bibfnamefont{K.}~\bibnamefont{Hornberger}},
  \bibinfo{year}{2012}, \bibinfo{journal}{J. Chem. Phys.}
  \textbf{\bibinfo{volume}{137}}, \bibinfo{pages}{044313}.

\bibitem{JankunasJCP15}
\bibinfo{author}{\bibnamefont{Jankunas}, \bibfnamefont{J.}},
  \bibinfo{author}{\bibfnamefont{K.}~\bibnamefont{Jachymski}},
  \bibinfo{author}{\bibfnamefont{M.}~\bibnamefont{Hapka}}, and
  \bibinfo{author}{\bibfnamefont{A.}~\bibnamefont{Osterwalder}},
  \bibinfo{year}{2015}, \bibinfo{journal}{J. Chem. Phys.}
  \textbf{\bibinfo{volume}{142}}(\bibinfo{number}{16}),
  \bibinfo{pages}{164305}.

\bibitem{JinYeCRev12}
\bibinfo{author}{\bibnamefont{Jin}, \bibfnamefont{D.~S.}}, and
  \bibinfo{author}{\bibfnamefont{J.}~\bibnamefont{Ye}}, \bibinfo{year}{2012},
  \bibinfo{journal}{Chem. Rev.} \textbf{\bibinfo{volume}{112}},
  \bibinfo{pages}{4801}.

\bibitem{JonesRMP06}
\bibinfo{author}{\bibnamefont{Jones}, \bibfnamefont{K.~M.}},
  \bibinfo{author}{\bibfnamefont{E.}~\bibnamefont{Tiesinga}},
  \bibinfo{author}{\bibfnamefont{P.~D.} \bibnamefont{Lett}}, and
  \bibinfo{author}{\bibfnamefont{P.~S.} \bibnamefont{Julienne}},
  \bibinfo{year}{2006}, \bibinfo{journal}{Rev. Mod. Phys.}
  \textbf{\bibinfo{volume}{78}}, \bibinfo{pages}{483}.

\bibitem{rabitz:1990}
\bibinfo{author}{\bibnamefont{Judson}, \bibfnamefont{R.}},
  \bibinfo{author}{\bibfnamefont{K.}~\bibnamefont{Lehmann}},
  \bibinfo{author}{\bibfnamefont{H.}~\bibnamefont{Rabitz}}, and
  \bibinfo{author}{\bibfnamefont{W.}~\bibnamefont{Warren}},
  \bibinfo{year}{1990}, \bibinfo{journal}{J. Mol. Struct.}
  \textbf{\bibinfo{volume}{223}}, \bibinfo{pages}{425}.

\bibitem{kamalov:2015}
\bibinfo{author}{\bibnamefont{Kamalov}, \bibfnamefont{A.}},
  \bibinfo{author}{\bibfnamefont{D.~W.} \bibnamefont{Broege}}, and
  \bibinfo{author}{\bibfnamefont{P.~H.} \bibnamefont{Bucksbaum}},
  \bibinfo{year}{2015}, \bibinfo{journal}{Phys. Rev. A}
  \textbf{\bibinfo{volume}{92}}, \bibinfo{pages}{013409}.

\bibitem{kanai:2005}
\bibinfo{author}{\bibnamefont{Kanai}, \bibfnamefont{T.}},
  \bibinfo{author}{\bibfnamefont{S.}~\bibnamefont{Minemoto}}, and
  \bibinfo{author}{\bibfnamefont{H.}~\bibnamefont{Sakai}},
  \bibinfo{year}{2005}, \bibinfo{journal}{Nature}
  \textbf{\bibinfo{volume}{435}}, \bibinfo{pages}{470}.

\bibitem{kanai:2007}
\bibinfo{author}{\bibnamefont{Kanai}, \bibfnamefont{T.}},
  \bibinfo{author}{\bibfnamefont{S.}~\bibnamefont{Minemoto}}, and
  \bibinfo{author}{\bibfnamefont{H.}~\bibnamefont{Sakai}},
  \bibinfo{year}{2007}, \bibinfo{journal}{Phys. Rev. Lett.}
  \textbf{\bibinfo{volume}{98}}, \bibinfo{pages}{053002}.

\bibitem{kanai:2001}
\bibinfo{author}{\bibnamefont{Kanai}, \bibfnamefont{T.}}, and
  \bibinfo{author}{\bibfnamefont{H.}~\bibnamefont{Sakai}},
  \bibinfo{year}{2001}, \bibinfo{journal}{J. Chem. Phys.}
  \textbf{\bibinfo{volume}{115}}, \bibinfo{pages}{5492}.

\bibitem{karczmarek:1999}
\bibinfo{author}{\bibnamefont{Karczmarek}, \bibfnamefont{J.}},
  \bibinfo{author}{\bibfnamefont{J.}~\bibnamefont{Wright}},
  \bibinfo{author}{\bibfnamefont{P.}~\bibnamefont{Corkum}}, and
  \bibinfo{author}{\bibfnamefont{M.}~\bibnamefont{Ivanov}},
  \bibinfo{year}{1999}, \bibinfo{journal}{Phys. Rev. Lett.}
  \textbf{\bibinfo{volume}{82}}, \bibinfo{pages}{3420}.

\bibitem{karman:2018}
\bibinfo{author}{\bibnamefont{Karman}, \bibfnamefont{T.}}, and
  \bibinfo{author}{\bibfnamefont{J.~M.} \bibnamefont{Hutson}},
  \bibinfo{year}{2018}, \bibinfo{journal}{Phys. Rev. Lett.}
  \textbf{\bibinfo{volume}{121}}, \bibinfo{pages}{163401}.

\bibitem{KarraJCP16}
\bibinfo{author}{\bibnamefont{Karra}, \bibfnamefont{M.}},
  \bibinfo{author}{\bibfnamefont{K.}~\bibnamefont{Sharma}},
  \bibinfo{author}{\bibfnamefont{B.}~\bibnamefont{Friedrich}},
  \bibinfo{author}{\bibfnamefont{S.}~\bibnamefont{Kais}}, and
  \bibinfo{author}{\bibfnamefont{D.}~\bibnamefont{Herschbach}},
  \bibinfo{year}{2016}, \bibinfo{journal}{J. Chem. Phys.}
  \textbf{\bibinfo{volume}{144}}, \bibinfo{pages}{094301}.

\bibitem{karras:2014}
\bibinfo{author}{\bibnamefont{Karras}, \bibfnamefont{G.}},
  \bibinfo{author}{\bibfnamefont{E.}~\bibnamefont{Hertz}},
  \bibinfo{author}{\bibfnamefont{F.}~\bibnamefont{Billard}},
  \bibinfo{author}{\bibfnamefont{B.}~\bibnamefont{Lavorel}},
  \bibinfo{author}{\bibfnamefont{J.-M.} \bibnamefont{Hartmann}}, and
  \bibinfo{author}{\bibfnamefont{O.}~\bibnamefont{Faucher}},
  \bibinfo{year}{2014}, \bibinfo{journal}{Phys. Rev. A}
  \textbf{\bibinfo{volume}{89}}, \bibinfo{pages}{063411}.

\bibitem{karras_echo}
\bibinfo{author}{\bibnamefont{Karras}, \bibfnamefont{G.}},
  \bibinfo{author}{\bibfnamefont{E.}~\bibnamefont{Hertz}},
  \bibinfo{author}{\bibfnamefont{F.}~\bibnamefont{Billard}},
  \bibinfo{author}{\bibfnamefont{B.}~\bibnamefont{Lavorel}},
  \bibinfo{author}{\bibfnamefont{J.-M.} \bibnamefont{Hartmann}},
  \bibinfo{author}{\bibfnamefont{O.}~\bibnamefont{Faucher}},
  \bibinfo{author}{\bibfnamefont{E.}~\bibnamefont{Gershnabel}},
  \bibinfo{author}{\bibfnamefont{Y.}~\bibnamefont{Prior}}, and
  \bibinfo{author}{\bibfnamefont{I.~S.} \bibnamefont{Averbukh}},
  \bibinfo{year}{2015}{\natexlab{a}}, \bibinfo{journal}{Phys. Rev. Lett.}
  \textbf{\bibinfo{volume}{114}}, \bibinfo{pages}{153601}.

\bibitem{karras:20016}
\bibinfo{author}{\bibnamefont{Karras}, \bibfnamefont{G.}},
  \bibinfo{author}{\bibfnamefont{E.}~\bibnamefont{Hertz}},
  \bibinfo{author}{\bibfnamefont{F.}~\bibnamefont{Billard}},
  \bibinfo{author}{\bibfnamefont{B.}~\bibnamefont{Lavorel}},
  \bibinfo{author}{\bibfnamefont{G.}~\bibnamefont{Siour}},
  \bibinfo{author}{\bibfnamefont{J.-M.} \bibnamefont{Hartmann}},
  \bibinfo{author}{\bibfnamefont{O.}~\bibnamefont{Faucher}},
  \bibinfo{author}{\bibfnamefont{E.}~\bibnamefont{Gershnabel}},
  \bibinfo{author}{\bibfnamefont{Y.}~\bibnamefont{Prior}}, and
  \bibinfo{author}{\bibfnamefont{I.~S.} \bibnamefont{Averbukh}},
  \bibinfo{year}{2016}, \bibinfo{journal}{Phys. Rev. A}
  \textbf{\bibinfo{volume}{94}}, \bibinfo{pages}{033404}.

\bibitem{karras:2015}
\bibinfo{author}{\bibnamefont{Karras}, \bibfnamefont{G.}},
  \bibinfo{author}{\bibfnamefont{M.}~\bibnamefont{Ndong}},
  \bibinfo{author}{\bibfnamefont{E.}~\bibnamefont{Hertz}},
  \bibinfo{author}{\bibfnamefont{D.}~\bibnamefont{Sugny}},
  \bibinfo{author}{\bibfnamefont{F.}~\bibnamefont{Billard}},
  \bibinfo{author}{\bibfnamefont{B.}~\bibnamefont{Lavorel}}, and
  \bibinfo{author}{\bibfnamefont{D.}~\bibnamefont{Sugny}},
  \bibinfo{year}{2015}{\natexlab{b}}, \bibinfo{journal}{Phys. Rev. Lett.}
  \textbf{\bibinfo{volume}{114}}, \bibinfo{pages}{113001}.

\bibitem{keller:2000}
\bibinfo{author}{\bibnamefont{Keller}, \bibfnamefont{A.}},
  \bibinfo{author}{\bibfnamefont{C.~M.} \bibnamefont{Dion}}, and
  \bibinfo{author}{\bibfnamefont{O.}~\bibnamefont{Atabek}},
  \bibinfo{year}{2000}, \bibinfo{journal}{Phys. Rev. A}
  \textbf{\bibinfo{volume}{61}}, \bibinfo{pages}{023409}.

\bibitem{KermanPRL04b}
\bibinfo{author}{\bibnamefont{Kerman}, \bibfnamefont{A.~J.}},
  \bibinfo{author}{\bibfnamefont{J.~M.} \bibnamefont{Sage}},
  \bibinfo{author}{\bibfnamefont{S.}~\bibnamefont{Sainis}},
  \bibinfo{author}{\bibfnamefont{T.}~\bibnamefont{Bergeman}}, and
  \bibinfo{author}{\bibfnamefont{D.}~\bibnamefont{DeMille}},
  \bibinfo{year}{2004}, \bibinfo{journal}{Phys. Rev. Lett.}
  \textbf{\bibinfo{volume}{92}}, \bibinfo{pages}{153001}.

\bibitem{KhodorkovskyNatCom15}
\bibinfo{author}{\bibnamefont{Khodorkovsky}, \bibfnamefont{Y.}},
  \bibinfo{author}{\bibfnamefont{U.}~\bibnamefont{Steinitz}},
  \bibinfo{author}{\bibfnamefont{J.-M.} \bibnamefont{Hartmann}}, and
  \bibinfo{author}{\bibfnamefont{I.~S.} \bibnamefont{Averbukh}},
  \bibinfo{year}{2015}, \bibinfo{journal}{Nature Commun.}
  \textbf{\bibinfo{volume}{6}}, \bibinfo{pages}{7791 EP }.

\bibitem{kienitz:2016}
\bibinfo{author}{\bibnamefont{Kienitz}, \bibfnamefont{J.~S.}},
  \bibinfo{author}{\bibfnamefont{S.}~\bibnamefont{Trippel}},
  \bibinfo{author}{\bibfnamefont{T.}~\bibnamefont{Mullins}},
  \bibinfo{author}{\bibfnamefont{K.}~\bibnamefont{Dlugolecki}},
  \bibinfo{author}{\bibfnamefont{R.}~\bibnamefont{Gonzalez-Ferez}}, and
  \bibinfo{author}{\bibfnamefont{J.}~\bibnamefont{K\"upper}},
  \bibinfo{year}{2016}, \bibinfo{journal}{ChemPhysChem}
  \textbf{\bibinfo{volume}{17}}(\bibinfo{number}{22}), \bibinfo{pages}{3740}.

\bibitem{kim:2016}
\bibinfo{author}{\bibnamefont{Kim}, \bibfnamefont{L.~Y.}},
  \bibinfo{author}{\bibfnamefont{J.~H.} \bibnamefont{Lee}},
  \bibinfo{author}{\bibfnamefont{H.~A.} \bibnamefont{Kim}},
  \bibinfo{author}{\bibfnamefont{S.~K.} \bibnamefont{Kwak}},
  \bibinfo{author}{\bibfnamefont{B.}~\bibnamefont{Friedrich}}, and
  \bibinfo{author}{\bibfnamefont{B.~S.} \bibnamefont{Zhao}},
  \bibinfo{year}{2016}, \bibinfo{journal}{Phys. Rev. A}
  \textbf{\bibinfo{volume}{94}}, \bibinfo{pages}{013428}.

\bibitem{kim:1996}
\bibinfo{author}{\bibnamefont{Kim}, \bibfnamefont{W.}}, and
  \bibinfo{author}{\bibfnamefont{P.~M.} \bibnamefont{Felker}},
  \bibinfo{year}{1996}, \bibinfo{journal}{J. Chem. Phys.}
  \textbf{\bibinfo{volume}{104}}(\bibinfo{number}{3}), \bibinfo{pages}{1147}.

\bibitem{kim:1997}
\bibinfo{author}{\bibnamefont{Kim}, \bibfnamefont{W.}}, and
  \bibinfo{author}{\bibfnamefont{P.~M.} \bibnamefont{Felker}},
  \bibinfo{year}{1997}, \bibinfo{journal}{J. Chem. Phys.}
  \textbf{\bibinfo{volume}{107}}(\bibinfo{number}{7}), \bibinfo{pages}{2193}.

\bibitem{kim:1998}
\bibinfo{author}{\bibnamefont{Kim}, \bibfnamefont{W.}}, and
  \bibinfo{author}{\bibfnamefont{P.~M.} \bibnamefont{Felker}},
  \bibinfo{year}{1998}, \bibinfo{journal}{J. Chem. Phys.}
  \textbf{\bibinfo{volume}{108}}(\bibinfo{number}{16}), \bibinfo{pages}{6763}.

\bibitem{kitano:2009}
\bibinfo{author}{\bibnamefont{Kitano}, \bibfnamefont{K.}},
  \bibinfo{author}{\bibfnamefont{H.}~\bibnamefont{Hasegawa}}, and
  \bibinfo{author}{\bibfnamefont{Y.}~\bibnamefont{Ohshima}},
  \bibinfo{year}{2009}, \bibinfo{journal}{Phys. Rev. Lett.}
  \textbf{\bibinfo{volume}{103}}, \bibinfo{pages}{223002}.

\bibitem{Kitano:11}
\bibinfo{author}{\bibnamefont{Kitano}, \bibfnamefont{K.}},
  \bibinfo{author}{\bibfnamefont{N.}~\bibnamefont{Ishii}}, and
  \bibinfo{author}{\bibfnamefont{J.}~\bibnamefont{Itatani}},
  \bibinfo{year}{2011}, \bibinfo{journal}{Phys. Rev. A}
  \textbf{\bibinfo{volume}{84}}, \bibinfo{pages}{053408}.

\bibitem{KleinNatPhys17}
\bibinfo{author}{\bibnamefont{Klein}, \bibfnamefont{A.}},
  \bibinfo{author}{\bibfnamefont{Y.}~\bibnamefont{Shagam}},
  \bibinfo{author}{\bibfnamefont{W.}~\bibnamefont{Skomorowski}},
  \bibinfo{author}{\bibfnamefont{P.~S.} \bibnamefont{Zuchowski}},
  \bibinfo{author}{\bibfnamefont{M.}~\bibnamefont{Pawlak}},
  \bibinfo{author}{\bibfnamefont{L.~M.~C.} \bibnamefont{Janssen}},
  \bibinfo{author}{\bibfnamefont{N.}~\bibnamefont{Moiseyev}},
  \bibinfo{author}{\bibfnamefont{S.~Y.~T.} \bibnamefont{van~de Meerakker}},
  \bibinfo{author}{\bibfnamefont{A.}~\bibnamefont{van~der Avoird}},
  \bibinfo{author}{\bibfnamefont{C.~P.} \bibnamefont{Koch}}, and
  \bibinfo{author}{\bibfnamefont{E.}~\bibnamefont{Narevicius}},
  \bibinfo{year}{2016}, \bibinfo{journal}{Nature Phys.}
  \textbf{\bibinfo{volume}{13}}, \bibinfo{pages}{35}.

\bibitem{KlosChapter18}
\bibinfo{author}{\bibnamefont{K\l{}os}, \bibfnamefont{J.}}, and
  \bibinfo{author}{\bibfnamefont{F.}~\bibnamefont{Lique}},
  \bibinfo{year}{2018}, in \emph{\bibinfo{booktitle}{Cold Chemistry: Molecular
  Scattering and Reactivity Near Absolute Zero}}, edited by
  \bibinfo{editor}{\bibfnamefont{O.}~\bibnamefont{Dulieu}} and
  \bibinfo{editor}{\bibfnamefont{A.}~\bibnamefont{Osterwalder}}
  (\bibinfo{publisher}{The Royal Society of Chemistry}),
  number~\bibinfo{number}{11} in \bibinfo{series}{Theoretical and Computational
  Chemistry Series}, pp. \bibinfo{pages}{46--91}.

\bibitem{kobzar:2005}
\bibinfo{author}{\bibnamefont{Kobzar}, \bibfnamefont{K.}},
  \bibinfo{author}{\bibfnamefont{B.}~\bibnamefont{Luy}},
  \bibinfo{author}{\bibfnamefont{N.}~\bibnamefont{Khaneja}}, and
  \bibinfo{author}{\bibfnamefont{S.~J.} \bibnamefont{Glaser}},
  \bibinfo{year}{2005}, \bibinfo{journal}{J. Magn. Reson.}
  \textbf{\bibinfo{volume}{173}}, \bibinfo{pages}{229}.

\bibitem{KochChapter18}
\bibinfo{author}{\bibnamefont{Koch}, \bibfnamefont{C.~P.}},
  \bibinfo{year}{2018}, in \emph{\bibinfo{booktitle}{Cold Chemistry: Molecular
  Scattering and Reactivity Near Absolute Zero}}, edited by
  \bibinfo{editor}{\bibfnamefont{O.}~\bibnamefont{Dulieu}} and
  \bibinfo{editor}{\bibfnamefont{A.}~\bibnamefont{Osterwalder}}
  (\bibinfo{publisher}{The Royal Society of Chemistry}),
  number~\bibinfo{number}{11} in \bibinfo{series}{Theoretical and Computational
  Chemistry Series}, pp. \bibinfo{pages}{496--536}.

\bibitem{KochChemRev12}
\bibinfo{author}{\bibnamefont{Koch}, \bibfnamefont{C.~P.}}, and
  \bibinfo{author}{\bibfnamefont{M.}~\bibnamefont{Shapiro}},
  \bibinfo{year}{2012}, \bibinfo{journal}{Chem. Rev.}
  \textbf{\bibinfo{volume}{212}}, \bibinfo{pages}{4928}.

\bibitem{KondovPRL18}
\bibinfo{author}{\bibnamefont{Kondov}, \bibfnamefont{S.~S.}},
  \bibinfo{author}{\bibfnamefont{C.-H.} \bibnamefont{Lee}},
  \bibinfo{author}{\bibfnamefont{M.}~\bibnamefont{McDonald}},
  \bibinfo{author}{\bibfnamefont{B.~H.} \bibnamefont{McGuyer}},
  \bibinfo{author}{\bibfnamefont{I.}~\bibnamefont{Majewska}},
  \bibinfo{author}{\bibfnamefont{R.}~\bibnamefont{Moszynski}}, and
  \bibinfo{author}{\bibfnamefont{T.}~\bibnamefont{Zelevinsky}},
  \bibinfo{year}{2018}, \bibinfo{journal}{Phys. Rev. Lett.}
  \textbf{\bibinfo{volume}{121}}, \bibinfo{pages}{143401}.

\bibitem{korech:2013}
\bibinfo{author}{\bibnamefont{Korech}, \bibfnamefont{O.}},
  \bibinfo{author}{\bibfnamefont{U.}~\bibnamefont{Steinitz}},
  \bibinfo{author}{\bibfnamefont{R.~J.} \bibnamefont{Gordon}},
  \bibinfo{author}{\bibfnamefont{I.~S.} \bibnamefont{Averbukh}}, and
  \bibinfo{author}{\bibfnamefont{Y.}~\bibnamefont{Prior}},
  \bibinfo{year}{2013}, \bibinfo{journal}{Nat. Photonics}
  \textbf{\bibinfo{volume}{7}}, \bibinfo{pages}{711}.

\bibitem{korobenko:2014}
\bibinfo{author}{\bibnamefont{Korobenko}, \bibfnamefont{A.}},
  \bibinfo{author}{\bibfnamefont{A.~A.} \bibnamefont{Milner}}, and
  \bibinfo{author}{\bibfnamefont{V.}~\bibnamefont{Milner}},
  \bibinfo{year}{2014}, \bibinfo{journal}{Phys. Rev. Lett.}
  \textbf{\bibinfo{volume}{112}}, \bibinfo{pages}{113004}.

\bibitem{KotochigovaPRA06}
\bibinfo{author}{\bibnamefont{Kotochigova}, \bibfnamefont{S.}}, and
  \bibinfo{author}{\bibfnamefont{E.}~\bibnamefont{Tiesinga}},
  \bibinfo{year}{2006}, \bibinfo{journal}{Phys. Rev. A}
  \textbf{\bibinfo{volume}{73}}, \bibinfo{pages}{041405}.

\bibitem{kozin:2003}
\bibinfo{author}{\bibnamefont{Kozin}, \bibfnamefont{I.~N.}}, and
  \bibinfo{author}{\bibfnamefont{R.~M.} \bibnamefont{Roberts}},
  \bibinfo{year}{2003}, \bibinfo{journal}{J. Chem. Phys.}
  \textbf{\bibinfo{volume}{118}}, \bibinfo{pages}{10523}.

\bibitem{KozyryevPRL17}
\bibinfo{author}{\bibnamefont{Kozyryev}, \bibfnamefont{I.}},
  \bibinfo{author}{\bibfnamefont{L.}~\bibnamefont{Baum}},
  \bibinfo{author}{\bibfnamefont{K.}~\bibnamefont{Matsuda}},
  \bibinfo{author}{\bibfnamefont{B.~L.} \bibnamefont{Augenbraun}},
  \bibinfo{author}{\bibfnamefont{L.}~\bibnamefont{Anderegg}},
  \bibinfo{author}{\bibfnamefont{A.~P.} \bibnamefont{Sedlack}}, and
  \bibinfo{author}{\bibfnamefont{J.~M.} \bibnamefont{Doyle}},
  \bibinfo{year}{2017}, \bibinfo{journal}{Phys. Rev. Lett.}
  \textbf{\bibinfo{volume}{118}}, \bibinfo{pages}{173201}.

\bibitem{KralPRL01}
\bibinfo{author}{\bibnamefont{Kr\'al}, \bibfnamefont{P.}}, and
  \bibinfo{author}{\bibfnamefont{M.}~\bibnamefont{Shapiro}},
  \bibinfo{year}{2001}, \bibinfo{journal}{Phys. Rev. Lett.}
  \textbf{\bibinfo{volume}{87}}, \bibinfo{pages}{183002}.

\bibitem{KralPRL03}
\bibinfo{author}{\bibnamefont{Kr\'al}, \bibfnamefont{P.}},
  \bibinfo{author}{\bibfnamefont{I.}~\bibnamefont{Thanopulos}},
  \bibinfo{author}{\bibfnamefont{M.}~\bibnamefont{Shapiro}}, and
  \bibinfo{author}{\bibfnamefont{D.}~\bibnamefont{Cohen}},
  \bibinfo{year}{2003}, \bibinfo{journal}{Phys. Rev. Lett.}
  \textbf{\bibinfo{volume}{90}}, \bibinfo{pages}{033001}.

\bibitem{kraus:14}
\bibinfo{author}{\bibnamefont{Kraus}, \bibfnamefont{P.~M.}},
  \bibinfo{author}{\bibfnamefont{D.}~\bibnamefont{Baykusheva}}, and
  \bibinfo{author}{\bibfnamefont{H.~J.} \bibnamefont{W\"orner}},
  \bibinfo{year}{2014}, \bibinfo{journal}{Phys. Rev. Lett.}
  \textbf{\bibinfo{volume}{113}}, \bibinfo{pages}{023001}.

\bibitem{kraus:2015}
\bibinfo{author}{\bibnamefont{Kraus}, \bibfnamefont{P.~M.}},
  \bibinfo{author}{\bibfnamefont{B.}~\bibnamefont{Mignolet}},
  \bibinfo{author}{\bibfnamefont{D.}~\bibnamefont{Baykusheva}},
  \bibinfo{author}{\bibfnamefont{A.}~\bibnamefont{Rupenyan}},
  \bibinfo{author}{\bibfnamefont{L.}~\bibnamefont{Horn{\'y}}},
  \bibinfo{author}{\bibfnamefont{E.~F.} \bibnamefont{Penka}},
  \bibinfo{author}{\bibfnamefont{G.}~\bibnamefont{Grassi}},
  \bibinfo{author}{\bibfnamefont{O.~I.} \bibnamefont{Tolstikhin}},
  \bibinfo{author}{\bibfnamefont{J.}~\bibnamefont{Schneider}},
  \bibinfo{author}{\bibfnamefont{F.}~\bibnamefont{Jensen}},
  \bibinfo{author}{\bibfnamefont{L.~B.} \bibnamefont{Madsen}},
  \bibinfo{author}{\bibfnamefont{A.~D.} \bibnamefont{Bandrauk}}, \emph{et~al.},
  \bibinfo{year}{2015}, \bibinfo{journal}{Science}
  \textbf{\bibinfo{volume}{350}}(\bibinfo{number}{6262}), \bibinfo{pages}{790}.

\bibitem{Kraus:12}
\bibinfo{author}{\bibnamefont{Kraus}, \bibfnamefont{P.~M.}},
  \bibinfo{author}{\bibfnamefont{A.}~\bibnamefont{Rupenyan}}, and
  \bibinfo{author}{\bibfnamefont{H.~J.} \bibnamefont{W\"orner}},
  \bibinfo{year}{2012}, \bibinfo{journal}{Phys. Rev. Lett.}
  \textbf{\bibinfo{volume}{109}}, \bibinfo{pages}{233903}.

\bibitem{KremsIRPC05}
\bibinfo{author}{\bibnamefont{Krems}, \bibfnamefont{R.~V.}},
  \bibinfo{year}{2005}, \bibinfo{journal}{Int. Rev. Phys. Chem.}
  \textbf{\bibinfo{volume}{24}}(\bibinfo{number}{1}), \bibinfo{pages}{99}.

\bibitem{KremsPCCP08}
\bibinfo{author}{\bibnamefont{Krems}, \bibfnamefont{R.~V.}},
  \bibinfo{year}{2008}, \bibinfo{journal}{Phys. Chem. Chem. Phys.}
  \textbf{\bibinfo{volume}{10}}, \bibinfo{pages}{4079}.

\bibitem{KremsBook18}
\bibinfo{author}{\bibnamefont{Krems}, \bibfnamefont{R.~V.}},
  \bibinfo{year}{2018}, \emph{\bibinfo{title}{Molecules in Electromagnetic
  Fields: From Ultracold Physics to Controlled Chemistry}}
  (\bibinfo{publisher}{Wiley}).

\bibitem{KreStwFrieColdMol}
\bibinfo{editor}{\bibnamefont{Krems}, \bibfnamefont{R.~V.}},
  \bibinfo{editor}{\bibfnamefont{W.~C.} \bibnamefont{Stwalley}}, and
  \bibinfo{editor}{\bibfnamefont{B.}~\bibnamefont{Friedrich}} (eds.),
  \bibinfo{year}{2009}, \emph{\bibinfo{title}{Cold molecules: theory,
  experiment, applications}} (\bibinfo{publisher}{Taylor\&Francis/CRC, Boca
  Raton, USA}).

\bibitem{KunsPRA11}
\bibinfo{author}{\bibnamefont{Kuns}, \bibfnamefont{K.~A.}},
  \bibinfo{author}{\bibfnamefont{A.~M.} \bibnamefont{Rey}}, and
  \bibinfo{author}{\bibfnamefont{A.~V.} \bibnamefont{Gorshkov}},
  \bibinfo{year}{2011}, \bibinfo{journal}{Phys. Rev. A}
  \textbf{\bibinfo{volume}{84}}, \bibinfo{pages}{063639}.

\bibitem{kupper:2014}
\bibinfo{author}{\bibnamefont{K\"upper}, \bibfnamefont{J.}},
  \bibinfo{author}{\bibfnamefont{S.}~\bibnamefont{Stern}},
  \bibinfo{author}{\bibfnamefont{L.}~\bibnamefont{Holmegaard}},
  \bibinfo{author}{\bibfnamefont{F.}~\bibnamefont{Filsinger}},
  \bibinfo{author}{\bibfnamefont{A.}~\bibnamefont{Rouz\'ee}},
  \bibinfo{author}{\bibfnamefont{A.}~\bibnamefont{Rudenko}},
  \bibinfo{author}{\bibfnamefont{P.}~\bibnamefont{Johnsson}},
  \bibinfo{author}{\bibfnamefont{A.~V.} \bibnamefont{Martin}},
  \bibinfo{author}{\bibfnamefont{M.}~\bibnamefont{Adolph}},
  \bibinfo{author}{\bibfnamefont{A.}~\bibnamefont{Aquila}},
  \bibinfo{author}{\bibfnamefont{S.~c.~v.} \bibnamefont{Bajt}},
  \bibinfo{author}{\bibfnamefont{A.}~\bibnamefont{Barty}}, \emph{et~al.},
  \bibinfo{year}{2014}, \bibinfo{journal}{Phys. Rev. Lett.}
  \textbf{\bibinfo{volume}{112}}, \bibinfo{pages}{083002}.

\bibitem{dirr:2012}
\bibinfo{author}{\bibnamefont{Kurniawan}, \bibfnamefont{I.}},
  \bibinfo{author}{\bibfnamefont{G.}~\bibnamefont{Dirr}}, and
  \bibinfo{author}{\bibfnamefont{U.}~\bibnamefont{Helmke}},
  \bibinfo{year}{2012}, \bibinfo{journal}{IEEE Trans. Autom. Control}
  \textbf{\bibinfo{volume}{57}}, \bibinfo{pages}{1984}.

\bibitem{KuznetsovaPRA08}
\bibinfo{author}{\bibnamefont{Kuznetsova}, \bibfnamefont{E.}},
  \bibinfo{author}{\bibfnamefont{R.}~\bibnamefont{C\^ot\'e}},
  \bibinfo{author}{\bibfnamefont{K.}~\bibnamefont{Kirby}}, and
  \bibinfo{author}{\bibfnamefont{S.~F.} \bibnamefont{Yelin}},
  \bibinfo{year}{2008}, \bibinfo{journal}{Phys. Rev. A}
  \textbf{\bibinfo{volume}{78}}, \bibinfo{pages}{012313}.

\bibitem{KuznetsovaPRA10}
\bibinfo{author}{\bibnamefont{Kuznetsova}, \bibfnamefont{E.}},
  \bibinfo{author}{\bibfnamefont{M.}~\bibnamefont{Gacesa}},
  \bibinfo{author}{\bibfnamefont{S.~F.} \bibnamefont{Yelin}}, and
  \bibinfo{author}{\bibfnamefont{R.}~\bibnamefont{C\^ot\'e}},
  \bibinfo{year}{2010}, \bibinfo{journal}{Phys. Rev. A}
  \textbf{\bibinfo{volume}{81}}, \bibinfo{pages}{030301}.

\bibitem{KuznetsovaPRA16}
\bibinfo{author}{\bibnamefont{Kuznetsova}, \bibfnamefont{E.}},
  \bibinfo{author}{\bibfnamefont{S.~T.} \bibnamefont{Rittenhouse}},
  \bibinfo{author}{\bibfnamefont{H.~R.} \bibnamefont{Sadeghpour}}, and
  \bibinfo{author}{\bibfnamefont{S.~F.} \bibnamefont{Yelin}},
  \bibinfo{year}{2016}, \bibinfo{journal}{Phys. Rev. A}
  \textbf{\bibinfo{volume}{94}}, \bibinfo{pages}{032325}.

\bibitem{KwasigrochPRA14}
\bibinfo{author}{\bibnamefont{Kwasigroch}, \bibfnamefont{M.~P.}}, and
  \bibinfo{author}{\bibfnamefont{N.~R.} \bibnamefont{Cooper}},
  \bibinfo{year}{2014}, \bibinfo{journal}{Phys. Rev. A}
  \textbf{\bibinfo{volume}{90}}, \bibinfo{pages}{021605}.

\bibitem{LaddNature10}
\bibinfo{author}{\bibnamefont{Ladd}, \bibfnamefont{T.~D.}},
  \bibinfo{author}{\bibfnamefont{F.}~\bibnamefont{Jelezko}},
  \bibinfo{author}{\bibfnamefont{R.}~\bibnamefont{Laflamme}},
  \bibinfo{author}{\bibfnamefont{Y.}~\bibnamefont{Nakamura}},
  \bibinfo{author}{\bibfnamefont{C.}~\bibnamefont{Monroe}}, and
  \bibinfo{author}{\bibfnamefont{J.~L.} \bibnamefont{O'Brien}},
  \bibinfo{year}{2010}, \bibinfo{journal}{Nature}
  \textbf{\bibinfo{volume}{464}}, \bibinfo{pages}{45 EP }.

\bibitem{landau}
\bibinfo{author}{\bibnamefont{Landau}, \bibfnamefont{L.~M.}}, and
  \bibinfo{author}{\bibfnamefont{E.~M.} \bibnamefont{Lifshitz}},
  \bibinfo{year}{1960}{\natexlab{a}}, \emph{\bibinfo{title}{Mechanics}}
  (\bibinfo{publisher}{Pergamon Press, Oxford}).

\bibitem{landauQM}
\bibinfo{author}{\bibnamefont{Landau}, \bibfnamefont{L.~M.}}, and
  \bibinfo{author}{\bibfnamefont{E.~M.} \bibnamefont{Lifshitz}},
  \bibinfo{year}{1960}{\natexlab{b}}, \emph{\bibinfo{title}{Quantum Mechanics}}
  (\bibinfo{publisher}{Pergamon Press, Oxford}).

\bibitem{LangPRL08}
\bibinfo{author}{\bibnamefont{Lang}, \bibfnamefont{F.}},
  \bibinfo{author}{\bibfnamefont{K.}~\bibnamefont{Winkler}},
  \bibinfo{author}{\bibfnamefont{C.}~\bibnamefont{Strauss}},
  \bibinfo{author}{\bibfnamefont{R.}~\bibnamefont{Grimm}}, and
  \bibinfo{author}{\bibfnamefont{J.~H.} \bibnamefont{Denschlag}},
  \bibinfo{year}{2008}, \bibinfo{journal}{Phys. Rev. Lett.}
  \textbf{\bibinfo{volume}{101}}, \bibinfo{pages}{133005}.

\bibitem{lapert:2009}
\bibinfo{author}{\bibnamefont{Lapert}, \bibfnamefont{M.}},
  \bibinfo{author}{\bibfnamefont{E.}~\bibnamefont{Hertz}},
  \bibinfo{author}{\bibfnamefont{S.}~\bibnamefont{Gu\'erin}}, and
  \bibinfo{author}{\bibfnamefont{D.}~\bibnamefont{Sugny}},
  \bibinfo{year}{2009}{\natexlab{a}}, \bibinfo{journal}{Phys. Rev. A}
  \textbf{\bibinfo{volume}{80}}, \bibinfo{pages}{051403}.

\bibitem{Lapert:12}
\bibinfo{author}{\bibnamefont{Lapert}, \bibfnamefont{M.}}, and
  \bibinfo{author}{\bibfnamefont{D.}~\bibnamefont{Sugny}},
  \bibinfo{year}{2012}, \bibinfo{journal}{Phys. Rev. A}
  \textbf{\bibinfo{volume}{85}}, \bibinfo{pages}{063418}.

\bibitem{lapert:2008}
\bibinfo{author}{\bibnamefont{Lapert}, \bibfnamefont{M.}},
  \bibinfo{author}{\bibfnamefont{R.}~\bibnamefont{Tehini}},
  \bibinfo{author}{\bibfnamefont{G.}~\bibnamefont{Turinici}}, and
  \bibinfo{author}{\bibfnamefont{D.}~\bibnamefont{Sugny}},
  \bibinfo{year}{2008}, \bibinfo{journal}{Phys. Rev. A}
  \textbf{\bibinfo{volume}{78}}, \bibinfo{pages}{023408}.

\bibitem{lapert:2009b}
\bibinfo{author}{\bibnamefont{Lapert}, \bibfnamefont{M.}},
  \bibinfo{author}{\bibfnamefont{R.}~\bibnamefont{Tehini}},
  \bibinfo{author}{\bibfnamefont{G.}~\bibnamefont{Turinici}}, and
  \bibinfo{author}{\bibfnamefont{D.}~\bibnamefont{Sugny}},
  \bibinfo{year}{2009}{\natexlab{b}}, \bibinfo{journal}{Phys. Rev. A}
  \textbf{\bibinfo{volume}{79}}, \bibinfo{pages}{063411}.

\bibitem{lapert:2010}
\bibinfo{author}{\bibnamefont{Lapert}, \bibfnamefont{M.}},
  \bibinfo{author}{\bibfnamefont{Y.}~\bibnamefont{Zhang}},
  \bibinfo{author}{\bibfnamefont{M.}~\bibnamefont{Braun}},
  \bibinfo{author}{\bibfnamefont{S.~J.} \bibnamefont{Glaser}}, and
  \bibinfo{author}{\bibfnamefont{D.}~\bibnamefont{Sugny}},
  \bibinfo{year}{2010}, \bibinfo{journal}{Phys. Rev. Lett.}
  \textbf{\bibinfo{volume}{104}}, \bibinfo{pages}{083001}.

\bibitem{larsen:2000}
\bibinfo{author}{\bibnamefont{Larsen}, \bibfnamefont{J.~J.}},
  \bibinfo{author}{\bibfnamefont{K.}~\bibnamefont{Hald}},
  \bibinfo{author}{\bibfnamefont{N.}~\bibnamefont{Bjerre}},
  \bibinfo{author}{\bibfnamefont{H.}~\bibnamefont{Stapelfeldt}}, and
  \bibinfo{author}{\bibfnamefont{T.}~\bibnamefont{Seideman}},
  \bibinfo{year}{2000}, \bibinfo{journal}{Phys. Rev. Lett.}
  \textbf{\bibinfo{volume}{85}}, \bibinfo{pages}{2470}.

\bibitem{larsen:1998}
\bibinfo{author}{\bibnamefont{Larsen}, \bibfnamefont{J.~J.}},
  \bibinfo{author}{\bibfnamefont{N.~J.} \bibnamefont{Mo/rkbak}},
  \bibinfo{author}{\bibfnamefont{J.}~\bibnamefont{Olesen}},
  \bibinfo{author}{\bibfnamefont{N.}~\bibnamefont{Bjerre}},
  \bibinfo{author}{\bibfnamefont{M.}~\bibnamefont{Machholm}},
  \bibinfo{author}{\bibfnamefont{S.~R.} \bibnamefont{Keiding}}, and
  \bibinfo{author}{\bibfnamefont{H.}~\bibnamefont{Stapelfeldt}},
  \bibinfo{year}{1998}, \bibinfo{journal}{J. Chem. Phys.}
  \textbf{\bibinfo{volume}{109}}(\bibinfo{number}{20}), \bibinfo{pages}{8857}.

\bibitem{larsen:1999}
\bibinfo{author}{\bibnamefont{Larsen}, \bibfnamefont{J.~J.}},
  \bibinfo{author}{\bibfnamefont{H.}~\bibnamefont{Sakai}},
  \bibinfo{author}{\bibfnamefont{C.~P.} \bibnamefont{Safvan}},
  \bibinfo{author}{\bibfnamefont{I.}~\bibnamefont{Wendt-Larsen}}, and
  \bibinfo{author}{\bibfnamefont{H.}~\bibnamefont{Stapelfeldt}},
  \bibinfo{year}{1999}{\natexlab{a}}, \bibinfo{journal}{J. Chem. Phys.}
  \textbf{\bibinfo{volume}{111}}, \bibinfo{pages}{7774}.

\bibitem{larsen99}
\bibinfo{author}{\bibnamefont{Larsen}, \bibfnamefont{J.~J.}},
  \bibinfo{author}{\bibfnamefont{I.}~\bibnamefont{Wendt-Larsen}}, and
  \bibinfo{author}{\bibfnamefont{H.}~\bibnamefont{Stapelfeldt}},
  \bibinfo{year}{1999}{\natexlab{b}}, \bibinfo{journal}{Phys. Rev. Lett.}
  \textbf{\bibinfo{volume}{83}}, \bibinfo{pages}{1123}.

\bibitem{Lassabliere:2018}
\bibinfo{author}{\bibnamefont{Lassabli\`ere}, \bibfnamefont{L.}}, and
  \bibinfo{author}{\bibfnamefont{G.}~\bibnamefont{Qu\'em\'ener}},
  \bibinfo{year}{2018}, \bibinfo{journal}{Phys. Rev. Lett.}
  \textbf{\bibinfo{volume}{121}}, \bibinfo{pages}{163402}.

\bibitem{Lavert-OfirNatChem2014}
\bibinfo{author}{\bibnamefont{Lavert-Ofir}, \bibfnamefont{E.}},
  \bibinfo{author}{\bibfnamefont{Y.}~\bibnamefont{Shagam}},
  \bibinfo{author}{\bibfnamefont{A.~B.} \bibnamefont{Henson}},
  \bibinfo{author}{\bibfnamefont{S.}~\bibnamefont{Gersten}},
  \bibinfo{author}{\bibfnamefont{J.}~\bibnamefont{K{\l}os}},
  \bibinfo{author}{\bibfnamefont{P.~S.} \bibnamefont{\.{Z}uchowski}},
  \bibinfo{author}{\bibfnamefont{J.}~\bibnamefont{Narevicius}}, and
  \bibinfo{author}{\bibfnamefont{E.}~\bibnamefont{Narevicius}},
  \bibinfo{year}{2014}, \bibinfo{journal}{Nature Chemistry}
  \textbf{\bibinfo{volume}{6}}, \bibinfo{pages}{332}.

\bibitem{lavorel:2016}
\bibinfo{author}{\bibnamefont{Lavorel}, \bibfnamefont{B.}},
  \bibinfo{author}{\bibfnamefont{P.}~\bibnamefont{Babilotte}},
  \bibinfo{author}{\bibfnamefont{G.}~\bibnamefont{Karras}},
  \bibinfo{author}{\bibfnamefont{F.}~\bibnamefont{Billard}},
  \bibinfo{author}{\bibfnamefont{E.}~\bibnamefont{Hertz}}, and
  \bibinfo{author}{\bibfnamefont{O.}~\bibnamefont{Faucher}},
  \bibinfo{year}{2016}, \bibinfo{journal}{Phys. Rev. A}
  \textbf{\bibinfo{volume}{94}}, \bibinfo{pages}{043422}.

\bibitem{le:2009}
\bibinfo{author}{\bibnamefont{Le}, \bibfnamefont{A.-T.}},
  \bibinfo{author}{\bibfnamefont{R.~R.} \bibnamefont{Lucchese}},
  \bibinfo{author}{\bibfnamefont{M.~T.} \bibnamefont{Lee}}, and
  \bibinfo{author}{\bibfnamefont{C.~D.} \bibnamefont{Lin}},
  \bibinfo{year}{2009}, \bibinfo{journal}{Phys. Rev. Lett.}
  \textbf{\bibinfo{volume}{102}}, \bibinfo{pages}{203001}.

\bibitem{LeePRA05}
\bibinfo{author}{\bibnamefont{Lee}, \bibfnamefont{C.}}, and
  \bibinfo{author}{\bibfnamefont{E.~A.} \bibnamefont{Ostrovskaya}},
  \bibinfo{year}{2005}, \bibinfo{journal}{Phys. Rev. A}
  \textbf{\bibinfo{volume}{72}}, \bibinfo{pages}{062321}.

\bibitem{LeePRL04}
\bibinfo{author}{\bibnamefont{Lee}, \bibfnamefont{K.~F.}},
  \bibinfo{author}{\bibfnamefont{D.~M.} \bibnamefont{Villeneuve}},
  \bibinfo{author}{\bibfnamefont{P.~B.} \bibnamefont{Corkum}}, and
  \bibinfo{author}{\bibfnamefont{E.~A.} \bibnamefont{Shapiro}},
  \bibinfo{year}{2004}, \bibinfo{journal}{Phys. Rev. Lett.}
  \textbf{\bibinfo{volume}{93}}, \bibinfo{pages}{233601}.

\bibitem{lee:2006}
\bibinfo{author}{\bibnamefont{Lee}, \bibfnamefont{K.~F.}},
  \bibinfo{author}{\bibfnamefont{D.~M.} \bibnamefont{Villeneuve}},
  \bibinfo{author}{\bibfnamefont{P.~B.} \bibnamefont{Corkum}},
  \bibinfo{author}{\bibfnamefont{A.}~\bibnamefont{Stolow}}, and
  \bibinfo{author}{\bibfnamefont{J.~G.} \bibnamefont{Underwood}},
  \bibinfo{year}{2006}, \bibinfo{journal}{Phys. Rev. Lett.}
  \textbf{\bibinfo{volume}{97}}, \bibinfo{pages}{173001}.

\bibitem{LevebvreBrionField}
\bibinfo{author}{\bibnamefont{Lefebvre-Brion}, \bibfnamefont{H.}}, and
  \bibinfo{author}{\bibfnamefont{R.~W.} \bibnamefont{Field}},
  \bibinfo{year}{2004}, \emph{\bibinfo{title}{The Spectra and Dynamics of
  Diatomic Molecules}} (\bibinfo{publisher}{Elsevier, New York}).

\bibitem{LehmannJCP18}
\bibinfo{author}{\bibnamefont{Lehmann}, \bibfnamefont{K.~K.}},
  \bibinfo{year}{2018}, \bibinfo{journal}{J. Chem. Phys.}
  \textbf{\bibinfo{volume}{149}}(\bibinfo{number}{9}), \bibinfo{pages}{094201}.

\bibitem{leibscher:2003}
\bibinfo{author}{\bibnamefont{Leibscher}, \bibfnamefont{M.}},
  \bibinfo{author}{\bibfnamefont{I.~S.} \bibnamefont{Averbukh}}, and
  \bibinfo{author}{\bibfnamefont{H.}~\bibnamefont{Rabitz}},
  \bibinfo{year}{2003}, \bibinfo{journal}{Phys. Rev. Lett.}
  \textbf{\bibinfo{volume}{90}}, \bibinfo{pages}{213001}.

\bibitem{leibscher:2004}
\bibinfo{author}{\bibnamefont{Leibscher}, \bibfnamefont{M.}},
  \bibinfo{author}{\bibfnamefont{I.~S.} \bibnamefont{Averbukh}}, and
  \bibinfo{author}{\bibfnamefont{H.}~\bibnamefont{Rabitz}},
  \bibinfo{year}{2004}, \bibinfo{journal}{Phys. Rev. A}
  \textbf{\bibinfo{volume}{69}}, \bibinfo{pages}{013402}.

\bibitem{LeibscherJCP19}
\bibinfo{author}{\bibnamefont{Leibscher}, \bibfnamefont{M.}},
  \bibinfo{author}{\bibfnamefont{T.~F.} \bibnamefont{Giesen}}, and
  \bibinfo{author}{\bibfnamefont{C.~P.} \bibnamefont{Koch}},
  \bibinfo{year}{2019}, \bibinfo{journal}{submitted to J. Chem. Phys.}
  \bibinfo{note}{ArXiv:1904.02208}.

\bibitem{LemeshkoDroplets16}
\bibinfo{author}{\bibnamefont{Lemeshko}, \bibfnamefont{M.}},
  \bibinfo{year}{2017}, \bibinfo{journal}{Phys. Rev. Lett.}
  \textbf{\bibinfo{volume}{118}}, \bibinfo{pages}{095301}.

\bibitem{LemeshkoJCP08}
\bibinfo{author}{\bibnamefont{Lemeshko}, \bibfnamefont{M.}}, and
  \bibinfo{author}{\bibfnamefont{B.}~\bibnamefont{Friedrich}},
  \bibinfo{year}{2008}, \bibinfo{journal}{J. Chem. Phys.}
  \textbf{\bibinfo{volume}{129}}(\bibinfo{number}{2}), \bibinfo{pages}{024301}.

\bibitem{LemeshkoPRA09}
\bibinfo{author}{\bibnamefont{Lemeshko}, \bibfnamefont{M.}}, and
  \bibinfo{author}{\bibfnamefont{B.}~\bibnamefont{Friedrich}},
  \bibinfo{year}{2009}{\natexlab{a}}, \bibinfo{journal}{Phys. Rev. A}
  \textbf{\bibinfo{volume}{79}}, \bibinfo{pages}{012718}.

\bibitem{LemeshkoJPCA09}
\bibinfo{author}{\bibnamefont{Lemeshko}, \bibfnamefont{M.}}, and
  \bibinfo{author}{\bibfnamefont{B.}~\bibnamefont{Friedrich}},
  \bibinfo{year}{2009}{\natexlab{b}}, \bibinfo{journal}{J. Phys. Chem. A}
  \textbf{\bibinfo{volume}{113}}(\bibinfo{number}{52}), \bibinfo{pages}{15055}.

\bibitem{LemeshkoPRL09}
\bibinfo{author}{\bibnamefont{Lemeshko}, \bibfnamefont{M.}}, and
  \bibinfo{author}{\bibfnamefont{B.}~\bibnamefont{Friedrich}},
  \bibinfo{year}{2009}{\natexlab{c}}, \bibinfo{journal}{Phys. Rev. Lett.}
  \textbf{\bibinfo{volume}{103}}, \bibinfo{pages}{053003}.

\bibitem{LemKreDoyKais13}
\bibinfo{author}{\bibnamefont{Lemeshko}, \bibfnamefont{M.}},
  \bibinfo{author}{\bibfnamefont{R.}~\bibnamefont{Krems}},
  \bibinfo{author}{\bibfnamefont{J.}~\bibnamefont{Doyle}}, and
  \bibinfo{author}{\bibfnamefont{S.}~\bibnamefont{Kais}}, \bibinfo{year}{2013},
  \bibinfo{journal}{Mol. Phys.} \textbf{\bibinfo{volume}{111}},
  \bibinfo{pages}{1648}.

\bibitem{LemeshkoPRL12}
\bibinfo{author}{\bibnamefont{Lemeshko}, \bibfnamefont{M.}},
  \bibinfo{author}{\bibfnamefont{R.~V.} \bibnamefont{Krems}}, and
  \bibinfo{author}{\bibfnamefont{H.}~\bibnamefont{Weimer}},
  \bibinfo{year}{2012}, \bibinfo{journal}{Phys. Rev. Lett.}
  \textbf{\bibinfo{volume}{109}}, \bibinfo{pages}{035301}.

\bibitem{lemeshko:2011}
\bibinfo{author}{\bibnamefont{Lemeshko}, \bibfnamefont{M.}},
  \bibinfo{author}{\bibfnamefont{M.}~\bibnamefont{Mustafa}},
  \bibinfo{author}{\bibfnamefont{S.}~\bibnamefont{Kais}}, and
  \bibinfo{author}{\bibfnamefont{B.}~\bibnamefont{Friedrich}},
  \bibinfo{year}{2011}, \bibinfo{journal}{Phys. Rev. A}
  \textbf{\bibinfo{volume}{83}}, \bibinfo{pages}{043415}.

\bibitem{Lemeshko_2016_book}
\bibinfo{author}{\bibnamefont{Lemeshko}, \bibfnamefont{M.}}, and
  \bibinfo{author}{\bibfnamefont{R.}~\bibnamefont{Schmidt}},
  \bibinfo{year}{2017}, \bibinfo{journal}{arXiv:1703.06753} .

\bibitem{levesque:2007}
\bibinfo{author}{\bibnamefont{Levesque}, \bibfnamefont{J.}},
  \bibinfo{author}{\bibfnamefont{Y.}~\bibnamefont{Mairesse}},
  \bibinfo{author}{\bibfnamefont{N.}~\bibnamefont{Dudovich}},
  \bibinfo{author}{\bibfnamefont{H.}~\bibnamefont{P\'epin}},
  \bibinfo{author}{\bibfnamefont{J.-C.} \bibnamefont{Kieffer}},
  \bibinfo{author}{\bibfnamefont{P.~B.} \bibnamefont{Corkum}}, and
  \bibinfo{author}{\bibfnamefont{D.~M.} \bibnamefont{Villeneuve}},
  \bibinfo{year}{2007}, \bibinfo{journal}{Phys. Rev. Lett.}
  \textbf{\bibinfo{volume}{99}}, \bibinfo{pages}{243001}.

\bibitem{LevineBook}
\bibinfo{author}{\bibnamefont{Levine}, \bibfnamefont{R.~D.}},
  \bibinfo{year}{2005}, \emph{\bibinfo{title}{Molecular Reaction Dynamics}}
  (\bibinfo{publisher}{Cambridge Univ. Press}).

\bibitem{LevinsenPRA11}
\bibinfo{author}{\bibnamefont{Levinsen}, \bibfnamefont{J.}},
  \bibinfo{author}{\bibfnamefont{N.~R.} \bibnamefont{Cooper}}, and
  \bibinfo{author}{\bibfnamefont{G.~V.} \bibnamefont{Shlyapnikov}},
  \bibinfo{year}{2011}, \bibinfo{journal}{Phys. Rev. A}
  \textbf{\bibinfo{volume}{84}}, \bibinfo{pages}{013603}.

\bibitem{li:2013}
\bibinfo{author}{\bibnamefont{Li}, \bibfnamefont{H.}},
  \bibinfo{author}{\bibfnamefont{W.}~\bibnamefont{Li}},
  \bibinfo{author}{\bibfnamefont{Y.}~\bibnamefont{Feng}},
  \bibinfo{author}{\bibfnamefont{H.}~\bibnamefont{Pan}}, and
  \bibinfo{author}{\bibfnamefont{H.}~\bibnamefont{Zeng}},
  \bibinfo{year}{2013}{\natexlab{a}}, \bibinfo{journal}{Phys. Rev. A}
  \textbf{\bibinfo{volume}{88}}, \bibinfo{pages}{013424}.

\bibitem{LiPRL13}
\bibinfo{author}{\bibnamefont{Li}, \bibfnamefont{S.}},
  \bibinfo{author}{\bibfnamefont{A.}~\bibnamefont{Yu}},
  \bibinfo{author}{\bibfnamefont{F.}~\bibnamefont{Toledo}},
  \bibinfo{author}{\bibfnamefont{Z.}~\bibnamefont{Han}},
  \bibinfo{author}{\bibfnamefont{H.}~\bibnamefont{Wang}},
  \bibinfo{author}{\bibfnamefont{H.~Y.} \bibnamefont{He}},
  \bibinfo{author}{\bibfnamefont{R.}~\bibnamefont{Wu}}, and
  \bibinfo{author}{\bibfnamefont{W.}~\bibnamefont{Ho}},
  \bibinfo{year}{2013}{\natexlab{b}}, \bibinfo{journal}{Phys. Rev. Lett.}
  \textbf{\bibinfo{volume}{111}}, \bibinfo{pages}{146102}.

\bibitem{LiPRL07}
\bibinfo{author}{\bibnamefont{Li}, \bibfnamefont{Y.}},
  \bibinfo{author}{\bibfnamefont{C.}~\bibnamefont{Bruder}}, and
  \bibinfo{author}{\bibfnamefont{C.~P.} \bibnamefont{Sun}},
  \bibinfo{year}{2007}, \bibinfo{journal}{Phys. Rev. Lett.}
  \textbf{\bibinfo{volume}{99}}, \bibinfo{pages}{130403}.

\bibitem{liao:2013}
\bibinfo{author}{\bibnamefont{Liao}, \bibfnamefont{S.-L.}},
  \bibinfo{author}{\bibfnamefont{T.-S.} \bibnamefont{Ho}},
  \bibinfo{author}{\bibfnamefont{H.}~\bibnamefont{Rabitz}}, and
  \bibinfo{author}{\bibfnamefont{S.-I.} \bibnamefont{Chu}},
  \bibinfo{year}{2013}, \bibinfo{journal}{Phys. Rev. A}
  \textbf{\bibinfo{volume}{87}}, \bibinfo{pages}{013429}.

\bibitem{LienNatCommun14}
\bibinfo{author}{\bibnamefont{Lien}, \bibfnamefont{C.-Y.}},
  \bibinfo{author}{\bibfnamefont{C.~M.} \bibnamefont{Seck}},
  \bibinfo{author}{\bibfnamefont{Y.-W.} \bibnamefont{Lin}},
  \bibinfo{author}{\bibfnamefont{J.~H.} \bibnamefont{Nguyen}},
  \bibinfo{author}{\bibfnamefont{D.~A.} \bibnamefont{Tabor}}, and
  \bibinfo{author}{\bibfnamefont{B.~C.} \bibnamefont{Odom}},
  \bibinfo{year}{2014}, \bibinfo{journal}{Nat. Commun.}
  \textbf{\bibinfo{volume}{5}}, \bibinfo{pages}{4783}.

\bibitem{LimPRL18}
\bibinfo{author}{\bibnamefont{Lim}, \bibfnamefont{J.}},
  \bibinfo{author}{\bibfnamefont{J.~R.} \bibnamefont{Almond}},
  \bibinfo{author}{\bibfnamefont{M.~A.} \bibnamefont{Trigatzis}},
  \bibinfo{author}{\bibfnamefont{J.~A.} \bibnamefont{Devlin}},
  \bibinfo{author}{\bibfnamefont{N.~J.} \bibnamefont{Fitch}},
  \bibinfo{author}{\bibfnamefont{B.~E.} \bibnamefont{Sauer}},
  \bibinfo{author}{\bibfnamefont{M.~R.} \bibnamefont{Tarbutt}}, and
  \bibinfo{author}{\bibfnamefont{E.~A.} \bibnamefont{Hinds}},
  \bibinfo{year}{2018}, \bibinfo{journal}{Phys. Rev. Lett.}
  \textbf{\bibinfo{volume}{120}}, \bibinfo{pages}{123201}.

\bibitem{LinPRA10}
\bibinfo{author}{\bibnamefont{Lin}, \bibfnamefont{C.-H.}},
  \bibinfo{author}{\bibfnamefont{Y.-T.} \bibnamefont{Hsu}},
  \bibinfo{author}{\bibfnamefont{H.}~\bibnamefont{Lee}}, and
  \bibinfo{author}{\bibfnamefont{D.-W.} \bibnamefont{Wang}},
  \bibinfo{year}{2010}, \bibinfo{journal}{Phys. Rev. A}
  \textbf{\bibinfo{volume}{81}}, \bibinfo{pages}{031601}.

\bibitem{lin:2016}
\bibinfo{author}{\bibnamefont{Lin}, \bibfnamefont{K.}},
  \bibinfo{author}{\bibfnamefont{P.}~\bibnamefont{Lu}},
  \bibinfo{author}{\bibfnamefont{J.}~\bibnamefont{Ma}},
  \bibinfo{author}{\bibfnamefont{X.}~\bibnamefont{Gong}},
  \bibinfo{author}{\bibfnamefont{Q.}~\bibnamefont{Song}},
  \bibinfo{author}{\bibfnamefont{Q.}~\bibnamefont{Ji}},
  \bibinfo{author}{\bibfnamefont{W.}~\bibnamefont{Zhang}},
  \bibinfo{author}{\bibfnamefont{H.}~\bibnamefont{Zeng}},
  \bibinfo{author}{\bibfnamefont{J.}~\bibnamefont{Wu}},
  \bibinfo{author}{\bibfnamefont{G.}~\bibnamefont{Karras}},
  \bibinfo{author}{\bibfnamefont{G.}~\bibnamefont{Siour}},
  \bibinfo{author}{\bibfnamefont{J.-M.} \bibnamefont{Hartmann}}, \emph{et~al.},
  \bibinfo{year}{2016}, \bibinfo{journal}{Phys. Rev. X}
  \textbf{\bibinfo{volume}{6}}, \bibinfo{pages}{041056}.

\bibitem{lin:2017}
\bibinfo{author}{\bibnamefont{Lin}, \bibfnamefont{K.}},
  \bibinfo{author}{\bibfnamefont{J.}~\bibnamefont{Ma}},
  \bibinfo{author}{\bibfnamefont{X.}~\bibnamefont{Gong}},
  \bibinfo{author}{\bibfnamefont{Q.}~\bibnamefont{Song}},
  \bibinfo{author}{\bibfnamefont{Q.}~\bibnamefont{Ji}},
  \bibinfo{author}{\bibfnamefont{W.}~\bibnamefont{Zhang}},
  \bibinfo{author}{\bibfnamefont{H.}~\bibnamefont{Li}},
  \bibinfo{author}{\bibfnamefont{P.}~\bibnamefont{Lu}},
  \bibinfo{author}{\bibfnamefont{H.}~\bibnamefont{Li}},
  \bibinfo{author}{\bibfnamefont{H.}~\bibnamefont{Zeng}},
  \bibinfo{author}{\bibfnamefont{J.}~\bibnamefont{Wu}},
  \bibinfo{author}{\bibfnamefont{J.-M.} \bibnamefont{Hartmann}}, \emph{et~al.},
  \bibinfo{year}{2017}, \bibinfo{journal}{Opt. Express}
  \textbf{\bibinfo{volume}{25}}, \bibinfo{pages}{24917}.

\bibitem{lin:2015}
\bibinfo{author}{\bibnamefont{Lin}, \bibfnamefont{K.}},
  \bibinfo{author}{\bibfnamefont{Q.}~\bibnamefont{Song}},
  \bibinfo{author}{\bibfnamefont{X.}~\bibnamefont{Gong}},
  \bibinfo{author}{\bibfnamefont{Q.}~\bibnamefont{Ji}},
  \bibinfo{author}{\bibfnamefont{H.}~\bibnamefont{Pan}},
  \bibinfo{author}{\bibfnamefont{J.}~\bibnamefont{Ding}},
  \bibinfo{author}{\bibfnamefont{H.}~\bibnamefont{Zeng}}, and
  \bibinfo{author}{\bibfnamefont{J.}~\bibnamefont{Wu}}, \bibinfo{year}{2015},
  \bibinfo{journal}{Phys. Rev. A} \textbf{\bibinfo{volume}{92}},
  \bibinfo{pages}{013410}.

\bibitem{kang:2018}
\bibinfo{author}{\bibnamefont{Lin}, \bibfnamefont{K.}},
  \bibinfo{author}{\bibfnamefont{I.}~\bibnamefont{Tutunnikov}},
  \bibinfo{author}{\bibfnamefont{J.}~\bibnamefont{Qiang}},
  \bibinfo{author}{\bibfnamefont{J.}~\bibnamefont{Ma}},
  \bibinfo{author}{\bibfnamefont{Q.}~\bibnamefont{Song}},
  \bibinfo{author}{\bibfnamefont{Q.}~\bibnamefont{Ji}},
  \bibinfo{author}{\bibfnamefont{W.}~\bibnamefont{Zhang}},
  \bibinfo{author}{\bibfnamefont{H.}~\bibnamefont{Li}},
  \bibinfo{author}{\bibfnamefont{F.}~\bibnamefont{Sun}},
  \bibinfo{author}{\bibfnamefont{X.}~\bibnamefont{Gong}},
  \bibinfo{author}{\bibfnamefont{H.}~\bibnamefont{Li}},
  \bibinfo{author}{\bibfnamefont{P.}~\bibnamefont{Lu}}, \emph{et~al.},
  \bibinfo{year}{2018}, \bibinfo{journal}{Nature Comm.}
  \textbf{\bibinfo{volume}{9}}, \bibinfo{pages}{5134}.

\bibitem{loesch:1990}
\bibinfo{author}{\bibnamefont{Loesch}, \bibfnamefont{H.~J.}}, and
  \bibinfo{author}{\bibfnamefont{A.}~\bibnamefont{Remscheid}},
  \bibinfo{year}{1990}, \bibinfo{journal}{J. Chem. Phys.}
  \textbf{\bibinfo{volume}{93}}(\bibinfo{number}{7}), \bibinfo{pages}{4779}.

\bibitem{LondonoPRA10}
\bibinfo{author}{\bibnamefont{Londo\~no}, \bibfnamefont{B.~E.}},
  \bibinfo{author}{\bibfnamefont{J.~E.} \bibnamefont{Mahecha}},
  \bibinfo{author}{\bibfnamefont{E.}~\bibnamefont{Luc-Koenig}}, and
  \bibinfo{author}{\bibfnamefont{A.}~\bibnamefont{Crubellier}},
  \bibinfo{year}{2010}, \bibinfo{journal}{Phys. Rev. A}
  \textbf{\bibinfo{volume}{82}}(\bibinfo{number}{1}), \bibinfo{pages}{012510}.

\bibitem{lu:2016}
\bibinfo{author}{\bibnamefont{Lu}, \bibfnamefont{J.}},
  \bibinfo{author}{\bibfnamefont{Y.}~\bibnamefont{Zhang}},
  \bibinfo{author}{\bibfnamefont{H.~Y.} \bibnamefont{Hwang}},
  \bibinfo{author}{\bibfnamefont{B.~K.} \bibnamefont{Ofori-Okai}},
  \bibinfo{author}{\bibfnamefont{S.}~\bibnamefont{Fleischer}}, and
  \bibinfo{author}{\bibfnamefont{K.~A.} \bibnamefont{Nelson}},
  \bibinfo{year}{2016}, \textbf{\bibinfo{volume}{113}}(\bibinfo{number}{42}),
  \bibinfo{pages}{11800}.

\bibitem{luo:2015}
\bibinfo{author}{\bibnamefont{Luo}, \bibfnamefont{S.}},
  \bibinfo{author}{\bibfnamefont{R.}~\bibnamefont{Zhu}},
  \bibinfo{author}{\bibfnamefont{L.}~\bibnamefont{He}},
  \bibinfo{author}{\bibfnamefont{W.}~\bibnamefont{Hu}},
  \bibinfo{author}{\bibfnamefont{X.}~\bibnamefont{Li}},
  \bibinfo{author}{\bibfnamefont{P.}~\bibnamefont{Ma}},
  \bibinfo{author}{\bibfnamefont{C.}~\bibnamefont{Wang}},
  \bibinfo{author}{\bibfnamefont{F.}~\bibnamefont{Liu}},
  \bibinfo{author}{\bibfnamefont{W.~G.} \bibnamefont{Roeterdink}},
  \bibinfo{author}{\bibfnamefont{S.}~\bibnamefont{Stolte}}, and
  \bibinfo{author}{\bibfnamefont{D.}~\bibnamefont{Ding}}, \bibinfo{year}{2015},
  \bibinfo{journal}{Phys. Rev. A} \textbf{\bibinfo{volume}{91}},
  \bibinfo{pages}{053408}.

\bibitem{LuxAngewandte12}
\bibinfo{author}{\bibnamefont{Lux}, \bibfnamefont{C.}},
  \bibinfo{author}{\bibfnamefont{M.}~\bibnamefont{Wollenhaupt}},
  \bibinfo{author}{\bibfnamefont{T.}~\bibnamefont{Bolze}},
  \bibinfo{author}{\bibfnamefont{Q.}~\bibnamefont{Liang}},
  \bibinfo{author}{\bibfnamefont{J.}~\bibnamefont{K\"ohler}},
  \bibinfo{author}{\bibfnamefont{C.}~\bibnamefont{Sarpe}}, and
  \bibinfo{author}{\bibfnamefont{T.}~\bibnamefont{Baumert}},
  \bibinfo{year}{2012}, \bibinfo{journal}{Angew. Chem. Int. Ed.}
  \textbf{\bibinfo{volume}{51}}(\bibinfo{number}{20}), \bibinfo{pages}{5001}.

\bibitem{maan:2016}
\bibinfo{author}{\bibnamefont{Maan}, \bibfnamefont{A.}},
  \bibinfo{author}{\bibfnamefont{D.~S.} \bibnamefont{Ahlawat}}, and
  \bibinfo{author}{\bibfnamefont{V.}~\bibnamefont{Prasad}},
  \bibinfo{year}{2016}, \bibinfo{journal}{Chemical Physics Letters}
  \textbf{\bibinfo{volume}{650}}, \bibinfo{pages}{29 }.

\bibitem{machholm:2001b}
\bibinfo{author}{\bibnamefont{Machholm}, \bibfnamefont{M.}},
  \bibinfo{year}{2001}, \bibinfo{journal}{J. Chem. Phys.}
  \textbf{\bibinfo{volume}{115}}, \bibinfo{pages}{10724}.

\bibitem{machholm:1999}
\bibinfo{author}{\bibnamefont{Machholm}, \bibfnamefont{M.}}, and
  \bibinfo{author}{\bibfnamefont{N.~E.} \bibnamefont{Henriksen}},
  \bibinfo{year}{1999}, \bibinfo{journal}{J. Chem. Phys.}
  \textbf{\bibinfo{volume}{111}}, \bibinfo{pages}{3051}.

\bibitem{machholm:2001}
\bibinfo{author}{\bibnamefont{Machholm}, \bibfnamefont{M.}}, and
  \bibinfo{author}{\bibfnamefont{N.~E.} \bibnamefont{Henriksen}},
  \bibinfo{year}{2001}, \bibinfo{journal}{Phys. Rev. Lett.}
  \textbf{\bibinfo{volume}{87}}, \bibinfo{pages}{193001}.

\bibitem{madsen:2009}
\bibinfo{author}{\bibnamefont{Madsen}, \bibfnamefont{C.~B.}},
  \bibinfo{author}{\bibfnamefont{L.~B.} \bibnamefont{Madsen}},
  \bibinfo{author}{\bibfnamefont{S.~S.} \bibnamefont{Viftrup}},
  \bibinfo{author}{\bibfnamefont{M.~P.} \bibnamefont{Johansson}},
  \bibinfo{author}{\bibfnamefont{T.~B.} \bibnamefont{Poulsen}},
  \bibinfo{author}{\bibfnamefont{L.}~\bibnamefont{Holmegaard}},
  \bibinfo{author}{\bibfnamefont{V.}~\bibnamefont{Kumarappan}},
  \bibinfo{author}{\bibfnamefont{K.~A.} \bibnamefont{J{\o}rgensen}}, and
  \bibinfo{author}{\bibfnamefont{H.}~\bibnamefont{Stapelfeldt}},
  \bibinfo{year}{2009}, \bibinfo{journal}{J. Chem. Phys.}
  \textbf{\bibinfo{volume}{130}}(\bibinfo{number}{23}),
  \bibinfo{pages}{234310}.

\bibitem{magann:2018}
\bibinfo{author}{\bibnamefont{Magann}, \bibfnamefont{A.}},
  \bibinfo{author}{\bibfnamefont{T.-S.} \bibnamefont{Ho}}, and
  \bibinfo{author}{\bibfnamefont{H.}~\bibnamefont{Rabitz}},
  \bibinfo{year}{2018}, \bibinfo{journal}{Phys. Rev. A}
  \textbf{\bibinfo{volume}{98}}, \bibinfo{pages}{043429}.

\bibitem{ManaiPRL12}
\bibinfo{author}{\bibnamefont{Manai}, \bibfnamefont{I.}},
  \bibinfo{author}{\bibfnamefont{R.}~\bibnamefont{Horchani}},
  \bibinfo{author}{\bibfnamefont{H.}~\bibnamefont{Lignier}},
  \bibinfo{author}{\bibfnamefont{P.}~\bibnamefont{Pillet}},
  \bibinfo{author}{\bibfnamefont{D.}~\bibnamefont{Comparat}},
  \bibinfo{author}{\bibfnamefont{A.}~\bibnamefont{Fioretti}}, and
  \bibinfo{author}{\bibfnamefont{M.}~\bibnamefont{Allegrini}},
  \bibinfo{year}{2012}, \bibinfo{journal}{Phys. Rev. Lett.}
  \textbf{\bibinfo{volume}{109}}, \bibinfo{pages}{183001}.

\bibitem{ManmanaPRA17}
\bibinfo{author}{\bibnamefont{Manmana}, \bibfnamefont{S.~R.}},
  \bibinfo{author}{\bibfnamefont{M.}~\bibnamefont{M\"oller}},
  \bibinfo{author}{\bibfnamefont{R.}~\bibnamefont{Gezzi}}, and
  \bibinfo{author}{\bibfnamefont{K.~R.~A.} \bibnamefont{Hazzard}},
  \bibinfo{year}{2017}, \bibinfo{journal}{Phys. Rev. A}
  \textbf{\bibinfo{volume}{96}}, \bibinfo{pages}{043618}.

\bibitem{ManmanaPRB13}
\bibinfo{author}{\bibnamefont{Manmana}, \bibfnamefont{S.~R.}},
  \bibinfo{author}{\bibfnamefont{E.~M.} \bibnamefont{Stoudenmire}},
  \bibinfo{author}{\bibfnamefont{K.~R.~A.} \bibnamefont{Hazzard}},
  \bibinfo{author}{\bibfnamefont{A.~M.} \bibnamefont{Rey}}, and
  \bibinfo{author}{\bibfnamefont{A.~V.} \bibnamefont{Gorshkov}},
  \bibinfo{year}{2013}, \bibinfo{journal}{Phys. Rev. B}
  \textbf{\bibinfo{volume}{87}}, \bibinfo{pages}{081106}.

\bibitem{DeMarco18}
\bibinfo{author}{\bibnamefont{Marco}, \bibfnamefont{L.~D.}},
  \bibinfo{author}{\bibfnamefont{G.}~\bibnamefont{Valtolina}},
  \bibinfo{author}{\bibfnamefont{K.}~\bibnamefont{Matsuda}},
  \bibinfo{author}{\bibfnamefont{W.~G.} \bibnamefont{Tobias}},
  \bibinfo{author}{\bibfnamefont{J.~P.} \bibnamefont{Covey}}, and
  \bibinfo{author}{\bibfnamefont{J.}~\bibnamefont{Ye}}, \bibinfo{year}{2018},
  \bibinfo{journal}{arXiv:1808.00028} .

\bibitem{matos:2003}
\bibinfo{author}{\bibnamefont{Matos-Abiague}, \bibfnamefont{A.}}, and
  \bibinfo{author}{\bibfnamefont{J.}~\bibnamefont{Berakdar}},
  \bibinfo{year}{2003}, \bibinfo{journal}{Phys. Rev. A}
  \textbf{\bibinfo{volume}{68}}, \bibinfo{pages}{063411}.

\bibitem{gonzales:2007}
\bibinfo{author}{\bibnamefont{Mayle}, \bibfnamefont{M.}},
  \bibinfo{author}{\bibfnamefont{R.}~\bibnamefont{Gonz\'alez-F\'erez}}, and
  \bibinfo{author}{\bibfnamefont{P.}~\bibnamefont{Schmelcher}},
  \bibinfo{year}{2007}, \bibinfo{journal}{Phys. Rev. A}
  \textbf{\bibinfo{volume}{75}}, \bibinfo{pages}{013421}.

\bibitem{McDonaldPRL18}
\bibinfo{author}{\bibnamefont{McDonald}, \bibfnamefont{M.}},
  \bibinfo{author}{\bibfnamefont{I.}~\bibnamefont{Majewska}},
  \bibinfo{author}{\bibfnamefont{C.-H.} \bibnamefont{Lee}},
  \bibinfo{author}{\bibfnamefont{S.~S.} \bibnamefont{Kondov}},
  \bibinfo{author}{\bibfnamefont{B.~H.} \bibnamefont{McGuyer}},
  \bibinfo{author}{\bibfnamefont{R.}~\bibnamefont{Moszynski}}, and
  \bibinfo{author}{\bibfnamefont{T.}~\bibnamefont{Zelevinsky}},
  \bibinfo{year}{2018}, \bibinfo{journal}{Phys. Rev. Lett.}
  \textbf{\bibinfo{volume}{120}}, \bibinfo{pages}{033201}.

\bibitem{McDonaldNat16}
\bibinfo{author}{\bibnamefont{McDonald}, \bibfnamefont{M.}},
  \bibinfo{author}{\bibfnamefont{B.~H.} \bibnamefont{McGuyer}},
  \bibinfo{author}{\bibfnamefont{F.}~\bibnamefont{Apfelbeck}},
  \bibinfo{author}{\bibfnamefont{C.~H.} \bibnamefont{Lee}},
  \bibinfo{author}{\bibfnamefont{I.}~\bibnamefont{Majewska}},
  \bibinfo{author}{\bibfnamefont{R.}~\bibnamefont{Moszynski}}, and
  \bibinfo{author}{\bibfnamefont{T.}~\bibnamefont{Zelevinsky}},
  \bibinfo{year}{2016}, \bibinfo{journal}{Nature}
  \textbf{\bibinfo{volume}{535}}, \bibinfo{pages}{122}.

\bibitem{MetcalfBook}
\bibinfo{author}{\bibnamefont{Metcalf}, \bibfnamefont{H.~J.}}, and
  \bibinfo{author}{\bibfnamefont{P.}~\bibnamefont{van~der Straten}},
  \bibinfo{year}{1999}, \emph{\bibinfo{title}{Laser Cooling and Trapping}}
  (\bibinfo{publisher}{Springer}, \bibinfo{address}{New York}).

\bibitem{MicheliNatPhys06}
\bibinfo{author}{\bibnamefont{Micheli}, \bibfnamefont{A.}},
  \bibinfo{author}{\bibfnamefont{G.~K.} \bibnamefont{Brennen}}, and
  \bibinfo{author}{\bibfnamefont{P.}~\bibnamefont{Zoller}},
  \bibinfo{year}{2006}, \bibinfo{journal}{Nature Phys.}
  \textbf{\bibinfo{volume}{2}}, \bibinfo{pages}{341 }.

\bibitem{MicheliPRA07}
\bibinfo{author}{\bibnamefont{Micheli}, \bibfnamefont{A.}},
  \bibinfo{author}{\bibfnamefont{G.}~\bibnamefont{Pupillo}},
  \bibinfo{author}{\bibfnamefont{H.~P.} \bibnamefont{B{\"u}chler}}, and
  \bibinfo{author}{\bibfnamefont{P.}~\bibnamefont{Zoller}},
  \bibinfo{year}{2007}, \bibinfo{journal}{Phys. Rev. A}
  \textbf{\bibinfo{volume}{76}}, \bibinfo{pages}{43604}.

\bibitem{BikashPRA16}
\bibinfo{author}{\bibnamefont{Midya}, \bibfnamefont{B.}},
  \bibinfo{author}{\bibfnamefont{M.}~\bibnamefont{Tomza}},
  \bibinfo{author}{\bibfnamefont{R.}~\bibnamefont{Schmidt}}, and
  \bibinfo{author}{\bibfnamefont{M.}~\bibnamefont{Lemeshko}},
  \bibinfo{year}{2016}, \bibinfo{journal}{Phys. Rev. A}
  \textbf{\bibinfo{volume}{94}}, \bibinfo{pages}{041601(R)}.

\bibitem{MilmanPRL07}
\bibinfo{author}{\bibnamefont{Milman}, \bibfnamefont{P.}},
  \bibinfo{author}{\bibfnamefont{A.}~\bibnamefont{Keller}},
  \bibinfo{author}{\bibfnamefont{E.}~\bibnamefont{Charron}}, and
  \bibinfo{author}{\bibfnamefont{O.}~\bibnamefont{Atabek}},
  \bibinfo{year}{2007}, \bibinfo{journal}{Phys. Rev. Lett.}
  \textbf{\bibinfo{volume}{99}}, \bibinfo{pages}{130405}.

\bibitem{milner:2015}
\bibinfo{author}{\bibnamefont{Milner}, \bibfnamefont{A.~A.}},
  \bibinfo{author}{\bibfnamefont{A.}~\bibnamefont{Korobenko}},
  \bibinfo{author}{\bibfnamefont{J.}~\bibnamefont{Flo\ss{}}},
  \bibinfo{author}{\bibfnamefont{I.~S.} \bibnamefont{Averbukh}}, and
  \bibinfo{author}{\bibfnamefont{V.}~\bibnamefont{Milner}},
  \bibinfo{year}{2015}{\natexlab{a}}, \bibinfo{journal}{Phys. Rev. Lett.}
  \textbf{\bibinfo{volume}{115}}, \bibinfo{pages}{033005}.

\bibitem{milner:2014}
\bibinfo{author}{\bibnamefont{Milner}, \bibfnamefont{A.~A.}},
  \bibinfo{author}{\bibfnamefont{A.}~\bibnamefont{Korobenko}},
  \bibinfo{author}{\bibfnamefont{J.~W.} \bibnamefont{Hepburn}}, and
  \bibinfo{author}{\bibfnamefont{V.}~\bibnamefont{Milner}},
  \bibinfo{year}{2014}, \bibinfo{journal}{Phys. Rev. Lett.}
  \textbf{\bibinfo{volume}{113}}, \bibinfo{pages}{043005}.

\bibitem{milner:2016}
\bibinfo{author}{\bibnamefont{Milner}, \bibfnamefont{A.~A.}},
  \bibinfo{author}{\bibfnamefont{A.}~\bibnamefont{Korobenko}}, and
  \bibinfo{author}{\bibfnamefont{V.}~\bibnamefont{Milner}},
  \bibinfo{year}{2016}, \bibinfo{journal}{Phys. Rev. A}
  \textbf{\bibinfo{volume}{93}}, \bibinfo{pages}{053408}.

\bibitem{MilnerPRL17}
\bibinfo{author}{\bibnamefont{Milner}, \bibfnamefont{A.~A.}},
  \bibinfo{author}{\bibfnamefont{A.}~\bibnamefont{Korobenko}}, and
  \bibinfo{author}{\bibfnamefont{V.}~\bibnamefont{Milner}},
  \bibinfo{year}{2017}, \bibinfo{journal}{Phys. Rev. Lett.}
  \textbf{\bibinfo{volume}{118}}, \bibinfo{pages}{243201}.

\bibitem{MilnerPRX15}
\bibinfo{author}{\bibnamefont{Milner}, \bibfnamefont{A.~A.}},
  \bibinfo{author}{\bibfnamefont{A.}~\bibnamefont{Korobenko}},
  \bibinfo{author}{\bibfnamefont{K.}~\bibnamefont{Rezaiezadeh}}, and
  \bibinfo{author}{\bibfnamefont{V.}~\bibnamefont{Milner}},
  \bibinfo{year}{2015}{\natexlab{b}}, \bibinfo{journal}{Phys. Rev. X}
  \textbf{\bibinfo{volume}{5}}, \bibinfo{pages}{031041}.

\bibitem{mirahmadi:2018}
\bibinfo{author}{\bibnamefont{Mirahmadi}, \bibfnamefont{M.}},
  \bibinfo{author}{\bibfnamefont{B.}~\bibnamefont{Schmidt}},
  \bibinfo{author}{\bibfnamefont{M.}~\bibnamefont{Karra}}, and
  \bibinfo{author}{\bibfnamefont{B.}~\bibnamefont{Friedrich}},
  \bibinfo{year}{2018}, \bibinfo{journal}{J. Chem. Phys.}
  \textbf{\bibinfo{volume}{149}}, \bibinfo{pages}{174109}.

\bibitem{MirandaNatPhys11}
\bibinfo{author}{\bibnamefont{de~Miranda}, \bibfnamefont{M.~H.~G.}},
  \bibinfo{author}{\bibfnamefont{A.}~\bibnamefont{Chotia}},
  \bibinfo{author}{\bibfnamefont{B.}~\bibnamefont{Neyenhuis}},
  \bibinfo{author}{\bibfnamefont{D.}~\bibnamefont{Wang}},
  \bibinfo{author}{\bibfnamefont{G.}~\bibnamefont{Quemener}},
  \bibinfo{author}{\bibfnamefont{S.}~\bibnamefont{Ospelkaus}},
  \bibinfo{author}{\bibfnamefont{J.~L.} \bibnamefont{Bohn}},
  \bibinfo{author}{\bibfnamefont{J.}~\bibnamefont{Ye}}, and
  \bibinfo{author}{\bibfnamefont{D.~S.} \bibnamefont{Jin}},
  \bibinfo{year}{2011}, \bibinfo{journal}{Nat. Phys.}
  \textbf{\bibinfo{volume}{7}}(\bibinfo{number}{6}), \bibinfo{pages}{502}.

\bibitem{rouchon:2004}
\bibinfo{author}{\bibnamefont{Mirrahimi}, \bibfnamefont{M.}}, and
  \bibinfo{author}{\bibfnamefont{P.}~\bibnamefont{Rouchon}},
  \bibinfo{year}{2004}, \bibinfo{journal}{IEEE Trans. A. C.}
  \textbf{\bibinfo{volume}{49}}, \bibinfo{pages}{745}.

\bibitem{MishimaChemPhys09}
\bibinfo{author}{\bibnamefont{Mishima}, \bibfnamefont{K.}}, and
  \bibinfo{author}{\bibfnamefont{K.}~\bibnamefont{Yamashita}},
  \bibinfo{year}{2009}, \bibinfo{journal}{Chemical Physics}
  \textbf{\bibinfo{volume}{361}}(\bibinfo{number}{1}), \bibinfo{pages}{106 },
  ISSN \bibinfo{issn}{0301-0104}.

\bibitem{mizuse:2015}
\bibinfo{author}{\bibnamefont{Mizuse}, \bibfnamefont{K.}},
  \bibinfo{author}{\bibfnamefont{K.}~\bibnamefont{Kitano}},
  \bibinfo{author}{\bibfnamefont{H.}~\bibnamefont{Hasegawa}}, and
  \bibinfo{author}{\bibfnamefont{Y.}~\bibnamefont{Ohshima}},
  \bibinfo{year}{2012}, \bibinfo{journal}{Sci. Adv.}
  \textbf{\bibinfo{volume}{1}}, \bibinfo{pages}{e1400185}.

\bibitem{MolhavePRA00}
\bibinfo{author}{\bibnamefont{M\o{}lhave}, \bibfnamefont{K.}}, and
  \bibinfo{author}{\bibfnamefont{M.}~\bibnamefont{Drewsen}},
  \bibinfo{year}{2000}, \bibinfo{journal}{Phys. Rev. A}
  \textbf{\bibinfo{volume}{62}}, \bibinfo{pages}{011401}.

\bibitem{MorrisonJPCA13}
\bibinfo{author}{\bibnamefont{Morrison}, \bibfnamefont{A.~M.}},
  \bibinfo{author}{\bibfnamefont{P.~L.} \bibnamefont{Raston}}, and
  \bibinfo{author}{\bibfnamefont{G.~E.} \bibnamefont{Douberly}},
  \bibinfo{year}{2013}, \bibinfo{journal}{J. Phys. Chem. A}
  \textbf{\bibinfo{volume}{117}}, \bibinfo{pages}{11640}.

\bibitem{MosesNatPhys17}
\bibinfo{author}{\bibnamefont{Moses}, \bibfnamefont{S.~A.}},
  \bibinfo{author}{\bibfnamefont{J.~P.} \bibnamefont{Covey}},
  \bibinfo{author}{\bibfnamefont{M.~T.} \bibnamefont{Miecnikowski}},
  \bibinfo{author}{\bibfnamefont{D.~S.} \bibnamefont{Jin}}, and
  \bibinfo{author}{\bibfnamefont{J.}~\bibnamefont{Ye}}, \bibinfo{year}{2017},
  \bibinfo{journal}{Nat. Phys.} \textbf{\bibinfo{volume}{13}},
  \bibinfo{pages}{13}.

\bibitem{MosesSci15}
\bibinfo{author}{\bibnamefont{Moses}, \bibfnamefont{S.~A.}},
  \bibinfo{author}{\bibfnamefont{J.~P.} \bibnamefont{Covey}},
  \bibinfo{author}{\bibfnamefont{M.~T.} \bibnamefont{Miecnikowski}},
  \bibinfo{author}{\bibfnamefont{B.}~\bibnamefont{Yan}},
  \bibinfo{author}{\bibfnamefont{B.}~\bibnamefont{Gadway}},
  \bibinfo{author}{\bibfnamefont{J.}~\bibnamefont{Ye}}, and
  \bibinfo{author}{\bibfnamefont{D.~S.} \bibnamefont{Jin}},
  \bibinfo{year}{2015}, \bibinfo{journal}{Science}
  \textbf{\bibinfo{volume}{350}}(\bibinfo{number}{6261}), \bibinfo{pages}{659}.

\bibitem{moskun_rotational_2006}
\bibinfo{author}{\bibnamefont{Moskun}, \bibfnamefont{A.~C.}},
  \bibinfo{author}{\bibfnamefont{A.~E.} \bibnamefont{Jailaubekov}},
  \bibinfo{author}{\bibfnamefont{S.~E.} \bibnamefont{Bradforth}},
  \bibinfo{author}{\bibfnamefont{G.}~\bibnamefont{Tao}}, and
  \bibinfo{author}{\bibfnamefont{R.~M.} \bibnamefont{Stratt}},
  \bibinfo{year}{2006}, \bibinfo{journal}{Science}
  \textbf{\bibinfo{volume}{311}}(\bibinfo{number}{5769}),
  \bibinfo{pages}{1907}.

\bibitem{MudrichIRPC14}
\bibinfo{author}{\bibnamefont{Mudrich}, \bibfnamefont{M.}}, and
  \bibinfo{author}{\bibfnamefont{F.}~\bibnamefont{Stienkemeier}},
  \bibinfo{year}{2014}, \bibinfo{journal}{Int. Rev. Phys. Chem.}
  \textbf{\bibinfo{volume}{33}}, \bibinfo{pages}{301}.

\bibitem{mun:2018}
\bibinfo{author}{\bibnamefont{Mun}, \bibfnamefont{J.~H.}}, and
  \bibinfo{author}{\bibfnamefont{H.}~\bibnamefont{Sakai}},
  \bibinfo{year}{2018}, \bibinfo{journal}{Phys. Rev. A}
  \textbf{\bibinfo{volume}{98}}, \bibinfo{pages}{013404}.

\bibitem{mun:2014}
\bibinfo{author}{\bibnamefont{Mun}, \bibfnamefont{J.~H.}},
  \bibinfo{author}{\bibfnamefont{D.}~\bibnamefont{Takei}},
  \bibinfo{author}{\bibfnamefont{S.}~\bibnamefont{Minemoto}}, and
  \bibinfo{author}{\bibfnamefont{H.}~\bibnamefont{Sakai}},
  \bibinfo{year}{2014}, \bibinfo{journal}{Phys. Rev. A}
  \textbf{\bibinfo{volume}{89}}, \bibinfo{pages}{051402}.

\bibitem{Mur-Petit2013}
\bibinfo{author}{\bibnamefont{Mur-Petit}, \bibfnamefont{J.}},
  \bibinfo{author}{\bibfnamefont{J.}~\bibnamefont{P{\'e}rez-R{\'\i}os}},
  \bibinfo{author}{\bibfnamefont{J.}~\bibnamefont{Campos-Mart{\'\i}nez}},
  \bibinfo{author}{\bibfnamefont{M.~I.} \bibnamefont{Hern{\'a}ndez}},
  \bibinfo{author}{\bibfnamefont{S.}~\bibnamefont{Willitsch}}, and
  \bibinfo{author}{\bibfnamefont{J.~J.} \bibnamefont{Garc{\'\i}a-Ripoll}},
  \bibinfo{year}{2013}, in \emph{\bibinfo{booktitle}{Architecture and Design of
  Molecule Logic Gates and Atom Circuits: Proceedings of the 2nd AtMol European
  Workshop}}, edited by
  \bibinfo{editor}{\bibfnamefont{N.}~\bibnamefont{Lorente}} and
  \bibinfo{editor}{\bibfnamefont{C.}~\bibnamefont{Joachim}}
  (\bibinfo{publisher}{Springer Berlin Heidelberg}, \bibinfo{address}{Berlin,
  Heidelberg}), ISBN \bibinfo{isbn}{978-3-642-33137-4}, pp.
  \bibinfo{pages}{267--277}.

\bibitem{muramatsu}
\bibinfo{author}{\bibnamefont{Muramatsu}, \bibfnamefont{M.}},
  \bibinfo{author}{\bibfnamefont{M.}~\bibnamefont{Hita}},
  \bibinfo{author}{\bibfnamefont{S.}~\bibnamefont{Minemoto}}, and
  \bibinfo{author}{\bibfnamefont{H.}~\bibnamefont{Sakai}},
  \bibinfo{year}{2009}, \bibinfo{journal}{Phys. Rev. A}
  \textbf{\bibinfo{volume}{79}}, \bibinfo{pages}{011403}.

\bibitem{nakagami:2008}
\bibinfo{author}{\bibnamefont{Nakagami}, \bibfnamefont{K.}},
  \bibinfo{author}{\bibfnamefont{Y.}~\bibnamefont{Mizumoto}}, and
  \bibinfo{author}{\bibfnamefont{Y.}~\bibnamefont{Ohtsuki}},
  \bibinfo{year}{2008}{\natexlab{a}}, \bibinfo{journal}{J. Chem. Phys.}
  \textbf{\bibinfo{volume}{129}}(\bibinfo{number}{19}),
  \bibinfo{pages}{194103}.

\bibitem{nakagami:2008b}
\bibinfo{author}{\bibnamefont{Nakagami}, \bibfnamefont{Y.}},
  \bibinfo{author}{\bibfnamefont{Y.}~\bibnamefont{Mizumoto}}, and
  \bibinfo{author}{\bibfnamefont{Y.}~\bibnamefont{Ohtsuki}},
  \bibinfo{year}{2008}{\natexlab{b}}, \bibinfo{journal}{J. Chem. Phys.}
  \textbf{\bibinfo{volume}{129}}, \bibinfo{pages}{194103}.

\bibitem{nakajima:2012}
\bibinfo{author}{\bibnamefont{Nakajima}, \bibfnamefont{K.}},
  \bibinfo{author}{\bibfnamefont{H.}~\bibnamefont{Abe}}, and
  \bibinfo{author}{\bibfnamefont{Y.}~\bibnamefont{Ohtsuki}},
  \bibinfo{year}{2012}, \bibinfo{journal}{J. Phys. Chem. A}
  \textbf{\bibinfo{volume}{116}}, \bibinfo{pages}{11219}.

\bibitem{nakajima:2016}
\bibinfo{author}{\bibnamefont{Nakajima}, \bibfnamefont{K.}},
  \bibinfo{author}{\bibfnamefont{M.}~\bibnamefont{Yoshida}},
  \bibinfo{author}{\bibfnamefont{T.}~\bibnamefont{Nakajima}}, and
  \bibinfo{author}{\bibfnamefont{Y.}~\bibnamefont{Ohtsuki}},
  \bibinfo{year}{2016}, \bibinfo{journal}{Mol. Phys.}
  \textbf{\bibinfo{volume}{115}}, \bibinfo{pages}{1}.

\bibitem{nalda:2004}
\bibinfo{author}{\bibnamefont{Nalda}, \bibfnamefont{R.~d.}},
  \bibinfo{author}{\bibfnamefont{E.}~\bibnamefont{Heesel}},
  \bibinfo{author}{\bibfnamefont{M.}~\bibnamefont{Lein}},
  \bibinfo{author}{\bibfnamefont{N.}~\bibnamefont{Hay}},
  \bibinfo{author}{\bibfnamefont{R.}~\bibnamefont{Velotta}},
  \bibinfo{author}{\bibfnamefont{E.}~\bibnamefont{Springate}},
  \bibinfo{author}{\bibfnamefont{M.}~\bibnamefont{Castillejo}}, and
  \bibinfo{author}{\bibfnamefont{J.~P.} \bibnamefont{Marangos}},
  \bibinfo{year}{2004}, \bibinfo{journal}{Phys. Rev. A}
  \textbf{\bibinfo{volume}{69}}, \bibinfo{pages}{031804}.

\bibitem{NattererPRL13}
\bibinfo{author}{\bibnamefont{Natterer}, \bibfnamefont{F.~D.}},
  \bibinfo{author}{\bibfnamefont{F.}~\bibnamefont{Patthey}}, and
  \bibinfo{author}{\bibfnamefont{H.}~\bibnamefont{Brune}},
  \bibinfo{year}{2013}, \bibinfo{journal}{Phys. Rev. Lett.}
  \textbf{\bibinfo{volume}{111}}, \bibinfo{pages}{175303}.

\bibitem{NdongJMO14}
\bibinfo{author}{\bibnamefont{Ndong}, \bibfnamefont{M.}},
  \bibinfo{author}{\bibfnamefont{C.~P.} \bibnamefont{Koch}}, and
  \bibinfo{author}{\bibfnamefont{D.}~\bibnamefont{Sugny}},
  \bibinfo{year}{2014}, \bibinfo{journal}{J. Mod. Opt.}
  \textbf{\bibinfo{volume}{61}}(\bibinfo{number}{10}), \bibinfo{pages}{857}.

\bibitem{Ndong:2013}
\bibinfo{author}{\bibnamefont{Ndong}, \bibfnamefont{M.}},
  \bibinfo{author}{\bibfnamefont{M.}~\bibnamefont{Lapert}},
  \bibinfo{author}{\bibfnamefont{C.~P.} \bibnamefont{Koch}}, and
  \bibinfo{author}{\bibfnamefont{D.}~\bibnamefont{Sugny}},
  \bibinfo{year}{2013}, \bibinfo{journal}{Phys. Rev. A}
  \textbf{\bibinfo{volume}{87}}, \bibinfo{pages}{043416}.

\bibitem{NiSci08}
\bibinfo{author}{\bibnamefont{Ni}, \bibfnamefont{K.-K.}},
  \bibinfo{author}{\bibfnamefont{S.}~\bibnamefont{Ospelkaus}},
  \bibinfo{author}{\bibfnamefont{M.~H.~G.} \bibnamefont{de~Miranda}},
  \bibinfo{author}{\bibfnamefont{A.}~\bibnamefont{Pe'er}},
  \bibinfo{author}{\bibfnamefont{B.}~\bibnamefont{Neyenhuis}},
  \bibinfo{author}{\bibfnamefont{J.~J.} \bibnamefont{Zirbel}},
  \bibinfo{author}{\bibfnamefont{S.}~\bibnamefont{Kotochigova}},
  \bibinfo{author}{\bibfnamefont{P.~S.} \bibnamefont{Julienne}},
  \bibinfo{author}{\bibfnamefont{D.~S.} \bibnamefont{Jin}}, and
  \bibinfo{author}{\bibfnamefont{J.}~\bibnamefont{Ye}}, \bibinfo{year}{2008},
  \bibinfo{journal}{Science}
  \textbf{\bibinfo{volume}{322}}(\bibinfo{number}{5899}), \bibinfo{pages}{231}.

\bibitem{nielsen:2012}
\bibinfo{author}{\bibnamefont{Nielsen}, \bibfnamefont{J.~H.}},
  \bibinfo{author}{\bibfnamefont{H.}~\bibnamefont{Stapelfeldt}},
  \bibinfo{author}{\bibfnamefont{J.}~\bibnamefont{K\"upper}},
  \bibinfo{author}{\bibfnamefont{B.}~\bibnamefont{Friedrich}},
  \bibinfo{author}{\bibfnamefont{J.~J.} \bibnamefont{Omiste}}, and
  \bibinfo{author}{\bibfnamefont{R.}~\bibnamefont{Gonz\'alez-F\'erez}},
  \bibinfo{year}{2012}, \bibinfo{journal}{Phys. Rev. Lett.}
  \textbf{\bibinfo{volume}{108}}, \bibinfo{pages}{193001}.

\bibitem{normand:1992}
\bibinfo{author}{\bibnamefont{Normand}, \bibfnamefont{D.}},
  \bibinfo{author}{\bibfnamefont{L.~A.} \bibnamefont{Lompre}}, and
  \bibinfo{author}{\bibfnamefont{C.}~\bibnamefont{Cornaggia}},
  \bibinfo{year}{1992}, \bibinfo{journal}{J. Phys. B}
  \textbf{\bibinfo{volume}{25}}, \bibinfo{pages}{L497}.

\bibitem{Oda:10}
\bibinfo{author}{\bibnamefont{Oda}, \bibfnamefont{K.}},
  \bibinfo{author}{\bibfnamefont{M.}~\bibnamefont{Hita}},
  \bibinfo{author}{\bibfnamefont{S.}~\bibnamefont{Minemoto}}, and
  \bibinfo{author}{\bibfnamefont{H.}~\bibnamefont{Sakai}},
  \bibinfo{year}{2010}, \bibinfo{journal}{Phys. Rev. Lett.}
  \textbf{\bibinfo{volume}{104}}, \bibinfo{pages}{213901}.

\bibitem{ohkubo:2004}
\bibinfo{author}{\bibnamefont{Ohkubo}, \bibfnamefont{J.}},
  \bibinfo{author}{\bibfnamefont{T.}~\bibnamefont{Kato}},
  \bibinfo{author}{\bibfnamefont{H.}~\bibnamefont{Kono}}, and
  \bibinfo{author}{\bibfnamefont{Y.}~\bibnamefont{Fujimura}},
  \bibinfo{year}{2004}, \bibinfo{journal}{J. Chem. Phys.}
  \textbf{\bibinfo{volume}{120}}(\bibinfo{number}{19}), \bibinfo{pages}{9123}.

\bibitem{ohshima:2010}
\bibinfo{author}{\bibnamefont{Ohshima}, \bibfnamefont{Y.}}, and
  \bibinfo{author}{\bibfnamefont{H.}~\bibnamefont{Hasegawa}},
  \bibinfo{year}{2010}, \bibinfo{journal}{Int. Rev. Phys. Chem.}
  \textbf{\bibinfo{volume}{29}}, \bibinfo{pages}{619}.

\bibitem{omiste:2018}
\bibinfo{author}{\bibnamefont{Omiste}, \bibfnamefont{J.~J.}},
  \bibinfo{year}{2018}, \bibinfo{journal}{Phys. Rev. A}
  \textbf{\bibinfo{volume}{97}}, \bibinfo{pages}{023407}.

\bibitem{omiste:2012}
\bibinfo{author}{\bibnamefont{Omiste}, \bibfnamefont{J.~J.}}, and
  \bibinfo{author}{\bibfnamefont{R.}~\bibnamefont{Gonz\'alez-F\'erez}},
  \bibinfo{year}{2012}, \bibinfo{journal}{Phys. Rev. A}
  \textbf{\bibinfo{volume}{86}}, \bibinfo{pages}{043437}.

\bibitem{Omiste:16}
\bibinfo{author}{\bibnamefont{Omiste}, \bibfnamefont{J.~J.}}, and
  \bibinfo{author}{\bibfnamefont{R.}~\bibnamefont{Gonz\'alez-F\'erez}},
  \bibinfo{year}{2016}, \bibinfo{journal}{Phys. Rev. A}
  \textbf{\bibinfo{volume}{94}}, \bibinfo{pages}{063408}.

\bibitem{omiste:2017}
\bibinfo{author}{\bibnamefont{Omiste}, \bibfnamefont{J.~J.}}, and
  \bibinfo{author}{\bibfnamefont{L.~B.} \bibnamefont{Madsen}},
  \bibinfo{year}{2017}, \bibinfo{journal}{Phys. Rev. A}
  \textbf{\bibinfo{volume}{95}}, \bibinfo{pages}{023402}.

\bibitem{Ordonez18}
\bibinfo{author}{\bibnamefont{Ordonez}, \bibfnamefont{A.~F.}}, and
  \bibinfo{author}{\bibfnamefont{O.}~\bibnamefont{Smirnova}},
  \bibinfo{year}{2018}, \bibinfo{journal}{arXiv:1802.06540} .

\bibitem{ortigoso:2004}
\bibinfo{author}{\bibnamefont{Ortigoso}, \bibfnamefont{J.}},
  \bibinfo{year}{2004}, \bibinfo{journal}{Phys. Rev. Lett.}
  \textbf{\bibinfo{volume}{93}}, \bibinfo{pages}{073001}.

\bibitem{ortigoso:2012}
\bibinfo{author}{\bibnamefont{Ortigoso}, \bibfnamefont{J.}},
  \bibinfo{year}{2012}, \bibinfo{journal}{J. Chem. Phys.}
  \textbf{\bibinfo{volume}{137}}, \bibinfo{pages}{044303}.

\bibitem{ortigoso:2013}
\bibinfo{author}{\bibnamefont{Ortigoso}, \bibfnamefont{J.}}, and
  \bibinfo{author}{\bibfnamefont{L.~H.} \bibnamefont{Coudert}},
  \bibinfo{year}{2013}, \bibinfo{journal}{Phys. Rev. A}
  \textbf{\bibinfo{volume}{87}}, \bibinfo{pages}{043403}.

\bibitem{ortigoso:1999}
\bibinfo{author}{\bibnamefont{Ortigoso}, \bibfnamefont{J.}},
  \bibinfo{author}{\bibfnamefont{M.}~\bibnamefont{Rodrigues}},
  \bibinfo{author}{\bibfnamefont{M.}~\bibnamefont{Gupta}}, and
  \bibinfo{author}{\bibfnamefont{B.~J.} \bibnamefont{Friedrich}},
  \bibinfo{year}{1999}, \bibinfo{journal}{J. Chem. Phys.}
  \textbf{\bibinfo{volume}{110}}, \bibinfo{pages}{3870}.

\bibitem{ortigoso:2010}
\bibinfo{author}{\bibnamefont{Ortigoso}, \bibfnamefont{J.}},
  \bibinfo{author}{\bibfnamefont{M.}~\bibnamefont{Rodr{\'\i}guez}},
  \bibinfo{author}{\bibfnamefont{J.}~\bibnamefont{Santos}},
  \bibinfo{author}{\bibfnamefont{A.}~\bibnamefont{Karpati}}, and
  \bibinfo{author}{\bibfnamefont{V.}~\bibnamefont{Szalay}},
  \bibinfo{year}{2010}, \bibinfo{journal}{J. Chem. Phys.}
  \textbf{\bibinfo{volume}{132}}(\bibinfo{number}{7}), \bibinfo{pages}{074105}.

\bibitem{OrtnerNJP09}
\bibinfo{author}{\bibnamefont{Ortner}, \bibfnamefont{M.}},
  \bibinfo{author}{\bibfnamefont{A.}~\bibnamefont{Micheli}},
  \bibinfo{author}{\bibfnamefont{G.}~\bibnamefont{Pupillo}}, and
  \bibinfo{author}{\bibfnamefont{P.}~\bibnamefont{Zoller}},
  \bibinfo{year}{2009}, \bibinfo{journal}{New Journal of Physics}
  \textbf{\bibinfo{volume}{11}}(\bibinfo{number}{5}), \bibinfo{pages}{055045}.

\bibitem{Ortner2011}
\bibinfo{author}{\bibnamefont{Ortner}, \bibfnamefont{M.}},
  \bibinfo{author}{\bibfnamefont{Y.~L.} \bibnamefont{Zhou}},
  \bibinfo{author}{\bibfnamefont{P.}~\bibnamefont{Rabl}}, and
  \bibinfo{author}{\bibfnamefont{P.}~\bibnamefont{Zoller}},
  \bibinfo{year}{2011}, \bibinfo{journal}{Quantum Information Processing}
  \textbf{\bibinfo{volume}{10}}, \bibinfo{pages}{793}.

\bibitem{OspelkausPRL10}
\bibinfo{author}{\bibnamefont{Ospelkaus}, \bibfnamefont{S.}},
  \bibinfo{author}{\bibfnamefont{K.-K.} \bibnamefont{Ni}},
  \bibinfo{author}{\bibfnamefont{G.}~\bibnamefont{Qu\'em\'ener}},
  \bibinfo{author}{\bibfnamefont{B.}~\bibnamefont{Neyenhuis}},
  \bibinfo{author}{\bibfnamefont{D.}~\bibnamefont{Wang}},
  \bibinfo{author}{\bibfnamefont{M.~H.~G.} \bibnamefont{de~Miranda}},
  \bibinfo{author}{\bibfnamefont{J.~L.} \bibnamefont{Bohn}},
  \bibinfo{author}{\bibfnamefont{J.}~\bibnamefont{Ye}}, and
  \bibinfo{author}{\bibfnamefont{D.~S.} \bibnamefont{Jin}},
  \bibinfo{year}{2010}{\natexlab{a}}, \bibinfo{journal}{Phys. Rev. Lett.}
  \textbf{\bibinfo{volume}{104}}, \bibinfo{pages}{030402}.

\bibitem{OspelkausSci10}
\bibinfo{author}{\bibnamefont{Ospelkaus}, \bibfnamefont{S.}},
  \bibinfo{author}{\bibfnamefont{K.-K.} \bibnamefont{Ni}},
  \bibinfo{author}{\bibfnamefont{D.}~\bibnamefont{Wang}},
  \bibinfo{author}{\bibfnamefont{M.~H.~G.} \bibnamefont{de~Miranda}},
  \bibinfo{author}{\bibfnamefont{B.}~\bibnamefont{Neyenhuis}},
  \bibinfo{author}{\bibfnamefont{G.}~\bibnamefont{Qu{\'e}m{\'e}ner}},
  \bibinfo{author}{\bibfnamefont{P.~S.} \bibnamefont{Julienne}},
  \bibinfo{author}{\bibfnamefont{J.~L.} \bibnamefont{Bohn}},
  \bibinfo{author}{\bibfnamefont{D.~S.} \bibnamefont{Jin}}, and
  \bibinfo{author}{\bibfnamefont{J.}~\bibnamefont{Ye}},
  \bibinfo{year}{2010}{\natexlab{b}}, \bibinfo{journal}{Science}
  \textbf{\bibinfo{volume}{327}}(\bibinfo{number}{5967}), \bibinfo{pages}{853}.

\bibitem{OspelkausNatPhys08}
\bibinfo{author}{\bibnamefont{Ospelkaus}, \bibfnamefont{S.}},
  \bibinfo{author}{\bibfnamefont{A.}~\bibnamefont{Pe'er}},
  \bibinfo{author}{\bibfnamefont{K.-K.} \bibnamefont{Ni}},
  \bibinfo{author}{\bibfnamefont{J.~J.} \bibnamefont{Zirbel}},
  \bibinfo{author}{\bibfnamefont{B.}~\bibnamefont{Neyenhuis}},
  \bibinfo{author}{\bibfnamefont{S.}~\bibnamefont{Kotochigova}},
  \bibinfo{author}{\bibfnamefont{P.~S.} \bibnamefont{Julienne}},
  \bibinfo{author}{\bibfnamefont{J.}~\bibnamefont{Ye}}, and
  \bibinfo{author}{\bibfnamefont{D.~S.} \bibnamefont{Jin}},
  \bibinfo{year}{2008}, \bibinfo{journal}{Nature Phys.}
  \textbf{\bibinfo{volume}{4}}, \bibinfo{pages}{622}.

\bibitem{OstendorfPRL06}
\bibinfo{author}{\bibnamefont{Ostendorf}, \bibfnamefont{A.}},
  \bibinfo{author}{\bibfnamefont{C.~B.} \bibnamefont{Zhang}},
  \bibinfo{author}{\bibfnamefont{M.~A.} \bibnamefont{Wilson}},
  \bibinfo{author}{\bibfnamefont{D.}~\bibnamefont{Offenberg}},
  \bibinfo{author}{\bibfnamefont{B.}~\bibnamefont{Roth}}, and
  \bibinfo{author}{\bibfnamefont{S.}~\bibnamefont{Schiller}},
  \bibinfo{year}{2006}, \bibinfo{journal}{Phys. Rev. Lett.}
  \textbf{\bibinfo{volume}{97}}, \bibinfo{pages}{243005}.

\bibitem{OsterwalderEPJTI2015}
\bibinfo{author}{\bibnamefont{Osterwalder}, \bibfnamefont{A.}},
  \bibinfo{year}{2015}, \bibinfo{journal}{EPJ Techniques and Instrumentation}
  \textbf{\bibinfo{volume}{2}}(\bibinfo{number}{1}), \bibinfo{pages}{10}, ISSN
  \bibinfo{issn}{2195-7045}.

\bibitem{OwschimikowJCP10}
\bibinfo{author}{\bibnamefont{Owschimikow}, \bibfnamefont{N.}},
  \bibinfo{author}{\bibfnamefont{F.}~\bibnamefont{K{\"o}nigsmann}},
  \bibinfo{author}{\bibfnamefont{J.}~\bibnamefont{Maurer}},
  \bibinfo{author}{\bibfnamefont{P.}~\bibnamefont{Giese}},
  \bibinfo{author}{\bibfnamefont{A.}~\bibnamefont{Ott}},
  \bibinfo{author}{\bibfnamefont{B.}~\bibnamefont{Schmidt}}, and
  \bibinfo{author}{\bibfnamefont{N.}~\bibnamefont{Schwentner}},
  \bibinfo{year}{2010}, \bibinfo{journal}{J. Chem. Phys.}
  \textbf{\bibinfo{volume}{133}}(\bibinfo{number}{4}), \bibinfo{pages}{044311}.

\bibitem{pabst:2013}
\bibinfo{author}{\bibnamefont{Pabst}, \bibfnamefont{S.}}, \bibinfo{year}{2013},
  \bibinfo{journal}{Eur. Phys. J. Spec. Top.} \textbf{\bibinfo{volume}{221}},
  \bibinfo{pages}{1}.

\bibitem{PalaoPRL02}
\bibinfo{author}{\bibnamefont{Palao}, \bibfnamefont{J.~P.}}, and
  \bibinfo{author}{\bibfnamefont{R.}~\bibnamefont{Kosloff}},
  \bibinfo{year}{2002}, \bibinfo{journal}{Phys. Rev. Lett.}
  \textbf{\bibinfo{volume}{89}}(\bibinfo{number}{18}), \bibinfo{pages}{188301}.

\bibitem{ParkPRL15}
\bibinfo{author}{\bibnamefont{Park}, \bibfnamefont{J.~W.}},
  \bibinfo{author}{\bibfnamefont{S.~A.} \bibnamefont{Will}}, and
  \bibinfo{author}{\bibfnamefont{M.~W.} \bibnamefont{Zwierlein}},
  \bibinfo{year}{2015}, \bibinfo{journal}{Phys. Rev. Lett.}
  \textbf{\bibinfo{volume}{114}}, \bibinfo{pages}{205302}.

\bibitem{PattersonPRA18}
\bibinfo{author}{\bibnamefont{Patterson}, \bibfnamefont{D.}},
  \bibinfo{year}{2018}, \bibinfo{journal}{Phys. Rev. A}
  \textbf{\bibinfo{volume}{97}}, \bibinfo{pages}{033403}.

\bibitem{PattersonPRL13}
\bibinfo{author}{\bibnamefont{Patterson}, \bibfnamefont{D.}}, and
  \bibinfo{author}{\bibfnamefont{J.~M.} \bibnamefont{Doyle}},
  \bibinfo{year}{2013}, \bibinfo{journal}{Phys. Rev. Lett.}
  \textbf{\bibinfo{volume}{111}}, \bibinfo{pages}{023008}.

\bibitem{PattersonNat13}
\bibinfo{author}{\bibnamefont{Patterson}, \bibfnamefont{D.}},
  \bibinfo{author}{\bibfnamefont{M.}~\bibnamefont{Schnell}}, and
  \bibinfo{author}{\bibfnamefont{J.~M.} \bibnamefont{Doyle}},
  \bibinfo{year}{2013}, \bibinfo{journal}{Nature}
  \textbf{\bibinfo{volume}{497}}, \bibinfo{pages}{475}.

\bibitem{pavicic:2007}
\bibinfo{author}{\bibnamefont{Pavicic}, \bibfnamefont{D.}},
  \bibinfo{author}{\bibfnamefont{K.~F.} \bibnamefont{Lee}},
  \bibinfo{author}{\bibfnamefont{D.~M.} \bibnamefont{Rayner}},
  \bibinfo{author}{\bibfnamefont{P.~B.} \bibnamefont{Corkum}}, and
  \bibinfo{author}{\bibfnamefont{D.~M.} \bibnamefont{Villeneuve}},
  \bibinfo{year}{2007}, \bibinfo{journal}{Phys. Rev. Lett.}
  \textbf{\bibinfo{volume}{98}}, \bibinfo{pages}{243001}.

\bibitem{PawlakJPCA17}
\bibinfo{author}{\bibnamefont{Pawlak}, \bibfnamefont{M.}},
  \bibinfo{author}{\bibfnamefont{Y.}~\bibnamefont{Shagam}},
  \bibinfo{author}{\bibfnamefont{A.}~\bibnamefont{Klein}},
  \bibinfo{author}{\bibfnamefont{E.}~\bibnamefont{Narevicius}}, and
  \bibinfo{author}{\bibfnamefont{N.}~\bibnamefont{Moiseyev}},
  \bibinfo{year}{2017}, \bibinfo{journal}{J. Phys. Chem. A}
  \textbf{\bibinfo{volume}{121}}(\bibinfo{number}{10}), \bibinfo{pages}{2194}.

\bibitem{PawlakJCP15}
\bibinfo{author}{\bibnamefont{Pawlak}, \bibfnamefont{M.}},
  \bibinfo{author}{\bibfnamefont{Y.}~\bibnamefont{Shagam}},
  \bibinfo{author}{\bibfnamefont{E.}~\bibnamefont{Narevicius}}, and
  \bibinfo{author}{\bibfnamefont{N.}~\bibnamefont{Moiseyev}},
  \bibinfo{year}{2015}, \bibinfo{journal}{J. Chem. Phys.}
  \textbf{\bibinfo{volume}{143}}(\bibinfo{number}{7}), \bibinfo{pages}{074114}.

\bibitem{pelzer:2007}
\bibinfo{author}{\bibnamefont{Pelzer}, \bibfnamefont{A.}},
  \bibinfo{author}{\bibfnamefont{S.}~\bibnamefont{Ramakrishna}}, and
  \bibinfo{author}{\bibfnamefont{T.}~\bibnamefont{Seideman}},
  \bibinfo{year}{2007}, \bibinfo{journal}{J. Chem. Phys.}
  \textbf{\bibinfo{volume}{126}}(\bibinfo{number}{3}), \bibinfo{pages}{034503}.

\bibitem{PentlehnerPRA13}
\bibinfo{author}{\bibnamefont{Pentlehner}, \bibfnamefont{D.}},
  \bibinfo{author}{\bibfnamefont{J.~H.} \bibnamefont{Nielsen}},
  \bibinfo{author}{\bibfnamefont{L.}~\bibnamefont{Christiansen}},
  \bibinfo{author}{\bibfnamefont{A.}~\bibnamefont{Slenczka}}, and
  \bibinfo{author}{\bibfnamefont{H.}~\bibnamefont{Stapelfeldt}},
  \bibinfo{year}{2013}{\natexlab{a}}, \bibinfo{journal}{Phys. Rev. A}
  \textbf{\bibinfo{volume}{87}}(\bibinfo{number}{6}), \bibinfo{pages}{063401}.

\bibitem{pentlehner:2013}
\bibinfo{author}{\bibnamefont{Pentlehner}, \bibfnamefont{D.}},
  \bibinfo{author}{\bibfnamefont{J.~H.} \bibnamefont{Nielsen}},
  \bibinfo{author}{\bibfnamefont{A.}~\bibnamefont{Slenczka}},
  \bibinfo{author}{\bibfnamefont{K.}~\bibnamefont{M\o{}lmer}}, and
  \bibinfo{author}{\bibfnamefont{H.}~\bibnamefont{Stapelfeldt}},
  \bibinfo{year}{2013}{\natexlab{b}}, \bibinfo{journal}{Phys. Rev. Lett.}
  \textbf{\bibinfo{volume}{110}}, \bibinfo{pages}{093002}.

\bibitem{PerezAngewandte17}
\bibinfo{author}{\bibnamefont{P\'erez}, \bibfnamefont{C.}},
  \bibinfo{author}{\bibfnamefont{A.~L.} \bibnamefont{Steber}},
  \bibinfo{author}{\bibfnamefont{S.~R.} \bibnamefont{Domingos}},
  \bibinfo{author}{\bibfnamefont{A.}~\bibnamefont{Krin}},
  \bibinfo{author}{\bibfnamefont{D.}~\bibnamefont{Schmitz}}, and
  \bibinfo{author}{\bibfnamefont{M.}~\bibnamefont{Schnell}},
  \bibinfo{year}{2017}, \bibinfo{journal}{Angew. Chem. Int. Ed.}
  \textbf{\bibinfo{volume}{56}}(\bibinfo{number}{41}), \bibinfo{pages}{12512}.

\bibitem{PeresRiosNJP10}
\bibinfo{author}{\bibnamefont{P{\'e}rez-R{\'\i}os}, \bibfnamefont{J.}},
  \bibinfo{author}{\bibfnamefont{F.}~\bibnamefont{Herrera}}, and
  \bibinfo{author}{\bibfnamefont{R.~V.} \bibnamefont{Krems}},
  \bibinfo{year}{2010}, \bibinfo{journal}{New Journal of Physics}
  \textbf{\bibinfo{volume}{12}}(\bibinfo{number}{10}), \bibinfo{pages}{103007}.

\bibitem{peronne:2003}
\bibinfo{author}{\bibnamefont{P\'eronne}, \bibfnamefont{E.}},
  \bibinfo{author}{\bibfnamefont{M.~D.} \bibnamefont{Poulsen}},
  \bibinfo{author}{\bibfnamefont{C.~Z.} \bibnamefont{Bisgaard}},
  \bibinfo{author}{\bibfnamefont{H.}~\bibnamefont{Stapelfeldt}}, and
  \bibinfo{author}{\bibfnamefont{T.}~\bibnamefont{Seideman}},
  \bibinfo{year}{2003}, \bibinfo{journal}{Phys. Rev. Lett.}
  \textbf{\bibinfo{volume}{91}}, \bibinfo{pages}{043003}.

\bibitem{peronne:2004}
\bibinfo{author}{\bibnamefont{P\'eronne}, \bibfnamefont{E.}},
  \bibinfo{author}{\bibfnamefont{M.~D.} \bibnamefont{Poulsen}},
  \bibinfo{author}{\bibfnamefont{H.}~\bibnamefont{Stapelfeldt}},
  \bibinfo{author}{\bibfnamefont{C.~Z.} \bibnamefont{Bisgaard}},
  \bibinfo{author}{\bibfnamefont{E.}~\bibnamefont{Hamilton}}, and
  \bibinfo{author}{\bibfnamefont{T.}~\bibnamefont{Seideman}},
  \bibinfo{year}{2004}, \bibinfo{journal}{Phys. Rev. A}
  \textbf{\bibinfo{volume}{70}}, \bibinfo{pages}{063410}.

\bibitem{peters:2011}
\bibinfo{author}{\bibnamefont{Peters}, \bibfnamefont{M.}},
  \bibinfo{author}{\bibfnamefont{T.~T.} \bibnamefont{Nguyen-Dang}},
  \bibinfo{author}{\bibfnamefont{C.}~\bibnamefont{Cornaggia}},
  \bibinfo{author}{\bibfnamefont{S.}~\bibnamefont{Saugout}},
  \bibinfo{author}{\bibfnamefont{E.}~\bibnamefont{Charron}},
  \bibinfo{author}{\bibfnamefont{A.}~\bibnamefont{Keller}}, and
  \bibinfo{author}{\bibfnamefont{O.}~\bibnamefont{Atabek}},
  \bibinfo{year}{2011}, \bibinfo{journal}{Phys. Rev. A}
  \textbf{\bibinfo{volume}{83}}, \bibinfo{pages}{051403}.

\bibitem{petretti:2010}
\bibinfo{author}{\bibnamefont{Petretti}, \bibfnamefont{S.}},
  \bibinfo{author}{\bibfnamefont{Y.~V.} \bibnamefont{Vanne}},
  \bibinfo{author}{\bibfnamefont{A.}~\bibnamefont{Saenz}},
  \bibinfo{author}{\bibfnamefont{A.}~\bibnamefont{Castro}}, and
  \bibinfo{author}{\bibfnamefont{P.}~\bibnamefont{Decleva}},
  \bibinfo{year}{2010}, \bibinfo{journal}{Phys. Rev. Lett.}
  \textbf{\bibinfo{volume}{104}}, \bibinfo{pages}{223001}.

\bibitem{PickeringPRL18}
\bibinfo{author}{\bibnamefont{Pickering}, \bibfnamefont{J.~D.}},
  \bibinfo{author}{\bibfnamefont{B.}~\bibnamefont{Shepperson}},
  \bibinfo{author}{\bibfnamefont{B.~A.~K.} \bibnamefont{H\"ubschmann}},
  \bibinfo{author}{\bibfnamefont{F.}~\bibnamefont{Thorning}}, and
  \bibinfo{author}{\bibfnamefont{H.}~\bibnamefont{Stapelfeldt}},
  \bibinfo{year}{2018}, \bibinfo{journal}{Phys. Rev. Lett.}
  \textbf{\bibinfo{volume}{120}}, \bibinfo{pages}{113202}.

\bibitem{PolletPRL10}
\bibinfo{author}{\bibnamefont{Pollet}, \bibfnamefont{L.}},
  \bibinfo{author}{\bibfnamefont{J.~D.} \bibnamefont{Picon}},
  \bibinfo{author}{\bibfnamefont{H.~P.} \bibnamefont{B\"uchler}}, and
  \bibinfo{author}{\bibfnamefont{M.}~\bibnamefont{Troyer}},
  \bibinfo{year}{2010}, \bibinfo{journal}{Phys. Rev. Lett.}
  \textbf{\bibinfo{volume}{104}}, \bibinfo{pages}{125302}.

\bibitem{poulsen:2006}
\bibinfo{author}{\bibnamefont{Poulsen}, \bibfnamefont{M.~D.}},
  \bibinfo{author}{\bibfnamefont{T.}~\bibnamefont{Ejdrup}},
  \bibinfo{author}{\bibfnamefont{H.}~\bibnamefont{Stapelfeldt}},
  \bibinfo{author}{\bibfnamefont{E.}~\bibnamefont{Hamilton}}, and
  \bibinfo{author}{\bibfnamefont{T.}~\bibnamefont{Seideman}},
  \bibinfo{year}{2006}, \bibinfo{journal}{Phys. Rev. A}
  \textbf{\bibinfo{volume}{73}}, \bibinfo{pages}{033405}.

\bibitem{PrehnPRL16}
\bibinfo{author}{\bibnamefont{Prehn}, \bibfnamefont{A.}},
  \bibinfo{author}{\bibfnamefont{M.}~\bibnamefont{Ibr\"ugger}},
  \bibinfo{author}{\bibfnamefont{R.}~\bibnamefont{Gl\"ockner}},
  \bibinfo{author}{\bibfnamefont{G.}~\bibnamefont{Rempe}}, and
  \bibinfo{author}{\bibfnamefont{M.}~\bibnamefont{Zeppenfeld}},
  \bibinfo{year}{2016}, \bibinfo{journal}{Phys. Rev. Lett.}
  \textbf{\bibinfo{volume}{116}}, \bibinfo{pages}{063005}.

\bibitem{prost:2017}
\bibinfo{author}{\bibnamefont{Prost}, \bibfnamefont{E.}},
  \bibinfo{author}{\bibfnamefont{H.}~\bibnamefont{Zhang}},
  \bibinfo{author}{\bibfnamefont{E.}~\bibnamefont{Hertz}},
  \bibinfo{author}{\bibfnamefont{F.}~\bibnamefont{Billard}},
  \bibinfo{author}{\bibfnamefont{B.}~\bibnamefont{Lavorel}},
  \bibinfo{author}{\bibfnamefont{P.}~\bibnamefont{Bejot}},
  \bibinfo{author}{\bibfnamefont{J.}~\bibnamefont{Zyss}},
  \bibinfo{author}{\bibfnamefont{I.~S.} \bibnamefont{Averbukh}}, and
  \bibinfo{author}{\bibfnamefont{O.}~\bibnamefont{Faucher}},
  \bibinfo{year}{2017}, \bibinfo{journal}{Phys. Rev. A}
  \textbf{\bibinfo{volume}{96}}, \bibinfo{pages}{043418}.

\bibitem{pupillo:2008}
\bibinfo{author}{\bibnamefont{Pupillo}, \bibfnamefont{G.}},
  \bibinfo{author}{\bibfnamefont{A.}~\bibnamefont{Griessner}},
  \bibinfo{author}{\bibfnamefont{A.}~\bibnamefont{Micheli}},
  \bibinfo{author}{\bibfnamefont{M.}~\bibnamefont{Ortner}},
  \bibinfo{author}{\bibfnamefont{D.-W.} \bibnamefont{Wang}}, and
  \bibinfo{author}{\bibfnamefont{P.}~\bibnamefont{Zoller}},
  \bibinfo{year}{2008}, \bibinfo{journal}{Phys. Rev. Lett.}
  \textbf{\bibinfo{volume}{100}}, \bibinfo{pages}{050402}.

\bibitem{purcell:2009}
\bibinfo{author}{\bibnamefont{Purcell}, \bibfnamefont{S.~M.}}, and
  \bibinfo{author}{\bibfnamefont{P.~F.} \bibnamefont{Barker}},
  \bibinfo{year}{2009}, \bibinfo{journal}{Phys. Rev. Lett.}
  \textbf{\bibinfo{volume}{103}}, \bibinfo{pages}{153001}.

\bibitem{QiCPC16}
\bibinfo{author}{\bibnamefont{Qi}, \bibfnamefont{W.}},
  \bibinfo{author}{\bibfnamefont{C.}~\bibnamefont{Yudong}},
  \bibinfo{author}{\bibfnamefont{K.}~\bibnamefont{Sabre}},
  \bibinfo{author}{\bibfnamefont{F.}~\bibnamefont{Bretislav}}, and
  \bibinfo{author}{\bibfnamefont{H.}~\bibnamefont{Dudley}},
  \bibinfo{year}{2016}, \bibinfo{journal}{ChemPhysChem}
  \textbf{\bibinfo{volume}{17}}, \bibinfo{pages}{3714}.

\bibitem{qin:2014}
\bibinfo{author}{\bibnamefont{Qin}, \bibfnamefont{C.}},
  \bibinfo{author}{\bibfnamefont{Y.}~\bibnamefont{Liu}},
  \bibinfo{author}{\bibfnamefont{X.}~\bibnamefont{Zhang}}, and
  \bibinfo{author}{\bibfnamefont{T.}~\bibnamefont{Gerber}},
  \bibinfo{year}{2014}, \bibinfo{journal}{Phys. Rev. A}
  \textbf{\bibinfo{volume}{90}}, \bibinfo{pages}{053429}.

\bibitem{qin:2012}
\bibinfo{author}{\bibnamefont{Qin}, \bibfnamefont{C.}},
  \bibinfo{author}{\bibfnamefont{Y.}~\bibnamefont{Tang}},
  \bibinfo{author}{\bibfnamefont{Y.}~\bibnamefont{Wang}}, and
  \bibinfo{author}{\bibfnamefont{B.}~\bibnamefont{Zhang}},
  \bibinfo{year}{2012}, \bibinfo{journal}{Phys. Rev. A}
  \textbf{\bibinfo{volume}{85}}, \bibinfo{pages}{053415}.

\bibitem{QuemenerChapter18}
\bibinfo{author}{\bibnamefont{Qu\'em\'ener}, \bibfnamefont{G.}},
  \bibinfo{year}{2018}, in \emph{\bibinfo{booktitle}{Cold Chemistry: Molecular
  Scattering and Reactivity Near Absolute Zero}}, edited by
  \bibinfo{editor}{\bibfnamefont{O.}~\bibnamefont{Dulieu}} and
  \bibinfo{editor}{\bibfnamefont{A.}~\bibnamefont{Osterwalder}}
  (\bibinfo{publisher}{The Royal Society of Chemistry}),
  number~\bibinfo{number}{11} in \bibinfo{series}{Theoretical and Computational
  Chemistry Series}, pp. \bibinfo{pages}{496--536}.

\bibitem{turinici:2007}
\bibinfo{author}{\bibnamefont{Rabitz}, \bibfnamefont{H.}}, and
  \bibinfo{author}{\bibfnamefont{G.}~\bibnamefont{Turinici}},
  \bibinfo{year}{2007}, \bibinfo{journal}{Phys. Rev. A}
  \textbf{\bibinfo{volume}{75}}, \bibinfo{pages}{043409}.

\bibitem{RablPRL06}
\bibinfo{author}{\bibnamefont{Rabl}, \bibfnamefont{P.}},
  \bibinfo{author}{\bibfnamefont{D.}~\bibnamefont{Demille}},
  \bibinfo{author}{\bibfnamefont{J.}~\bibnamefont{Doyle}},
  \bibinfo{author}{\bibfnamefont{M.}~\bibnamefont{Lukin}},
  \bibinfo{author}{\bibfnamefont{R.}~\bibnamefont{Schoelkopf}}, and
  \bibinfo{author}{\bibfnamefont{P.}~\bibnamefont{Zoller}},
  \bibinfo{year}{2006}, \bibinfo{journal}{Phys. Rev. Lett.}
  \textbf{\bibinfo{volume}{97}}, \bibinfo{pages}{033003}.

\bibitem{RamakrishnaSeideman2005}
\bibinfo{author}{\bibnamefont{Ramakrishna}, \bibfnamefont{S.}}, and
  \bibinfo{author}{\bibfnamefont{T.}~\bibnamefont{Seideman}},
  \bibinfo{year}{2005}, \bibinfo{journal}{Phys. Rev. Lett.}
  \textbf{\bibinfo{volume}{95}}, \bibinfo{pages}{113001}.

\bibitem{ramakrishna:2006jcp}
\bibinfo{author}{\bibnamefont{Ramakrishna}, \bibfnamefont{S.}}, and
  \bibinfo{author}{\bibfnamefont{T.}~\bibnamefont{Seideman}},
  \bibinfo{year}{2006}, \bibinfo{journal}{J. Chem. Phys.}
  \textbf{\bibinfo{volume}{124}}(\bibinfo{number}{3}), \bibinfo{pages}{034101}.

\bibitem{Ramakrishna:07}
\bibinfo{author}{\bibnamefont{Ramakrishna}, \bibfnamefont{S.}}, and
  \bibinfo{author}{\bibfnamefont{T.}~\bibnamefont{Seideman}},
  \bibinfo{year}{2007}{\natexlab{a}}, \bibinfo{journal}{Phys. Rev. Lett.}
  \textbf{\bibinfo{volume}{99}}, \bibinfo{pages}{113901}.

\bibitem{ramakrishna:2007}
\bibinfo{author}{\bibnamefont{Ramakrishna}, \bibfnamefont{S.}}, and
  \bibinfo{author}{\bibfnamefont{T.}~\bibnamefont{Seideman}},
  \bibinfo{year}{2007}{\natexlab{b}}, \bibinfo{journal}{Phys. Rev. Lett.}
  \textbf{\bibinfo{volume}{99}}, \bibinfo{pages}{103001}.

\bibitem{ramakrishna:2008}
\bibinfo{author}{\bibnamefont{Ramakrishna}, \bibfnamefont{S.}}, and
  \bibinfo{author}{\bibfnamefont{T.}~\bibnamefont{Seideman}},
  \bibinfo{year}{2008}, \bibinfo{journal}{Phys. Rev. A}
  \textbf{\bibinfo{volume}{77}}, \bibinfo{pages}{053411}.

\bibitem{ramakrishna:2013}
\bibinfo{author}{\bibnamefont{Ramakrishna}, \bibfnamefont{S.}}, and
  \bibinfo{author}{\bibfnamefont{T.}~\bibnamefont{Seideman}},
  \bibinfo{year}{2013}, \bibinfo{journal}{Phys. Rev. A}
  \textbf{\bibinfo{volume}{87}}, \bibinfo{pages}{023411}.

\bibitem{Ramakrishna:10}
\bibinfo{author}{\bibnamefont{Ramakrishna}, \bibfnamefont{S.}},
  \bibinfo{author}{\bibfnamefont{P.~A.~J.} \bibnamefont{Sherratt}},
  \bibinfo{author}{\bibfnamefont{A.~D.} \bibnamefont{Dutoi}}, and
  \bibinfo{author}{\bibfnamefont{T.}~\bibnamefont{Seideman}},
  \bibinfo{year}{2010}, \bibinfo{journal}{Phys. Rev. A}
  \textbf{\bibinfo{volume}{81}}, \bibinfo{pages}{021802}.

\bibitem{RanganPRL04}
\bibinfo{author}{\bibnamefont{Rangan}, \bibfnamefont{C.}},
  \bibinfo{author}{\bibfnamefont{A.~M.} \bibnamefont{Bloch}},
  \bibinfo{author}{\bibfnamefont{C.}~\bibnamefont{Monroe}}, and
  \bibinfo{author}{\bibfnamefont{P.~H.} \bibnamefont{Bucksbaum}},
  \bibinfo{year}{2004}, \bibinfo{journal}{Phys. Rev. Lett.}
  \textbf{\bibinfo{volume}{92}}, \bibinfo{pages}{113004}.

\bibitem{reckenthaeler:2009}
\bibinfo{author}{\bibnamefont{Reckenthaeler}, \bibfnamefont{P.}},
  \bibinfo{author}{\bibfnamefont{M.}~\bibnamefont{Centurion}},
  \bibinfo{author}{\bibfnamefont{W.}~\bibnamefont{Fu\ss{}}},
  \bibinfo{author}{\bibfnamefont{S.~A.} \bibnamefont{Trushin}},
  \bibinfo{author}{\bibfnamefont{F.}~\bibnamefont{Krausz}}, and
  \bibinfo{author}{\bibfnamefont{E.~E.} \bibnamefont{Fill}},
  \bibinfo{year}{2009}, \bibinfo{journal}{Phys. Rev. Lett.}
  \textbf{\bibinfo{volume}{102}}, \bibinfo{pages}{213001}.

\bibitem{ReichsollnerPRL17}
\bibinfo{author}{\bibnamefont{Reichs\"ollner}, \bibfnamefont{L.}},
  \bibinfo{author}{\bibfnamefont{A.}~\bibnamefont{Schindewolf}},
  \bibinfo{author}{\bibfnamefont{T.}~\bibnamefont{Takekoshi}},
  \bibinfo{author}{\bibfnamefont{R.}~\bibnamefont{Grimm}}, and
  \bibinfo{author}{\bibfnamefont{H.-C.} \bibnamefont{N\"agerl}},
  \bibinfo{year}{2017}, \bibinfo{journal}{Phys. Rev. Lett.}
  \textbf{\bibinfo{volume}{118}}, \bibinfo{pages}{073201}.

\bibitem{ren:2012}
\bibinfo{author}{\bibnamefont{Ren}, \bibfnamefont{X.}},
  \bibinfo{author}{\bibfnamefont{V.}~\bibnamefont{Makhija}}, and
  \bibinfo{author}{\bibfnamefont{V.}~\bibnamefont{Kumarappan}},
  \bibinfo{year}{2012}, \bibinfo{journal}{Phys. Rev. A}
  \textbf{\bibinfo{volume}{85}}, \bibinfo{pages}{033405}.

\bibitem{ren:2014}
\bibinfo{author}{\bibnamefont{Ren}, \bibfnamefont{X.}},
  \bibinfo{author}{\bibfnamefont{V.}~\bibnamefont{Makhija}}, and
  \bibinfo{author}{\bibfnamefont{V.}~\bibnamefont{Kumarappan}},
  \bibinfo{year}{2014}{\natexlab{a}}, \bibinfo{journal}{Phys. Rev. Lett.}
  \textbf{\bibinfo{volume}{112}}, \bibinfo{pages}{173602}.

\bibitem{ren:2014b}
\bibinfo{author}{\bibnamefont{Ren}, \bibfnamefont{X.}},
  \bibinfo{author}{\bibfnamefont{V.}~\bibnamefont{Makhija}},
  \bibinfo{author}{\bibfnamefont{H.}~\bibnamefont{Li}},
  \bibinfo{author}{\bibfnamefont{M.~F.} \bibnamefont{Kling}}, and
  \bibinfo{author}{\bibfnamefont{V.}~\bibnamefont{Kumarappan}},
  \bibinfo{year}{2014}{\natexlab{b}}, \bibinfo{journal}{Phys. Rev. A}
  \textbf{\bibinfo{volume}{90}}, \bibinfo{pages}{013419}.

\bibitem{renard:2005}
\bibinfo{author}{\bibnamefont{Renard}, \bibfnamefont{M.}},
  \bibinfo{author}{\bibfnamefont{E.}~\bibnamefont{Hertz}},
  \bibinfo{author}{\bibfnamefont{S.}~\bibnamefont{Gu\'erin}},
  \bibinfo{author}{\bibfnamefont{H.~R.} \bibnamefont{Jauslin}},
  \bibinfo{author}{\bibfnamefont{B.}~\bibnamefont{Lavorel}}, and
  \bibinfo{author}{\bibfnamefont{O.}~\bibnamefont{Faucher}},
  \bibinfo{year}{2005}, \bibinfo{journal}{Phys. Rev. A}
  \textbf{\bibinfo{volume}{72}}, \bibinfo{pages}{025401}.

\bibitem{renard:2004}
\bibinfo{author}{\bibnamefont{Renard}, \bibfnamefont{M.}},
  \bibinfo{author}{\bibfnamefont{E.}~\bibnamefont{Hertz}},
  \bibinfo{author}{\bibfnamefont{B.}~\bibnamefont{Lavorel}}, and
  \bibinfo{author}{\bibfnamefont{O.}~\bibnamefont{Faucher}},
  \bibinfo{year}{2004}{\natexlab{a}}, \bibinfo{journal}{Phys. Rev. A}
  \textbf{\bibinfo{volume}{69}}, \bibinfo{pages}{043401}.

\bibitem{renard:2003}
\bibinfo{author}{\bibnamefont{Renard}, \bibfnamefont{V.}},
  \bibinfo{author}{\bibfnamefont{M.}~\bibnamefont{Renard}},
  \bibinfo{author}{\bibfnamefont{S.}~\bibnamefont{Gu\'erin}},
  \bibinfo{author}{\bibfnamefont{Y.~T.} \bibnamefont{Pashayan}},
  \bibinfo{author}{\bibfnamefont{B.}~\bibnamefont{Lavorel}},
  \bibinfo{author}{\bibfnamefont{O.}~\bibnamefont{Faucher}}, and
  \bibinfo{author}{\bibfnamefont{H.~R.} \bibnamefont{Jauslin}},
  \bibinfo{year}{2003}, \bibinfo{journal}{Phys. Rev. Lett.}
  \textbf{\bibinfo{volume}{90}}, \bibinfo{pages}{153601}.

\bibitem{renard:2004b}
\bibinfo{author}{\bibnamefont{Renard}, \bibfnamefont{V.}},
  \bibinfo{author}{\bibfnamefont{M.}~\bibnamefont{Renard}},
  \bibinfo{author}{\bibfnamefont{A.}~\bibnamefont{Rouz\'ee}},
  \bibinfo{author}{\bibfnamefont{S.}~\bibnamefont{Gu\'erin}},
  \bibinfo{author}{\bibfnamefont{H.~R.} \bibnamefont{Jauslin}},
  \bibinfo{author}{\bibfnamefont{B.}~\bibnamefont{Lavorel}}, and
  \bibinfo{author}{\bibfnamefont{O.}~\bibnamefont{Faucher}},
  \bibinfo{year}{2004}{\natexlab{b}}, \bibinfo{journal}{Phys. Rev. A}
  \textbf{\bibinfo{volume}{70}}, \bibinfo{pages}{033420}.

\bibitem{reuter:2008}
\bibinfo{author}{\bibnamefont{Reuter}, \bibfnamefont{M.~G.}},
  \bibinfo{author}{\bibfnamefont{M.}~\bibnamefont{Sukharev}}, and
  \bibinfo{author}{\bibfnamefont{T.}~\bibnamefont{Seideman}},
  \bibinfo{year}{2008}, \bibinfo{journal}{Phys. Rev. Lett.}
  \textbf{\bibinfo{volume}{101}}, \bibinfo{pages}{208303}.

\bibitem{RodriguesIRPC16}
\bibinfo{author}{\bibnamefont{Rodr{\'\i}guez-Cantano}, \bibfnamefont{R.}},
  \bibinfo{author}{\bibfnamefont{T.}~\bibnamefont{Gonz{\'a}lez-Lezana}}, and
  \bibinfo{author}{\bibfnamefont{P.}~\bibnamefont{Villarreal}},
  \bibinfo{year}{2016}, \bibinfo{journal}{Int. Rev. Phys. Chem.}
  \textbf{\bibinfo{volume}{35}}, \bibinfo{pages}{37}.

\bibitem{rosca:2001}
\bibinfo{author}{\bibnamefont{Rosca-Pruna}, \bibfnamefont{F.}}, and
  \bibinfo{author}{\bibfnamefont{M.~J.~J.} \bibnamefont{Vrakking}},
  \bibinfo{year}{2001}, \bibinfo{journal}{Phys. Rev. Lett.}
  \textbf{\bibinfo{volume}{87}}, \bibinfo{pages}{153902}.

\bibitem{rosca:2002a}
\bibinfo{author}{\bibnamefont{Rosca-Pruna}, \bibfnamefont{F.}}, and
  \bibinfo{author}{\bibfnamefont{M.~J.~J.} \bibnamefont{Vrakking}},
  \bibinfo{year}{2002}{\natexlab{a}}, \bibinfo{journal}{J. Chem. Phys.}
  \textbf{\bibinfo{volume}{116}}(\bibinfo{number}{15}), \bibinfo{pages}{6567}.

\bibitem{rosca:2002b}
\bibinfo{author}{\bibnamefont{Rosca-Pruna}, \bibfnamefont{F.}}, and
  \bibinfo{author}{\bibfnamefont{M.~J.~J.} \bibnamefont{Vrakking}},
  \bibinfo{year}{2002}{\natexlab{b}}, \bibinfo{journal}{J. Chem. Phys.}
  \textbf{\bibinfo{volume}{116}}(\bibinfo{number}{15}), \bibinfo{pages}{6579}.

\bibitem{rosenberg:2017}
\bibinfo{author}{\bibnamefont{Rosenberg}, \bibfnamefont{D.}},
  \bibinfo{author}{\bibfnamefont{D.}~\bibnamefont{Damari}},
  \bibinfo{author}{\bibfnamefont{S.}~\bibnamefont{Kallush}}, and
  \bibinfo{author}{\bibfnamefont{S.}~\bibnamefont{Fleischer}},
  \bibinfo{year}{2017}, \bibinfo{journal}{J. Phys. Chem. Lett.}
  \textbf{\bibinfo{volume}{8}}, \bibinfo{pages}{5128}.

\bibitem{rosenberg:2018}
\bibinfo{author}{\bibnamefont{Rosenberg}, \bibfnamefont{D.}},
  \bibinfo{author}{\bibfnamefont{R.}~\bibnamefont{Damari}}, and
  \bibinfo{author}{\bibfnamefont{S.}~\bibnamefont{Fleischer}},
  \bibinfo{year}{2018}, \bibinfo{journal}{Phys. Rev. Lett.}
  \textbf{\bibinfo{volume}{121}}, \bibinfo{pages}{234101}.

\bibitem{rouzee:2011}
\bibinfo{author}{\bibnamefont{Rouz\'ee}, \bibfnamefont{A.}},
  \bibinfo{author}{\bibfnamefont{O.}~\bibnamefont{Ghafur}},
  \bibinfo{author}{\bibfnamefont{K.}~\bibnamefont{Vidma}},
  \bibinfo{author}{\bibfnamefont{A.}~\bibnamefont{Gijsbertsen}},
  \bibinfo{author}{\bibfnamefont{O.~M.} \bibnamefont{Shir}},
  \bibinfo{author}{\bibfnamefont{T.}~\bibnamefont{B\"ack}},
  \bibinfo{author}{\bibfnamefont{A.}~\bibnamefont{Meijer}},
  \bibinfo{author}{\bibfnamefont{W.~J.} \bibnamefont{van~der Zande}},
  \bibinfo{author}{\bibfnamefont{D.}~\bibnamefont{Parker}}, and
  \bibinfo{author}{\bibfnamefont{M.~J.~J.} \bibnamefont{Vrakking}},
  \bibinfo{year}{2011}, \bibinfo{journal}{Phys. Rev. A}
  \textbf{\bibinfo{volume}{84}}, \bibinfo{pages}{033415}.

\bibitem{rouzee:2009}
\bibinfo{author}{\bibnamefont{Rouz\'ee}, \bibfnamefont{A.}},
  \bibinfo{author}{\bibfnamefont{A.}~\bibnamefont{Gijsbertsen}},
  \bibinfo{author}{\bibfnamefont{O.}~\bibnamefont{Ghafur}},
  \bibinfo{author}{\bibfnamefont{O.~M.} \bibnamefont{Shir}},
  \bibinfo{author}{\bibfnamefont{T.}~\bibnamefont{B{\"a}ck}},
  \bibinfo{author}{\bibfnamefont{S.}~\bibnamefont{Stolte}}, and
  \bibinfo{author}{\bibfnamefont{M.~J.~J.} \bibnamefont{Vrakking}},
  \bibinfo{year}{2009}, \bibinfo{journal}{New Journal of Physics}
  \textbf{\bibinfo{volume}{11}}(\bibinfo{number}{10}), \bibinfo{pages}{105040}.

\bibitem{rouzee:2006}
\bibinfo{author}{\bibnamefont{Rouz\'ee}, \bibfnamefont{A.}},
  \bibinfo{author}{\bibfnamefont{S.}~\bibnamefont{Gu\'erin}},
  \bibinfo{author}{\bibfnamefont{V.}~\bibnamefont{Boudon}},
  \bibinfo{author}{\bibfnamefont{B.}~\bibnamefont{Lavorel}}, and
  \bibinfo{author}{\bibfnamefont{O.}~\bibnamefont{Faucher}},
  \bibinfo{year}{2006}, \bibinfo{journal}{Phys. Rev. A}
  \textbf{\bibinfo{volume}{73}}, \bibinfo{pages}{033418}.

\bibitem{rouzee:2008}
\bibinfo{author}{\bibnamefont{Rouz\'ee}, \bibfnamefont{A.}},
  \bibinfo{author}{\bibfnamefont{S.}~\bibnamefont{Gu\'erin}},
  \bibinfo{author}{\bibfnamefont{O.}~\bibnamefont{Faucher}}, and
  \bibinfo{author}{\bibfnamefont{B.}~\bibnamefont{Lavorel}},
  \bibinfo{year}{2008}, \bibinfo{journal}{Phys. Rev. A}
  \textbf{\bibinfo{volume}{77}}, \bibinfo{pages}{043412}.

\bibitem{rupenyan:2012}
\bibinfo{author}{\bibnamefont{Rupenyan}, \bibfnamefont{A.}},
  \bibinfo{author}{\bibfnamefont{J.~B.} \bibnamefont{Bertrand}},
  \bibinfo{author}{\bibfnamefont{D.~M.} \bibnamefont{Villeneuve}}, and
  \bibinfo{author}{\bibfnamefont{H.~J.} \bibnamefont{W\"orner}},
  \bibinfo{year}{2012}, \bibinfo{journal}{Phys. Rev. Lett.}
  \textbf{\bibinfo{volume}{108}}, \bibinfo{pages}{033903}.

\bibitem{sadovskii:1990}
\bibinfo{author}{\bibnamefont{Sadovskii}, \bibfnamefont{D.~A.}},
  \bibinfo{author}{\bibfnamefont{B.~I.} \bibnamefont{Zhilinskii}},
  \bibinfo{author}{\bibfnamefont{J.-P.} \bibnamefont{Champion}}, and
  \bibinfo{author}{\bibfnamefont{G.}~\bibnamefont{Pierre}},
  \bibinfo{year}{1990}, \bibinfo{journal}{J. Chem. Phys.}
  \textbf{\bibinfo{volume}{92}}, \bibinfo{pages}{1523}.

\bibitem{SaffmanMolmerRMP10}
\bibinfo{author}{\bibnamefont{Saffman}, \bibfnamefont{M.}},
  \bibinfo{author}{\bibfnamefont{T.~G.} \bibnamefont{Walker}}, and
  \bibinfo{author}{\bibfnamefont{K.}~\bibnamefont{M{\o}lmer}},
  \bibinfo{year}{2010}, \bibinfo{journal}{Rev. Mod. Phys.}
  \textbf{\bibinfo{volume}{82}}, \bibinfo{pages}{2313}.

\bibitem{sakai:2003}
\bibinfo{author}{\bibnamefont{Sakai}, \bibfnamefont{H.}},
  \bibinfo{author}{\bibfnamefont{S.}~\bibnamefont{Minemoto}},
  \bibinfo{author}{\bibfnamefont{H.}~\bibnamefont{Nanjo}},
  \bibinfo{author}{\bibfnamefont{H.}~\bibnamefont{Tanji}}, and
  \bibinfo{author}{\bibfnamefont{T.}~\bibnamefont{Suzuki}},
  \bibinfo{year}{2003}, \bibinfo{journal}{Phys. Rev. Lett.}
  \textbf{\bibinfo{volume}{90}}, \bibinfo{pages}{083001}.

\bibitem{sakai:1999}
\bibinfo{author}{\bibnamefont{Sakai}, \bibfnamefont{H.}},
  \bibinfo{author}{\bibfnamefont{C.~P.} \bibnamefont{Safvan}},
  \bibinfo{author}{\bibfnamefont{J.~J.} \bibnamefont{Larsen}},
  \bibinfo{author}{\bibfnamefont{K.~M.} \bibnamefont{Hilligsoe}},
  \bibinfo{author}{\bibfnamefont{K.}~\bibnamefont{Hald}}, and
  \bibinfo{author}{\bibfnamefont{H.}~\bibnamefont{Stapelfeldt}},
  \bibinfo{year}{1999}, \bibinfo{journal}{J. Chem. Phys.}
  \textbf{\bibinfo{volume}{110}}, \bibinfo{pages}{10235}.

\bibitem{salomon:2005}
\bibinfo{author}{\bibnamefont{Salomon}, \bibfnamefont{J.}},
  \bibinfo{author}{\bibfnamefont{C.~M.} \bibnamefont{Dion}}, and
  \bibinfo{author}{\bibfnamefont{G.}~\bibnamefont{Turinici}},
  \bibinfo{year}{2005}, \bibinfo{journal}{J. Chem. Phys.}
  \textbf{\bibinfo{volume}{123}}(\bibinfo{number}{14}),
  \bibinfo{pages}{144310}.

\bibitem{schatz:2018}
\bibinfo{author}{\bibnamefont{Schatz}, \bibfnamefont{K.}},
  \bibinfo{author}{\bibfnamefont{B.}~\bibnamefont{Friedrich}},
  \bibinfo{author}{\bibfnamefont{S.}~\bibnamefont{Becker}}, and
  \bibinfo{author}{\bibfnamefont{B.}~\bibnamefont{Schmidt}},
  \bibinfo{year}{2018}, \bibinfo{journal}{Phys. Rev. A}
  \textbf{\bibinfo{volume}{97}}, \bibinfo{pages}{053417}.

\bibitem{SchillerPRL14}
\bibinfo{author}{\bibnamefont{Schiller}, \bibfnamefont{S.}},
  \bibinfo{author}{\bibfnamefont{D.}~\bibnamefont{Bakalov}}, and
  \bibinfo{author}{\bibfnamefont{V.~I.} \bibnamefont{Korobov}},
  \bibinfo{year}{2014}, \bibinfo{journal}{Phys. Rev. Lett.}
  \textbf{\bibinfo{volume}{113}}, \bibinfo{pages}{023004}.

\bibitem{SchillerPRA05}
\bibinfo{author}{\bibnamefont{Schiller}, \bibfnamefont{S.}}, and
  \bibinfo{author}{\bibfnamefont{V.}~\bibnamefont{Korobov}},
  \bibinfo{year}{2005}, \bibinfo{journal}{Phys. Rev. A}
  \textbf{\bibinfo{volume}{71}}, \bibinfo{pages}{032505}.

\bibitem{SchinkeBook}
\bibinfo{author}{\bibnamefont{Schinke}, \bibfnamefont{R.}},
  \bibinfo{year}{1993}, \emph{\bibinfo{title}{Photodissociation Dynamics}},
  Cambridge Monographs on Atomic, Molecular and Chemical Physics
  (\bibinfo{publisher}{Cambridge University Press}).

\bibitem{schmidt:2014}
\bibinfo{author}{\bibnamefont{Schmidt}, \bibfnamefont{B.}}, and
  \bibinfo{author}{\bibfnamefont{B.}~\bibnamefont{Friedrich}},
  \bibinfo{year}{2014}, \bibinfo{journal}{J. Chem. Phys.}
  \textbf{\bibinfo{volume}{140}}(\bibinfo{number}{6}), \bibinfo{pages}{064317}.

\bibitem{schmidt:2015}
\bibinfo{author}{\bibnamefont{Schmidt}, \bibfnamefont{B.}}, and
  \bibinfo{author}{\bibfnamefont{B.}~\bibnamefont{Friedrich}},
  \bibinfo{year}{2015}, \bibinfo{journal}{Phys. Rev. A}
  \textbf{\bibinfo{volume}{91}}, \bibinfo{pages}{022111}.

\bibitem{SchmidtPRL08}
\bibinfo{author}{\bibnamefont{Schmidt}, \bibfnamefont{K.~P.}},
  \bibinfo{author}{\bibfnamefont{J.}~\bibnamefont{Dorier}}, and
  \bibinfo{author}{\bibfnamefont{A.~M.} \bibnamefont{L\"auchli}},
  \bibinfo{year}{2008}, \bibinfo{journal}{Phys. Rev. Lett.}
  \textbf{\bibinfo{volume}{101}}, \bibinfo{pages}{150405}.

\bibitem{SchmidtLem15}
\bibinfo{author}{\bibnamefont{Schmidt}, \bibfnamefont{R.}}, and
  \bibinfo{author}{\bibfnamefont{M.}~\bibnamefont{Lemeshko}},
  \bibinfo{year}{2015}, \bibinfo{journal}{Phys. Rev. Lett.}
  \textbf{\bibinfo{volume}{114}}, \bibinfo{pages}{203001}.

\bibitem{SchmidtLem16}
\bibinfo{author}{\bibnamefont{Schmidt}, \bibfnamefont{R.}}, and
  \bibinfo{author}{\bibfnamefont{M.}~\bibnamefont{Lemeshko}},
  \bibinfo{year}{2016}, \bibinfo{journal}{Phys. Rev. X}
  \textbf{\bibinfo{volume}{6}}, \bibinfo{pages}{011012}.

\bibitem{SchneiderNatPhys10}
\bibinfo{author}{\bibnamefont{Schneider}, \bibfnamefont{T.}},
  \bibinfo{author}{\bibfnamefont{B.}~\bibnamefont{Roth}},
  \bibinfo{author}{\bibfnamefont{H.}~\bibnamefont{Duncker}},
  \bibinfo{author}{\bibfnamefont{I.}~\bibnamefont{Ernsting}}, and
  \bibinfo{author}{\bibfnamefont{S.}~\bibnamefont{Schiller}},
  \bibinfo{year}{2010}, \bibinfo{journal}{Nat. Phys.}
  \textbf{\bibinfo{volume}{6}}, \bibinfo{pages}{275}.

\bibitem{SchusterPRA11}
\bibinfo{author}{\bibnamefont{Schuster}, \bibfnamefont{D.~I.}},
  \bibinfo{author}{\bibfnamefont{L.~S.} \bibnamefont{Bishop}},
  \bibinfo{author}{\bibfnamefont{I.~L.} \bibnamefont{Chuang}},
  \bibinfo{author}{\bibfnamefont{D.}~\bibnamefont{DeMille}}, and
  \bibinfo{author}{\bibfnamefont{R.~J.} \bibnamefont{Schoelkopf}},
  \bibinfo{year}{2011}, \bibinfo{journal}{Phys. Rev. A}
  \textbf{\bibinfo{volume}{83}}, \bibinfo{pages}{012311}.

\bibitem{Segev2019}
\bibinfo{author}{\bibnamefont{Segev}, \bibfnamefont{Y.}},
  \bibinfo{author}{\bibfnamefont{M.}~\bibnamefont{Pitzer}},
  \bibinfo{author}{\bibfnamefont{M.}~\bibnamefont{Karpov}},
  \bibinfo{author}{\bibfnamefont{N.}~\bibnamefont{Akerman}},
  \bibinfo{author}{\bibfnamefont{J.}~\bibnamefont{Narevicius}}, and
  \bibinfo{author}{\bibfnamefont{E.}~\bibnamefont{Narevicius}},
  \bibinfo{year}{2019}, \bibinfo{journal}{arXiv:1902.04549} .

\bibitem{seideman:1995}
\bibinfo{author}{\bibnamefont{Seideman}, \bibfnamefont{T.}},
  \bibinfo{year}{1995}, \bibinfo{journal}{J. Chem. Phys.}
  \textbf{\bibinfo{volume}{103}}, \bibinfo{pages}{7887}.

\bibitem{seideman:1997a}
\bibinfo{author}{\bibnamefont{Seideman}, \bibfnamefont{T.}},
  \bibinfo{year}{1997}{\natexlab{a}}, \bibinfo{journal}{J. Chem. Phys.}
  \textbf{\bibinfo{volume}{106}}, \bibinfo{pages}{2881}.

\bibitem{seideman:1997b}
\bibinfo{author}{\bibnamefont{Seideman}, \bibfnamefont{T.}},
  \bibinfo{year}{1997}{\natexlab{b}}, \bibinfo{journal}{J. Chem. Phys.}
  \textbf{\bibinfo{volume}{107}}, \bibinfo{pages}{10420}.

\bibitem{seideman:1997c}
\bibinfo{author}{\bibnamefont{Seideman}, \bibfnamefont{T.}},
  \bibinfo{year}{1997}{\natexlab{c}}, \bibinfo{journal}{Phys. Rev. A}
  \textbf{\bibinfo{volume}{56}}, \bibinfo{pages}{R17}.

\bibitem{seideman:1999b}
\bibinfo{author}{\bibnamefont{Seideman}, \bibfnamefont{T.}},
  \bibinfo{year}{1999}{\natexlab{a}}, \bibinfo{journal}{J. Chem. Phys.}
  \textbf{\bibinfo{volume}{111}}, \bibinfo{pages}{4397}.

\bibitem{seideman:1999}
\bibinfo{author}{\bibnamefont{Seideman}, \bibfnamefont{T.}},
  \bibinfo{year}{1999}{\natexlab{b}}, \bibinfo{journal}{Phys. Rev. Lett.}
  \textbf{\bibinfo{volume}{83}}, \bibinfo{pages}{4971}.

\bibitem{seideman:2001}
\bibinfo{author}{\bibnamefont{Seideman}, \bibfnamefont{T.}},
  \bibinfo{year}{2001}, \bibinfo{journal}{J. Chem. Phys.}
  \textbf{\bibinfo{volume}{115}}(\bibinfo{number}{13}), \bibinfo{pages}{5965}.

\bibitem{Seideman:05}
\bibinfo{author}{\bibnamefont{Seideman}, \bibfnamefont{T.}}, and
  \bibinfo{author}{\bibfnamefont{E.}~\bibnamefont{Hamilton}},
  \bibinfo{year}{2005}, \bibinfo{journal}{Adv. At. Mol. Opt.}
  \textbf{\bibinfo{volume}{52}}, \bibinfo{pages}{289 }.

\bibitem{ShagamNatChem15}
\bibinfo{author}{\bibnamefont{Shagam}, \bibfnamefont{Y.}},
  \bibinfo{author}{\bibfnamefont{A.}~\bibnamefont{Klein}},
  \bibinfo{author}{\bibfnamefont{W.}~\bibnamefont{Skomorowski}},
  \bibinfo{author}{\bibfnamefont{R.}~\bibnamefont{Yun}},
  \bibinfo{author}{\bibfnamefont{V.}~\bibnamefont{Averbukh}},
  \bibinfo{author}{\bibfnamefont{C.~P.} \bibnamefont{Koch}}, and
  \bibinfo{author}{\bibfnamefont{E.}~\bibnamefont{Narevicius.}},
  \bibinfo{year}{2015}, \bibinfo{journal}{Nature Chem.}
  \textbf{\bibinfo{volume}{7}}(\bibinfo{number}{11}), \bibinfo{pages}{921}.

\bibitem{ShagamJPCC13}
\bibinfo{author}{\bibnamefont{Shagam}, \bibfnamefont{Y.}}, and
  \bibinfo{author}{\bibfnamefont{E.}~\bibnamefont{Narevicius}},
  \bibinfo{year}{2013}, \bibinfo{journal}{J. Phys. Chem. C}
  \textbf{\bibinfo{volume}{117}}(\bibinfo{number}{43}), \bibinfo{pages}{22454}.

\bibitem{shapiro:2003}
\bibinfo{author}{\bibnamefont{Shapiro}, \bibfnamefont{E.~A.}},
  \bibinfo{author}{\bibfnamefont{I.}~\bibnamefont{Khavkine}},
  \bibinfo{author}{\bibfnamefont{M.}~\bibnamefont{Spanner}}, and
  \bibinfo{author}{\bibfnamefont{M.~Y.} \bibnamefont{Ivanov}},
  \bibinfo{year}{2003}, \bibinfo{journal}{Phys. Rev. A}
  \textbf{\bibinfo{volume}{67}}, \bibinfo{pages}{013406}.

\bibitem{ShapiroBook2}
\bibinfo{author}{\bibnamefont{Shapiro}, \bibfnamefont{M.}}, and
  \bibinfo{author}{\bibfnamefont{P.}~\bibnamefont{Brumer}},
  \bibinfo{year}{2012}, \emph{\bibinfo{title}{Quantum Control of Molecular
  Processes}} (\bibinfo{publisher}{Wiley Interscience}), \bibinfo{edition}{2nd,
  revised and enlarged edition} edition.

\bibitem{sharma:2015}
\bibinfo{author}{\bibnamefont{Sharma}, \bibfnamefont{K.}}, and
  \bibinfo{author}{\bibfnamefont{B.}~\bibnamefont{Friedrich}},
  \bibinfo{year}{2015}, \bibinfo{journal}{New J. Phys.}
  \textbf{\bibinfo{volume}{17}}, \bibinfo{pages}{045017}.

\bibitem{SheppersonPRA18}
\bibinfo{author}{\bibnamefont{Shepperson}, \bibfnamefont{B.}},
  \bibinfo{author}{\bibfnamefont{A.~S.} \bibnamefont{Chatterley}},
  \bibinfo{author}{\bibfnamefont{L.}~\bibnamefont{Christiansen}},
  \bibinfo{author}{\bibfnamefont{A.~A.} \bibnamefont{S\o{}ndergaard}}, and
  \bibinfo{author}{\bibfnamefont{H.}~\bibnamefont{Stapelfeldt}},
  \bibinfo{year}{2018}, \bibinfo{journal}{Phys. Rev. A}
  \textbf{\bibinfo{volume}{97}}, \bibinfo{pages}{013427}.

\bibitem{Shepperson17}
\bibinfo{author}{\bibnamefont{Shepperson}, \bibfnamefont{B.}},
  \bibinfo{author}{\bibfnamefont{A.~S.} \bibnamefont{Chatterley}},
  \bibinfo{author}{\bibfnamefont{A.~A.} \bibnamefont{S{\o}ndergaard}},
  \bibinfo{author}{\bibfnamefont{L.}~\bibnamefont{Christiansen}},
  \bibinfo{author}{\bibfnamefont{M.}~\bibnamefont{Lemeshko}}, and
  \bibinfo{author}{\bibfnamefont{H.}~\bibnamefont{Stapelfeldt}},
  \bibinfo{year}{2017}{\natexlab{a}}, \bibinfo{journal}{J. Chem. Phys.}
  \textbf{\bibinfo{volume}{147}}, \bibinfo{pages}{013946}.

\bibitem{Shepperson16}
\bibinfo{author}{\bibnamefont{Shepperson}, \bibfnamefont{B.}},
  \bibinfo{author}{\bibfnamefont{A.~A.} \bibnamefont{S{\o}ndergaard}},
  \bibinfo{author}{\bibfnamefont{L.}~\bibnamefont{Christiansen}},
  \bibinfo{author}{\bibfnamefont{J.}~\bibnamefont{Kaczmarczyk}},
  \bibinfo{author}{\bibfnamefont{R.~E.} \bibnamefont{Zillich}},
  \bibinfo{author}{\bibfnamefont{M.}~\bibnamefont{Lemeshko}}, and
  \bibinfo{author}{\bibfnamefont{H.}~\bibnamefont{Stapelfeldt}},
  \bibinfo{year}{2017}{\natexlab{b}}, \bibinfo{journal}{Phys. Rev. Lett.}
  \textbf{\bibinfo{volume}{118}}, \bibinfo{pages}{203203}.

\bibitem{ShiNJP13}
\bibinfo{author}{\bibnamefont{Shi}, \bibfnamefont{M.}},
  \bibinfo{author}{\bibfnamefont{P.~F.} \bibnamefont{Herskind}},
  \bibinfo{author}{\bibfnamefont{M.}~\bibnamefont{Drewsen}}, and
  \bibinfo{author}{\bibfnamefont{I.~L.} \bibnamefont{Chuang}},
  \bibinfo{year}{2013}, \bibinfo{journal}{New J. Phys.}
  \textbf{\bibinfo{volume}{15}}(\bibinfo{number}{11}), \bibinfo{pages}{113019}.

\bibitem{ShioyaMolPhys07}
\bibinfo{author}{\bibnamefont{Shioya}, \bibfnamefont{K.}},
  \bibinfo{author}{\bibfnamefont{K.}~\bibnamefont{Mishima}}, and
  \bibinfo{author}{\bibfnamefont{K.}~\bibnamefont{Yamashita}},
  \bibinfo{year}{2007}, \bibinfo{journal}{Molecular Physics}
  \textbf{\bibinfo{volume}{105}}(\bibinfo{number}{9}), \bibinfo{pages}{1283}.

\bibitem{shir:2008}
\bibinfo{author}{\bibnamefont{Shir}, \bibfnamefont{O.~M.}},
  \bibinfo{author}{\bibfnamefont{V.}~\bibnamefont{Beltrani}},
  \bibinfo{author}{\bibfnamefont{T.}~\bibnamefont{B{\"a}ck}},
  \bibinfo{author}{\bibfnamefont{H.}~\bibnamefont{Rabitz}}, and
  \bibinfo{author}{\bibfnamefont{M.~J.~J.} \bibnamefont{Vrakking}},
  \bibinfo{year}{2008}, \bibinfo{journal}{Journal of Physics B: Atomic,
  Molecular and Optical Physics}
  \textbf{\bibinfo{volume}{41}}(\bibinfo{number}{7}), \bibinfo{pages}{074021}.

\bibitem{chuan:2013}
\bibinfo{author}{\bibnamefont{Shu}, \bibfnamefont{C.-C.}}, and
  \bibinfo{author}{\bibfnamefont{N.~E.} \bibnamefont{Henriksen}},
  \bibinfo{year}{2013}, \bibinfo{journal}{Phys. Rev. A}
  \textbf{\bibinfo{volume}{87}}, \bibinfo{pages}{013408}.

\bibitem{shu:2016}
\bibinfo{author}{\bibnamefont{Shu}, \bibfnamefont{C.-C.}},
  \bibinfo{author}{\bibfnamefont{T.-S.} \bibnamefont{Ho}}, and
  \bibinfo{author}{\bibfnamefont{H.}~\bibnamefont{Rabitz}},
  \bibinfo{year}{2016}, \bibinfo{journal}{Phys. Rev. A}
  \textbf{\bibinfo{volume}{93}}, \bibinfo{pages}{053418}.

\bibitem{shu:2009}
\bibinfo{author}{\bibnamefont{Shu}, \bibfnamefont{C.-C.}},
  \bibinfo{author}{\bibfnamefont{K.-J.} \bibnamefont{Yuan}},
  \bibinfo{author}{\bibfnamefont{W.-H.} \bibnamefont{Hu}}, and
  \bibinfo{author}{\bibfnamefont{S.-L.} \bibnamefont{Cong}},
  \bibinfo{year}{2009}, \bibinfo{journal}{Phys. Rev. A}
  \textbf{\bibinfo{volume}{80}}, \bibinfo{pages}{011401}.

\bibitem{shu:2008}
\bibinfo{author}{\bibnamefont{Shu}, \bibfnamefont{C.-C.}},
  \bibinfo{author}{\bibfnamefont{K.-J.} \bibnamefont{Yuan}},
  \bibinfo{author}{\bibfnamefont{W.-H.} \bibnamefont{Hu}},
  \bibinfo{author}{\bibfnamefont{J.}~\bibnamefont{Yang}}, and
  \bibinfo{author}{\bibfnamefont{S.-L.} \bibnamefont{Cong}},
  \bibinfo{year}{2008}, \bibinfo{journal}{Phys. Rev. A}
  \textbf{\bibinfo{volume}{78}}, \bibinfo{pages}{055401}.

\bibitem{ShubertACIE2014}
\bibinfo{author}{\bibnamefont{Shubert}, \bibfnamefont{V.~A.}},
  \bibinfo{author}{\bibfnamefont{D.}~\bibnamefont{Schmitz}},
  \bibinfo{author}{\bibfnamefont{D.}~\bibnamefont{Patterson}},
  \bibinfo{author}{\bibfnamefont{J.~M.} \bibnamefont{Doyle}}, and
  \bibinfo{author}{\bibfnamefont{M.}~\bibnamefont{Schnell}},
  \bibinfo{year}{2014}, \bibinfo{journal}{Angew. Chem., Int. Ed.}
  \textbf{\bibinfo{volume}{53}}, \bibinfo{pages}{1152}.

\bibitem{ShubertJPCL16}
\bibinfo{author}{\bibnamefont{Shubert}, \bibfnamefont{V.~A.}},
  \bibinfo{author}{\bibfnamefont{D.}~\bibnamefont{Schmitz}},
  \bibinfo{author}{\bibfnamefont{C.}~\bibnamefont{P{\'e}rez}},
  \bibinfo{author}{\bibfnamefont{C.}~\bibnamefont{Medcraft}},
  \bibinfo{author}{\bibfnamefont{A.}~\bibnamefont{Krin}},
  \bibinfo{author}{\bibfnamefont{S.~R.} \bibnamefont{Domingos}},
  \bibinfo{author}{\bibfnamefont{D.}~\bibnamefont{Patterson}}, and
  \bibinfo{author}{\bibfnamefont{M.}~\bibnamefont{Schnell}},
  \bibinfo{year}{2016}, \bibinfo{journal}{J. Phys. Chem. Lett.}
  \textbf{\bibinfo{volume}{7}}(\bibinfo{number}{2}), \bibinfo{pages}{341}.

\bibitem{ShumanNature10}
\bibinfo{author}{\bibnamefont{Shuman}, \bibfnamefont{E.~S.}},
  \bibinfo{author}{\bibfnamefont{J.~F.} \bibnamefont{Barry}}, and
  \bibinfo{author}{\bibfnamefont{D.}~\bibnamefont{DeMille}},
  \bibinfo{year}{2010}, \bibinfo{journal}{Nature}
  \textbf{\bibinfo{volume}{467}}(\bibinfo{number}{7317}), \bibinfo{pages}{820}.

\bibitem{ShumanPRL09}
\bibinfo{author}{\bibnamefont{Shuman}, \bibfnamefont{E.~S.}},
  \bibinfo{author}{\bibfnamefont{J.~F.} \bibnamefont{Barry}},
  \bibinfo{author}{\bibfnamefont{D.~R.} \bibnamefont{Glenn}}, and
  \bibinfo{author}{\bibfnamefont{D.}~\bibnamefont{DeMille}},
  \bibinfo{year}{2009}, \bibinfo{journal}{Phys. Rev. Lett.}
  \textbf{\bibinfo{volume}{103}}, \bibinfo{pages}{223001}.

\bibitem{SiskaRMP93}
\bibinfo{author}{\bibnamefont{Siska}, \bibfnamefont{P.~E.}},
  \bibinfo{year}{1993}, \bibinfo{journal}{Rev. Mod. Phys.}
  \textbf{\bibinfo{volume}{65}}, \bibinfo{pages}{337}.

\bibitem{skinner2010}
\bibinfo{author}{\bibnamefont{Skinner}, \bibfnamefont{T.~E.}}, and
  \bibinfo{author}{\bibfnamefont{N.~I.} \bibnamefont{Gershenzon}},
  \bibinfo{year}{2010}, \bibinfo{journal}{Journal of Magnetic Resonance}
  \textbf{\bibinfo{volume}{204}}(\bibinfo{number}{2}), \bibinfo{pages}{248 },
  ISSN \bibinfo{issn}{1090-7807}.

\bibitem{skinner:2012}
\bibinfo{author}{\bibnamefont{Skinner}, \bibfnamefont{T.~E.}},
  \bibinfo{author}{\bibfnamefont{N.~I.} \bibnamefont{Gershenzon}},
  \bibinfo{author}{\bibfnamefont{N.}~\bibnamefont{Nimbalkar}}, and
  \bibinfo{author}{\bibfnamefont{S.~J.} \bibnamefont{Glaser}},
  \bibinfo{year}{2012}, \bibinfo{journal}{J. Magn. Reson.}
  \textbf{\bibinfo{volume}{217}}, \bibinfo{pages}{53}.

\bibitem{SkomorowskiJPCA16}
\bibinfo{author}{\bibnamefont{Skomorowski}, \bibfnamefont{W.}},
  \bibinfo{author}{\bibfnamefont{Y.}~\bibnamefont{Shagam}},
  \bibinfo{author}{\bibfnamefont{E.}~\bibnamefont{Narevicius}}, and
  \bibinfo{author}{\bibfnamefont{C.~P.} \bibnamefont{Koch}},
  \bibinfo{year}{2016}, \bibinfo{journal}{J. Phys. Chem. A}
  \textbf{\bibinfo{volume}{120}}(\bibinfo{number}{19}), \bibinfo{pages}{3309}.

\bibitem{Slipchenko2005}
\bibinfo{author}{\bibnamefont{Slipchenko}, \bibfnamefont{M.~N.}}, and
  \bibinfo{author}{\bibfnamefont{A.~F.} \bibnamefont{Vilesov}},
  \bibinfo{year}{2005}, \bibinfo{journal}{Chem. Phys. Lett.}
  \textbf{\bibinfo{volume}{412}}, \bibinfo{pages}{176}.

\bibitem{SowinskiPRL12}
\bibinfo{author}{\bibnamefont{Sowi\'{n}ski}, \bibfnamefont{T.}},
  \bibinfo{author}{\bibfnamefont{O.}~\bibnamefont{Dutta}},
  \bibinfo{author}{\bibfnamefont{P.}~\bibnamefont{Hauke}},
  \bibinfo{author}{\bibfnamefont{L.}~\bibnamefont{Tagliacozzo}}, and
  \bibinfo{author}{\bibfnamefont{M.}~\bibnamefont{Lewenstein}},
  \bibinfo{year}{2012}, \bibinfo{journal}{Phys. Rev. Lett.}
  \textbf{\bibinfo{volume}{108}}, \bibinfo{pages}{115301}.

\bibitem{spanner:2012}
\bibinfo{author}{\bibnamefont{Spanner}, \bibfnamefont{M.}},
  \bibinfo{author}{\bibfnamefont{S.}~\bibnamefont{Patchkovskii}},
  \bibinfo{author}{\bibfnamefont{E.}~\bibnamefont{Frumker}}, and
  \bibinfo{author}{\bibfnamefont{P.}~\bibnamefont{Corkum}},
  \bibinfo{year}{2012}, \bibinfo{journal}{Phys. Rev. Lett.}
  \textbf{\bibinfo{volume}{109}}, \bibinfo{pages}{113001}.

\bibitem{spanner:2004}
\bibinfo{author}{\bibnamefont{Spanner}, \bibfnamefont{M.}},
  \bibinfo{author}{\bibfnamefont{E.~A.} \bibnamefont{Shapiro}}, and
  \bibinfo{author}{\bibfnamefont{M.}~\bibnamefont{Ivanov}},
  \bibinfo{year}{2004}, \bibinfo{journal}{Phys. Rev. Lett.}
  \textbf{\bibinfo{volume}{92}}, \bibinfo{pages}{093001}.

\bibitem{StaanumPRL08}
\bibinfo{author}{\bibnamefont{Staanum}, \bibfnamefont{P.~F.}},
  \bibinfo{author}{\bibfnamefont{K.}~\bibnamefont{H\o{}jbjerre}},
  \bibinfo{author}{\bibfnamefont{R.}~\bibnamefont{Wester}}, and
  \bibinfo{author}{\bibfnamefont{M.}~\bibnamefont{Drewsen}},
  \bibinfo{year}{2008}, \bibinfo{journal}{Phys. Rev. Lett.}
  \textbf{\bibinfo{volume}{100}}, \bibinfo{pages}{243003}.

\bibitem{StaanumNatPhys10}
\bibinfo{author}{\bibnamefont{Staanum}, \bibfnamefont{P.~F.}},
  \bibinfo{author}{\bibfnamefont{K.~H.} \bibnamefont{jbjerre}},
  \bibinfo{author}{\bibfnamefont{P.~S.} \bibnamefont{Skyt}},
  \bibinfo{author}{\bibfnamefont{A.~K.} \bibnamefont{Hansen}}, and
  \bibinfo{author}{\bibfnamefont{M.}~\bibnamefont{Drewsen}},
  \bibinfo{year}{2010}, \bibinfo{journal}{Nature Phys.}
  \textbf{\bibinfo{volume}{6}}, \bibinfo{pages}{271}.

\bibitem{Stapelfeldt:03}
\bibinfo{author}{\bibnamefont{Stapelfeldt}, \bibfnamefont{H.}}, and
  \bibinfo{author}{\bibfnamefont{T.}~\bibnamefont{Seideman}},
  \bibinfo{year}{2003}, \bibinfo{journal}{Rev. Mod. Phys.}
  \textbf{\bibinfo{volume}{75}}, \bibinfo{pages}{543}.

\bibitem{steinitz:2012}
\bibinfo{author}{\bibnamefont{Steinitz}, \bibfnamefont{U.}},
  \bibinfo{author}{\bibfnamefont{Y.}~\bibnamefont{Prior}}, and
  \bibinfo{author}{\bibfnamefont{I.~S.} \bibnamefont{Averbukh}},
  \bibinfo{year}{2012}, \bibinfo{journal}{Phys. Rev. Lett.}
  \textbf{\bibinfo{volume}{109}}, \bibinfo{pages}{033001}.

\bibitem{steinitz:2014}
\bibinfo{author}{\bibnamefont{Steinitz}, \bibfnamefont{U.}},
  \bibinfo{author}{\bibfnamefont{Y.}~\bibnamefont{Prior}}, and
  \bibinfo{author}{\bibfnamefont{I.~S.} \bibnamefont{Averbukh}},
  \bibinfo{year}{2014}, \bibinfo{journal}{Phys. Rev. Lett.}
  \textbf{\bibinfo{volume}{112}}, \bibinfo{pages}{013004}.

\bibitem{StienkemeierJPB06}
\bibinfo{author}{\bibnamefont{Stienkemeier}, \bibfnamefont{F.}}, and
  \bibinfo{author}{\bibfnamefont{K.~K.} \bibnamefont{Lehmann}},
  \bibinfo{year}{2006}, \bibinfo{journal}{J. Phys. B}
  \textbf{\bibinfo{volume}{39}}, \bibinfo{pages}{R127}.

\bibitem{stolte:1988}
\bibinfo{author}{\bibnamefont{Stolte}, \bibfnamefont{S.}},
  \bibinfo{year}{1988}, \emph{\bibinfo{title}{Atomic and molecular beams
  methods}} (\bibinfo{publisher}{Oxford University, New York}).

\bibitem{StuhlNature12}
\bibinfo{author}{\bibnamefont{Stuhl}, \bibfnamefont{B.~K.}},
  \bibinfo{author}{\bibfnamefont{M.~T.} \bibnamefont{Hummon}},
  \bibinfo{author}{\bibfnamefont{M.}~\bibnamefont{Yeo}},
  \bibinfo{author}{\bibfnamefont{G.}~\bibnamefont{Quemener}},
  \bibinfo{author}{\bibfnamefont{J.~L.} \bibnamefont{Bohn}}, and
  \bibinfo{author}{\bibfnamefont{J.}~\bibnamefont{Ye}}, \bibinfo{year}{2012},
  \bibinfo{journal}{Nature}
  \textbf{\bibinfo{volume}{492}}(\bibinfo{number}{7429}), \bibinfo{pages}{396}.

\bibitem{sugita:2000}
\bibinfo{author}{\bibnamefont{Sugita}, \bibfnamefont{A.}},
  \bibinfo{author}{\bibfnamefont{M.}~\bibnamefont{Mashino}},
  \bibinfo{author}{\bibfnamefont{M.}~\bibnamefont{Kawasaki}},
  \bibinfo{author}{\bibfnamefont{Y.}~\bibnamefont{Matsumi}},
  \bibinfo{author}{\bibfnamefont{R.~J.} \bibnamefont{Gordon}}, and
  \bibinfo{author}{\bibfnamefont{R.}~\bibnamefont{Bersohn}},
  \bibinfo{year}{2000}, \bibinfo{journal}{J. Chem. Phys.}
  \textbf{\bibinfo{volume}{112}}, \bibinfo{pages}{2164}.

\bibitem{sugny:2004}
\bibinfo{author}{\bibnamefont{Sugny}, \bibfnamefont{D.}},
  \bibinfo{author}{\bibfnamefont{A.}~\bibnamefont{Keller}},
  \bibinfo{author}{\bibfnamefont{O.}~\bibnamefont{Atabek}},
  \bibinfo{author}{\bibfnamefont{D.}~\bibnamefont{Daems}},
  \bibinfo{author}{\bibfnamefont{C.~M.} \bibnamefont{Dion}},
  \bibinfo{author}{\bibfnamefont{S.}~\bibnamefont{Gu\'erin}}, and
  \bibinfo{author}{\bibfnamefont{H.~R.} \bibnamefont{Jauslin}},
  \bibinfo{year}{2004}{\natexlab{a}}, \bibinfo{journal}{Phys. Rev. A}
  \textbf{\bibinfo{volume}{69}}, \bibinfo{pages}{033402}.

\bibitem{sugny:2005b}
\bibinfo{author}{\bibnamefont{Sugny}, \bibfnamefont{D.}},
  \bibinfo{author}{\bibfnamefont{A.}~\bibnamefont{Keller}},
  \bibinfo{author}{\bibfnamefont{O.}~\bibnamefont{Atabek}},
  \bibinfo{author}{\bibfnamefont{D.}~\bibnamefont{Daems}},
  \bibinfo{author}{\bibfnamefont{C.~M.} \bibnamefont{Dion}},
  \bibinfo{author}{\bibfnamefont{S.}~\bibnamefont{Gu\'erin}}, and
  \bibinfo{author}{\bibfnamefont{H.~R.} \bibnamefont{Jauslin}},
  \bibinfo{year}{2005}{\natexlab{a}}, \bibinfo{journal}{Phys. Rev. A}
  \textbf{\bibinfo{volume}{72}}, \bibinfo{pages}{032704}.

\bibitem{sugny:2005}
\bibinfo{author}{\bibnamefont{Sugny}, \bibfnamefont{D.}},
  \bibinfo{author}{\bibfnamefont{A.}~\bibnamefont{Keller}},
  \bibinfo{author}{\bibfnamefont{O.}~\bibnamefont{Atabek}},
  \bibinfo{author}{\bibfnamefont{D.}~\bibnamefont{Daems}},
  \bibinfo{author}{\bibfnamefont{C.~M.} \bibnamefont{Dion}},
  \bibinfo{author}{\bibfnamefont{S.}~\bibnamefont{Gu\'erin}}, and
  \bibinfo{author}{\bibfnamefont{H.~R.} \bibnamefont{Jauslin}},
  \bibinfo{year}{2005}{\natexlab{b}}, \bibinfo{journal}{Phys. Rev. A}
  \textbf{\bibinfo{volume}{71}}, \bibinfo{pages}{063402}.

\bibitem{sugny:2004b}
\bibinfo{author}{\bibnamefont{Sugny}, \bibfnamefont{D.}},
  \bibinfo{author}{\bibfnamefont{A.}~\bibnamefont{Keller}},
  \bibinfo{author}{\bibfnamefont{O.}~\bibnamefont{Atabek}},
  \bibinfo{author}{\bibfnamefont{D.}~\bibnamefont{Daems}},
  \bibinfo{author}{\bibfnamefont{S.}~\bibnamefont{Gu\'erin}}, and
  \bibinfo{author}{\bibfnamefont{H.~R.} \bibnamefont{Jauslin}},
  \bibinfo{year}{2004}{\natexlab{b}}, \bibinfo{journal}{Phys. Rev. A}
  \textbf{\bibinfo{volume}{69}}, \bibinfo{pages}{043407}.

\bibitem{Sugny:14}
\bibinfo{author}{\bibnamefont{Sugny}, \bibfnamefont{D.}},
  \bibinfo{author}{\bibfnamefont{S.}~\bibnamefont{Vranckx}},
  \bibinfo{author}{\bibfnamefont{M.}~\bibnamefont{Ndong}},
  \bibinfo{author}{\bibfnamefont{N.}~\bibnamefont{Vaeck}}, and
  \bibinfo{author}{\bibfnamefont{M.}~\bibnamefont{Desouter-Lecomte}},
  \bibinfo{year}{2014}, \bibinfo{journal}{Phys. Rev. A}
  \textbf{\bibinfo{volume}{90}}.

\bibitem{SundarSciRep18}
\bibinfo{author}{\bibnamefont{Sundar}, \bibfnamefont{B.}},
  \bibinfo{author}{\bibfnamefont{B.}~\bibnamefont{Gadway}}, and
  \bibinfo{author}{\bibfnamefont{K.~R.~A.} \bibnamefont{Hazzard}},
  \bibinfo{year}{2018}, \bibinfo{journal}{Scientific Reports}
  \textbf{\bibinfo{volume}{8}}(\bibinfo{number}{1}), \bibinfo{pages}{3422}.

\bibitem{suzuki:2008}
\bibinfo{author}{\bibnamefont{Suzuki}, \bibfnamefont{T.}},
  \bibinfo{author}{\bibfnamefont{Y.}~\bibnamefont{Sugawara}},
  \bibinfo{author}{\bibfnamefont{S.}~\bibnamefont{Minemoto}}, and
  \bibinfo{author}{\bibfnamefont{H.}~\bibnamefont{Sakai}},
  \bibinfo{year}{2008}, \bibinfo{journal}{Phys. Rev. Lett.}
  \textbf{\bibinfo{volume}{100}}, \bibinfo{pages}{033603}.

\bibitem{SyzranovNatComm14}
\bibinfo{author}{\bibnamefont{Syzranov}, \bibfnamefont{S.~V.}},
  \bibinfo{author}{\bibfnamefont{M.~L.} \bibnamefont{Wall}},
  \bibinfo{author}{\bibfnamefont{V.}~\bibnamefont{Gurarie}}, and
  \bibinfo{author}{\bibfnamefont{A.~M.} \bibnamefont{Rey}},
  \bibinfo{year}{2014}, \bibinfo{journal}{Nature Comm.}
  \textbf{\bibinfo{volume}{5}}, \bibinfo{pages}{5391}.

\bibitem{SzalewiczIRPC08}
\bibinfo{author}{\bibnamefont{Szalewicz}, \bibfnamefont{K.}},
  \bibinfo{year}{2008}, \bibinfo{journal}{Int. Rev. Phys. Chem.}
  \textbf{\bibinfo{volume}{27}}, \bibinfo{pages}{273}.

\bibitem{takei:2016}
\bibinfo{author}{\bibnamefont{Takei}, \bibfnamefont{D.}},
  \bibinfo{author}{\bibfnamefont{J.~H.} \bibnamefont{Mun}},
  \bibinfo{author}{\bibfnamefont{S.}~\bibnamefont{Minemoto}}, and
  \bibinfo{author}{\bibfnamefont{H.}~\bibnamefont{Sakai}},
  \bibinfo{year}{2016}, \bibinfo{journal}{Phys. Rev. A}
  \textbf{\bibinfo{volume}{94}}, \bibinfo{pages}{013401}.

\bibitem{tanji:2005}
\bibinfo{author}{\bibnamefont{Tanji}, \bibfnamefont{H.}},
  \bibinfo{author}{\bibfnamefont{S.}~\bibnamefont{Minemoto}}, and
  \bibinfo{author}{\bibfnamefont{H.}~\bibnamefont{Sakai}},
  \bibinfo{year}{2005}, \bibinfo{journal}{Phys. Rev. A}
  \textbf{\bibinfo{volume}{72}}, \bibinfo{pages}{063401}.

\bibitem{tao_molecular_2006}
\bibinfo{author}{\bibnamefont{Tao}, \bibfnamefont{G.}}, and
  \bibinfo{author}{\bibfnamefont{R.~M.} \bibnamefont{Stratt}},
  \bibinfo{year}{2006}, \bibinfo{journal}{J. Chem. Phys.}
  \textbf{\bibinfo{volume}{125}}(\bibinfo{number}{11}),
  \bibinfo{pages}{114501}.

\bibitem{Tehini:2012}
\bibinfo{author}{\bibnamefont{Tehini}, \bibfnamefont{R.}},
  \bibinfo{author}{\bibfnamefont{M.~Z.} \bibnamefont{Hoque}},
  \bibinfo{author}{\bibfnamefont{O.}~\bibnamefont{Faucher}}, and
  \bibinfo{author}{\bibfnamefont{D.}~\bibnamefont{Sugny}},
  \bibinfo{year}{2012}, \bibinfo{journal}{Phys. Rev. A}
  \textbf{\bibinfo{volume}{85}}, \bibinfo{pages}{043423}.

\bibitem{Tehini:08}
\bibinfo{author}{\bibnamefont{Tehini}, \bibfnamefont{R.}}, and
  \bibinfo{author}{\bibfnamefont{D.}~\bibnamefont{Sugny}},
  \bibinfo{year}{2008}, \bibinfo{journal}{Phys. Rev. A}
  \textbf{\bibinfo{volume}{77}}, \bibinfo{pages}{023407}.

\bibitem{tenney:2016}
\bibinfo{author}{\bibnamefont{Tenney}, \bibfnamefont{I.~F.}},
  \bibinfo{author}{\bibfnamefont{M.}~\bibnamefont{Artamonov}},
  \bibinfo{author}{\bibfnamefont{T.}~\bibnamefont{Seideman}}, and
  \bibinfo{author}{\bibfnamefont{P.~H.} \bibnamefont{Bucksbaum}},
  \bibinfo{year}{2016}, \bibinfo{journal}{Phys. Rev. A}
  \textbf{\bibinfo{volume}{93}}, \bibinfo{pages}{013421}.

\bibitem{TeschPRL02}
\bibinfo{author}{\bibnamefont{Tesch}, \bibfnamefont{C.}}, and
  \bibinfo{author}{\bibfnamefont{R.}~\bibnamefont{de~Vivie-Riedle}},
  \bibinfo{year}{2002}, \bibinfo{journal}{Phys. Rev. Lett.}
  \textbf{\bibinfo{volume}{89}}, \bibinfo{pages}{157901}.

\bibitem{ACMEedm}
\bibinfo{author}{\bibnamefont{{The ACME Collaboration}}},
  \bibinfo{author}{\bibfnamefont{J.}~\bibnamefont{Baron}},
  \bibinfo{author}{\bibfnamefont{W.~C.} \bibnamefont{Campbell}},
  \bibinfo{author}{\bibfnamefont{D.}~\bibnamefont{DeMille}},
  \bibinfo{author}{\bibfnamefont{J.~M.} \bibnamefont{Doyle}},
  \bibinfo{author}{\bibfnamefont{G.}~\bibnamefont{Gabrielse}},
  \bibinfo{author}{\bibfnamefont{Y.~V.} \bibnamefont{Gurevich}},
  \bibinfo{author}{\bibfnamefont{P.~W.} \bibnamefont{Hess}},
  \bibinfo{author}{\bibfnamefont{N.~R.} \bibnamefont{Hutzler}},
  \bibinfo{author}{\bibfnamefont{E.}~\bibnamefont{Kirilov}},
  \bibinfo{author}{\bibfnamefont{I.}~\bibnamefont{Kozyryev}},
  \bibinfo{author}{\bibfnamefont{B.~R.} \bibnamefont{O'Leary}}, \emph{et~al.},
  \bibinfo{year}{2014}, \bibinfo{journal}{Science}
  \textbf{\bibinfo{volume}{343}}(\bibinfo{number}{6168}), \bibinfo{pages}{269}.

\bibitem{thesing:2017}
\bibinfo{author}{\bibnamefont{Thesing}, \bibfnamefont{L.~V.}},
  \bibinfo{author}{\bibfnamefont{J.}~\bibnamefont{K{\"u}pper}}, and
  \bibinfo{author}{\bibfnamefont{R.}~\bibnamefont{Gonz{\'a}lez-F{\'e}rez}},
  \bibinfo{year}{2017}, \bibinfo{journal}{J. Chem. Phys.}
  \textbf{\bibinfo{volume}{146}}(\bibinfo{number}{24}),
  \bibinfo{pages}{244304}.

\bibitem{thomas:2018}
\bibinfo{author}{\bibnamefont{Thomas}, \bibfnamefont{E.~F.}},
  \bibinfo{author}{\bibfnamefont{A.~A.} \bibnamefont{S\o{}ndergaard}},
  \bibinfo{author}{\bibfnamefont{B.}~\bibnamefont{Shepperson}},
  \bibinfo{author}{\bibfnamefont{N.~E.} \bibnamefont{Henriksen}}, and
  \bibinfo{author}{\bibfnamefont{H.}~\bibnamefont{Stapelfeldt}},
  \bibinfo{year}{2018}, \bibinfo{journal}{Phys. Rev. Lett.}
  \textbf{\bibinfo{volume}{120}}, \bibinfo{pages}{163202}.

\bibitem{ToenniesAngChem04}
\bibinfo{author}{\bibnamefont{Toennies}, \bibfnamefont{J.}}, and
  \bibinfo{author}{\bibfnamefont{A.}~\bibnamefont{Vilesov}},
  \bibinfo{year}{2004}, \bibinfo{journal}{Ang. Chem. Int. Ed.}
  \textbf{\bibinfo{volume}{43}}, \bibinfo{pages}{2622}.

\bibitem{ToenniesMolPhys13}
\bibinfo{author}{\bibnamefont{Toennies}, \bibfnamefont{J.~P.}},
  \bibinfo{year}{2013}, \bibinfo{journal}{Molecular Physics}
  \textbf{\bibinfo{volume}{111}}(\bibinfo{number}{12-13}),
  \bibinfo{pages}{1879}.

\bibitem{ToenniesARPC98}
\bibinfo{author}{\bibnamefont{Toennies}, \bibfnamefont{J.~P.}}, and
  \bibinfo{author}{\bibfnamefont{A.~F.} \bibnamefont{Vilesov}},
  \bibinfo{year}{1998}, \bibinfo{journal}{Annu. Rev. Phys. Chem.}
  \textbf{\bibinfo{volume}{49}}, \bibinfo{pages}{1}.

\bibitem{TomzaPRL14}
\bibinfo{author}{\bibnamefont{Tomza}, \bibfnamefont{M.}},
  \bibinfo{author}{\bibfnamefont{R.}~\bibnamefont{Gonz\'alez-F\'erez}},
  \bibinfo{author}{\bibfnamefont{C.~P.} \bibnamefont{Koch}}, and
  \bibinfo{author}{\bibfnamefont{R.}~\bibnamefont{Moszynski}},
  \bibinfo{year}{2014}, \bibinfo{journal}{Phys. Rev. Lett.}
  \textbf{\bibinfo{volume}{112}}, \bibinfo{pages}{113201}.

\bibitem{TongPRL10}
\bibinfo{author}{\bibnamefont{Tong}, \bibfnamefont{X.}},
  \bibinfo{author}{\bibfnamefont{A.~H.} \bibnamefont{Winney}}, and
  \bibinfo{author}{\bibfnamefont{S.}~\bibnamefont{Willitsch}},
  \bibinfo{year}{2010}, \bibinfo{journal}{Phys. Rev. Lett.}
  \textbf{\bibinfo{volume}{105}}, \bibinfo{pages}{143001}.

\bibitem{TordrupPRA08}
\bibinfo{author}{\bibnamefont{Tordrup}, \bibfnamefont{K.}}, and
  \bibinfo{author}{\bibfnamefont{K.}~\bibnamefont{M\o{}lmer}},
  \bibinfo{year}{2008}, \bibinfo{journal}{Phys. Rev. A}
  \textbf{\bibinfo{volume}{77}}, \bibinfo{pages}{020301}.

\bibitem{torres:2007}
\bibinfo{author}{\bibnamefont{Torres}, \bibfnamefont{R.}},
  \bibinfo{author}{\bibfnamefont{N.}~\bibnamefont{Kajumba}},
  \bibinfo{author}{\bibfnamefont{J.~G.} \bibnamefont{Underwood}},
  \bibinfo{author}{\bibfnamefont{J.~S.} \bibnamefont{Robinson}},
  \bibinfo{author}{\bibfnamefont{S.}~\bibnamefont{Baker}},
  \bibinfo{author}{\bibfnamefont{J.~W.~G.} \bibnamefont{Tisch}},
  \bibinfo{author}{\bibfnamefont{R.}~\bibnamefont{de~Nalda}},
  \bibinfo{author}{\bibfnamefont{W.~A.} \bibnamefont{Bryan}},
  \bibinfo{author}{\bibfnamefont{R.}~\bibnamefont{Velotta}},
  \bibinfo{author}{\bibfnamefont{C.}~\bibnamefont{Altucci}},
  \bibinfo{author}{\bibfnamefont{I.~C.~E.} \bibnamefont{Turcu}}, and
  \bibinfo{author}{\bibfnamefont{J.~P.} \bibnamefont{Marangos}},
  \bibinfo{year}{2007}, \bibinfo{journal}{Phys. Rev. Lett.}
  \textbf{\bibinfo{volume}{98}}, \bibinfo{pages}{203007}.

\bibitem{torres:2005}
\bibinfo{author}{\bibnamefont{Torres}, \bibfnamefont{R.}},
  \bibinfo{author}{\bibfnamefont{R.}~\bibnamefont{de~Nalda}}, and
  \bibinfo{author}{\bibfnamefont{J.~P.} \bibnamefont{Marangos}},
  \bibinfo{year}{2005}, \bibinfo{journal}{Phys. Rev. A}
  \textbf{\bibinfo{volume}{72}}, \bibinfo{pages}{023420}.

\bibitem{TownesSchawlow}
\bibinfo{author}{\bibnamefont{Townes}, \bibfnamefont{C.~H.}}, and
  \bibinfo{author}{\bibfnamefont{A.~L.} \bibnamefont{Schawlow}},
  \bibinfo{year}{1975}, \emph{\bibinfo{title}{Microwave Spectroscopy}}
  (\bibinfo{publisher}{Dover, New York}).

\bibitem{trippel:2015}
\bibinfo{author}{\bibnamefont{Trippel}, \bibfnamefont{S.}},
  \bibinfo{author}{\bibfnamefont{T.}~\bibnamefont{Mullins}},
  \bibinfo{author}{\bibfnamefont{N.~L.~M.} \bibnamefont{M\"uller}},
  \bibinfo{author}{\bibfnamefont{J.~S.} \bibnamefont{Kienitz}},
  \bibinfo{author}{\bibfnamefont{R.}~\bibnamefont{Gonz\'alez-F\'erez}}, and
  \bibinfo{author}{\bibfnamefont{J.}~\bibnamefont{K\"upper}},
  \bibinfo{year}{2015}, \bibinfo{journal}{Phys. Rev. Lett.}
  \textbf{\bibinfo{volume}{114}}, \bibinfo{pages}{103003}.

\bibitem{trippel:2014}
\bibinfo{author}{\bibnamefont{Trippel}, \bibfnamefont{S.}},
  \bibinfo{author}{\bibfnamefont{T.}~\bibnamefont{Mullins}},
  \bibinfo{author}{\bibfnamefont{N.~L.~M.} \bibnamefont{M\"uller}},
  \bibinfo{author}{\bibfnamefont{J.~S.} \bibnamefont{Kienitz}},
  \bibinfo{author}{\bibfnamefont{J.~J.} \bibnamefont{Omiste}},
  \bibinfo{author}{\bibfnamefont{H.}~\bibnamefont{Stapelfeldt}},
  \bibinfo{author}{\bibfnamefont{R.}~\bibnamefont{Gonz\'alez-F\'erez}}, and
  \bibinfo{author}{\bibfnamefont{J.}~\bibnamefont{K\"upper}},
  \bibinfo{year}{2014}, \bibinfo{journal}{Phys. Rev. A}
  \textbf{\bibinfo{volume}{89}}, \bibinfo{pages}{051401}.

\bibitem{TruppeNatPhys17}
\bibinfo{author}{\bibnamefont{Truppe}, \bibfnamefont{S.}},
  \bibinfo{author}{\bibfnamefont{H.~J.} \bibnamefont{Williams}},
  \bibinfo{author}{\bibfnamefont{M.}~\bibnamefont{Hambach}},
  \bibinfo{author}{\bibfnamefont{L.}~\bibnamefont{Caldwell}},
  \bibinfo{author}{\bibfnamefont{N.~J.} \bibnamefont{Fitch}},
  \bibinfo{author}{\bibfnamefont{E.~A.} \bibnamefont{Hinds}},
  \bibinfo{author}{\bibfnamefont{B.~E.} \bibnamefont{Sauer}}, and
  \bibinfo{author}{\bibfnamefont{M.~R.} \bibnamefont{Tarbutt}},
  \bibinfo{year}{2017}, \bibinfo{journal}{Nature Phys.}
  \textbf{\bibinfo{volume}{13}}, \bibinfo{pages}{1173}.

\bibitem{turinici:2004}
\bibinfo{author}{\bibnamefont{Turinici}, \bibfnamefont{G.}}, and
  \bibinfo{author}{\bibfnamefont{H.}~\bibnamefont{Rabitz}},
  \bibinfo{year}{2004}, \bibinfo{journal}{Phys. Rev. A}
  \textbf{\bibinfo{volume}{70}}, \bibinfo{pages}{063412}.

\bibitem{TutunnikovJPCL18}
\bibinfo{author}{\bibnamefont{Tutunnikov}, \bibfnamefont{I.}},
  \bibinfo{author}{\bibfnamefont{E.}~\bibnamefont{Gershnabel}},
  \bibinfo{author}{\bibfnamefont{S.}~\bibnamefont{Gold}}, and
  \bibinfo{author}{\bibfnamefont{I.~S.} \bibnamefont{Averbukh}},
  \bibinfo{year}{2018}, \bibinfo{journal}{J. Phys. Chem. Lett.}
  \textbf{\bibinfo{volume}{9}}(\bibinfo{number}{5}), \bibinfo{pages}{1105}.

\bibitem{UnderwoodRSI15}
\bibinfo{author}{\bibnamefont{Underwood}, \bibfnamefont{J.~G.}},
  \bibinfo{author}{\bibfnamefont{I.}~\bibnamefont{Procino}},
  \bibinfo{author}{\bibfnamefont{L.}~\bibnamefont{Christiansen}},
  \bibinfo{author}{\bibfnamefont{J.}~\bibnamefont{Maurer}}, and
  \bibinfo{author}{\bibfnamefont{H.}~\bibnamefont{Stapelfeldt}},
  \bibinfo{year}{2015}, \bibinfo{journal}{Rev. Sci. Instrum.}
  \textbf{\bibinfo{volume}{86}}(\bibinfo{number}{7}), \bibinfo{pages}{073101}.

\bibitem{underwood:2003}
\bibinfo{author}{\bibnamefont{Underwood}, \bibfnamefont{J.~G.}},
  \bibinfo{author}{\bibfnamefont{M.}~\bibnamefont{Spanner}},
  \bibinfo{author}{\bibfnamefont{M.~Y.} \bibnamefont{Ivanov}},
  \bibinfo{author}{\bibfnamefont{J.}~\bibnamefont{Mottershead}},
  \bibinfo{author}{\bibfnamefont{B.~J.} \bibnamefont{Sussman}}, and
  \bibinfo{author}{\bibfnamefont{A.}~\bibnamefont{Stolow}},
  \bibinfo{year}{2003}, \bibinfo{journal}{Phys. Rev. Lett.}
  \textbf{\bibinfo{volume}{90}}, \bibinfo{pages}{223001}.

\bibitem{vandamme:SR}
\bibinfo{author}{\bibnamefont{Van~Damme}, \bibfnamefont{L.}},
  \bibinfo{author}{\bibfnamefont{D.}~\bibnamefont{Leiner}},
  \bibinfo{author}{\bibfnamefont{P.}~\bibnamefont{Mardesic}}, and
  \bibinfo{author}{\bibfnamefont{S.~J.} \bibnamefont{Glaser}},
  \bibinfo{year}{2017}{\natexlab{a}}, \bibinfo{journal}{Sci. Rep.}
  \textbf{\bibinfo{volume}{3998}}, \bibinfo{pages}{7}.

\bibitem{vandamme:2017}
\bibinfo{author}{\bibnamefont{Van~Damme}, \bibfnamefont{L.}},
  \bibinfo{author}{\bibfnamefont{P.}~\bibnamefont{Mardesic}}, and
  \bibinfo{author}{\bibfnamefont{D.}~\bibnamefont{Sugny}},
  \bibinfo{year}{2017}{\natexlab{b}}, \bibinfo{journal}{Physica D}
  \textbf{\bibinfo{volume}{338}}, \bibinfo{pages}{17}.

\bibitem{VanhaeckeMolPhys07}
\bibinfo{author}{\bibnamefont{Vanhaecke}, \bibfnamefont{N.}}, and
  \bibinfo{author}{\bibfnamefont{O.}~\bibnamefont{Dulieu}},
  \bibinfo{year}{2007}, \bibinfo{journal}{Molecular Physics}
  \textbf{\bibinfo{volume}{105}}(\bibinfo{number}{11-12}),
  \bibinfo{pages}{1723}.

\bibitem{velotta:2001}
\bibinfo{author}{\bibnamefont{Velotta}, \bibfnamefont{R.}},
  \bibinfo{author}{\bibfnamefont{N.}~\bibnamefont{Hay}},
  \bibinfo{author}{\bibfnamefont{M.~B.} \bibnamefont{Mason}},
  \bibinfo{author}{\bibfnamefont{M.}~\bibnamefont{Castillejo}}, and
  \bibinfo{author}{\bibfnamefont{J.~P.} \bibnamefont{Marangos}},
  \bibinfo{year}{2001}, \bibinfo{journal}{Phys. Rev. Lett.}
  \textbf{\bibinfo{volume}{87}}, \bibinfo{pages}{183901}.

\bibitem{vidal:2014}
\bibinfo{author}{\bibnamefont{Vidal}, \bibfnamefont{S.}},
  \bibinfo{author}{\bibfnamefont{J.}~\bibnamefont{Degert}},
  \bibinfo{author}{\bibfnamefont{M.}~\bibnamefont{Tondusson}},
  \bibinfo{author}{\bibfnamefont{E.}~\bibnamefont{Freysz}}, and
  \bibinfo{author}{\bibfnamefont{J.}~\bibnamefont{Oberl\'e}},
  \bibinfo{year}{2014}, \bibinfo{journal}{J. Opt. Soc. Am. B}
  \textbf{\bibinfo{volume}{31}}, \bibinfo{pages}{149}.

\bibitem{Viellard:13}
\bibinfo{author}{\bibnamefont{Viellard}, \bibfnamefont{T.}},
  \bibinfo{author}{\bibfnamefont{F.}~\bibnamefont{Chaussard}},
  \bibinfo{author}{\bibfnamefont{F.}~\bibnamefont{Billard}},
  \bibinfo{author}{\bibfnamefont{D.}~\bibnamefont{Sugny}},
  \bibinfo{author}{\bibfnamefont{O.}~\bibnamefont{Faucher}},
  \bibinfo{author}{\bibfnamefont{S.}~\bibnamefont{Ivanov}},
  \bibinfo{author}{\bibfnamefont{J.-M.} \bibnamefont{Hartmann}},
  \bibinfo{author}{\bibfnamefont{C.}~\bibnamefont{Boulet}}, and
  \bibinfo{author}{\bibfnamefont{B.}~\bibnamefont{Lavorel}},
  \bibinfo{year}{2013}, \bibinfo{journal}{Phys. Rev. A}
  \textbf{\bibinfo{volume}{87}}.

\bibitem{viellard:2008}
\bibinfo{author}{\bibnamefont{Viellard}, \bibfnamefont{T.}},
  \bibinfo{author}{\bibfnamefont{F.}~\bibnamefont{Chaussard}},
  \bibinfo{author}{\bibfnamefont{D.}~\bibnamefont{Sugny}},
  \bibinfo{author}{\bibfnamefont{B.}~\bibnamefont{Lavorel}}, and
  \bibinfo{author}{\bibfnamefont{O.}~\bibnamefont{Faucher}},
  \bibinfo{year}{2008}, \bibinfo{journal}{J. Raman Spec.}
  \textbf{\bibinfo{volume}{39}}, \bibinfo{pages}{694}.

\bibitem{viftrup:2009}
\bibinfo{author}{\bibnamefont{Viftrup}, \bibfnamefont{S.~S.}},
  \bibinfo{author}{\bibfnamefont{V.}~\bibnamefont{Kumarappan}},
  \bibinfo{author}{\bibfnamefont{L.}~\bibnamefont{Holmegaard}},
  \bibinfo{author}{\bibfnamefont{C.~Z.} \bibnamefont{Bisgaard}},
  \bibinfo{author}{\bibfnamefont{H.}~\bibnamefont{Stapelfeldt}},
  \bibinfo{author}{\bibfnamefont{M.}~\bibnamefont{Artamonov}},
  \bibinfo{author}{\bibfnamefont{E.}~\bibnamefont{Hamilton}}, and
  \bibinfo{author}{\bibfnamefont{T.}~\bibnamefont{Seideman}},
  \bibinfo{year}{2009}, \bibinfo{journal}{Phys. Rev. A}
  \textbf{\bibinfo{volume}{79}}, \bibinfo{pages}{023404}.

\bibitem{viftrup:2007}
\bibinfo{author}{\bibnamefont{Viftrup}, \bibfnamefont{S.~S.}},
  \bibinfo{author}{\bibfnamefont{V.}~\bibnamefont{Kumarappan}},
  \bibinfo{author}{\bibfnamefont{S.}~\bibnamefont{Trippel}},
  \bibinfo{author}{\bibfnamefont{H.}~\bibnamefont{Stapelfeldt}},
  \bibinfo{author}{\bibfnamefont{E.}~\bibnamefont{Hamilton}}, and
  \bibinfo{author}{\bibfnamefont{T.}~\bibnamefont{Seideman}},
  \bibinfo{year}{2007}, \bibinfo{journal}{Phys. Rev. Lett.}
  \textbf{\bibinfo{volume}{99}}, \bibinfo{pages}{143602}.

\bibitem{villeneuve:2000}
\bibinfo{author}{\bibnamefont{Villeneuve}, \bibfnamefont{D.~M.}},
  \bibinfo{author}{\bibfnamefont{S.~A.} \bibnamefont{Aseyev}},
  \bibinfo{author}{\bibfnamefont{P.}~\bibnamefont{Dietrich}},
  \bibinfo{author}{\bibfnamefont{M.}~\bibnamefont{Spanner}},
  \bibinfo{author}{\bibfnamefont{M.~Y.} \bibnamefont{Ivanov}}, and
  \bibinfo{author}{\bibfnamefont{P.~B.} \bibnamefont{Corkum}},
  \bibinfo{year}{2000}, \bibinfo{journal}{Phys. Rev. Lett.}
  \textbf{\bibinfo{volume}{85}}, \bibinfo{pages}{542}.

\bibitem{ViteauSci08}
\bibinfo{author}{\bibnamefont{Viteau}, \bibfnamefont{M.}},
  \bibinfo{author}{\bibfnamefont{A.}~\bibnamefont{Chotia}},
  \bibinfo{author}{\bibfnamefont{M.}~\bibnamefont{Allegrini}},
  \bibinfo{author}{\bibfnamefont{N.}~\bibnamefont{Bouloufa}},
  \bibinfo{author}{\bibfnamefont{O.}~\bibnamefont{Dulieu}},
  \bibinfo{author}{\bibfnamefont{D.}~\bibnamefont{Comparat}}, and
  \bibinfo{author}{\bibfnamefont{P.}~\bibnamefont{Pillet}},
  \bibinfo{year}{2008}, \bibinfo{journal}{Science}
  \textbf{\bibinfo{volume}{321}}(\bibinfo{number}{5886}), \bibinfo{pages}{232}.

\bibitem{VogeliusPRL02}
\bibinfo{author}{\bibnamefont{Vogelius}, \bibfnamefont{I.~S.}},
  \bibinfo{author}{\bibfnamefont{L.~B.} \bibnamefont{Madsen}}, and
  \bibinfo{author}{\bibfnamefont{M.}~\bibnamefont{Drewsen}},
  \bibinfo{year}{2002}, \bibinfo{journal}{Phys. Rev. Lett.}
  \textbf{\bibinfo{volume}{89}}, \bibinfo{pages}{173003}.

\bibitem{VogeliusJPB06}
\bibinfo{author}{\bibnamefont{Vogelius}, \bibfnamefont{I.~S.}},
  \bibinfo{author}{\bibfnamefont{L.~B.} \bibnamefont{Madsen}}, and
  \bibinfo{author}{\bibfnamefont{M.}~\bibnamefont{Drewsen}},
  \bibinfo{year}{2006}, \bibinfo{journal}{J. Phys. B}
  \textbf{\bibinfo{volume}{39}}(\bibinfo{number}{19}), \bibinfo{pages}{S1259}.

\bibitem{vrakking:1997}
\bibinfo{author}{\bibnamefont{Vrakking}, \bibfnamefont{M.~J.~J.}}, and
  \bibinfo{author}{\bibfnamefont{S.}~\bibnamefont{Stolte}},
  \bibinfo{year}{1997}, \bibinfo{journal}{Chem. Phys. Lett.}
  \textbf{\bibinfo{volume}{271}}, \bibinfo{pages}{209}.

\bibitem{WallAnnPhys13}
\bibinfo{author}{\bibnamefont{Wall}, \bibfnamefont{M.~L.}},
  \bibinfo{author}{\bibfnamefont{K.}~\bibnamefont{Maeda}}, and
  \bibinfo{author}{\bibfnamefont{L.~D.} \bibnamefont{Carr}},
  \bibinfo{year}{2013}, \bibinfo{journal}{Annalen der Physik}
  \textbf{\bibinfo{volume}{525}}(\bibinfo{number}{10-11}),
  \bibinfo{pages}{845}.

\bibitem{WallNJP15}
\bibinfo{author}{\bibnamefont{Wall}, \bibfnamefont{M.~L.}},
  \bibinfo{author}{\bibfnamefont{K.}~\bibnamefont{Maeda}}, and
  \bibinfo{author}{\bibfnamefont{L.~D.} \bibnamefont{Carr}},
  \bibinfo{year}{2015}, \bibinfo{journal}{New J. Phys.}
  \textbf{\bibinfo{volume}{17}}, \bibinfo{pages}{025001}.

\bibitem{WallPRA17}
\bibinfo{author}{\bibnamefont{Wall}, \bibfnamefont{M.~L.}},
  \bibinfo{author}{\bibfnamefont{N.~P.} \bibnamefont{Mehta}},
  \bibinfo{author}{\bibfnamefont{R.}~\bibnamefont{Mukherjee}},
  \bibinfo{author}{\bibfnamefont{S.~S.} \bibnamefont{Alam}}, and
  \bibinfo{author}{\bibfnamefont{K.~R.~A.} \bibnamefont{Hazzard}},
  \bibinfo{year}{2017}, \bibinfo{journal}{Phys. Rev. A}
  \textbf{\bibinfo{volume}{95}}, \bibinfo{pages}{043635}.

\bibitem{WallquistPS09}
\bibinfo{author}{\bibnamefont{Wallquist}, \bibfnamefont{M.}},
  \bibinfo{author}{\bibfnamefont{K.}~\bibnamefont{Hammerer}},
  \bibinfo{author}{\bibfnamefont{P.}~\bibnamefont{Rabl}},
  \bibinfo{author}{\bibfnamefont{M.}~\bibnamefont{Lukin}}, and
  \bibinfo{author}{\bibfnamefont{P.}~\bibnamefont{Zoller}},
  \bibinfo{year}{2009}, \bibinfo{journal}{Physica Scripta}
  \textbf{\bibinfo{volume}{T137}}, \bibinfo{pages}{014001}.

\bibitem{weber:2013}
\bibinfo{author}{\bibnamefont{Weber}, \bibfnamefont{S.~J.}},
  \bibinfo{author}{\bibfnamefont{M.}~\bibnamefont{Oppermann}}, and
  \bibinfo{author}{\bibfnamefont{J.~P.} \bibnamefont{Marangos}},
  \bibinfo{year}{2013}, \bibinfo{journal}{Phys. Rev. Lett.}
  \textbf{\bibinfo{volume}{111}}, \bibinfo{pages}{263601}.

\bibitem{WeiJCP11}
\bibinfo{author}{\bibnamefont{Wei}, \bibfnamefont{Q.}},
  \bibinfo{author}{\bibfnamefont{S.}~\bibnamefont{Kais}},
  \bibinfo{author}{\bibfnamefont{B.}~\bibnamefont{Friedrich}}, and
  \bibinfo{author}{\bibfnamefont{D.}~\bibnamefont{Herschbach}},
  \bibinfo{year}{2011}{\natexlab{a}}, \bibinfo{journal}{J. Chem. Phys.}
  \textbf{\bibinfo{volume}{134}}, \bibinfo{pages}{124107}.

\bibitem{WeiJCP11b}
\bibinfo{author}{\bibnamefont{Wei}, \bibfnamefont{Q.}},
  \bibinfo{author}{\bibfnamefont{S.}~\bibnamefont{Kais}},
  \bibinfo{author}{\bibfnamefont{B.}~\bibnamefont{Friedrich}}, and
  \bibinfo{author}{\bibfnamefont{D.}~\bibnamefont{Herschbach}},
  \bibinfo{year}{2011}{\natexlab{b}}, \bibinfo{journal}{J. Chem. Phys.}
  \textbf{\bibinfo{volume}{135}}, \bibinfo{pages}{154102}.

\bibitem{WeimerMolPhys13}
\bibinfo{author}{\bibnamefont{Weimer}, \bibfnamefont{H.}},
  \bibinfo{year}{2013}, \bibinfo{journal}{Molecular Physics}
  \textbf{\bibinfo{volume}{111}}(\bibinfo{number}{12-13}),
  \bibinfo{pages}{1753}.

\bibitem{weiner:2000}
\bibinfo{author}{\bibnamefont{Weiner}, \bibfnamefont{A.~M.}},
  \bibinfo{year}{2000}, \bibinfo{journal}{Review of Scientific Instruments}
  \textbf{\bibinfo{volume}{71}}(\bibinfo{number}{5}), \bibinfo{pages}{1929}.

\bibitem{WillPRL16}
\bibinfo{author}{\bibnamefont{Will}, \bibfnamefont{S.~A.}},
  \bibinfo{author}{\bibfnamefont{J.~W.} \bibnamefont{Park}},
  \bibinfo{author}{\bibfnamefont{Z.~Z.} \bibnamefont{Yan}},
  \bibinfo{author}{\bibfnamefont{H.}~\bibnamefont{Loh}}, and
  \bibinfo{author}{\bibfnamefont{M.~W.} \bibnamefont{Zwierlein}},
  \bibinfo{year}{2016}, \bibinfo{journal}{Phys. Rev. Lett.}
  \textbf{\bibinfo{volume}{116}}, \bibinfo{pages}{225306}.

\bibitem{WillitschIRPC12}
\bibinfo{author}{\bibnamefont{Willitsch}, \bibfnamefont{S.}},
  \bibinfo{year}{2012}, \bibinfo{journal}{Int. Rev. Phys. Chem.}
  \textbf{\bibinfo{volume}{31}}(\bibinfo{number}{2}), \bibinfo{pages}{175}.

\bibitem{WolfNature16}
\bibinfo{author}{\bibnamefont{Wolf}, \bibfnamefont{F.}},
  \bibinfo{author}{\bibfnamefont{Y.}~\bibnamefont{Wan}},
  \bibinfo{author}{\bibfnamefont{J.~C.} \bibnamefont{Heip}},
  \bibinfo{author}{\bibfnamefont{F.}~\bibnamefont{Gebert}},
  \bibinfo{author}{\bibfnamefont{C.}~\bibnamefont{Shi}}, and
  \bibinfo{author}{\bibfnamefont{P.~O.} \bibnamefont{Schmidt}},
  \bibinfo{year}{2016}, \bibinfo{journal}{Nature}
  \textbf{\bibinfo{volume}{530}}, \bibinfo{pages}{457}.

\bibitem{WollenhauptAnnuRevPhysChem05}
\bibinfo{author}{\bibnamefont{Wollenhaupt}, \bibfnamefont{M.}},
  \bibinfo{author}{\bibfnamefont{V.}~\bibnamefont{Engel}}, and
  \bibinfo{author}{\bibfnamefont{T.}~\bibnamefont{Baumert}},
  \bibinfo{year}{2005}, \bibinfo{journal}{Annu. Rev. Phys. Chem.}
  \textbf{\bibinfo{volume}{56}}(\bibinfo{number}{1}), \bibinfo{pages}{25}.

\bibitem{Wolter:2016}
\bibinfo{author}{\bibnamefont{Wolter}, \bibfnamefont{B.}},
  \bibinfo{author}{\bibfnamefont{M.~G.} \bibnamefont{Pullen}},
  \bibinfo{author}{\bibfnamefont{A.-T.} \bibnamefont{Le}},
  \bibinfo{author}{\bibfnamefont{M.}~\bibnamefont{Baudisch}},
  \bibinfo{author}{\bibfnamefont{K.}~\bibnamefont{Doblhoff-Dier}},
  \bibinfo{author}{\bibfnamefont{A.}~\bibnamefont{Senftleben}},
  \bibinfo{author}{\bibfnamefont{M.}~\bibnamefont{Hemmer}},
  \bibinfo{author}{\bibfnamefont{C.~D.} \bibnamefont{Schr{\"o}ter}},
  \bibinfo{author}{\bibfnamefont{J.}~\bibnamefont{Ullrich}},
  \bibinfo{author}{\bibfnamefont{T.}~\bibnamefont{Pfeifer}},
  \bibinfo{author}{\bibfnamefont{R.}~\bibnamefont{Moshammer}},
  \bibinfo{author}{\bibfnamefont{S.}~\bibnamefont{Gr{\"a}fe}}, \emph{et~al.},
  \bibinfo{year}{2016}, \bibinfo{journal}{Science}
  \textbf{\bibinfo{volume}{354}}(\bibinfo{number}{6310}), \bibinfo{pages}{308}.

\bibitem{zeng:2010}
\bibinfo{author}{\bibnamefont{Wu}, \bibfnamefont{J.}}, and
  \bibinfo{author}{\bibfnamefont{H.}~\bibnamefont{Zeng}}, \bibinfo{year}{2010},
  \bibinfo{journal}{Phys. Rev. A} \textbf{\bibinfo{volume}{81}},
  \bibinfo{pages}{053401}.

\bibitem{wu:1994}
\bibinfo{author}{\bibnamefont{Wu}, \bibfnamefont{M.}},
  \bibinfo{author}{\bibfnamefont{R.~J.} \bibnamefont{Bemish}}, and
  \bibinfo{author}{\bibfnamefont{R.~E.} \bibnamefont{Miller}},
  \bibinfo{year}{1994}, \bibinfo{journal}{J. Chem. Phys.}
  \textbf{\bibinfo{volume}{101}}(\bibinfo{number}{11}), \bibinfo{pages}{9447}.

\bibitem{XiangPRA12}
\bibinfo{author}{\bibnamefont{Xiang}, \bibfnamefont{P.}},
  \bibinfo{author}{\bibfnamefont{M.}~\bibnamefont{Litinskaya}}, and
  \bibinfo{author}{\bibfnamefont{R.~V.} \bibnamefont{Krems}},
  \bibinfo{year}{2012}, \bibinfo{journal}{Phys. Rev. A}
  \textbf{\bibinfo{volume}{85}}, \bibinfo{pages}{061401}.

\bibitem{xie:2014}
\bibinfo{author}{\bibnamefont{Xie}, \bibfnamefont{X.}},
  \bibinfo{author}{\bibfnamefont{K.}~\bibnamefont{Doblhoff-Dier}},
  \bibinfo{author}{\bibfnamefont{H.}~\bibnamefont{Xu}},
  \bibinfo{author}{\bibfnamefont{S.}~\bibnamefont{Roither}},
  \bibinfo{author}{\bibfnamefont{M.~S.} \bibnamefont{Sch\"offler}},
  \bibinfo{author}{\bibfnamefont{D.}~\bibnamefont{Kartashov}},
  \bibinfo{author}{\bibfnamefont{S.}~\bibnamefont{Erattupuzha}},
  \bibinfo{author}{\bibfnamefont{T.}~\bibnamefont{Rathje}},
  \bibinfo{author}{\bibfnamefont{G.~G.} \bibnamefont{Paulus}},
  \bibinfo{author}{\bibfnamefont{K.}~\bibnamefont{Yamanouchi}},
  \bibinfo{author}{\bibfnamefont{A.}~\bibnamefont{Baltu\ifmmode~\check{s}\else
  \v{s}\fi{}ka}}, \bibinfo{author}{\bibfnamefont{S.}~\bibnamefont{Gr\"afe}},
  \emph{et~al.}, \bibinfo{year}{2014}, \bibinfo{journal}{Phys. Rev. Lett.}
  \textbf{\bibinfo{volume}{112}}, \bibinfo{pages}{163003}.

\bibitem{XuNJP15}
\bibinfo{author}{\bibnamefont{Xu}, \bibfnamefont{T.}}, and
  \bibinfo{author}{\bibfnamefont{R.~V.} \bibnamefont{Krems}},
  \bibinfo{year}{2015}, \bibinfo{journal}{New J. Phys.}
  \textbf{\bibinfo{volume}{17}}, \bibinfo{pages}{065014}.

\bibitem{yachmenev:2016}
\bibinfo{author}{\bibnamefont{Yachmenev}, \bibfnamefont{A.}}, and
  \bibinfo{author}{\bibfnamefont{S.~N.} \bibnamefont{Yurchenko}},
  \bibinfo{year}{2016}{\natexlab{a}}, \bibinfo{journal}{Phys. Rev. Lett.}
  \textbf{\bibinfo{volume}{117}}, \bibinfo{pages}{033001}.

\bibitem{YachmenevPRL16}
\bibinfo{author}{\bibnamefont{Yachmenev}, \bibfnamefont{A.}}, and
  \bibinfo{author}{\bibfnamefont{S.~N.} \bibnamefont{Yurchenko}},
  \bibinfo{year}{2016}{\natexlab{b}}, \bibinfo{journal}{Phys. Rev. Lett.}
  \textbf{\bibinfo{volume}{117}}, \bibinfo{pages}{033001}.

\bibitem{YanNature13}
\bibinfo{author}{\bibnamefont{Yan}, \bibfnamefont{B.}},
  \bibinfo{author}{\bibfnamefont{S.~A.} \bibnamefont{Moses}},
  \bibinfo{author}{\bibfnamefont{B.}~\bibnamefont{Gadway}},
  \bibinfo{author}{\bibfnamefont{J.~P.} \bibnamefont{Covey}},
  \bibinfo{author}{\bibfnamefont{K.~R.~A.} \bibnamefont{Hazzard}},
  \bibinfo{author}{\bibfnamefont{A.~M.} \bibnamefont{Rey}},
  \bibinfo{author}{\bibfnamefont{D.~S.} \bibnamefont{Jin}}, and
  \bibinfo{author}{\bibfnamefont{J.}~\bibnamefont{Ye}}, \bibinfo{year}{2013},
  \bibinfo{journal}{Nature} \textbf{\bibinfo{volume}{501}},
  \bibinfo{pages}{521}.

\bibitem{YaoPRL13}
\bibinfo{author}{\bibnamefont{Yao}, \bibfnamefont{N.~Y.}},
  \bibinfo{author}{\bibfnamefont{A.~V.} \bibnamefont{Gorshkov}},
  \bibinfo{author}{\bibfnamefont{C.~R.} \bibnamefont{Laumann}},
  \bibinfo{author}{\bibfnamefont{A.~M.} \bibnamefont{L\"auchli}},
  \bibinfo{author}{\bibfnamefont{J.}~\bibnamefont{Ye}}, and
  \bibinfo{author}{\bibfnamefont{M.~D.} \bibnamefont{Lukin}},
  \bibinfo{year}{2013}, \bibinfo{journal}{Phys. Rev. Lett.}
  \textbf{\bibinfo{volume}{110}}, \bibinfo{pages}{185302}.

\bibitem{YaoNatPhys18}
\bibinfo{author}{\bibnamefont{Yao}, \bibfnamefont{N.~Y.}},
  \bibinfo{author}{\bibfnamefont{M.~P.} \bibnamefont{Zaletel}},
  \bibinfo{author}{\bibfnamefont{D.~M.} \bibnamefont{Stamper-Kurn}}, and
  \bibinfo{author}{\bibfnamefont{A.}~\bibnamefont{Vishwanath}},
  \bibinfo{year}{2018}, \bibinfo{journal}{Nature Physics}
  \textbf{\bibinfo{volume}{14}}(\bibinfo{number}{4}), \bibinfo{pages}{405}.

\bibitem{YelinPRA06}
\bibinfo{author}{\bibnamefont{Yelin}, \bibfnamefont{S.~F.}},
  \bibinfo{author}{\bibfnamefont{K.}~\bibnamefont{Kirby}}, and
  \bibinfo{author}{\bibfnamefont{R.}~\bibnamefont{C\^ot\'e}},
  \bibinfo{year}{2006}, \bibinfo{journal}{Phys. Rev. A}
  \textbf{\bibinfo{volume}{74}}, \bibinfo{pages}{050301}.

\bibitem{YeoPRL15}
\bibinfo{author}{\bibnamefont{Yeo}, \bibfnamefont{M.}},
  \bibinfo{author}{\bibfnamefont{M.~T.} \bibnamefont{Hummon}},
  \bibinfo{author}{\bibfnamefont{A.~L.} \bibnamefont{Collopy}},
  \bibinfo{author}{\bibfnamefont{B.}~\bibnamefont{Yan}},
  \bibinfo{author}{\bibfnamefont{B.}~\bibnamefont{Hemmerling}},
  \bibinfo{author}{\bibfnamefont{E.}~\bibnamefont{Chae}},
  \bibinfo{author}{\bibfnamefont{J.~M.} \bibnamefont{Doyle}}, and
  \bibinfo{author}{\bibfnamefont{J.}~\bibnamefont{Ye}}, \bibinfo{year}{2015},
  \bibinfo{journal}{Phys. Rev. Lett.} \textbf{\bibinfo{volume}{114}},
  \bibinfo{pages}{223003}.

\bibitem{Yoshida:14}
\bibinfo{author}{\bibnamefont{Yoshida}, \bibfnamefont{M.}}, and
  \bibinfo{author}{\bibfnamefont{Y.}~\bibnamefont{Ohtsuki}},
  \bibinfo{year}{2014}, \bibinfo{journal}{Phys. Rev. A}
  \textbf{\bibinfo{volume}{90}}, \bibinfo{pages}{013415}.

\bibitem{yoshida:2015}
\bibinfo{author}{\bibnamefont{Yoshida}, \bibfnamefont{M.}}, and
  \bibinfo{author}{\bibfnamefont{Y.}~\bibnamefont{Ohtsuki}},
  \bibinfo{year}{2015}, \bibinfo{journal}{Chemical Physics Letters}
  \textbf{\bibinfo{volume}{633}}, \bibinfo{pages}{169 }.

\bibitem{yoshii:2008}
\bibinfo{author}{\bibnamefont{Yoshii}, \bibfnamefont{K.}},
  \bibinfo{author}{\bibfnamefont{G.}~\bibnamefont{Miyaji}}, and
  \bibinfo{author}{\bibfnamefont{K.}~\bibnamefont{Miyazaki}},
  \bibinfo{year}{2008}, \bibinfo{journal}{Phys. Rev. Lett.}
  \textbf{\bibinfo{volume}{101}}, \bibinfo{pages}{183902}.

\bibitem{YuPCCP18}
\bibinfo{author}{\bibnamefont{Yu}, \bibfnamefont{H.}},
  \bibinfo{author}{\bibfnamefont{T.-S.} \bibnamefont{Ho}}, and
  \bibinfo{author}{\bibfnamefont{H.}~\bibnamefont{Rabitz}},
  \bibinfo{year}{2018}, \bibinfo{journal}{Phys. Chem. Chem. Phys.}
  \textbf{\bibinfo{volume}{20}}, \bibinfo{pages}{13008}.

\bibitem{yuan:2011}
\bibinfo{author}{\bibnamefont{Yuan}, \bibfnamefont{L.}},
  \bibinfo{author}{\bibfnamefont{S.~W.} \bibnamefont{Teitelbaum}},
  \bibinfo{author}{\bibfnamefont{A.}~\bibnamefont{Robinson}}, and
  \bibinfo{author}{\bibfnamefont{A.~S.} \bibnamefont{Mullin}},
  \bibinfo{year}{2011}, \bibinfo{journal}{Proceedings of the National Academy
  of Sciences} \textbf{\bibinfo{volume}{108}}(\bibinfo{number}{17}),
  \bibinfo{pages}{6872}.

\bibitem{yun:2011}
\bibinfo{author}{\bibnamefont{Yun}, \bibfnamefont{H.}},
  \bibinfo{author}{\bibfnamefont{H.~T.} \bibnamefont{Kim}},
  \bibinfo{author}{\bibfnamefont{C.~M.} \bibnamefont{Kim}},
  \bibinfo{author}{\bibfnamefont{C.~H.} \bibnamefont{Nam}}, and
  \bibinfo{author}{\bibfnamefont{J.}~\bibnamefont{Lee}}, \bibinfo{year}{2011},
  \bibinfo{journal}{Phys. Rev. A} \textbf{\bibinfo{volume}{84}},
  \bibinfo{pages}{065401}.

\bibitem{YzombardPRL15}
\bibinfo{author}{\bibnamefont{Yzombard}, \bibfnamefont{P.}},
  \bibinfo{author}{\bibfnamefont{M.}~\bibnamefont{Hamamda}},
  \bibinfo{author}{\bibfnamefont{S.}~\bibnamefont{Gerber}},
  \bibinfo{author}{\bibfnamefont{M.}~\bibnamefont{Doser}}, and
  \bibinfo{author}{\bibfnamefont{D.}~\bibnamefont{Comparat}},
  \bibinfo{year}{2015}, \bibinfo{journal}{Phys. Rev. Lett.}
  \textbf{\bibinfo{volume}{114}}, \bibinfo{pages}{213001}.

\bibitem{ZahedpourPRL14}
\bibinfo{author}{\bibnamefont{Zahedpour}, \bibfnamefont{S.}},
  \bibinfo{author}{\bibfnamefont{J.~K.} \bibnamefont{Wahlstrand}}, and
  \bibinfo{author}{\bibfnamefont{H.~M.} \bibnamefont{Milchberg}},
  \bibinfo{year}{2014}, \bibinfo{journal}{Phys. Rev. Lett.}
  \textbf{\bibinfo{volume}{112}}, \bibinfo{pages}{143601}.

\bibitem{Zare:88}
\bibinfo{author}{\bibnamefont{Zare}, \bibfnamefont{R.~N.}},
  \bibinfo{year}{1988}, \emph{\bibinfo{title}{Angular momentum: understanding
  spatial aspects in chemistry and physics}}
  (\bibinfo{publisher}{Wiley-Interscience Publication}).

\bibitem{ZeppenfeldNature12}
\bibinfo{author}{\bibnamefont{Zeppenfeld}, \bibfnamefont{M.}},
  \bibinfo{author}{\bibfnamefont{B.~G.~U.} \bibnamefont{Englert}},
  \bibinfo{author}{\bibfnamefont{R.}~\bibnamefont{Glockner}},
  \bibinfo{author}{\bibfnamefont{A.}~\bibnamefont{Prehn}},
  \bibinfo{author}{\bibfnamefont{M.}~\bibnamefont{Mielenz}},
  \bibinfo{author}{\bibfnamefont{C.}~\bibnamefont{Sommer}},
  \bibinfo{author}{\bibfnamefont{L.~D.} \bibnamefont{van Buuren}},
  \bibinfo{author}{\bibfnamefont{M.}~\bibnamefont{Motsch}}, and
  \bibinfo{author}{\bibfnamefont{G.}~\bibnamefont{Rempe}},
  \bibinfo{year}{2012}, \bibinfo{journal}{Nature}
  \textbf{\bibinfo{volume}{491}}, \bibinfo{pages}{7425}.

\bibitem{ZeppenfeldPRA09}
\bibinfo{author}{\bibnamefont{Zeppenfeld}, \bibfnamefont{M.}},
  \bibinfo{author}{\bibfnamefont{M.}~\bibnamefont{Motsch}},
  \bibinfo{author}{\bibfnamefont{P.~W.~H.} \bibnamefont{Pinkse}}, and
  \bibinfo{author}{\bibfnamefont{G.}~\bibnamefont{Rempe}},
  \bibinfo{year}{2009}, \bibinfo{journal}{Phys. Rev. A}
  \textbf{\bibinfo{volume}{80}}, \bibinfo{pages}{041401}.

\bibitem{ZhangChapter18}
\bibinfo{author}{\bibnamefont{Zhang}, \bibfnamefont{D.}}, and
  \bibinfo{author}{\bibfnamefont{S.}~\bibnamefont{Willitsch}},
  \bibinfo{year}{2018}, in \emph{\bibinfo{booktitle}{Cold Chemistry: Molecular
  Scattering and Reactivity Near Absolute Zero}}, edited by
  \bibinfo{editor}{\bibfnamefont{O.}~\bibnamefont{Dulieu}} and
  \bibinfo{editor}{\bibfnamefont{A.}~\bibnamefont{Osterwalder}}
  (\bibinfo{publisher}{The Royal Society of Chemistry}),
  number~\bibinfo{number}{11} in \bibinfo{series}{Theoretical and Computational
  Chemistry Series}, pp. \bibinfo{pages}{496--536}.

\bibitem{ZhangJCP18}
\bibinfo{author}{\bibnamefont{Zhang}, \bibfnamefont{H.}},
  \bibinfo{author}{\bibfnamefont{F.}~\bibnamefont{Billard}},
  \bibinfo{author}{\bibfnamefont{X.}~\bibnamefont{Yu}},
  \bibinfo{author}{\bibfnamefont{O.}~\bibnamefont{Faucher}}, and
  \bibinfo{author}{\bibfnamefont{B.}~\bibnamefont{Lavorel}},
  \bibinfo{year}{2018}, \bibinfo{journal}{J. Chem. Phys.}
  \textbf{\bibinfo{volume}{148}}(\bibinfo{number}{12}),
  \bibinfo{pages}{124303}.

\bibitem{zhang2019}
\bibinfo{author}{\bibnamefont{Zhang}, \bibfnamefont{H.}},
  \bibinfo{author}{\bibfnamefont{B.}~\bibnamefont{Lavorel}},
  \bibinfo{author}{\bibfnamefont{F.}~\bibnamefont{Billard}},
  \bibinfo{author}{\bibfnamefont{J.-M.} \bibnamefont{Hartmann}},
  \bibinfo{author}{\bibfnamefont{E.}~\bibnamefont{Hertz}},
  \bibinfo{author}{\bibfnamefont{O.}~\bibnamefont{Faucher}},
  \bibinfo{author}{\bibfnamefont{J.}~\bibnamefont{Ma}},
  \bibinfo{author}{\bibfnamefont{J.}~\bibnamefont{Wu}},
  \bibinfo{author}{\bibfnamefont{E.}~\bibnamefont{Gershnabel}},
  \bibinfo{author}{\bibfnamefont{Y.}~\bibnamefont{Prior}}, and
  \bibinfo{author}{\bibfnamefont{I.~S.} \bibnamefont{Averbukh}},
  \bibinfo{year}{2019}, \bibinfo{journal}{Phys. Rev. Lett.}
  \textbf{\bibinfo{volume}{122}}, \bibinfo{pages}{193401}.

\bibitem{zhang:2011}
\bibinfo{author}{\bibnamefont{Zhang}, \bibfnamefont{S.}},
  \bibinfo{author}{\bibfnamefont{C.}~\bibnamefont{Lu}},
  \bibinfo{author}{\bibfnamefont{T.}~\bibnamefont{Jia}},
  \bibinfo{author}{\bibfnamefont{Z.}~\bibnamefont{Wang}}, and
  \bibinfo{author}{\bibfnamefont{Z.}~\bibnamefont{Sun}},
  \bibinfo{year}{2011}{\natexlab{a}}, \bibinfo{journal}{Phys. Rev. A}
  \textbf{\bibinfo{volume}{83}}, \bibinfo{pages}{043410}.

\bibitem{zhang:2011b}
\bibinfo{author}{\bibnamefont{Zhang}, \bibfnamefont{S.}},
  \bibinfo{author}{\bibfnamefont{J.}~\bibnamefont{Shi}},
  \bibinfo{author}{\bibfnamefont{H.}~\bibnamefont{Zhang}},
  \bibinfo{author}{\bibfnamefont{T.}~\bibnamefont{Jia}},
  \bibinfo{author}{\bibfnamefont{Z.}~\bibnamefont{Wang}}, and
  \bibinfo{author}{\bibfnamefont{Z.}~\bibnamefont{Sun}},
  \bibinfo{year}{2011}{\natexlab{b}}, \bibinfo{journal}{Phys. Rev. A}
  \textbf{\bibinfo{volume}{83}}, \bibinfo{pages}{023416}.

\bibitem{ZhangSciRep17}
\bibinfo{author}{\bibnamefont{Zhang}, \bibfnamefont{Z.-Y.}}, and
  \bibinfo{author}{\bibfnamefont{J.-M.} \bibnamefont{Liu}},
  \bibinfo{year}{2017}, \bibinfo{journal}{Scientific Reports}
  \textbf{\bibinfo{volume}{7}}(\bibinfo{number}{1}), \bibinfo{pages}{17822}.

\bibitem{zhdanov:2011}
\bibinfo{author}{\bibnamefont{Zhdanov}, \bibfnamefont{D.}}, and
  \bibinfo{author}{\bibfnamefont{H.}~\bibnamefont{Rabitz}},
  \bibinfo{year}{2011}, \bibinfo{journal}{Phys. Rev. A}
  \textbf{\bibinfo{volume}{83}}, \bibinfo{pages}{061402}.

\bibitem{zhdanov:2015}
\bibinfo{author}{\bibnamefont{Zhdanov}, \bibfnamefont{D.~V.}}, and
  \bibinfo{author}{\bibfnamefont{T.}~\bibnamefont{Seideman}},
  \bibinfo{year}{2015}, \bibinfo{journal}{Phys. Rev. A}
  \textbf{\bibinfo{volume}{92}}, \bibinfo{pages}{012129}.

\bibitem{zhdanov:2008}
\bibinfo{author}{\bibnamefont{Zhdanov}, \bibfnamefont{D.~V.}}, and
  \bibinfo{author}{\bibfnamefont{V.~N.} \bibnamefont{Zadkov}},
  \bibinfo{year}{2008}, \bibinfo{journal}{Phys. Rev. A}
  \textbf{\bibinfo{volume}{77}}, \bibinfo{pages}{011401}.

\bibitem{zhdanovich:2012}
\bibinfo{author}{\bibnamefont{Zhdanovich}, \bibfnamefont{S.}},
  \bibinfo{author}{\bibfnamefont{C.}~\bibnamefont{Bloomquist}},
  \bibinfo{author}{\bibfnamefont{J.}~\bibnamefont{Flo\ss{}}},
  \bibinfo{author}{\bibfnamefont{I.~S.} \bibnamefont{Averbukh}},
  \bibinfo{author}{\bibfnamefont{J.~W.} \bibnamefont{Hepburn}}, and
  \bibinfo{author}{\bibfnamefont{V.}~\bibnamefont{Milner}},
  \bibinfo{year}{2012}, \bibinfo{journal}{Phys. Rev. Lett.}
  \textbf{\bibinfo{volume}{109}}, \bibinfo{pages}{043003}.

\bibitem{zhdanovich:2011}
\bibinfo{author}{\bibnamefont{Zhdanovich}, \bibfnamefont{S.}},
  \bibinfo{author}{\bibfnamefont{A.~A.} \bibnamefont{Milner}},
  \bibinfo{author}{\bibfnamefont{C.}~\bibnamefont{Bloomquist}},
  \bibinfo{author}{\bibfnamefont{J.}~\bibnamefont{Flo\ss{}}},
  \bibinfo{author}{\bibfnamefont{I.~S.} \bibnamefont{Averbukh}},
  \bibinfo{author}{\bibfnamefont{J.~W.} \bibnamefont{Hepburn}}, and
  \bibinfo{author}{\bibfnamefont{V.}~\bibnamefont{Milner}},
  \bibinfo{year}{2011}, \bibinfo{journal}{Phys. Rev. Lett.}
  \textbf{\bibinfo{volume}{107}}, \bibinfo{pages}{243004}.

\bibitem{ZhouPRA11}
\bibinfo{author}{\bibnamefont{Zhou}, \bibfnamefont{Y.~L.}},
  \bibinfo{author}{\bibfnamefont{M.}~\bibnamefont{Ortner}}, and
  \bibinfo{author}{\bibfnamefont{P.}~\bibnamefont{Rabl}}, \bibinfo{year}{2011},
  \bibinfo{journal}{Phys. Rev. A} \textbf{\bibinfo{volume}{84}},
  \bibinfo{pages}{052332}.

\bibitem{ZhuJCP13}
\bibinfo{author}{\bibnamefont{Zhu}, \bibfnamefont{J.}},
  \bibinfo{author}{\bibfnamefont{S.}~\bibnamefont{Kais}},
  \bibinfo{author}{\bibfnamefont{Q.}~\bibnamefont{Wei}},
  \bibinfo{author}{\bibfnamefont{D.}~\bibnamefont{Herschbach}}, and
  \bibinfo{author}{\bibfnamefont{B.}~\bibnamefont{Friedrich}},
  \bibinfo{year}{2013}, \bibinfo{journal}{J. Chem. Phys.}
  \textbf{\bibinfo{volume}{138}}(\bibinfo{number}{2}), \bibinfo{pages}{024104}.

\bibitem{znakovskaya:2014}
\bibinfo{author}{\bibnamefont{Znakovskaya}, \bibfnamefont{I.}},
  \bibinfo{author}{\bibfnamefont{M.}~\bibnamefont{Spanner}},
  \bibinfo{author}{\bibfnamefont{S.}~\bibnamefont{De}},
  \bibinfo{author}{\bibfnamefont{H.}~\bibnamefont{Li}},
  \bibinfo{author}{\bibfnamefont{D.}~\bibnamefont{Ray}},
  \bibinfo{author}{\bibfnamefont{P.}~\bibnamefont{Corkum}},
  \bibinfo{author}{\bibfnamefont{I.~V.} \bibnamefont{Litvinyuk}},
  \bibinfo{author}{\bibfnamefont{C.~L.} \bibnamefont{Cocke}}, and
  \bibinfo{author}{\bibfnamefont{M.~F.} \bibnamefont{Kling}},
  \bibinfo{year}{2014}, \bibinfo{journal}{Phys. Rev. Lett.}
  \textbf{\bibinfo{volume}{112}}, \bibinfo{pages}{113005}.

\bibitem{zon:1975}
\bibinfo{author}{\bibnamefont{Zon}, \bibfnamefont{B.~A.}}, and
  \bibinfo{author}{\bibfnamefont{B.~G.} \bibnamefont{Katsnelson}},
  \bibinfo{year}{1975}, \bibinfo{journal}{Sov. Phys. JETP}
  \textbf{\bibinfo{volume}{42}}, \bibinfo{pages}{595}.

\end{thebibliography}


\end{document}